\pdfoutput=1
\RequirePackage{ifpdf}
\ifpdf % We~are running pdfTeX in pdf mode
\documentclass[pdftex]{sigma}
\else
\documentclass{sigma}
\fi

\usepackage{mathtools}
\usepackage[dvipsnames]{xcolor}
\usepackage{tikz}
\usetikzlibrary{arrows.meta,decorations.markings,decorations.pathmorphing}
\usepackage{pgfplots}
\usepgfplotslibrary{polar}

\DeclareMathOperator{\tr}{tr}

\DeclareMathOperator{\CC}{\mathbb{C}}
\DeclareMathOperator{\RR}{\mathbb{R}}

\DeclareMathOperator{\OO}{\mathcal{O}}

\DeclareMathOperator{\Ai}{Ai}

\DeclareMathOperator*{\Ress}{Res}
\newcommand{\Res}{\displaystyle\Ress}

\DeclareMathOperator{\YY}{{\bf Y}}
\DeclareMathOperator{\XX}{{\bf X}}
\DeclareMathOperator{\TT}{{\bf T}}
\DeclareMathOperator{\WW}{{\bf W}}
\DeclareMathOperator{\UU}{{\bf U}}

\DeclareMathOperator{\GG}{{\bf G}}
\DeclareMathOperator{\boldS}{{\bf S}}
\DeclareMathOperator{\boldR}{{\bf R}}
\newcommand{\eps}{\varepsilon}

\numberwithin{equation}{section}

\newtheorem{Theorem}{Theorem}[section]
\newtheorem*{Theorem*}{Theorem}
\newtheorem{Corollary}[Theorem]{Corollary}
\newtheorem{Lemma}[Theorem]{Lemma}
\newtheorem{Proposition}[Theorem]{Proposition}
 { \theoremstyle{definition}
\newtheorem{Definition}[Theorem]{Definition}

\newtheorem{Remark}[Theorem]{Remark} }

\begin{document}

\allowdisplaybreaks

\newcommand{\arXivNumber}{2405.03259}

\renewcommand{\PaperNumber}{079}

\FirstPageHeading

\ShortArticleName{The Ising Model Coupled to 2D Gravity: Genus Zero Partition Function}

\ArticleName{The Ising Model Coupled to 2D Gravity:\\ Genus Zero Partition Function}

\Author{Maurice DUITS~$^{\rm a}$, Nathan HAYFORD~$^{\rm a}$ and Seung-Yeop LEE~$^{\rm b}$}

\AuthorNameForHeading{M.~Duits, N.~Hayford and S.-Y.~Lee}

\Address{$^{\rm a)}$~Department of Mathematics, Royal Institute of Technology (KTH), Stockholm, Sweden}
\EmailD{\href{mailto:duits@kth.se}{duits@kth.se}, \href{mailto:nhayford@kth.se}{nhayford@kth.se}}

\Address{$^{\rm b)}$~Department of Mathematics and Statistics, University of South Florida, Tampa, FL, USA}
\EmailD{\href{mailto:lees3@usf.edu}{lees3@usf.edu}}

\ArticleDates{Received January 31, 2025, in final form September 03, 2025; Published online September 24, 2025}

\Abstract{We compute the genus $0$ free energy for the $2$-matrix model with quartic interactions, which acts as a generating function for the Ising model's partition function on a random, $4$-regular, planar graph. This is consistent with the predictions of Kazakov and Boulatov on this model, as well as subsequent confirmation of this formula using combinatorial methods. We also provide a new parametric formula for the free energy and give a characterization of the phase space. Our analysis is based on a steepest descent Riemann--Hilbert analysis of the associated biorthogonal polynomials and the corresponding isomonodromic $\tau$-function. A key ingredient in the analysis is a parametrization of the spectral curve. This analysis lays the groundwork for the subsequent study of the multicritical point, which we will study in a forthcoming work.}


\Classification{30E15; 30E10; 30E25; 41A60; 05C30; 05C10}

\tableofcontents

\section{Introduction}

The 2-matrix model is the probability measure on the space of pairs of $n \times n$ Hermitian matrices~$X$,~$Y$ defined by
\begin{equation} \label{eq:twomatrixmodel}
 \frac{1}{Z_{n}} \exp N \tr (\tau X Y- V_1(X) -V_2(Y) ) {\rm d}X {\rm d}Y,
\end{equation}
where ${\rm d}X$, ${\rm d}Y$ are the Haar measures on the space of $n\times n$ Hermitian matrices, $N>0$ is a~large parameter (which will eventually be set to $n$ the size of the matrix), $\tau>0$ is constant called the coupling constant, and $V_j$ are two polynomials of even degree and with positive leading coefficient. The normalizing constant $Z_{n}$, which depends on $n$, $N$, $\tau$, and any parameters present in $V_1$, $V_2$, is called the \textit{partition function}, and will essentially be the main object of study in this work.

An important motivation for studying the partition function of the 2-matrix model comes from its connection to 2-dimensional quantum gravity~\cite{KB,DS,GM1,GM2}. It is conjectured in the physics literature that, by tuning the parameters in the polynomials $V_1$ and $V_2$, the 2-matrix model contains all multi-critical models of type $(q,p)$ with arbitrary $q$, whereas the one matrix models ($\tau=0$) only contains critical models with $q=2$. Perhaps the most important first example is the case of two quartic potentials, which is related to the Ising model coupled to 2-dimensional gravity and the type $(3,4)$ critical model. The richer structure also makes the 2-matrix model harder to analyze when compared to the well-understood one matrix model and a rigorous treatment of the asymptotic behavior as $N\to \infty$ of the general 2-matrix model is an important open problem. To date, the only rigorous results in the literature are in~\cite{DG,DK1,DKM}, but these results do not cover the case of quantum gravity coupled to the Ising model. The current writing is the first in a series of three papers where we discuss this multi-critical point. More precisely, we will study the limiting behavior of its partition function $Z_n$ as $n=N \to \infty$ in case of the quartic potentials:
\[
V_1(x) = V\bigl(x;{\rm e}^H t\bigr):=\frac12 x^2 + \frac{{\rm e}^H t}{4} x^4 \qquad \text{and} \qquad V_2(y)=V\bigl(y;{\rm e}^{-H} t\bigr):=\frac12 y^2 + \frac{{\rm e}^{-H} t}{4} y^4,
\]
where $H \in \mathbb R $ and $t$ are parameters. In this case, we write the partition function as
 \begin{equation*}
 Z_n(\tau,t,H;N) := \iint \exp N \tr \bigl(\tau X Y- V\bigl(X;{\rm e}^H t\bigr) -V\bigl(Y;{\rm e}^{-H} t\bigr)\bigr) {\rm d}X {\rm d}Y,
 \end{equation*}
where the integration is carried out over pairs of $n\times n$ Hermitian matrices.
Note that to ensure that the measure in \eqref{eq:twomatrixmodel} is finite we need that $t>0$. However, for fixed $N$ the partition function $Z_N(\tau,t,H;N)$ depends analytically on $t$ and can be analytically extended to complex~$t$ with a~branch point at $t=0$. In fact, we are mainly interested in $t<0$ and discuss this continuation in detail below.
Our main results are for the \textit{genus zero free energy} defined as
 \begin{equation}\label{free-energy-2-matrix-model}
 F(\tau,t,H) := \lim_{N\to \infty} \frac{1}{N^2} \log \frac{Z_{N}(\tau,t,H;N)}{Z_{N}(\tau,0,0;N)}.
 \end{equation}
In this paper, will show that, up to some critical value, the analytic continuation for $t<0$ the genus zero free energy exists, and we derive an analytic expression for it, confirming predictions from the physics literature~\cite{Kazakov2, Kazakov1}. We will also see that at the multi-critical point
 \begin{equation}\label{multicritical-point}
 \tau = \frac{1}{4},\qquad t = -\frac{5}{72}, \qquad H=0,
 \end{equation}
 the free energy undergoes a phase transition. In two forthcoming works~\cite{DHL3, DHL2}, we will address this phase transition by means of a multi-scaling limit and show that it gives rise to the $(3,4)$ minimal model coupled to gravity, as conjectured. Before we come to the statement of our main results and strategy of the proof, we start with a more detailed historical discussion on this model.

\subsection{The Ising model coupled to gravity and the 2-matrix model}
The 2-dimensional Ising model has long been a source of interest in statistical physics, as it is an exactly solvable lattice model which exhibits a $2^{\rm nd}$
order phase transition at finite temperature. The model describes a ferromagnet with only nearest-neighbor interactions, and can be defined on
any graph\footnote{Throughout the present work, by graph we actually mean \textit{multigraph}, i.e., we allow for loops and multiple edges between vertices.} $G := (V,E)$ with vertices $V$ and edges $E$ as follows. The Hamiltonian for the ferromagnetic Ising model is a functional on maps $\psi\colon V\to \{\pm 1\}$, and is defined as
 \begin{equation*}
 \mathcal{H}(\psi;h) = -\sum_{(x,y)\in E} \psi(x)\psi(y) - h \sum_{x\in V} \psi(x).
 \end{equation*}
The parameter $h \in \mathbb{R}$ is called the \textit{magnetic field}.
The partition function for this model is defined to be
 \begin{equation*} %\label{Ising-partition}
 Z_G(\beta;h) := \sum_{\psi} {\rm e}^{-\beta \mathcal{H}(\psi;h)},
 \end{equation*}
where $\beta > 0$ is a parameter called the \textit{inverse temperature}, and the sum is taken over all $2^{|V|}$ maps $\psi\colon V \to \{\pm 1\}$. In other words, we are considering the Boltzmann distribution on the system at temperature $\beta^{-1}$. Typically, one is interested in calculating the \textit{free energy} of this model, defined as
 \begin{equation*}
 F_G(\beta;h) = -\frac{1}{\beta} \log Z_G(\beta;h).
 \end{equation*}
If the graph is of infinite size (i.e., the number of vertices is infinite), one is instead interested in the free energy \textit{per unit site};
that is, to say
 \begin{equation*}
 f(\beta;h) := \lim_{|V|\to \infty} \frac{1}{|V|} F_G(\beta;h),
 \end{equation*}
where $|V|$ denotes the number of vertices in $G$, and the limit is taken in an appropriate sense.

The model was introduced by Ising~\cite{Ising}, although he only studied the 1-dimensional model, and incorrectly
conjectured that the model in general did \textit{not} exhibit a phase transition. It was not until over 20 years later that Onsager~\cite{Onsager} announced that the
2-dimensional model indeed exhibited a phase transition.\footnote{In fact, Onsager himself did not provide a proof, instead only furnishing the expression for the free energy. It took until 1952 for a fully rigorous proof to be published by Yang~\cite{Yang}.} The universality of the critical exponents appearing in the 2-dimensional model were verified
for various lattices (cf.~\cite{MW}, or \cite[Chapter~11 and references therein]{Baxter}), but these considerations were markedly limited by the fact that computations could be performed explicitly only
for choices of fairly regular lattices (e.g., square lattice, triangular lattice, etc.). The next breakthrough in the study of critical phenomena came in the 1960s, when Kadanoff applied quantum field-theoretic techniques (the renormalization group, or RG) to describe the Ising phase transition~\cite{Kadanoff}; these ideas were subsequently further developed by Wilson~\cite{Wilson1,Wilson2}.
One of the features of the RG approach was the ability to explain the phenomenon of universality: seemingly very different physical systems end up having identical critical exponents. Fixed points of the RG flow exhibit scale invariance; furthermore, many of the statistical systems of interest also enjoy translation invariance. These observations led Belavin, Polyakov and Zamolodchikov~\cite{BPZ} in 1984 to postulate that \textit{conformal invariance} should manifest at these fixed points. Their work resulted in the characterization of 2D critical phenomena by the celebrated \textit{minimal models} of conformal field theory (CFT). The~Ising critical point itself is the so-called $(3,4)$ minimal model.

Happening concurrently with these developments in statistical mechanics were the first developments in the study of 2-dimensional quantum gravity. Part of the hope of this program was that an exact solution of a 2D theory of gravity could shed light on how higher-dimensional versions might function; another motivating factor was a number of recent advances in string theory, in which the techniques of 2D gravity played an integral role. One of the first major works on this subject was Polyakov's work on bosonic string theory~\cite{Polyakov1}, which demonstrated that the Liouville (also called continuum) approach to 2D gravity could be exactly solved. The~drawback of this theory was that explicit calculations proved difficult, and hindered the theory for the next decade or so. Indeed, the aforementioned work of Belavin, Polyakov, and Zamolodchikov~\cite{BPZ} was later described
by Polyakov as ``an unsuccessful attempt to solve the Liouville theory''~\cite{Polyakov2}. An~alternative approach to this problem was put forth by the Saclay school of theoretical physics: to replace the functional integral over geometries by a sum over discretized surfaces, who began to develop this approach in the late 70s~\cite{BIPZ,IZ}. This sum could be calculated with the help of matrix integrals~\cite{DKKM, David,Kazakov4, KMK}. This method was quite successful in the case of a pure theory of gravity, as the model was indeed completely integrable, and was closely linked to the theory of orthogonal polynomials, and of Painlev\'{e} equations~\cite{FIK1,FIK2}. The~next natural step in this program was to try an compute how matter interacted with a nontrivial gravitational background.

These developments together prompted Kazakov to consider the Ising model coupled to 2-dimensional gravity: that is, the Ising model on a random lattice. More precisely, he found that the 2D Ising model on a random $4$-regular planar graph could be described by the large $N$-limit of a
2-matrix model. Here \textit{$4$-regular} means that each vertex is connected to four edges. The partition function for the Ising model on a random $4$-regular planar graph with $n$ vertices is defined as
 \begin{equation}\label{Ising-random-partition-n} % RS:
 \mathcal{Z}_n(\beta;h) = \sum_{\substack{G:|V| = n,\\ G \ {\textrm{planar}}}} Z_G(\beta;h),
 \end{equation}
where the sum is taken over all $4$-regular, planar graphs with $n$ vertices. In~\cite{Kazakov1}, Kazakov considered the formal generating function
 \begin{equation}\label{Ising-random-partition-generating}
 \mathcal{Z}(\tau,t;H) = \sum_{n \in \mathbb{N}} \biggl(\frac{-t\tau}{4(1-\tau^2)^2}\biggr)^n \mathcal{Z}_n(\beta;h),
 \end{equation}
where we identify
 \begin{equation*}
 \tau := {\rm e}^{-2\beta},\qquad H := \beta h,
 \end{equation*}
and $t$ is a parameter. Kazakov demonstrated that the generating function \eqref{Ising-random-partition-generating} is equivalent to the planar $(n \to \infty)$ limit of the free energy of the 2-matrix model \eqref{free-energy-2-matrix-model}:
\begin{equation*}
 F(\tau,t,H) = \mathcal{Z}(\tau,t,H).
\end{equation*}

In~\cite{Kazakov2, Kazakov1}, a formula was derived for this quantity, which subsequently allowed for the~prediction of the shift of the critical exponents of the Ising model when coupled
to gravity. Kazakov's description of the critical point turned out to be in direct agreement with the newly predicted results of Kniznik, Polyakov and Zamolodchikov~\cite{KPZ} arising from coupling certain CFTs to matter; this is the so-called KPZ formula.
Subsequent analysis for this model and closely related ones was performed in~\cite{Kazakov2,Kazakov3} (see also the work~\cite{Zamolodchikov-Ishimoto}, which describes a generalization of the Ising/massless case to massive Majorana fermions). The culmination of this work in the physics literature were the papers of
Douglas and Shenker~\cite{DS}, Br\'{e}zin and Kazakov~\cite{KB}, and Gross and Migdal~\cite{GM1,GM2}. These works identified the critical points of the $2$-matrix model (more generally, the matrix-chain model) with the CFT minimal models coupled to gravity. These critical points were characterized in terms of so-called string equations, which come from additional symmetries of reductions of the KP hierarchy. This work completed the picture of pure gravity provided by the $1$-matrix model (which described the $(2,2g+1)$ critical points) to describe all $(p,q)$ minimal models, for $(p,q)$ coprime. For an overview of this story from the physics perspective, see
the excellent review~\cite{PGZ} and references therein.

In the mathematics literature, the ``pure gravity'' situation, in which no matter is present, stemmed from the works~\cite{DS,GM1,KM}. This was subsequently described rigorously by Fokas, Its and Kitaev in~\cite{FIK1,FIK2}, and more recently (in line with the language of this work) by Duits and Kuijlaars~\cite{DK0} (the quartic model), and Bleher and Dea\~{n}o~\cite{BleherDeano} (the cubic model). Similarly, the other critical points of the $1$-matrix model are now widely accepted to be well-understood (see, for example,~\cite{CV}). However, the same cannot be said for the $2$-matrix model, due in part to the lack of orthogonal polynomial techniques that had been core to the study of the $1$-matrix model. This changed with the work of Kuijlaars and McLaughlin~\cite{KM}, in which they found a~characterization of this model in terms of multiple orthogonal polynomials. Since there was already a wide mathematical literature on the asymptotics of such polynomials~\cite{Aptekarev,NikishinSorokin,VanAssche}, this opened the doors for the possibility of
the analysis of critical phenomena in the $2$-matrix model, the first physically interesting critical point being the one studied originally by Kazakov~\cite{Kazakov1} corresponding to the Ising model. Some progress in this direction was made by Kuijlaars, Duits, and their collaborators~\cite{DG,DGK,DK1,DKM}, but no direct analytic derivation of the Ising critical point has been
written down. In this series of works (see also the forthcoming~\cite{DHL3, DHL2}), we provide a rigorous analysis of the critical point appearing in Kazakov's work, using steepest descent analysis.

As was mentioned before, the partition function \eqref{Ising-random-partition-generating} is a generating function for the Ising model on $4$-regular planar maps. This
generating function (along with its analog for $3$-regular planar maps) has been studied using purely combinatorial techniques, bypassing the use of
matrix integrals. The first rigorous derivation of the results of Kazakov and Boulatov for
both the~$3$ and~$4$-regular generating functions have been obtained by Bousquet-M\'{e}lou and Schaeffer in~\cite{MelouSchaeffer}. Alternative bijective approaches were found by Bouttier, Di Francesco and Guitter in~\cite{BFG,BDFG}, and later by Bernardi and Bousquet-M\'{e}lou in~\cite{BBM} using functional equations. More recently still, the groups of Albenque, M\'{e}nard and Schaeffer~\cite{AM22, AMS} and the group of
Chen and Turunen~\cite{CT1,CT2,Tur} constructed local limits of various families of random maps coupled to the Ising model. These local limits correspond to infinite random maps and correspond to a~\textit{thermodynamic limit}. These groups were able to rigorously establish the existence of
the Ising phase transition coupled to a random background, and calculate some of the critical exponents, in agreement with the predictions of Boulatov and Kazakov. In the present work, we study the (analytic continuation of) the associated matrix integral originally put forth by Kazakov, and study it using analytical techniques, i.e., Riemann--Hilbert analysis of biorthogonal polynomials. These techniques will allow us in subsequent works~\cite{DHL3, DHL2} to study the multi-scaling limit of the Ising critical point, and prove that the partition function of this model converges under the appropriate scaling to a $\tau$-function for the so-called $(3,4)$ string equation, thus rigorously proving the results of~\cite{Kazakov3,CGM}.

\subsection{A family of biorthogonal polynomials}
The 2-matrix model considered above, with partition function \eqref{eq:twomatrixmodel}, can be studied using orthogonal polynomial-type methods. For convenience, we will first introduce the parameter
 \begin{equation} \label{eq:qeH}
 q := {\rm e}^H.
 \end{equation}
Using the Itzykson--Zuber integral over the unitary group (cf., for example,~\cite{IZ,Mehta,ZJZ}), one can convert the integral over the spaces of formal hermitian
matrices to an integral over eigenvalue coordinates; the partition function in the eigenvalue coordinates reads\footnote{Note that we have changed from $H$ in \eqref{free-energy-2-matrix-model} to $q$, by a slight abuse of notation. We hope that this will not cause confusion further in this article.}
 \begin{gather*} %\label{eigenvalue-measure}
 Z_{n}(\tau,t,q;N) = \frac{C_{n,N}}{\tau^{\frac{n(n-1)}{2}}} \\
 \hphantom{Z_{n}(\tau,t,q;N) =}{}\times\!
 \iint\! \Delta(x) \Delta(y) \exp\Biggl\{ N\sum_{i=1}^n \bigl(\tau x_iy_i - V(x_i;q t) - V\bigl(y_i;q^{-1}t\bigr) \bigr)\Biggr\}\prod_{i=1}^n {\rm d}x_i {\rm d}y_i,
 \end{gather*}
where
\begin{equation*}
 V(z;t) = \frac{1}{2}z^2 + \frac{t}{4} z^4,
\end{equation*}
$\Delta(x)$, $\Delta(y)$ denote the usual Vandermonde determinants in the variables $\{x_i\}$, $\{y_i\}$, respectively, and $C_{n,N}$ is a constant independent of the parameters $\tau$,~$t$:
 \begin{equation*}
 C_{n,N} = \frac{(2\pi)^{n(n-1)}}{\bigl(\prod_{p=1}^n p!\bigr)^2} \Biggl(\prod_{p=1}^{n-1} p!\Biggr) N^{-\frac{n(n-1)}{2}}
 = \frac{1}{(n!)^2}\frac{(2\pi)^{n(n-1)}}{\prod_{p=1}^{n-1} p!}N^{-\frac{n(n-1)}{2}}.
 \end{equation*}

By performing elementary row/column transformations on the Vandermonde determinants $\Delta(x)$, $\Delta(y)$, and applying Andr\'eief's identity, we can rewrite the partition function in terms of the following family of biorthogonal polynomials~\cite{Mehta}
 \begin{equation} \label{biorthogonal-polynomials}
 \int_\Gamma \int_{\Gamma} p_k(z)q_j(w)\exp \bigl[ N\bigl(\tau z w - V(z;q t) - V\bigl(w;q^{-1}t\bigr) \bigr)\bigr] {\rm d}z {\rm d}w = h_j \delta_{kj},
 \end{equation}
where $\Gamma := \mathbb{R}$ is taken to be the real line.
In this language, the partition function can be expressed in terms of the orthogonality coefficients:
 \begin{equation}\label{partition-function-orthogonality-coeff}
 Z_{n}(\tau,t,q;N) = n!\frac{C_{n,N}}{\tau^{\frac{-n(n-1)}{2}}}\prod_{j=0}^{n-1} h_j(\tau,t,q;N).
 \end{equation}
 The study of the large-$n$ asymptotic behavior of the partition function thus reduces to the study of the large $n$ asymptotic behavior of the associated biorthogonal polynomials.

\subsection{Analytic continuation}
The critical point considered by Kazakov (cf.~\cite{Kazakov3}, after formula (1)) corresponds equation \eqref{multicritical-point} in the present work.
This leads to an immediate issue: the partition function \eqref{partition-function-orthogonality-coeff} does not converge for $t<0$. We therefore must interpret the partition function as an analytic continuation of the $t>0$ partition function. For finite $n$, we make this continuation as follows. Note that since all relevant quantities (the biorthogonal polynomials, the partition function, etc.) can be written in terms of moments of the biorthogonality measure, it is sufficient to consider
analytic continuation of the moments. For $k,j\geq 0$, we define
 \begin{gather*}
 m_{jk}(\tau,t,q;N) := \int_{\Gamma_{t}}\int_{\Gamma_{t}} z^{j}w^{k}\exp\biggl\{Ng[\tau zw - \frac{1}{2}z^2 - \frac{1}{2}w^2 - \frac{tqz^4}{4} - \frac{tq^{-1}w^4}{4}\biggr]\biggr\}{\rm d}z{\rm d}w,
 \end{gather*}
where the contour $\Gamma_t := \bigl\{x {\rm e}^{{\rm i}\theta}\mid x\in \RR,\, \theta = \arg \bigl(t^{-1/4}\bigr)\bigr\}$, and is oriented in the direction of increasing $x$. This definition coincides with the standard definition of the moments when~${t>0}$, and is analytic in $t$, and so acts as
a suitable analytic continuation of the functions $m_{jk}(\tau,t,q;N)$. Evidently, the moments have a branch point $0$, and so there are
two possible analytic continuations we can make: one where we analytically continue $t$ through the upper half plane, and one where we
continue through the lower half plane. This would in principle require two separate analyses. However, due to the Schwarz symmetry (for real $\tau$, $q$, $N$,
and $|{\arg t}| < \pi$),
 \begin{equation*}%\label{Schwarz-symmetry-moments}
 \overline{m_{jk}(\tau,\bar{t},q;N)} = m_{jk}(\tau,t,q;N),
 \end{equation*}
we can obtain one analytic continuation from the other by reflection. Thus, it is sufficient to consider an analytic continuation through
the \textit{upper half plane}. For $t<0$, the contour $\Gamma$ is defined as starting from ${\rm e}^{3\pi {\rm i}/4}\cdot \infty$, and ending at
${\rm e}^{-\pi {\rm i}/4}\cdot\infty$. We remark that we have the freedom to later redefine the contour $\Gamma$ locally, as long as we retain its
asymptotic properties. We shall indeed redefine $\Gamma$ in a more precise manner in the next section, in order to guarantee that certain inequalities are satisfied on it.

From here on, we thus consider the contour $\Gamma$ as chosen before, and assume that the parameters of the model all have fixed sign:
 \begin{equation*}
 \tau >0,\quad t < 0, \quad q >0, \qquad H\in \RR.
 \end{equation*}

\subsection{Riemann--Hilbert problem}

As observed in~\cite{KM}, the biorthogonal polynomials defined by \eqref{biorthogonal-polynomials} admit a Riemann--Hilbert formulation, given as follows.
Consider the following Riemann--Hilbert problem (RHP) on $\Gamma$ (recall that the contour $\Gamma$ is defined as starting from
${\rm e}^{3\pi {\rm i}/4}\cdot \infty$, and ending at ${\rm e}^{-\pi {\rm i}/4}\cdot\infty$):
 \begin{equation*} %\label{RHP-1}
 \YY_{+}(z) = \YY_{-}(z) \left[\mathbb{I} + {\rm e}^{-NV(z;q t)}\begin{pmatrix}
 0 & f(z) & \frac{f'(z)}{N\tau} & \frac{f''(z)}{(N\tau)^2} \\
 0 & 0 & 0 & 0 \\
 0 & 0 & 0 & 0 \\
 0 & 0 & 0 & 0
 \end{pmatrix} \right]
, \qquad z \in \Gamma,
 \end{equation*}
where
 \begin{equation*}
 f(z) = \int_{\Gamma} \exp\bigl[ N\bigl(\tau z w - V\bigl(w;q^{-1}t\bigr) \bigr)\bigr] {\rm d}w.
 \end{equation*}
The asymptotics of $\YY(z)$ are chosen to be
 \begin{equation*}
 \YY(z) = \biggl[ \mathbb{I} + \OO\biggl(\frac{1}{z}\biggr) \biggr]
 \begin{pmatrix}
 z^n & 0 & 0 & 0 \\
 0 & z^{-n/3} & 0 & 0 \\
 0 & 0 & z^{-n/3} & 0 \\
 0 & 0 & 0 & z^{-n/3}
 \end{pmatrix}, \qquad |z| \to \infty,
 \end{equation*}
where $n$ is a multiple of $3$. \emph{This restriction on $n$ is taken for simplicity of exposition; the assumption is not essential} (see also~\cite{DK1,DKM}). This RHP admits a unique exact solution, with the 1-1 entry of $\YY(z)$ being the degree $n$ monic biorthogonal polynomial\footnote{Note that, in general, the relation \eqref{biorthogonal-polynomials} in fact defines \textit{two} sets of polynomials $\{p_n(z)\}$, $\{q_n(w)\}$, which in general do not coincide, except in the special case that the potentials $V_ 1(z) = V_2(z)$ are the same. However, because the potentials in question in this work are related by $V_1(z;qt) = V_2\bigl(z,q^{-1}t\bigr)$, the corresponding polynomials are related by $q_n(z;\tau,t,q;N) = p_n\bigl(z;\tau,t,q^{-1};N\bigr)$. } defined by the relations~\eqref{biorthogonal-polynomials}:
 \begin{equation*}
 [\YY]_{11}(z) = p_n(z).
 \end{equation*}

The partition function of the matrix model defined by \eqref{eq:twomatrixmodel} (more precisely, its analytic continuation) can be related to an isomonodromic $\boldsymbol{\tau}$-function, given in terms of $\YY(z)$~\cite{Bertola-Marchal}. The $\boldsymbol{\tau}$-function is defined to be
 \begin{equation*} %\label{tau-definition}
 {\rm d}\log \boldsymbol{\tau}_n := \Res_{z=\infty}\tr\bigl[ \YY^{-1}(z) \YY'(z) {\rm d}\Psi_0(z) \Psi_0^{-1}(z)\bigr],
 \end{equation*}
where $\Psi_0$ is an explicit matrix function, and the differential is in the variables of deformation~$t$,~$\tau$,~$q$. Its definition will be given more precisely in the succeeding text. The differential of the partition function
considered by Kazakov is proportional to this isomonodromic-$\boldsymbol{\tau}$ function:
 \begin{equation*} %\label{partition-2}
 {\rm d}\log \frac{Z_{n}(\tau,t,q;N)}{\tau^{\frac{-n(n-1)}{2}}C_{n,N}} = {\rm d}\log \biggl[\biggl(\frac{\tau}{t^2}\biggr)^{\frac{n}{2}(\frac{n}{3}-1)}\boldsymbol{\tau}_n\biggr].
 \end{equation*}
Furthermore, one may calculate $Z_{n}(\tau,0,0;N)$ explicitly (cf.~\cite[Appendix A.49]{Mehta-Book}):
 \begin{equation*} %\label{partition-frozen}
 Z_{n}(\tau,0,0;N) = \tau^{\frac{-n(n-1)}{2}}C_{n,N} \times N^{-\frac{n(n+1)}{2}}\frac{(2\pi)^{n}\tau^{n(n-1)/2}}{(1-\tau^2)^{n^2/2}}\prod_{p=1}^{n-1}p!.
 \end{equation*}
Thus, by computing the isomonodromic $\boldsymbol{\tau}$-function, we can compute all quantities of interest related to the original partition function of the $2$-matrix model. In particular, we have that
 \begin{equation*}
 {\rm d}\log \frac{Z_{n}(\tau,t,q;N)}{Z_{n}(\tau,0,0;N)} = {\rm d}\log \biggl[\frac{\bigl(1-\tau^2\bigr)^{n^2/2}}{\tau^{n^2/3}t^{n(\frac{n}{3}-1)}}\boldsymbol{\tau}_n\biggr].
 \end{equation*}
The remainder of this work is devoted to the calculation of this $\boldsymbol{\tau}$-function.

\subsection{Spectral curve}

In recent decades, the Riemann--Hilbert approach has emerged as a powerful tool for computing the asymptotic behavior of polynomials satisfying orthogonality relations. Its breakthrough occurred with the seminal work~\cite{DKMVZ}, where the authors successfully applied the Deift--Zhou steepest descent method~\cite{Deift-Zhou} to the Riemann--Hilbert problem for orthogonal polynomials on the real line with a varying weight ${\rm e}^{-nV(x)}{\rm d}x$. An important contribution to this analysis came from potential theory: the limiting zero distribution of the polynomials is given by the~equilibrium measure $\mu_V$ for the Coulomb gas in the external field $V$:
 \begin{equation*}
 \mu_{V} := \arg \min_{\nu\in \mathcal{M}_1(\RR)} \iint \log \frac{1}{|x-y|} {\rm d}\nu(x){\rm d}\nu(y) + \int V(x) {\rm d}\nu(x),
 \end{equation*}
where $\mathcal{M}_1(\RR)$ is the set of all unit Borel measures supported on $\RR$.
By introducing the equilibrium measure into the Riemann--Hilbert problem (via a transformation involving the $g$-function constructed from the equilibrium measure) and employing the Deift--Zhou steepest descent method, one can obtain a full asymptotic description of the polynomials and their properties. When $V$ is a polynomial, this equilibrium measure can be characterized in terms of hyperelliptic curve. Specifically, there exists a polynomial $Q$ such that
\begin{equation} \label{eq:1myx}
 y(x)=V'(x)- \int \frac{1}{t-x}{\rm d}\mu_V(t)
\end{equation}
satisfies the algebraic equation
\begin{equation*}
 y^2-V'(x)y+Q(x)=0.
\end{equation*}
This defines a curve known as the spectral curve, which plays a crucial role in the analysis. Indeed, if $Q(x)$ is known explicitly, then the spectral curve can be used to perform the steepest descent analysis without knowledge of the equilibrium measure. The defining feature relevant for steepest descent analysis is the so-called \textit{S-property} of the harmonic function \smash{$\phi(x) :=\text{Re} \int y(x){\rm d}x$}~\cite{KS}:
 \begin{equation*}
 \frac{\partial}{\partial n_+} \phi(x) = \frac{\partial}{\partial n_-} \phi(x), \qquad x\in \operatorname{supp}\mu,
 \end{equation*}
where $\frac{\partial}{\partial n_{\pm}}$ denote the normal derivatives on either side of $\operatorname{supp}\mu$.
 It is precisely this property which allows for the opening of lenses, one of the standard transformations in the Deift--Zhou analysis. The polynomial $Q$ is determined in terms of the first few moments of the equilibrium measure. In general, without prior knowledge of this measure, these moments are difficult to compute. Without the aid of potential theory, one would need to prove that such an $S$-property holds by other means.

For special potentials $V_1$ and $V_2$, a Deift--Zhou steepest descent for the Riemann--Hilbert problem characterizing biorthogonal polynomials has been studied in~\cite{DG, DK1,DKM}. These works demonstrated that the limiting zero distribution for the polynomials can be characterized by a~vector equilibrium problem (see also~\cite{DGK}). This problem involves three measures that simultaneously minimize an energy functional. Since its full description is rather complicated, we omit it here to maintain the flow of presentation. Similar to the case of orthogonal polynomials on the real line described above, the minimizing vector of measures can be characterized by an algebraic curve. This aligns with predictions made in~\cite{Eynard,Orantin} based on loop equations. It is important to note that the vector equilibrium problem requires the potentials to be such that the matrix integrals converge as integrals over the real line. It does not trivially extend to the case $t<0$, which is of interest to us. Nevertheless, from the loop equation approach one still expects that, in the problem at hand, the limiting zero distribution is given by an algebraic curve of the form
\begin{align*}
 \tau q X^4 + \tau q^{-1}Y^4 -tX^3Y^3 - qX^3Y - q^{-1}Y^3X + t\tau^{-1}X^2Y^2
 + a X^2 + b Y^2 + c XY + d=0,
\end{align*}
for some constants $a$, $b$, $c$, and $d$. Here $Y(X)$, $X(Y)$ are the analogs of \eqref{eq:1myx}:
 \begin{align*} %\label{eq:eq2m1m}
& \tau Y=X + t q X^3 + \int \frac{{\rm d}\mu(\zeta)}{\zeta - X},\\
& \tau X=Y + t q^{-1} Y^3 + \int \frac{{\rm d}\tilde \mu(\zeta)}{\zeta - Y},
 \end{align*}
and $\mu$, $\tilde{\mu}$ represent the limiting zero distributions of $p_n$ and $q_n$, respectively.

 However, finding the constants $a$, $b$, $c$ and $d$ explicitly is challenging. One important aspect of our analysis is the computation of the spectral curve for certain parameter values with $t<0$, and we are able to prove that the function $Y(X)$ satisfies an $S$-property of its own (cf.\ Lemma~\ref{global-inequalities}, and the proof of Proposition~\ref{lensing-generic}).
We will use this as input in the Deift--Zhou steepest descent analysis.

\section{Statement of results}

\subsection{Asymptotic behavior of the partition function}
We will now state our main result: an exact expression for $F(\tau,t,H)$ as given in \eqref{free-energy-2-matrix-model}.
To this end, we first consider the equation
 \begin{align}
 0 = \mathfrak{I}(\sigma;\tau,t,q) :={}& {-}t -\frac{1}{9}\tau^2\sigma \bigl(\sigma^2 - 3\bigr) -\frac{1}{3}\frac{\sigma}{(1+\sigma)^2} \nonumber\\
 \hphantom{0 = \mathfrak{I}(\sigma;\tau,t,q) :=}{}
 &{} + \frac{2}{3}\biggl(\frac{\sigma}{1-\sigma^2}\biggr)^2\Biggl[\underbrace{\frac{1}{2}\bigl(q+q^{-1}\bigr)}_{\cosh H} -1\Biggr].\label{critical-surface-ideal}
 \end{align}
Note that for fixed $q={\rm e}^{H}$ and $0< \tau <1$, this is an algebraic curve in the $(t,\sigma)$ plane. By the implicit function theorem, since
 \begin{equation*}
 \frac{\partial \mathfrak{I}}{\partial \sigma}(\sigma;\tau,t,q)\big|_{t=0,\sigma=0} = \frac{1}{3}\bigl(\tau^2-1\bigr),
 \end{equation*}
for $t=0$ there is a unique solution $\sigma(\tau,t, H)$ of \eqref{critical-surface-ideal} which is analytic in a neighborhood of~${t=0}$, and such that $\sigma(\tau,0,H)=0$, for any $H\in \RR$, $0<\tau<1$. The solution can be analytically continued along the negative part of the real-$t$ axis. Either one hits a branch point at some critical value $t_{\rm cr}$ such that $\sigma(\tau,t_{\rm cr},H)$ is a double root of \eqref{critical-surface-ideal}, or $\sigma(\tau,t,H)$ is well defined for all $t<0$. In the latter case, we set $t_{\rm cr}=-\infty$ (this case in practice does not happen for the values of $\tau$, $H$ we consider here).
\begin{Definition} \label{def:sigmaDtcrx}
For fixed $H$ and $0<\tau <1$, we denote with $t_{\rm cr}(\tau,H)$ the largest negative value such that the solution $\sigma(\tau,t,H)$, characterized by $\sigma(\tau,0, H)=0$, is real analytic for $t\in (t_{\rm cr},0].$ Moreover, we define
\[
\mathcal D=\{(\tau,t,H) \mid 0<\tau<1,\, H \in \mathbb{R},\, t_{\rm cr}(\tau,H)< t <0\},
\]
and refer to $\mathcal{D}$ as the phase space.
\end{Definition}
Note that $\mathcal{D}$ is somewhat implicitly defined, as we have not indicated how $t_{\rm cr}$ can be found in terms of $\tau$ and $H$. As we will show below, for $H=0$ it is possible to give $t_{\rm cr}$ as an explicit function of $\tau$. For $H\neq 0$, we will see that $\mathcal D$ has a very convenient parametrization. That parametrization is still somewhat complicated and we have not been able to derive an explicit expression for $t_{\rm cr}$ as a function of both $\tau$ and $H$. We will postpone this discussion for now and first present the main result of the paper.
\begin{Theorem}\label{main-theorem}
 Let $(\tau,t,H)$ belong to the region $\mathcal D$. Then, as $n\to \infty$,
 \begin{align}\label{Kazakov-Free-Energy}
 F(\tau,t,H):={}&\lim_{n\to \infty} \frac{1}{n^2}\log \frac{ Z_{n}(\tau,t,H;n) }{ Z_{n}(\tau,0,0;n)} \nonumber \\
={}& \frac{3}{4} + \frac{1}{2}\log \frac{\bigl(1-\tau^2\bigr)\sigma(\tau,t,H)}{-3t} - \int_{0}^{\sigma(\tau,t,H)}\biggl(\lambda(u)-\frac{1}{2}\lambda(u)^2\biggr) \frac{{\rm d}u}{u},
\end{align}
 where $\lambda(u)$ is the rational function
 \begin{equation*}
 \lambda(u) = -\frac{1}{t}\biggl[\frac{1}{9}\tau^2u\bigl(u^2-3\bigr) + \frac{1}{3}\frac{u}{(u+1)^2} - \frac{2}{3}\biggl(\frac{u}{u^2-1}\biggr)^2[\cosh H -1]\biggr],
 \end{equation*}
 and $\sigma(\tau,t,H)$ from Definition~$\ref{def:sigmaDtcrx}$.
 \end{Theorem}

 This result was first formally computed in~\cite{Kazakov2, Kazakov1}, based partially on the work~\cite{Mehta}. Our results are consistent with this formula. We remark that an alternative form of this result was derived using combinatorial methods, cf.~\cite{BDFG}, or~\cite{Eynard-book}
 and references therein.
 \begin{Remark}
 One should compare \eqref{critical-surface-ideal} to formula (17) in \cite{Kazakov2}, with the translation of notations
$$\renewcommand{\arraystretch}{1.2}
 \begin{array}{|c||c|c|c|c|}
 \hline
 \text{ours} & \tau & t & q & \sigma\\
 \hline
 \text{theirs} & c & g = g(z) & B = \bigl[q+q^{-1}-2\bigr] & -z\\
 \hline
 \end{array}
$$
 \end{Remark}
 \subsection{Structure of the phase space}
 We will now discuss the structure of the phase space $\mathcal D$ in greater detail. We will start by discussing the $H=0$ first, as it is somewhat simpler, and then discuss the general situation.

\subsubsection[$H=0$]{$\boldsymbol{H=0}$}
We begin with a description of the phase space when $H=0$, as this region can be presented in a more transparent manner. All of the proofs of statements in this section follow from the more general results of the generic case ($H\neq 0$).
\begin{figure}
 \centering
 \includegraphics{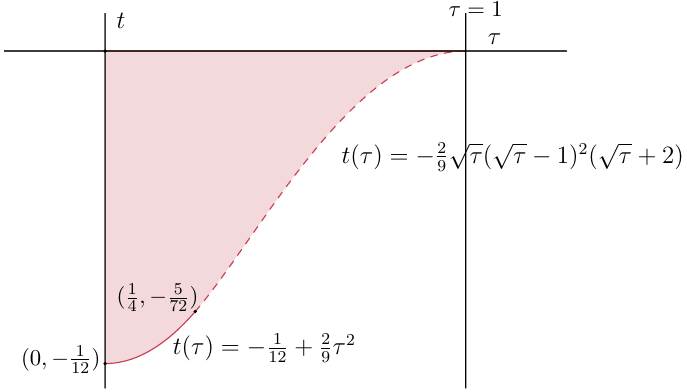}
% \begin{overpic}[scale=.15]{Pictures/PhaseDiagram.pdf}
% \put (79,66) {$\tau = 1$}
% \put (86,61) {\large $\tau$}
% \put (20,64) {\large $t$}
% \put (3,4) {$(0,-\frac{1}{12})$}
% \put (20,15) {$(\frac{1}{4},-\frac{5}{72})$}
% \put (30,6) {\large $t(\tau) = -\frac{1}{12} + \frac{2}{9}\tau^2$ }
% \put (60,40) {\large $t(\tau) = -\frac{2}{9} \sqrt{\tau}(\sqrt{\tau} - 1)^2(\sqrt{\tau} +2)$}
% \end{overpic}
 \caption{The phase portrait for the $2$-matrix model with quartic interactions in the plane $H=0$. The low-temperature critical curve is represented by the solid red line; the high-temperature critical curve is represented by the dashed red line.} \label{fig:PhaseDiagram}
\end{figure}
%\textcolor{orange}{Maybe do not treat $H=0$ first and move this down?}
 For $H = 0$, $0<\tau<1$ we have the following values for $t_{\rm cr}(\tau,0)$:
 \begin{equation*}
 t_{\rm cr}(\tau,0) =
 \begin{cases}
 -\dfrac{1}{12} +\dfrac{2}{9}\tau^2,& 0<\tau \leq \dfrac{1}{4},\\
 - \dfrac{2}{9} \sqrt{\tau}(\sqrt{\tau} + 1)^2(\sqrt{\tau} +2),& \dfrac{1}{4} < \tau < 1.
 \end{cases}
 \end{equation*}
 See also Figure~\ref{fig:PhaseDiagram}. These curves can be obtained as the most negative real solution to the system of equations $\mathfrak{I}(\sigma;\tau,t,0) = 0$, $\frac{\partial \mathfrak{I}}{\partial \sigma}(\sigma;\tau,t,0) = 0$. The intersection of $\mathcal D$ and the $H=0$ plane is the region enclosed by $t=0$, $\tau=0$ and a curve connecting $(0,-\frac{1}{12},0)$ to $(1,0,0)$. This curve consist of two parts: one part connecting $\bigl(0,-\frac{1}{12},0\bigr)$ to the multicritcial point $\bigl(\frac{1}{4},-\frac{5}{72},0\bigr)$, and the other connects $\bigl(\frac{1}{4},-\frac{5}{72},0\bigr)$ to $(1,0,0)$. We will refer to the first curve as the low-temperature critical curve and denote it by $\gamma_{{\rm low},0}$. The second, denoted by $\gamma_{{\rm high},0}$, will be referred to as the high-temperature critical curve. Note that these are precisely the critical curves that appeared in Kazakov's original work~\cite{Kazakov1}. Note that, when $\tau\to 0$, the partition function $Z_N(\tau,t,0;n)$ decouples into two independent partition functions for the $1$-matrix model. The point $\bigl(0,-\frac{1}{12},0\bigr)$ is the critical point for the 1-matrix model that was studied in~\cite{DK0, FIK1,FIK2}. In this case, the phase transition can be described in terms of special solutions to the Painlev\'e~I equation. At~the~multi-critical point, $\sigma(\tau,t,0)$ is a triple root and the description of this phase transition is the topic of a forthcoming paper.

 %Finally, we mention that \textcolor{orange}{I guess this is proved from the parametrization? It is not so important, so I am fine with removing it too}
 %\begin{equation}
 % \sigma(\tau,t,0) =
 % \begin{cases}
 % 1, & (\tau,t,0) \in \gamma_{{\rm low},0},\\
 % \tau^{-1/2} - 1, & (\tau,t,0) \in \gamma_{{\rm high},0}.
 % \end{cases}
 %\end{equation}
 We now proceed with general case.
 \subsubsection{The general case}

 Due to the aforementioned symmetry in $H<0$ and $H>0$, it will be convenient to use $q={\rm e}^{H}$ as in \eqref{eq:qeH} and only consider $0< q \leq 1$. We will then set
\[
 D=\{(\tau,t,q) \mid t<0, \, \tau>0,\, 0<q \leq 1, \, (t,\tau,\log q) \in \mathcal{D}\}.
 \]
One of the important ingredients in our analysis is a particular parametrization of the phase space $\mathcal D$. This parametrization has not been used before in the literature, to the best of our knowledge.

 We start by defining the set $R$ of all possible values for our parameters
 \begin{equation} \label{eq:phasespaceabc}
 R := \bigl\{(a,b,c)\mid 1 \leq a \leq b^{-1},\, 0 < c \leq b\bigr\}.
 \end{equation}
 Then on $R$ we define
 \begin{align*}
 \Pi\colon \ R\longrightarrow \mathbb{R}^3, \qquad
 (a,b,c) \longmapsto (\tau,t,q),
 \end{align*}
 where
 $\tau$, $t$ and $q$ are defined as
 \begin{align*}
 &\tau= \tau(a,b,c) = \frac{1}{\sqrt{(a^4c^2 + a^2b^2c^2 + a^2 + c^2)(a^4b^2 + a^2b^2c^2 + a^2 + b^2)}}, \\
 &t= t(a,b,c) = -\frac{a^2bc\bigl(a^4b^2c^2 + 3a^4 + 3a^2b^2 + 3a^2c^2 + 3b^2c^2 - 3\bigr)}{9(a^4c^2 + a^2b^2c^2 + a^2 + c^2)(a^4b^2 + a^2b^2c^2 + a^2 + b^2)},\\
 &q= q(a,b,c) = \frac{c\bigl(a^4b^2 + a^2b^2c^2 + a^2 + b^2\bigr)}{b(a^4c^2 + a^2b^2c^2 + a^2 + c^2)}.
 \end{align*}

 The Jacobian of this map is given by
 \begin{gather}\label{eq:jacobian}
 \frac{\partial(\tau,t,q)}{\partial(a,b,c)} = -\frac{4ac\bigl(a^2-1\bigr)\bigl(a^2+1\bigr)\bigl(a^2b^2-1\bigr)\bigl(a^2c^2-1\bigr)\bigl(b^2c^2-1\bigr)\bigl(a^2-b^2\bigr)\bigl(a^2-c^2\bigr)}{3b(a^2b^2c^2 + a^4b^2+a^2+b^2)^{3/2}(a^2b^2c^2+a^4c^2 + a^2+c^2)^{7/2}},
 \end{gather}
and this does not vanish in the interior of $R$. It does vanish on the surfaces defined by $a=1$ and $a=b^{-1}$, and this is the reason that they are of particular importance to us. We define these components below.

\begin{figure}[t]
 \centering
 \includegraphics{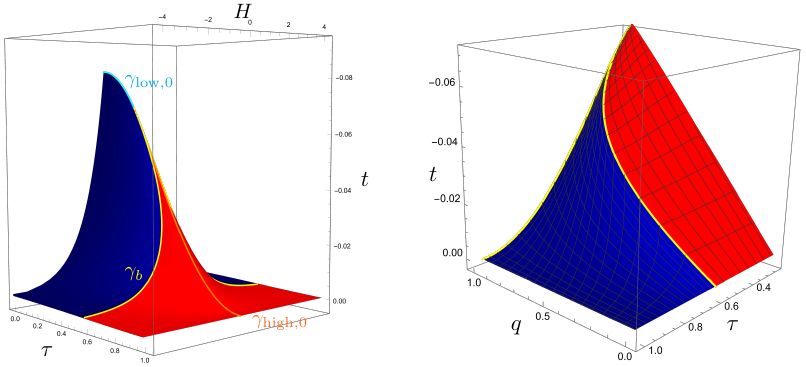}
% \begin{overpic}[scale=.5]{Pictures/PhaseDiagram3D.pdf}
% \put (12,3) {$\tau$}
% \put (100,50) {$t$}
% \put (65,96) {\small $H$}
% \put (35,78) {\small \textcolor{Cyan}{$\gamma_{{\rm low},0}$}}
% \put (70,12) {\small \textcolor{Orange}{$\gamma_{{\rm high},0}$}}
% \put (35,25) {\small \textcolor{Yellow}{$\gamma_{b}$}}
% \end{overpic}\quad \qquad
% \begin{overpic}[scale=.5]{Pictures/criticalsurfaceq.pdf}
% \put (0,50) {$t$}
% \put (22,10) {$q$}
% \put (80,10) {$\tau$}
% \end{overpic}
 \caption{The left picture illustrates the phase portrait for the $2$-matrix model with quartic interactions in the $(\tau,t,H)$ plane. The region where the spectral curve is of genus zero analyzed in this work lies in the half space where $\tau>0$, $t<0$, and is bounded by the red and blue surfaces. The genus zero partition function $F(\tau,t,H)$ is continuous on this surface, and in the genus zero region; on the surface, $F$ is analytic away from the low-temperature critical curve, shown here in light blue. The multicritical point $\bigl(\frac{1}{4},-\frac{5}{72},0\bigr)$ is the meeting point of the high-temperature critical curve (orange), the two phase boundaries joining the low and high temperature phases (yellow), and the low-temperature critical curve. The picture on the right represents the phase diagram in the coordinates $\tau$, $t$, $q$ and only the part $0<q \leq 1$.}
 \label{fig:PhaseDiagram3D}
\end{figure}

 \begin{Definition} \label{defn:criticalsurfaces} \quad
 \begin{enumerate}\itemsep=0pt
 \item[$(1)$] The low-temperature critical surface $S_{{\rm low}}$, is the image under $\Pi$ of ${a=b^{-1}}$, with $0<b <1$, $0<c<b$.
 \item[$(2)$] The high-temperature critical surface $S_{{\rm high}}$, is given by setting $a=1$, with $0 < b < 1$, $0< c < b$.
 \item[$(3)$] The $H\neq 0$ $\tau$-critical curve $\gamma_b$, is given by setting $a=b=1$, with $0 < c <1$.
 \end{enumerate}
 Moreover, the intersection of these critical surface with $H=0$ can be obtained by further setting~${b=c}$ and we thus define:
 \begin{enumerate}\itemsep=0pt
 \item[$(4)$] The low-temperature $H=0$ critical curve $\gamma_{{\rm low},0}$ is given by setting $a=b^{-1}$, $c=b$.
 \item[$(5)$] The high-temperature $H=0$ critical curve $\gamma_{{\rm high},0}$ is given by setting $a=1$, $c=b$.
 \item[$(6)$] The multicritical point is given by setting $a = b = c = 1$.
 \end{enumerate}
 \end{Definition}
 The critical curve $\gamma_b$ is the interface between the low-temperature and high-temperature critical surfaces. A graphical representation of these critical surfaces is given in Figure~\ref{fig:PhaseDiagram3D}. From the parametrization of the critical surfaces, it is not hard to compute the normal to these surfaces. It is interesting to note that the normal is continuous when passing from the low-temperature to the high-temperature critical surface.

 Before we comment further on these critical surface, we first mention the necessary technical result that our map $\Pi$ indeed parametrizes the phase space $ D$ and that the critical surfaces indeed are the sets defined by $t=t_{\rm cr}$.

 \begin{Proposition} \label{prop:phase} The set $\Pi(R)$ is the region enclosed by $\tau=0$, $\tau=1$, $q=0$, $q=1$, $t=0$, and the critical surfaces $S_{{\rm low}}\cup S_{{\rm high}}\cup \gamma_b$. Moreover, $D=\Pi(R)$ and $\sigma(t,\tau,H)=a^2bc$.
 \end{Proposition}
 The proof of this proposition is given in Appendix~\ref{app:param}.

The critical surfaces $S_{{\rm low}}$, $S_{{\rm high}}$ are implicit in the subsequent works of Boulatov and Kazakov~\cite{Kazakov2}, although they are never explicitly derived. Our parametrizations based on \eqref{tau-param}--\eqref{q-param} of these surfaces is new and turn out to be very useful in our analysis of the Riemann--Hilbert problem.

\begin{Remark}
When $(a,b,c)$ tend to a point on the critical surface, the Jacobian
 \smash{\raisebox{0.8pt}{$\frac{\partial(\tau,t,q)}{\partial(a,b,c)} \to 0$}}. In~terms of the equation $0 = \mathfrak{I}(\sigma;\tau,t,q)$, this may be interpreted as the root \raisebox{-0.4pt}{$\sigma = \sigma(\tau,t,q)$} becoming a double root, i.e.,
 \begin{equation*}
 \mathfrak{I}(\sigma;\tau,t,q) = \frac{\partial \mathfrak{I} }{\partial \sigma}(\sigma;\tau,t,q) = 0,
 \end{equation*}
 for $(a,b,c)$ on the critical surface and away from the multicritical point. At the multicritical point $(\tau,t,q) = \bigl(\frac{1}{4},-\frac{5}{72},1\bigr)$, the
 root $\sigma$ becomes a \textit{triple root} of the equation $0 = \mathfrak{I}(\sigma;\tau,t,q)$. One can thus characterize the critical surface as a subset of the zero locus of the discriminant of equation \eqref{critical-surface-ideal}. In Appendix~\ref{Critical-Surface-Appendix}, we write an
 an explicit formula for this discriminant, and thus furnish an implicit formula for the critical surface.
\end{Remark}

 \subsection{Spectral curve}

 An important ingredient in our analysis is the spectral curve for the limiting zero distribution of the polynomials, which we will now introduce. First
 observe that, for $(a,b,c)\in R$, the function~$\sigma$ is bounded:
 \begin{equation*}
 0 \leq \sigma \leq 1,
 \end{equation*}
 with $\sigma = 1$ only if $a=b^{-1}$, $c=b$ (this is precisely the low-temperature $H=0$ critical curve).
 Then we define the rational functions
 \begin{align} \label{Xv}
 &X(v) = \sqrt{-\frac{\tau \sigma}{3t}}\biggl(v + \frac{\sigma q^{-1} -1}{\tau(\sigma^2-1)} v^{-1} - \frac{\sigma}{3q}v^{-3}\biggr),\\
 &Y(v) = \sqrt{-\frac{\tau \sigma}{3t}}\biggl(v^{-1} + \frac{\sigma q -1}{\tau(\sigma^2-1)} v - \frac{\sigma q}{3}v^{3}\biggr).\label{Yv}
 \end{align}
One can eliminate the parameter $v$ from the pair of functions $(X(v),Y(v))$ to obtain an implicit formula for the Riemann surface these functions parametrize
 \begin{equation*}
 \mathfrak{S}(X(v),Y(v)) = 0.
 \end{equation*}
Here $\mathfrak{S}(X,Y)$ is a degree $6$ polynomial in $X$ and $Y$, with rational coefficients in the variables~$\tau$,~$t$,~$q$, and $\sigma$ (recall that $\sigma$ is defined as a special solution to the algebraic equation \eqref{critical-surface-ideal}). Explicitly, this polynomial is
 \begin{align*}
 \mathfrak{S}(X,Y)={}& \tau q X^4 + \tau q^{-1}Y^4 -tX^3Y^3 - qX^3Y - q^{-1}Y^3X + t\tau^{-1}X^2Y^2\\
 &{}{+}\, \mathfrak{s}_2(\sigma;\tau,t,q) X^2 + \mathfrak{s}_2\bigl(\sigma;\tau,t,q^{-1}\bigr) Y^2 + \mathfrak{s}_1(\sigma;\tau,t,q) XY + \mathfrak{s}_0(\sigma;\tau,t,q).\nonumber
 \end{align*}
where $\mathfrak{s}_i(\sigma;\tau,t,q)$ are given by
 \begin{gather*}
 \mathfrak{s}_2(\sigma;\tau,t,q)= \frac{1-6\sigma-3\sigma^2}{27\tau t (\sigma+1)^3} - \frac{1}{27\tau t}\bigl(\sigma^{3} \tau^{2}+3 \sigma^{2} \tau^{2}-9 \sigma \tau^{2}-27 \tau^{2}+1\bigr) \nonumber\\
 \hphantom{\mathfrak{s}_2(\sigma;\tau,t,q)=}{}
 -\frac{(q-1)\sigma}{27\tau t(\sigma^2 - 1)^3}\bigl[\bigl(9 \tau^{2}-9\bigr)+9(q+1) \sigma +\bigl(4q^{-1}-\bigl(28 \tau^{2}+6\bigr) \bigr) \sigma^{2} \\
 \hphantom{\mathfrak{s}_2(\sigma;\tau,t,q)=-\frac{(q-1)\sigma}{27\tau t(\sigma^2 - 1)^3}\bigl[}{}
 -6 (q+1 ) \sigma^{3}\!+\bigl(30 \tau^{2}+3\bigr)\sigma^{4}\!+(q+1) \sigma^{5}\!-12\sigma^{6} \tau^{2}\!+\sigma^{8} \tau^{2}\bigr]\nonumber,\\
 \mathfrak{s}_1(\sigma;\tau,t,q) = -\frac{8 \bigl(5 \sigma +3\bigr)}{81 \bigl(\sigma +1\bigr)^{2} t} -\frac{\sigma^{6} \tau^{2}-15 \sigma^{4} \tau^{2}+27 \sigma^{2} \tau^{2}-3 \sigma^{2}+243 \tau^{2}+24 \sigma +171}{243 t}\nonumber\\
 \hphantom{\mathfrak{s}_1(\sigma;\tau,t,q) =}{}
 -\frac{4\sigma^3\bigl(\sigma^2+3\bigr)}{81t(\sigma^2 - 1)^2}\bigl[q+q^{-1}-2\bigr],\\
 \mathfrak{s}_0(\sigma;\tau,t,q) = -\frac{\sigma}{19683t^2\tau q^2(\sigma^2 - 1)^4} \nonumber \\
 \hphantom{\mathfrak{s}_0(\sigma;\tau,t,q) =}{}
 \times\biggl[15 q \sigma^{6} \tau^{2}+9 q \sigma^{5} t +\bigl(12-111 \tau^{2}\bigr) q\sigma^{4} +15\biggl(q^{2}-\frac{6}{5}t q +1\biggr) \sigma^{3}\nonumber\\
 \hphantom{\mathfrak{s}_0(\sigma;\tau,t,q) =\times\bigl[}{}
 +\bigl(177 \tau^{2}-15\bigr) q \sigma^{2}-54\biggl(q^{2}-\frac{1}{6}t q +1\biggr) \sigma
 + 81\bigl(1-\tau^{2}\bigr) q\biggr]^2.
 \end{gather*}
 These constants are given for completeness and we will not use their exact expression in this paper. Indeed, it is the uniformization that we will be working with. Note also, that since the curve has uniformization in terms of rational functions, it has genus zero.

 The spectral curve provides us with a 4-sheeted Riemann surface. In the interior of the phase space, it is easy to see from the uniformization that there are four branch points (given by $X(v)$ with $v$ a solution to $X'(v)=0$) denoted by $X=\pm \alpha, \pm \beta$ with $0<\alpha<\beta$. The sheet structure is represented in Figure~\ref{fig:Noncritical-Curve}. The first sheet has a branch cut at $[-\alpha,\alpha]$ at which it is glued to the second sheet in a crosswise manner. The second sheet has additional cuts at $(-\infty,-\beta]$ and~${[\beta,\infty)}$ at which it is glued in a crosswise manner to the third and fourth sheet respectively. On~each sheet $j=1,2,3,4$, we define the uniformization coordinate~$v_j(X)$ as the map such that~$X(v_j(X))$
 is the identity map on the $j^{\rm th}$ sheet (these maps are uniquely determined by their large $X$ behavior; the expansions of rescaled versions of the functions $v_j(X)$ for $X\to\infty$ are given in Appendix~\ref{Appendix-A}. See also Remark~\ref{uv-relation-remark} for the exact relation between the coordinates $u$ and $v$). Then, we can define the function $Y(X)$ on our Riemann surface
 by setting its value on the $j^{\rm th}$ sheet, $Y_j(X)$, to be $Y_j(X) = Y(v_j(X))$.

 Although we do not prove this explicitly, it can be shown from our analysis that zeros of the polynomials $p_n(x)$ accumulate on the interval $(-\alpha,\alpha)$ and have limiting distribution, which we denote by $\mu$. We can then recover the measure $\mu$ from the boundary behavior of $Y_1$ on the cut,
 \begin{equation} \label{eq:definitionmu}
 {\rm d}\mu(s) = \frac{\tau}{2\pi {\rm i}}[Y_{1,+}(s) - Y_{1,-}(s)]{\bf 1}_{[-\alpha,\alpha]}(s){\rm d}s.
 \end{equation}
 There is also a second measure that plays a role for us. Its defined by taking boundary values on the other cuts,
 \begin{align}\label{eq:definitionnu}
 &{\rm d}\nu(s)= -\frac{\tau}{2\pi {\rm i}}[Y_{3,+}(s) - Y_{3,-}(s)]{\bf 1}_{(-\infty,-\beta]}(s){\rm d}s -\frac{\tau}{2\pi {\rm i}}[Y_{4,+}(s) - Y_{4,-}(s)]{\bf 1}_{[\beta,\infty)}(s){\rm d}s.
 \end{align}
 The interpretation of $\nu$ for the asymptotic behavior of the polynomials is not as obvious as it is for $\mu$ and it serves mostly as an auxiliary measure in our analysis. Note that from the representations \eqref{eq:definitionmu} and \eqref{eq:definitionnu} it is far from obvious that these measures are positive and we will put significant effort in proving this. The positivity is important for our lensing inequalities in the Riemann--Hilbert analysis.

 Both measures $\mu$ and $\nu$ are absolutely continuous and have an analytic density. Away from the critical surfaces, these densities vanish as a square root near the endpoints. On the critical surfaces, one or both of these measures will have different behaviors at their endpoints and this will break down the limiting behavior of the partition function.

 \begin{Remark}
 When changing roles of $X$ and $Y$, the $Y$ coordinate has branch points at ${Y=\pm \tilde \alpha, \pm \tilde \beta}$ with $0<\tilde \alpha\leq \tilde \beta$. The zeros of the polynomials $q_n(x)$ accumulate on an interval $(-\tilde \alpha, \tilde \alpha)$ and have limiting distribution, which we denote by $\tilde \mu$. Moreover, $(X,Y)$ given by \begin{align*}
 \tau X = Y + t q^{-1} Y^3 + \int \frac{{\rm d}\tilde{\mu}(\zeta)}{\zeta - Y}
 \end{align*}
 are on the spectral curve and defines $X$ as a function on $Y$ on the first sheet with an analytic continuation to the other sheets of the surface.
\end{Remark}

\subsection{Phase transitions and special points on the spectral curve} \label{phase-transitions-section}
We will comment now on the phase transitions on the critical surfaces of the parameter space.

When the parameters approach the critical surface, the measures will observe different behavior of $\mu$ and $\nu$ at their endpoints:\footnote{Throughout, $\rho,\tilde{\rho} > 0$ represent irrelevant constant factors. All limits
 are taken from inside the support of the measure, i.e., $X\nearrow \alpha$, $X\searrow -\alpha$, $X\nearrow -\beta$, and $X\searrow \beta$, respectively.}
\begin{enumerate}\itemsep=0pt
 \item \textit{The generic (non-critical) case.} %$1<a<\frac{1}{b}, 0< b < 1, 0 < c < b$.
 The measures $\mu$, $\nu$ vanish as square roots
 at their finite endpoints:
 \begin{align*}
 &\frac{{\rm d}\mu}{{\rm d}s}= \rho |X\mp\alpha|^{1/2}[1 + \OO(|X\mp\alpha|)], \qquad X\to \pm\alpha,\\
 &\frac{{\rm d}\nu}{{\rm d}s}= \tilde{\rho} |X\mp\beta|^{1/2}[1 + \OO(|X\mp\beta|)], \qquad X\to \pm\beta.
 \end{align*}
 \item \textit{The low-temperature critical surface $S_{{\rm low}}$.}
 %($a=\frac{1}{b}, 0 < b < 1$, $0< c < b$)
 An extra zero at the endpoint of $\nu$:
 \begin{align*}
 &\frac{{\rm d}\mu}{{\rm d}s}= \rho |X\mp\alpha|^{1/2}[1 + \OO(|X\mp\alpha|)], \qquad X\to \pm\alpha,\\
 &\frac{{\rm d}\nu}{{\rm d}s}= \tilde{\rho} |X\mp\beta|^{3/2}[1 + \OO(|X\mp\beta|)], \qquad X\to \pm\beta.
 \end{align*}
 \item \textit{The high-temperature critical surface and curve $S_{{\rm high}} \cup\gamma_{{\rm high},0}$.}
 %$ ($a=1, 0 < b < 1$, $0< c \leq b$)
 An extra zero at the endpoint
 of $\mu$:
 \begin{align*}
 &\frac{{\rm d}\mu}{{\rm d}s}= \rho |X\mp\alpha|^{3/2}[1 + \OO(|X\mp\alpha|)], \qquad X\to \pm\alpha,\\
 &\frac{{\rm d}\nu}{{\rm d}s}= \tilde{\rho} |X\mp\beta|^{1/2}[1 + \OO(|X\mp\beta|)], \qquad X\to \pm\beta.
 \end{align*}
 \item \textit{The low-temperature $H=0$ critical curve.}
 %$\gamma_{{\rm low},0}$ ($a=1/b, 0<b<1, b=c$)
 Extra zeros at the endpoints of both measures:
 \begin{align*}
 &\frac{{\rm d}\mu}{{\rm d}s}= \rho |X\mp\alpha|^{3/2}[1 + \OO(|X\mp\alpha|)], \qquad X\to \pm\alpha,\\
 &\frac{{\rm d}\nu}{{\rm d}s}= \tilde{\rho} |X\mp\beta|^{3/2}[1 + \OO(|X\mp\beta|)], \qquad X\to \pm\beta.
 \end{align*}
 \item \textit{The $H\neq 0$ $\tau$-critical curve.}
 %$\gamma_b$ ($a=b=1$, $0<c<1$)
 The supports of $\mu$ and $\nu$ touch, and vanish locally as a cube root:
 \begin{align*}
 &\frac{{\rm d}\mu}{{\rm d}s}= \rho |X\mp\alpha|^{1/3}\bigl[1 + \OO\bigl(|X\mp\alpha|^{1/3}\bigr)\bigr], \qquad X\to \pm\alpha,\\
 &\frac{{\rm d}\nu}{{\rm d}s}= \tilde{\rho} |X\mp\alpha|^{1/3}\bigl[1 + \OO\bigl(|X\mp\beta|^{1/3}\bigr)\bigr], \qquad X\to \pm\alpha.
 \end{align*}
 \item \textit{The multicritical point.}
 %($a=b=c=1$)
 The supports of $\mu$ and $\nu$ touch, and an extra zero at touching point:
 \begin{align*}
 &\frac{{\rm d}\mu}{{\rm d}s}= \rho |X\mp\alpha|^{4/3}\bigl[1 + \OO\bigl(|X\mp\alpha|^{1/3}\bigr)\bigr], \qquad X\to \pm\alpha,\\
 &\frac{{\rm d}\nu}{{\rm d}s}= \tilde{\rho} |X\mp\alpha|^{4/3}\bigl[1 + \OO\bigl(|X\mp\beta|^{1/3}\bigr)\bigr], \qquad X\to \pm\alpha.
 \end{align*}
 \end{enumerate}

\begin{figure}[!ht]
 \centering
 \includegraphics[scale=1.03]{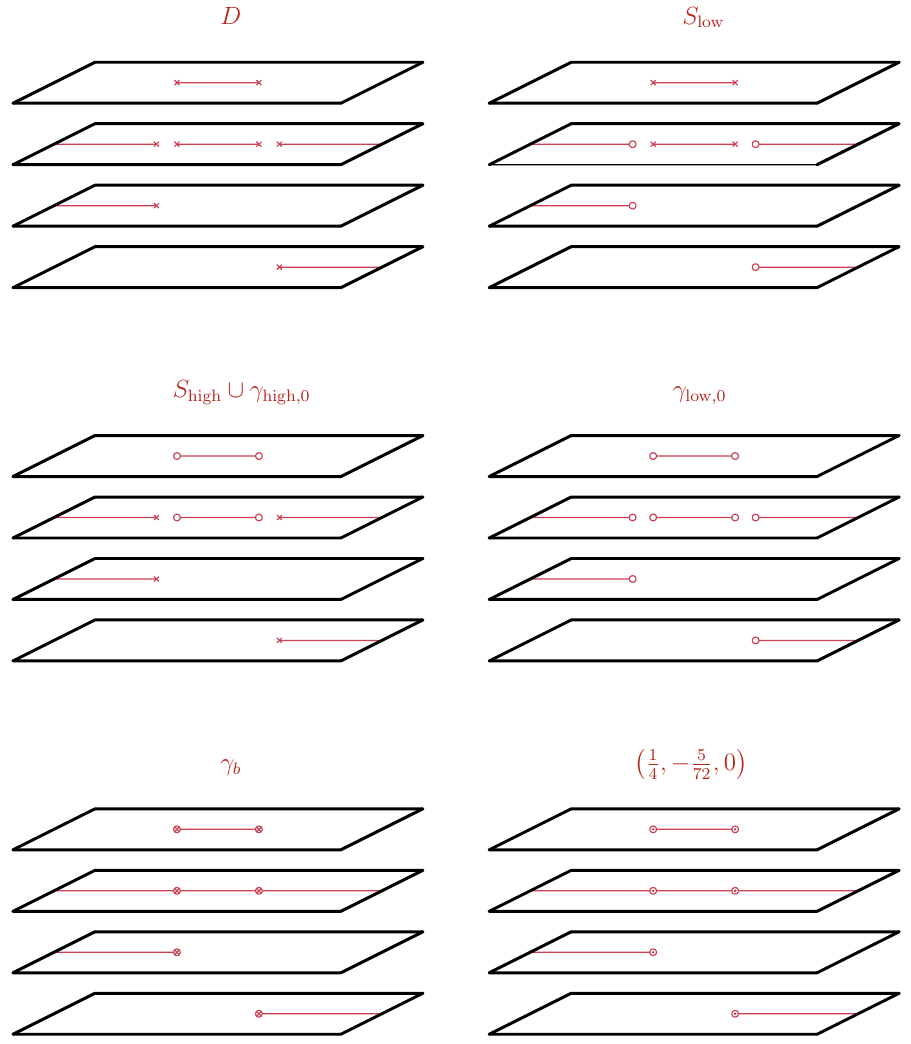}
% \begin{subfigure}{0.4\textwidth}
% \begin{overpic}[scale=.38]{Pictures/Branching-Behavior/Subcritical.pdf}
% \put (50,85) {\large \textcolor{BrickRed}{$D$}}
% \end{overpic}
% \end{subfigure}
% \hfill \vspace{-25pt}
% \begin{subfigure}{0.4\textwidth}
% \begin{overpic}[scale=.38]{Pictures/Branching-Behavior/SLow.pdf}
% \put (47,85) {\large \textcolor{BrickRed}{$S_{{\rm low}}$}}
% \end{overpic}
% \end{subfigure}
% \hfill \vspace{-25pt}
% \begin{subfigure}{0.4\textwidth}
% \begin{overpic}[scale=.38]{Pictures/Branching-Behavior/SHigh.pdf}
% \put (40,85) {\large \textcolor{BrickRed}{$S_{{\rm high}} \cup \gamma_{{\rm high},0}$}}
% \end{overpic}
% \end{subfigure}
% \hfill \vspace{-25pt}
% \begin{subfigure}{0.4\textwidth}
% \begin{overpic}[scale=.38]{Pictures/Branching-Behavior/Low0.pdf}
% \put (45,85) {\large \textcolor{BrickRed}{$\gamma_{{\rm low},0}$}}
% \end{overpic}
% \end{subfigure}
% \hfill \vspace{-25pt}
% \begin{subfigure}{0.4\textwidth}
% \begin{overpic}[scale=.38]{Pictures/Branching-Behavior/gammab.pdf}
% \put (50,85) {\large \textcolor{BrickRed}{$\gamma_b$}}
% \end{overpic}
% \end{subfigure}
% \hfill \vspace{-60pt}
% \begin{subfigure}{0.4\textwidth}
% \begin{overpic}[scale=.38]{Pictures/Branching-Behavior/Multicritical.pdf}
% \put (37,85) {\large \textcolor{BrickRed}{$\left(\frac{1}{4},-\frac{5}{72},0\right)$}}
% \end{overpic}
% \end{subfigure}
% \hfill \vspace{15pt}
 \caption{Possible endpoint behaviors of the measures $\mu$ and $\nu$ in various regions of the phase space. Each region is characterized by the behavior of $\mu$, $\nu$ at their endpoints. The symbols $\textcolor{BrickRed}{\times},\textcolor{BrickRed}{\bigcirc}, \textcolor{BrickRed}{\otimes}, \textcolor{BrickRed}{\odot}$ correspond to the following vanishing exponents of $\mu$, $\nu$ near their endpoints: $\textcolor{BrickRed}{\times} \leftrightarrow \frac{1}{2}$, $\textcolor{BrickRed}{\bigcirc} \leftrightarrow \frac{3}{2}$, $\textcolor{BrickRed}{\otimes} \leftrightarrow \frac{1}{3}$, $\textcolor{BrickRed}{\odot} \leftrightarrow \frac{4}{3}$.}
 \label{tab:SpectralCurveCriticalPhenomenon}
\end{figure}

The behaviors of the above measures are depicted in Figure~\ref{tab:SpectralCurveCriticalPhenomenon}. All these critical behaviors break our analysis in a crucial way and will lead to different scaling limits for the partition function.

The situation on the $S_{{\rm high}}$ and $S_{{\rm low}}$ is similar to the 1-matrix model critical point. Indeed also in that case, there is a merging of critical point to a branch point. This means that in a~double scaling limit, the partition function is described by a special solution to the Painlev\'e~I equation. For the Riemann--Hilbert analysis, it means that one should be able to treat the local parametrices in terms of Painlev\'{e}~I parametrices, as has been
 performed in previous works~\cite{BleherDeano, DK0} in the $1$-matrix model.

The situation on the critical curve $\gamma_b$ is intriguing. The situation is different from the rest for the critical surface. Nevertheless, we still expect that the phase transition is described by the same Painlev\'e~I equation as on the critical surfaces. We believe that a $3\times 3$ Painlev\'{e}~I parametrix (similar to the one arising from the Lax pair in~\cite{JKT}) should be used. We hope to address this in a future work.

Finally, the multi-critical point is the most important point of the critical surface and we shall treat this in the sequel to this work~\cite{DHL3, DHL2}.

\subsection{Strategy of the proof and an associated Riemann--Hilbert problem}

The strategy for the proof of Theorem~\ref{main-theorem} is a Deift--Zhou steepest descent analysis~\cite{Deift-Zhou} of $\YY$, i.e., a sequence of explicit invertible transformations
 \begin{equation*}
 \YY\longmapsto\XX\longmapsto\UU\longmapsto\TT\longmapsto\boldS\longmapsto\boldR,
 \end{equation*}
with the final Riemann--Hilbert problem for $\boldR$ having jumps that tend
to the identity matrix as $n\to \infty$, uniformly and in $L^2$, with
normalized behavior at infinity. Standard argument then allow us to obtain an asymptotic expansion for $\boldR$, and thus, utilizing the invertibility of the transformations, an asymptotic expansion of $\YY$. This can be used to write an expression for the partition function \eqref{Ising-random-partition-n}.

In Section~\ref{section3}, we analyze the spectral curve of this model. Section~\ref{section4} is devoted to the first and second transformations $\YY\mapsto\XX\mapsto\UU$. The transformations $\UU\mapsto\TT\mapsto\boldS$ which correspond to the opening of lenses are completed in Section~\ref{section5}. In Section~\ref{section6}, we construct the parametrices, and perform the final transformation $\boldS\mapsto\boldR$. We conclude with Section~\ref{section7}, in which we furnish a proof of Theorem~\ref{main-theorem}.

We will make use of some notations frequently; we establish them here for the convenience of the reader.
 \begin{itemize}\itemsep=0pt
 \item Throughout, $\omega := {\rm e}^{\frac{2\pi {\rm i}}{3}} = -\frac{1}{2} + {\rm i}\frac{\sqrt{3}}{2}$ is the principal third root of unity,
 \item We denote the $4\times 4$ matrix with a 1 in the $(i,j)^{\rm th}$ entry and zeros elsewhere by $E_{ij}$,
 \item The Pauli matrices
 \begin{equation*}
 \sigma_{1} := \begin{pmatrix} 0 & 1\\ 1 & 0\end{pmatrix}, \qquad
 \sigma_{2} := \begin{pmatrix} 0 & -{\rm i}\\ {\rm i} & \hphantom{-} 0\end{pmatrix},\qquad
 \sigma_{3} := \begin{pmatrix} 1 & \hphantom{-} 0\\ 0 & -1 \end{pmatrix}.
 \end{equation*}
 In particular, for the third Pauli matrix $\sigma_3$, we will often write expressions such as $z^{C \sigma_3}$, or ${\rm e}^{f(z)\sigma_3}$. These expressions are defined to be
 \begin{equation*}
 z^{C \sigma_3} := \begin{pmatrix}
 z^{C} & 0\\
 0 & z^{-C}
 \end{pmatrix},\qquad
 {\rm e}^{f(z)\sigma_3} :=
 \begin{pmatrix}
 {\rm e}^{f(z)} & 0\\
 0 & {\rm e}^{-f(z)}
 \end{pmatrix}.
 \end{equation*}
 \item The matrix $\hat{\sigma}_{ij}$ is defined to be the $4\times 4$ matrix which permutes the $i^{\rm th}$ and $j^{\rm th}$ row/column. For example, the matrix $\hat{\sigma}_{24}$ would be
 \begin{equation*}
 \hat{\sigma}_{24} =
 \begin{pmatrix}
 1 & 0 & 0 & 0\\
 0 & 0 & 0 & 1\\
 0 & 0 & 1 & 0\\
 0 & 1 & 0 & 0
 \end{pmatrix}.
 \end{equation*}
 The matrix $\hat{\sigma}_{ij}$ permutes the $i^{\rm th}$ and $j^{\rm th}$ row and column of a given matrix $A$ by conjugation; again using our example of $\hat{\sigma}_{24}$,
 \begin{equation*}
 \hat{\sigma}_{24} A \hat{\sigma}_{24} =
 \hat{\sigma}_{24}\begin{pmatrix}
 a_{11} & a_{12} & a_{13} & a_{14}\\
 a_{21} & a_{22} & a_{23} & a_{24}\\
 a_{31} & a_{32} & a_{33} & a_{34}\\
 a_{41} & a_{42} & a_{43} & a_{44}
 \end{pmatrix}\hat{\sigma}_{24}
 = \begin{pmatrix}
 a_{11} & a_{14} & a_{13} & a_{12}\\
 a_{41} & a_{44} & a_{43} & a_{42}\\
 a_{31} & a_{34} & a_{33} & a_{32}\\
 a_{21} & a_{24} & a_{23} & a_{22}
 \end{pmatrix}.
 \end{equation*}
 \item For readability purposes, blocks of zeros in matrices will be denoted simply by zero, where there is no cause for ambiguity. For example, the if $A$ is a $3\times 3$ matrix, then the expressions
 \begin{equation*}
 \begin{pmatrix}
 1 & 0 & 0 & 0\\
 0 & & & & \\
 0 & & A & \\
 0 & & &
 \end{pmatrix}=
 \begin{pmatrix}
 1 & 0_{3\times 1}\\
 0_{1\times 3} & A
 \end{pmatrix}=
 \begin{pmatrix}
 1 & 0 \\
 0 & A
 \end{pmatrix}
 \end{equation*}
 all have identical meaning.
 \item If $A$ is an $n\times n$ matrix and $B$ is an $m\times m$ matrix, we define the matrix $A\oplus B$ to be the block diagonal matrix
 \begin{equation*}
 A\oplus B :=
 \begin{pmatrix}
 A & 0\\
 0 & B
 \end{pmatrix}.
 \end{equation*}
 \item If $X(z)$ is the solution to a Riemann Hilbert problem defined on the contour $\gamma$, we write $J_X(z)\colon \gamma \to \CC$ as its jump matrix: i.e.,
 \begin{equation*}
 X_+(z) = X_-(z) J_X(z), \qquad z\in \gamma.
 \end{equation*}
 We will also sometimes write $\gamma_X$ or $\Gamma_X$ to denote the contour $\gamma$ corresponding to the Riemann--Hilbert problem for $X(z)$.
 \end{itemize}

\section{Definition and analysis of the spectral curve}\label{section3}
In this section, we define the spectral curve and an associated set of functions, which we will later use in the second
transformation. These functions are typically referred to as ``$g$-functions'' (or collectively as \textit{the} $g$-function) in the Riemann--Hilbert community, see~\cite{DKMVZ}. This function is used to transform the original Riemann--Hilbert problem, which contains exponentially growing terms, into a Riemann--Hilbert problem with oscillatory terms. It is required that this special function satisfy certain inequalities, which we refer to as \textit{lensing inequalities}, so that we can eventually transform the resulting Riemann--Hilbert problem with oscillations
into one which is exponentially close to a RHP with constant jumps; this transform is often referred to as the \textit{opening of lenses}. We also prove a number of inequalities necessary for the later ``lensing'' transformations. In what follows, the role of the $g$-function will be
played by a collection of functions $\Omega_j(z)$.

We begin by constructing the spectral curve, and give a basic analysis of the spectral curve for
all values of the parameters $(\tau,t,q)$ in the region $D$ bounded by $\tau=0$, $t=0$, and $q=1$ planes, the \textit{infinite temperature plane} $\tau =1$, and a ``critical surface'', which comprises the remaining boundary components of the region $D$.

\subsection{Definition of the spectral curve}
The analysis of the spectral curve is based on the ansatz that the curve is of genus zero, and thus rationally parameterized. Let us momentarily discuss the situation formally; once we have formulated a workable object, we shall prove that it is the correct one. Consider the partition function
 \begin{gather*}
 Z= \iint \Delta(x) \Delta(y) \exp\Biggl\{-n\sum_{i=1}^n \biggl(\frac{1}{2}x_i^2 + \frac{1}{2}y_i^2 + \frac{t q}{4} x_i^4 + \frac{t q^{-1}}{4} y_i^4 - \tau x_iy_i\biggr)\Biggr\} \prod_{i=1}^n {\rm d}x_i {\rm d}y_i\\
 \hphantom{Z}{}
= \iint \exp - \Bigg\{ \sum_{i < j} \log\frac{1}{|x_i - x_j|} + \sum_{i < j} \log\frac{1}{|y_i - y_j|} \\
\hphantom{Z=\iint \exp - \Bigg\{}{}
+ n\sum_{i=1}^n \biggl(\frac{1}{2}x_i^2 + \frac{1}{2}y_i^2 + \frac{t q}{4} x_i^4 + \frac{t q^{-1}}{4} y_i^4 - \tau x_iy_i\biggr)\Bigg\} \prod_{i=1}^n {\rm d}x_i {\rm d}y_i.
 \end{gather*}
In the large $n$ limit, under the appropriate scaling, we expect the $x_i$, $y_i$ to accumulate to a~limiting distributions that together minimize the term in the exponent; these distributions, $\mu$ and $\tilde \mu$ should be subject to the stationarity conditions that are given by
 \begin{align*}
 &X + t q X^3 + \int \frac{{\rm d}\mu(\zeta)}{\zeta - X} - \tau Y= 0,\\
 &Y + t q^{-1} Y^3 + \int \frac{{\rm d}\tilde{\mu}(\zeta)}{\zeta - Y} - \tau X= 0.
 \end{align*}
Expanding at infinity, we find that
 \begin{align}
 &X + t qX^3 - \frac{1}{X} - \tau Y= \OO\bigl(X^{-2}\bigr), \label{stationarity-X}\\
 &Y + t q^{-1} Y^3 - \frac{1}{Y} - \tau X= \OO\bigl(Y^{-2}\bigr).\label{stationarity-Y}
 \end{align}

We will construct solutions to these equations from an algebraic curve $\mathfrak{S}(X,Y) = 0$, by making the ansatz that it has genus $0$, i.e., that it is rationally parameterized.
\begin{Proposition}
 A $3$-parameter family of solutions to the stationarity equations \eqref{stationarity-X} and~\eqref{stationarity-Y} is given by the rational functions
 \begin{align}
 &X(u;a,b,c)= A \int^u \frac{\bigl(u^2-a^2\bigr)\bigl(u^2-b^2\bigr)}{u^4} {\rm d}u = A\biggl(u + \frac{a^2+b^2}{u} - \frac{a^2b^2}{3u^3}\biggr), \label{X-parameterization}\\
 &Y(u;a,b,c)= X\bigl(u^{-1};a,c,b\bigr) = B\biggl(\frac{1}{u} + \bigl(a^2+c^2\bigr)u - \frac{a^2c^2}{3}u^3\biggr),\nonumber%\label{Y-parametrization}
 \end{align}
 with $\tau$, $t$, $q$, $A$, and $B$ defined parametrically in terms of $a$, $b$, and $c$ as
 \begin{align}
 &\tau= \tau(a,b,c) = \frac{1}{\sqrt{\bigl(a^4c^2 + a^2b^2c^2 + a^2 + c^2\bigr)\bigl(a^4b^2 + a^2b^2c^2 + a^2 + b^2\bigr)}}, \label{tau-param}\\
 &t= t(a,b,c) = -\frac{a^2bc\bigl(a^4b^2c^2 + 3a^4 + 3a^2b^2 + 3a^2c^2 + 3b^2c^2 - 3\bigr)}{9\bigl(a^4c^2 + a^2b^2c^2 + a^2 + c^2\bigr)\bigl(a^4b^2 + a^2b^2c^2 + a^2 + b^2\bigr)},\label{t-param}\\
 &q= q(a,b,c) = \frac{c\bigl(a^4b^2 + a^2b^2c^2 + a^2 + b^2\bigr)}{b\bigl(a^4c^2 + a^2b^2c^2 + a^2 + c^2\bigr)},\label{q-param}\\
 &A= A(a,b,c) = \sqrt{\frac{3\bigl(a^4c^2 + a^2b^2c^2 + a^2 + c^2\bigr)}{a^4b^2c^2 + 3a^4 + 3a^2b^2 + 3a^2c^2 + 3b^2c^2 - 3}}\label{A-param}, \\
 &B = B(a,b,c) = A(a,c,b).\nonumber
 \end{align}
\end{Proposition}
\begin{Remark}\label{uv-relation-remark} Before we give the proof, we first make a slight abuse of notation. Note that we have defined $X(v)$ and $Y(v)$ already in \eqref{Xv} and \eqref{Yv}. It would be more prudent to use notations $\hat X(u;a,b,c)$ and $\hat Y(u;a,b,c)$. However, this is simply a scaling of the variables: $\hat{X}(u) = X\bigl(v\sqrt{A/B}\bigr)$ and $\hat{Y}(u) = Y\bigl(v\sqrt{A/B}\bigr)$. Thus the $u$ and $v$ variable give different parametrizations of the same objects. To avoid cumbersome notation, we will simply write~$X(u;a,b,c)$ and $Y(u;a,b,c)$. A convenient property of the $abc$-parametrization is that the branch points of $X(u)$ and $Y(u)$ are at $u=a,b$, and $u=b,c$, respectively.
\end{Remark}
\begin{proof}
 We can generically set
 \begin{align*}
 &X(u) = A \bigl[ u + \alpha_0 + \alpha_1 u^{-1} + \alpha_2 u^{-2} + \alpha_3 u^{-3}\bigr],\\
 &Y(u) = B \bigl[ u^{-1} + \beta_0 + \beta_1 u + \beta_2 u^{2} + \beta_3 u^{3}\bigr].
 \end{align*}
This fixes the positions of the infinities of the $X$- and $Y$-coordinates in the uniformizing plane as~$u=\infty$, $u=0$, respectively.
One then sees that this is the most general form of a rational function with poles only at $u=0,\infty$ which can possibly be a solution to the stationarity equations. This is because the leading terms on the right hand side have equal order. For example, as $u\to \infty$, $tX(u)^3 = \OO\bigl(u^3\bigr)$, and $Y(u) = X\bigl(u^{-1}\bigr) = \OO\bigl(u^{3}\bigr)$, with all other terms of order $\OO\bigl(u^2\bigr)$ or lower. The free parameters $A$, $B$, $\{\alpha_k\}$,
and $\{\beta_k\}$ must then be chosen so that the asymptotics
 \begin{align*}
 &X(u) + tqX(u)^3 - \frac{1}{X(u)} - \tau Y(u) = \OO\bigl(u^{-2}\bigr),\qquad u\to \infty,\\
 &Y(u) + tq^{-1}Y(u)^3 - \frac{1}{Y(u)} - \tau X(u) = \OO\bigl(u^{2}\bigr),\qquad u \to 0,
 \end{align*}
hold. Since the stationarity equations are odd functions of $X$, $Y$, we can assume (without loss of generality) that
 \begin{equation*}
 X(u) = -X(-u), \qquad Y(u) = -Y(-u),
 \end{equation*}
this implies that the branch points of $X$ (resp.\ $Y$) will be symmetric about $u=0$. This fixes the form of $X(u)$, $Y(u)$ as
 \begin{equation*}
 X(u) = A\biggl(u + \frac{a^2+b^2}{u} -\frac{a^2b^2}{3u^3}\biggr),\qquad Y(u) = B\biggl(\frac{1}{u} + \bigl(c^2+d^2\bigr)u -\frac{c^2d^2}{3}u^3\biggr),
 \end{equation*}
for some parameters $a$, $b$, $c$, $d$ (note that the branch points here are at $u=\pm a,b$, $u=\pm c,d$, respectively). We also have a degree of freedom available arising from reparametrization: ${u\to \lambda u}$. We use this freedom to fix $d=a$. Inserting these expressions for $X(u)$, $Y(u)$ into the stationarity equation \eqref{stationarity-X}, we obtain that
 \begin{gather*}
 \biggl(tA^3 +\frac{1}{3}\tau Ba^2c^2\biggr) u^3 + \bigl(A + 3 q tA^3\bigl(a^2+b^2\bigr) - \tau B \bigl(a^2+c^2\bigr)\bigr) u \\
 \qquad\quad{}+ \biggl( A\bigl(a^2 + b^2\bigr) + t A^3\bigl(-a^2b^2 + \bigl(a^2 + b^2\bigr)^2 + \bigl(a^2 + b^2\bigr)\bigl(2a^2 + 2b^2\bigr)\bigr) - \frac{1}{A} - \tau B \biggr) u^{-1} \\
 \qquad{}= \OO\bigl(u^{-2}\bigr).
 \end{gather*}
A similar equation at $u=0$ holds upon interchanging $u\to u^{-1}$, $b\leftrightarrow c$, $A\leftrightarrow B$, $q\leftrightarrow q^{-1}$. The requirement that the above vanishes to order $\OO\bigl(u^{-2}\bigr)$ (and the equivalent condition at $u=0$) yield a total of $6$ equations on the coefficients; one of these equations is redundant, and so we can solve the remaining equations for $A$, $B$, $t$, $\tau$, $q$ in terms of the parameters $a$, $b$, $c$. The unique solution (up to a sign, and possibly a relabelling $b\leftrightarrow c$) is given by equations \eqref{tau-param}--\eqref{A-param}. This completes the proof.
\end{proof}

We have now obtained a $3$-parameter family of solutions to the stationarity equations up to terms of order $\OO\bigl(X^{-2}\bigr)$. Each fixed triple $(a,b,c)$ parametrizes a Riemann surface. However, if one considers all possible values $(a,b,c)$, not all of them are relevant or suited for steepest descent analysis. Indeed, it will also be important that the measures $\mu$ and $\nu$ as given in \eqref{eq:definitionmu} and \eqref{eq:definitionnu} are positive when we are to open lenses. These constrains will be satisfied for all the parameters in our phase space $R$ as defined in \eqref{eq:phasespaceabc}.

\subsection{Sheet structure of the spectral curve}

The associated Riemann surface given by the parametrization $(X(u),Y(u))$ is called the \textit{spectral curve}. Let us study the structure of the spectral curve. We shall treat the parametric curve~$(X,Y)$ as a branched covering of the sphere over the $X$-coordinate; by construction, the $X$-coordinate has branch points ($X'(u) = 0$) at $u = \pm a,\pm b$, and $\infty$. This discussion is repeated from the introduction, for sake of clarity, and because of our slight abuse of notation on the relation between the coordinates $u$, $v$.

\begin{figure}[t]
 \centering
 \includegraphics{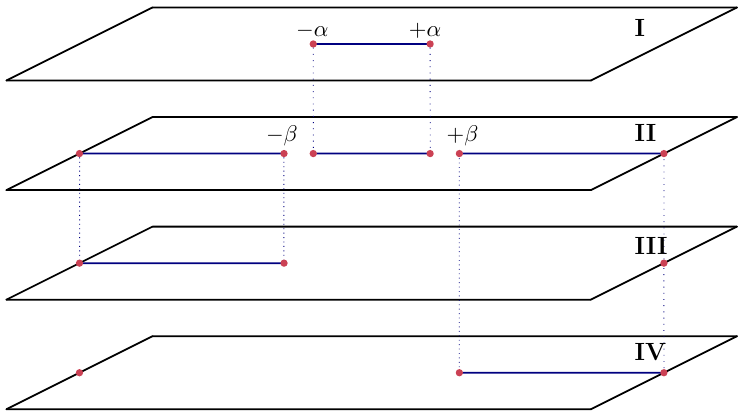}
% \begin{overpic}[scale=.3]{Pictures/NonCriticalSurface.pdf}
% \put (85,54) {\large \bf I}
% \put (85,40) {\large \bf II}
% \put (85,25) {\large \bf III}
% \put (85,11) {\large \bf IV}
% \put (55,54) {$+\alpha$}
% \put (40,54) {$-\alpha$}
% \put (60,40) {$+\beta$}
% \put (36,40) {$-\beta$}
% \end{overpic}
 \caption{A representative example of a the spectral curve in the physical plane, when the associated parameters $(\tau,t,H) \notin \gamma_b \cup \bigl\{\bigl(\frac{1}{4},-\frac{5}{72},0\bigr)\bigr\}$; the sheets are labelled ${\rm I},\dots,{\rm IV}$.}
 \label{fig:Noncritical-Curve}
\end{figure}

\begin{figure}[t]
 \centering
 \includegraphics{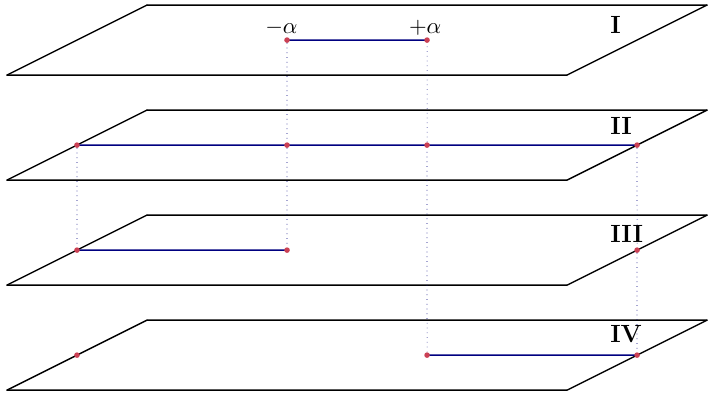}
% \begin{overpic}[scale=.23]{Pictures/CriticalSurface.pdf}
% \put (85,54) {\large \bf I}
% \put (85,40) {\large \bf II}
% \put (85,25) {\large \bf III}
% \put (85,11) {\large \bf IV}
% \put (57,54) {$+\alpha$}
% \put (37,54) {$-\alpha$}
% \end{overpic}
 \caption{The spectral curve, when the associated parameters $(\tau,t,H) \in \gamma_b \cup \bigl\{\bigl(\frac{1}{4},-\frac{5}{72},0\bigr)\bigr\}$. Note that the branch points at $z = \alpha$ and $z=\beta$ have merged here.}
 \label{fig:Critical-Curve}
\end{figure}

Away from the multicritical point and curve $\gamma_{b}$, the spectral curve is $4$-sheeted; this family of spectral curves have generically the same structure, and are shown in Figure~\ref{fig:Noncritical-Curve}. There are $4$ branch points, all of which lie on the real axis: at $\pm \alpha := X(\pm a)$, and $\pm \beta := X(\pm b)$. We have the inequalities $0 < \alpha < \beta < \infty$. The structure of the curve is as follows: Sheets $1$ and $2$ are glued along $[-\alpha,\alpha]$, sheet $2$ is glued to sheet $3$ along the interval $(-\infty,-\beta]$, and finally sheets~$2$ and~$4$ are glued along $[\beta,\infty)$. At the multicritical point $a=b=c=1$, and on the curve $\gamma_b$, the curve further degenerates, and the branch points $\pm \beta \to \pm \alpha$. This family of spectral curves is shown in Figure~\ref{fig:Critical-Curve}. In this case, sheets
$1$ and $2$ are glued along the interval $[-\alpha,\alpha]$, sheet $2$ is glued to sheet $3$ along the interval $(-\infty,-\alpha]$, and finally sheet $2$ is glued to sheet $4$ along the interval $[\alpha,\infty)$.

\begin{figure}[t]
 \centering
 \includegraphics{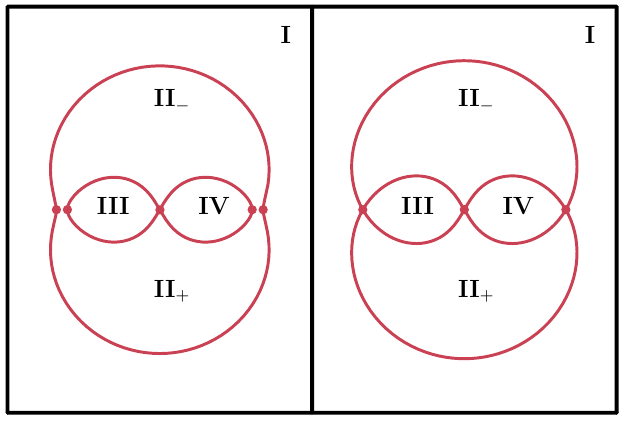}
 %\begin{overpic}[scale=.5]{Pictures/UniformizingPlane.pdf}
 %% generic
 %\put (25,20) {\large {\bf $\text{II}_+$}}
 %\put (25,50) {\large {\bf $\text{II}_-$}}
 %\put (45,60) {\large {\bf $\text{I}$}}
 %\put (16,33){\large {\bf $\text{III}$}}
 %\put (32,33){\large {\bf $\text{IV}$}}
 %% multicritical
 %\put (73,20) {\large {\bf $\text{II}_+$}}
 %\put (73,50) {\large {\bf $\text{II}_-$}}
 %\put (93,60) {\large {\bf $\text{I}$}}
 %\put (64,33){\large {\bf $\text{III}$}}
 %\put (80,33){\large {\bf $\text{IV}$}}
 %\end{overpic}

 \caption{The branch cuts in the uniformizing plane $u = x+{\rm i}y$ away from the multicritical point (left) and at the multicritical point (right). The noncritical case has $a=1.0184$, ${b=c=0.9100}$. The images of each sheet under the uniformizing map are labelled ${\rm I},\dots,{\rm IV}$. Note that, at criticality, the second sheet is split into two connected components; this is consistent with the picture in the physical plane, depicted in Figures~\ref{fig:Noncritical-Curve} and~\ref{fig:Critical-Curve}.}
 \label{fig:UniformizingPlane-Comparison}
\end{figure}

In the uniformization plane, the spectral curve is shown at and away from the multicritical point $(1,1,1)$ in Figure~\ref{fig:UniformizingPlane-Comparison}.

\subsection[The functions $\Omega$]{The functions $\boldsymbol{\Omega}$}

Of particular importance is the following (multi-valued) function:
 \begin{equation*}
 \tau\Omega(X) = \tau\int Y {\rm d}X.
 \end{equation*}
This function will play the role of what is often referred to as the $g$-function in our Deift/Zhou steepest descent analysis of the Riemann--Hilbert problem. The function $\tau\Omega(X)$ is in general a~multivalued function in $\CC$; its real part, however, can be defined in a single-valued manner on the spectral curve. The function $\Omega(u) := \int Y(u) X'(u) {\rm d}u$ in the uniformization coordinate is~explicit:
 \begin{align}
 \tau \Omega(u)={}& \tau \int Y {\rm d}X = \tau \int Y(u) X'(u) {\rm d}u \nonumber\\
={}& \frac{1}{a^2b^4c^2 + 3a^2b^2 + 3a^2c^2 + 3b^4 + 3b^2c^2 - 3}\biggl(-\frac{1}{4} b^2c^2 u^4 - \frac{3}{4}a^2b^2 u^{-4}\label{Omega-uniformizer}\\
 &{}{+} \frac{1}{2}\bigl(a^2b^2c^2 + b^4c^2 + 3b^2 + 3c^2\bigr)u^2- \frac{3}{2}\bigl( a^2b^4 + a^2b^2c^2 - a^2 - b^2\bigr)u^{-2}\biggr) - \log(u).\nonumber
 \end{align}
From here on, we shall relabel $X(u)$ as
 \begin{equation*}
 X(u) =: z(u),
 \end{equation*}
as the notation is more convenient and consistent later on. We also define the uniformization mappings $u_j(z)$, $j=1,2,3,4$, as the inverse functions to $z(u)$, such that $z(u_j(z))$ is the identity mapping in sheet $j$, away from the branch cuts.

We can expand $\Omega$ on each sheet of the spectral curve, by inverting the uniformization coordinate ($u_j(z)$, $j=1,2,3,4$), and inserting these expansions into the expression above for $\Omega$. Since~${z=\infty}$ (corresponding to $u=0$ on the lower three sheets, $u=\infty$ on the first sheet) is a~common branch point for all curves in the family parameterized by $R$, we can write an~expression for the expansion of $\Omega_j(z) := \Omega(u_j(z))$ at infinity that holds for all $(a,b,c) \in R$. We have the following proposition.

 \begin{Proposition}[Expansion of $\Omega_j(z)$ at $z=\infty$] For any fixed $(a,b,c)\in R$, with $t(a,b,c)$, $\tau(a,b,c)$, $q(a,b,c)$ defined as in
 \eqref{tau-param}--\eqref{q-param}, we have the following expansions of the functions~$\Omega_j(z)$:
 \begin{align}
 &\tau \Omega_1(z) = \frac{t q}{4}z^4 + \frac{1}{2}z^2 - \log z + \ell_0 + \frac{C_0}{z^2} + \OO\big(z^{-4}\big), \label{Omega-Sheet-1-infty}\\
 &\tau \Omega_{2}(z) = \begin{cases}
 -\dfrac{3\omega^2}{4}\dfrac{\tau^{4/3}}{(-tq^{-1})^{1/3}}z^{4/3} - \dfrac{\omega}{2} \dfrac{\tau^{2/3}}{(-tq^{-1})^{2/3}} z^{2/3} & \\
 \quad{}+ \dfrac{1}{3}\log z+ \ell_1 + \dfrac{\omega^2 C_1}{z^{2/3}} + \OO\left(\dfrac{1}{z^{4/3}}\right), & \operatorname{Im} z > 0, \vspace{1mm}\\
 -\dfrac{3\omega}{4}\dfrac{\tau^{4/3}}{(-tq^{-1})^{1/3}}z^{4/3} - \dfrac{\omega^2}{2} \dfrac{\tau^{2/3}}{(-tq^{-1})^{2/3}} z^{2/3} & \\
 \quad{} + \dfrac{1}{3}\log z+ \ell_1 + \dfrac{\omega C_1}{z^{2/3}}+ \OO\bigl(\frac{1}{z^{4/3}}\bigr), & \operatorname{Im} z < 0,
 \end{cases} \label{Omega-Sheet-2-infty}\\
 &\tau \Omega_{3}(z) = -\frac{3}{4}\frac{\tau^{4/3}}{(-tq^{-1})^{1/3}}z^{4/3} - \frac{1}{2} \frac{\tau^{2/3}}{(-tq^{-1})^{2/3}} z^{2/3} + \frac{1}{3}\log z + \ell_1 + \frac{C_1}{z^{2/3}} + \OO\biggl(\frac{1}{z^{4/3}}\biggr),\label{Omega-Sheet-3-infty}\\
 &\tau \Omega_{4}(z) = \begin{cases}
 -\dfrac{3\omega}{4}\dfrac{\tau^{4/3}}{(-tq^{-1})^{1/3}}z^{4/3} - \dfrac{\omega^2}{2} \dfrac{\tau^{2/3}}{(-tq^{-1})^{2/3}} z^{2/3} & \\
 \quad{} + \dfrac{1}{3}\log z+ \ell_1 + \dfrac{\omega C_1}{z^{2/3}}+ \OO\left(\dfrac{1}{z^{4/3}}\right), & \operatorname{Im} z > 0, \vspace{1mm}\\
 -\dfrac{3\omega^2}{4}\dfrac{\tau^{4/3}}{(-tq^{-1})^{1/3}}z^{4/3} - \dfrac{\omega}{2} \dfrac{\tau^{2/3}}{(-tq^{-1})^{2/3}} z^{2/3} &\\
 \quad{} + \dfrac{1}{3}\log z+ \ell_1 + \dfrac{\omega^2 C_1}{z^{2/3}} + \OO\left(\dfrac{1}{z^{4/3}}\right),& \operatorname{Im} z < 0.
 \end{cases}\label{Omega-Sheet-4-infty}
 \end{align}
 Here the constants $\ell_0 := \ell_0(a,b,c)$, $\ell_1 := \ell_1(a,b,c)$ are defined as
 \begin{gather*}
 \ell_0(a,b,c) = \frac{9a^6c^2 + 20a^4b^2c^2 + 9a^2b^4c^2 + 18a^4 + 18a^2b^2 + 18a^2c^2 + 18b^2c^2}{6(a^4b^2c^2 + 3a^4 + 3a^2b^2 + 3a^2c^2 + 3b^2c^2 - 3)}\\
 \hphantom{\ell_0(a,b,c) =}{}
 -\log[A(a,b,c)],\\
 \ell_1(a,b,c) = -\frac{3\bigl(2a^6b^2 + 2a^4b^4 + 2a^4b^2c^2 + 2a^2b^4c^2 + a^4 + 4a^2b^2 + b^4\bigr)}{2a^2b^2(a^4b^2c^2 + 3a^4 + 3a^2b^2 + 3a^2c^2 + 3b^2c^2 - 3)}\\
 \hphantom{\ell_1(a,b,c) =}{}
 - \frac{1}{3}\log \biggl[\frac{1}{3}a^2b^2 A(a,b,c)\biggr].
 \end{gather*}
 and the constant $C_1 := C_1(a,b,c)$ is defined as
 \begin{gather}
 C_1(a,b,c) = \frac{c^{3/2}\tau^{7/3}}{18b^{3/2}(-t)^{4/3}q^{1/6}}\bigl(3a^8b^4c^2 + 3a^6b^6c^2 + 3a^8b^2 + 3a^6b^4 + 3a^6b^2c^2\label{C1-constant}\\
 \hphantom{C_1(a,b,c) = \frac{c^{3/2}\tau^{7/3}}{18b^{3/2}(-t)^{4/3}q^{1/6}}\bigl(}{}
 + 3a^4b^6 + 3a^4b^4c^2 + 3a^2b^6c^2 - a^6 - 9a^4b^2 - 9a^2b^4 - b^6\bigr).\nonumber
 \end{gather}
 \textit{The constant $C_0$ has an explicit expression in terms of $a$, $b$, $c$, but it is irrelevant in further calculations.}
 \end{Proposition}
 \begin{proof}
 The proof of this proposition is a straightforward calculation; as we have alluded to, one must first expand the uniformization coordinate at $z=\infty$ on each of the
 sheets, then insert this expansion into the expression for $\Omega(u)$ in the uniformization coordinate (see equation \eqref{Omega-uniformizer}). The expansions of the uniformization coordinate on each sheet near the branch points are given in Appendix~\ref{Appendix-A}.
 \end{proof}

\subsection[Definition of the contours $\Gamma$, $\Gamma_1$, and $\Gamma_2$]{Definition of the contours $\boldsymbol{\Gamma}$, $\boldsymbol{\Gamma_1}$, and $\boldsymbol{\Gamma_2}$}
\begin{figure}[t]\centering
\includegraphics{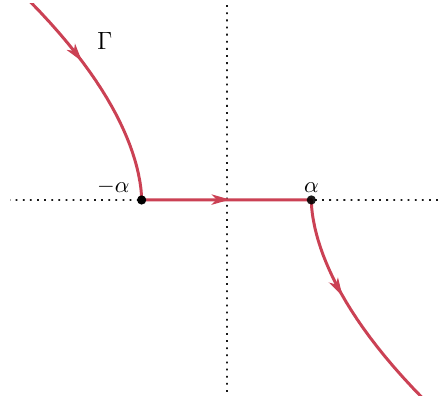}

% \begin{overpic}[scale=.38]{Pictures/GammaContours.pdf}
% \put (20,47) {$-\alpha$}
% \put (68,47) {$\alpha$}
% \put (20,80) {\large $\Gamma$}
% \end{overpic}

\caption{The contour $\Gamma$, which starts at ${\rm e}^{3\pi {\rm i}/4}\cdot \infty$, passes through $-\alpha$, $+\alpha$ on the real axis, and then goes off to ${\rm e}^{-\pi {\rm i}/4}\cdot\infty$.}
 \label{fig:Gamma-RHP}
\end{figure}

Our next step is to define a number of contour which our Riemann--Hilbert problem will rely on.
These contours will be chosen so that the functions $\Omega_j(z)$ satisfy certain inequalities on them. The first such contour is $\Gamma$, on which the
matrix-valued function $\YY(z)$ has jumps. We redefine~$\Gamma$ to be the contour starting at \smash{${\rm e}^{\frac{3\pi {\rm i}}{4}}\cdot\infty$}, passes through $z=-\alpha$,
then continues along the real axis until it reaches $z= + \alpha$, then goes off again to infinity in the direction \smash{${\rm e}^{-\frac{\pi {\rm i}}{4}}$}. The modified
contour $\Gamma$ is depicted in Figure~\ref{fig:Gamma-RHP}.

We also define two new contours, $\Gamma_1$ and $\Gamma_2$, which will appear in the first transformation. The contour $\Gamma_2$ is defined to start at
$z= -\infty$, then travel along the real axis until it reaches $z=-\beta$; the contour then goes off to infinity in the sector $-\frac{\pi}{4} <\arg (z+ \beta) < 0$.
The exact direction of approach to infinity will be established in the next section, when we will require certain inequalities to hold on $\Gamma_2$.

\begin{figure}[t]\centering
\includegraphics{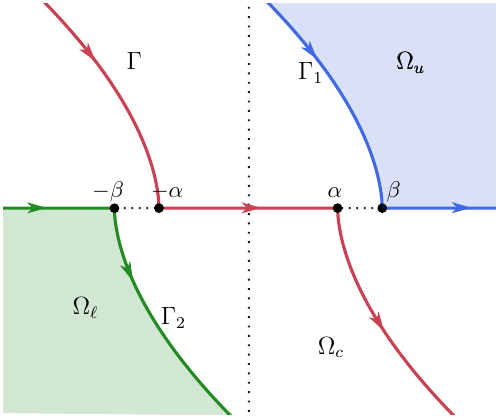}
% \begin{overpic}[scale=.4]{Pictures/GammaContours2-color.pdf}
% \put (30,44) {$-\alpha$}
% \put (66,44) {$\alpha$}
% \put (25,70) {\large $\Gamma$}
% \put (18,44) {$-\beta$}
% \put (78,44) {$\beta$}
% \put (60,68) {\large $\Gamma_1$}
% \put (32,18) {\large $\Gamma_2$}
% \put (14,20) {\large $\Omega_\ell$}
% \put (80,70) {\large $\Omega_u$}
% \put (80,70) {\large $\Omega_u$}
% \put (64,12) {\large $\Omega_c$}
% \end{overpic}
%		\end{center}
 \caption{The contours $\Gamma$, $\Gamma_1$, and $\Gamma_2$. Also depicted are the regions $\Omega_{\ell}$ and $\Omega_u$ (shaded), as well as $\Omega_c$ (unshaded).} \label{fig:Gamma-ABC}
\end{figure}

Similarly, we define $\Gamma_1$ to be the contour starting at infinity in the sector \smash{$0 < \arg (z- \beta ) < \frac{3\pi}{4}$}, and approaches the point
$z=\beta$; then, $\Gamma_1$ goes off to infinity again along the positive real axis. Again, the exact specifications of $\Gamma_1$ will be established in the
next section to guarantee that certain inequalities hold. The family of contours $\Gamma$, $\Gamma_1$, and $\Gamma_2$ are depicted in Figure~\ref{fig:Gamma-ABC}. We label the regions above $\Gamma_1$, below $\Gamma_2$, and bounded between $\Gamma_1$ and $\Gamma_2$ by $\Omega_u$, $\Omega_{\ell}$, an $\Omega_c$,
respectively.

\subsection{Lensing inequalities}\label{ssec:SecondTransformation}
We now begin the work of proving that the function $\Omega$ we constructed indeed satisfies the~inequalities necessary for lensing. The main idea of this
section is contained in this subsection; it explains how we can extend local inequalities near the branch points to inequalities that hold on the full
branch cuts. Most of the work that remains in the subsequent subsections involves expanding $\Omega$ near the branch points, and checking that the correct
inequalities hold there.
This check differs in the generic, critical, and multicritical cases, as the expansions of $\Omega$ near the~branch points
are different in each case. However, we stress that the proof is essentially the~same, and relies only on the following lemma. We will first need some setup
in the uniformizing plane. Define the polar functions
 \begin{align*}
 &r_+(\theta;a,b)= \sqrt{\frac{1}{2}\bigl(a^2+b^2\bigr) + \sqrt{\frac{4}{3} a^2b^2 \sin^2\theta + \frac{1}{4}\bigl(a^2-b^2\bigr)^2}}, \qquad \theta \in (-\pi,0)\cup (0,\pi),\\
 &r_-(\theta;a,b)= \sqrt{\frac{1}{2}\bigl(a^2+b^2\bigr) - \sqrt{\frac{4}{3} a^2b^2 \sin^2\theta + \frac{1}{4}\bigl(a^2-b^2\bigr)^2}},\\
 &\qquad\theta\in \Bigl(-\frac{\pi}{3}, 0\Bigr)\cup\Bigl(0,\frac{\pi}{3}\Bigr) \cup \biggl(\frac{2\pi}{3},\pi\biggr)\cup \biggl(\pi,\frac{4\pi}{3}\biggr) \nonumber.
 \end{align*}
 These functions define the preimages of the branch cuts (without the branch points) in the~uniformizing plane in polar form: $r_+(\theta;a,b)$ defines the
 preimage of the cut connecting sheets~$1$ and~$2$, and $r_-(\theta;a,b)$ defines the preimages of the cuts between sheets $2$ and $4$ \big(for $\theta\in\bigl(-\frac{\pi}{3},\frac{\pi}{3}\bigr)\setminus\{0\}$\big) and sheets $2$ and $3$ \big(for $\theta\in \bigl(\frac{2\pi}{3},\frac{4\pi}{3}\bigr) \setminus\{\pi\}$\big). One can obtain these expressions by solving the equation
 $\operatorname{Im} z \bigl(r{\rm e}^{{\rm i}\theta}\bigr) = 0$ for $r$ (see equation \eqref{X-parameterization}). Note that $r_{\pm}(\theta;a,b) = r_{\pm}(\theta;b,a)$.
 There are six connected components of these curves, corresponding to the $+/-$ sides of each branch cut. We now can state the main lemma.

\begin{figure}[t]\centering
\includegraphics{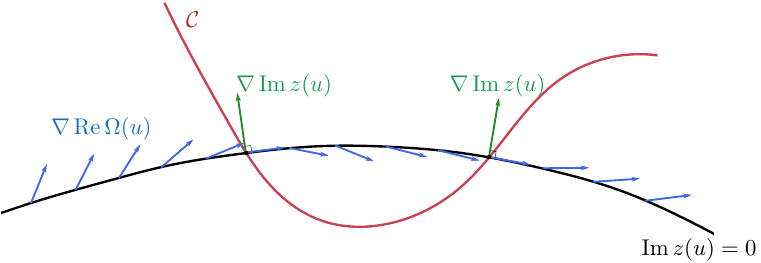}
% \begin{overpic}[scale=.8]{Pictures/Intersections.pdf}
% \put (7,22) {\textcolor{RoyalBlue}{$\nabla \operatorname{Re} \Omega (u)$}}
% \put (90,5) {$\operatorname{Im} z(u) = 0$}
% \put (26,37) {\textcolor{BrickRed}{$\mathcal{C}$}}
% \put (33,28) {\textcolor{ForestGreen}{$\nabla \operatorname{Im} z(u)$}}
% \put (63,28) {\textcolor{ForestGreen}{$\nabla \operatorname{Im} z(u)$}}
% \end{overpic}
 \caption{The curve $\operatorname{Im} z =0$, and the curve $\mathcal{C}$, which characterizes the places that $\nabla \operatorname{Re} \Omega (u)$ and $\nabla \operatorname{Im} z(u)$ are perpendicular. At the places where $\mathcal{C}$ and $\operatorname{Im} z =0$ intersect, the normal vectors to these curves are perpendicular, implying that the direction of steepest descent may change sign, as shown above.} %\label{fig:Intersections}
\end{figure}

\begin{Lemma}\label{global-inequalities}
 Define the vector field
 \begin{equation*}
 \hat{{\bf n}} := \frac{\nabla \operatorname{Im}z(u)}{||\nabla \operatorname{Im}z(u)||}
 \end{equation*}
$($note that this is the normal vector to the preimages of the branch cuts in the uniformizing plane under the mapping $z(u))$.
 For any $(a,b,c) \in R = \bigl\{(a,b,c)\mid 0 < b\leq 1,\, 1 \leq a \leq b^{-1},\, 0 < c \leq b\bigr\}$, the function
 \begin{equation*}
 \nabla \operatorname{Re}\Omega(u) \cdot \hat{{\bf n}} = \frac{\partial}{\partial n} \operatorname{Re}\Omega(u)
 \end{equation*}
 is of constant sign on each connected component of the preimages of the branch cuts.
\end{Lemma}

\begin{proof}
Let $\mathcal{C}$ denote the curve where $\frac{\partial}{\partial n} \operatorname{Re}\Omega(u) = 0$. In order to prove that the direction of steepest descent is constant on each component of the branch cuts, it is sufficient to check that the curves $\operatorname{Im} z(u) = 0$ and $\mathcal{C}$ do not intersect.\footnote{Here it is important to note that we have omitted the branch points from the definition of these curves. Indeed, $\operatorname{Im} z(u) = 0$ and $\mathcal{C}$ may intersect at the branch points: this occurs in the case of criticality. However, it does not affect any of the arguments that follow.}

First, since $\Omega(u)$ and $z(u)$ are analytic functions, the Cauchy--Riemann equations yield that $\partial[ \operatorname{Re}\Omega(u)] = \frac{1}{2}\partial \Omega(u)$,
and similarly $\partial [\operatorname{Im} z(u)] = \frac{1}{2{\rm i}}\partial z(u)$ (here $\partial$ denotes the holomorphic derivative in $u$). It follows that the $\nabla \operatorname{Re}\Omega(u)$ is perpendicular to $\nabla \operatorname{Im} z(u)$
if and only if the quotient of these two expressions is purely imaginary,
 \begin{equation*}
 \frac{\partial \Omega(u)}{{\rm i} \partial z(u)} \in {\rm i}\mathbb{R}.
 \end{equation*}

It follows that the curve $\mathcal{C}$ is characterized by the condition
 \begin{equation*}
 \operatorname{Im}\frac{\partial \Omega(u)}{\partial z(u)} = 0,
 \end{equation*}
which is equivalent to the condition \smash{$\operatorname{Im}\frac{\partial \Omega}{\partial z}(u) = \operatorname{Im}Y (u) = 0$} (cf.\ equation \eqref{Omega-uniformizer}).
We thus must prove that the components of the branch cuts (a subset of $\operatorname{Im} z(u) = 0$ as defined before) and the corresponding components of
$\operatorname{Im} Y(u) =0$ do not intersect.
We have already parameterized the branch cuts in the uniformizing plane in polar form; we can similarly parameterize the corresponding components of $\operatorname{Im}Y (u) = 0$. Since \smash{$X(u;a,b) = \overline{Y\bigl(u^{-1},a,b\bigr)}$}, we have that
 \begin{equation*}
 \operatorname{Im} Y\bigl(r{\rm e}^{{\rm i}\theta}\bigr) = 0 \ \Longrightarrow \ r = \frac{1}{r_{\pm}(\theta;a,c)}.
 \end{equation*}
Therefore, we must prove that the following equations have no solution, for any value of ${(a,b,c) \in R}$, and any $\theta$ in the corresponding parameter ranges
for $r_+$, $r_-$:
 \begin{enumerate}\itemsep=0pt
 \item[$(1)$] $r_{+}(\theta;a,b) = \frac{1}{r_{+}(\theta;a,c)}$,
 \item[$(2)$] $r_{-}(\theta;a,b) = \frac{1}{r_{-}(\theta;a,c)}$,
 \item[$(3)$] $r_{+}(\theta;a,b) = \frac{1}{r_{-}(\theta;a,c)}$,
 \item[$(4)$] $r_{-}(\theta;a,b) = \frac{1}{r_{+}(\theta;a,c)}$.
 \end{enumerate}

To show this, we state a sequence of inequalities we shall make use of. Provided $0 < c \leq b \leq 1$, for any $\theta$ in the parameter ranges for $r_{\pm}$, and any $1 \leq a \leq b^{-1}$, we claim that the following inequalities hold:
 \begin{itemize}\itemsep=0pt\samepage
 \item[$(i)$] $r_+(\theta;a,b) \geq 1$, with equality only if $a=1$,
 \item[$(ii)$] $r_-(\theta;a,b) \leq 1$, with equality only if $a=b=1$,
 \item[$(iii)$] $r_{+}(\theta;a,b) \geq r_{+}(\theta;a,c)$, with equality only if $b=c$,
 \item[$(iv)$] $r_{-}(\theta;a,b) \geq r_{-}(\theta;a,c)$ with equality only if $b=c$.
 \end{itemize}

Let us momentarily assume that $(i)$--$(iv)$ hold, and see how they imply $(1)$--$(4)$. For the equation~$(1)$, we have that
 \begin{equation*}
 (1) \ \Longleftrightarrow \ r_{+}(\theta;a,b)r_{+}(\theta;a,c) = 1,
 \end{equation*}
and since
 \begin{equation*}
 r_{+}(\theta;a,b)r_{+}(\theta;a,c) \overset{(iii)}{\geq} r_{+}(\theta;a,c)^2 \overset{(i)}{\geq} 1,
 \end{equation*}
we have only to check that the equation $r_{+}(\theta;a,b)^2 = 1$ has no solutions, for $0< b \leq 1$, $1 \leq a \leq b^{-1}$.
Similarly, for equation $(2)$,
 \begin{equation*}
 (2) \ \Longleftrightarrow \ r_{-}(\theta;a,b)r_{-}(\theta;a,c) = 1,
 \end{equation*}
and so
 \begin{equation*}
 r_{-}(\theta;a,b)r_{-}(\theta;a,c) \overset{(iv)}{\leq} r_{-}(\theta;a,b)^2 \overset{(ii)}{\leq} 1.
 \end{equation*}
 Thus, we have only to check that the equation $r_{-}(\theta;a,b)^2 = 1$ has no solution, for $0< b \leq 1$, $1 \leq a \leq b^{-1}$.
 For equations $(3)$ and $(4)$, we obtain that
 \begin{align*}
 &r_+(\theta;a,b)r_-(\theta;a,c)\overset{(iv)}{\leq} r_+(\theta;a,b) r_-(\theta;a,b),\\
 &r_{-}(\theta;a,b)r_+(\theta;a,c)\overset{(iii)}{\leq} r_+(\theta;a,b) r_-(\theta;a,b),
 \end{align*}
and so we have only to show that $r_+(\theta;a,b) r_-(\theta;a,b) \leq 1$, for $0< b \leq 1$, $1 \leq a \leq b^{-1}$. In~summary, our original problem involving equations $(1)$--$(4)$ has been reduced to showing that the following equations:
 \begin{equation}\label{simplified-intersection-equations}
 r_+(\theta;a,b)^2 = 1, \qquad r_-(\theta;a,b)^2 = 1,\qquad r_+(\theta;a,b)r_-(\theta;a,b) = 1,
 \end{equation}
have no solutions, provided $0< b \leq 1$, $1 \leq a \leq b^{-1}$.
Now, let us check that the first two equations \eqref{simplified-intersection-equations} have no solutions. We have that
 \begin{align*}
 1 =r_{\pm}(\theta)^2 \ \Longleftrightarrow \ {}& 1-\frac{1}{2}\bigl(a^2+b^2\bigr) = \sqrt{\frac{4}{3} a^2b^2 \sin^2\theta + \frac{1}{4}\bigl(a^2-b^2\bigr)^2}\\
 \Longleftrightarrow \ {}& a^2b^2- a^2 -b^2 + 1 = \frac{4}{3} a^2b^2 \sin^2\theta\\
 \Longleftrightarrow \ {}& \frac{3}{4}\biggl(1 -\frac{1}{b^2} - \frac{1}{a^2} + \frac{1}{a^2b^2}\biggr) = \sin^2\theta.
 \end{align*}
Since $\theta = 0,\pi$ are excluded from the range of $r_{\pm}$, we need that the left hand side of the above equation to satisfy one of the inequalities
 \begin{equation*}
 \frac{3}{4}\biggl(1 -\frac{1}{b^2} - \frac{1}{a^2} + \frac{1}{a^2b^2}\biggr) \leq 0 \qquad \text{or} \qquad
 \frac{3}{4}\biggl(1 -\frac{1}{b^2} - \frac{1}{a^2} + \frac{1}{a^2b^2}\biggr) > 1.
 \end{equation*}
Indeed, if $0< b \leq 1$, then
 \begin{equation*}
 \frac{1}{b^2} + \frac{1}{a^2} - \frac{1}{a^2b^2} \geq 1 + \frac{1}{a^2} - \frac{1}{a^2} = 1,
 \end{equation*}
so that
 \begin{equation*}
 \frac{3}{4}\biggl(1 -\frac{1}{b^2} - \frac{1}{a^2} + \frac{1}{a^2b^2}\biggr) \leq 0,
 \end{equation*}
and the necessary inequality holds in the region $R$. On the other hand, for the other intersection equation, we have that
 \begin{align*}
 1 > r_{+}(\theta)r_{-}(\theta) \ \Longleftrightarrow \ {}& 1 > \frac{1}{4}\bigl(a^2+b^2\bigr)^2 - \frac{4}{3}a^2b^2\sin^2\theta - \frac{1}{4}\bigl(a^2-b^2\bigr)^2\\
 \Longleftrightarrow \ {}& a^2b^2 - 1 < \frac{4}{3} a^2b^2\sin^2\theta\\
 \Longleftrightarrow \ {}& \frac{3}{4}\biggl(1 - \frac{1}{a^2b^2}\biggr) < \sin^2\theta,
 \end{align*}
since $1 \leq a \leq b^{-1}$, \smash{$\frac{3}{4}\bigl(1 - \frac{1}{a^2b^2}\bigr) < 0$}, and the desired inequality holds.

What remains to be shown is that the inequalities $(i)$--$(iv)$ hold. Indeed, we have that
 \begin{align*}
 r_+(\theta;a,b) &{}= \sqrt{\frac{1}{2}\bigl(a^2 + b^2\bigr) + \sqrt{\frac{4}{3}a^2b^2\sin^2\theta + \frac{1}{4}\bigl(a^2-b^2\bigr)^2}} \\
 &{}\geq \sqrt{\frac{1}{2}\bigl(a^2 + b^2\bigr) + \sqrt{0 + \frac{1}{4}\bigl(a^2-b^2\bigr)^2}} \\
 &{}= \sqrt{\frac{1}{2}\bigl(a^2 + b^2\bigr) + \frac{1}{2}\bigl(a^2-b^2\bigr)} = a \geq 1,
 \end{align*}
which proves $(i)$. Similarly,
 \begin{align*}
 r_-(\theta;a,b) &{}= \sqrt{\frac{1}{2}\bigl(a^2 + b^2\bigr) - \sqrt{\frac{4}{3}a^2b^2\sin^2\theta + \frac{1}{4}\bigl(a^2-b^2\bigr)^2}} \\
 &{}\leq \sqrt{\frac{1}{2}\bigl(a^2 + b^2\bigr) - \sqrt{0 + \frac{1}{4}\bigl(a^2-b^2\bigr)^2}} \\
 &{}= \sqrt{\frac{1}{2}\bigl(a^2 + b^2\bigr) - \frac{1}{2}\bigl(a^2-b^2\bigr)} = b \leq 1,
 \end{align*}
proving $(ii)$.

Let us now prove the inequalities $(iii)$, $(iv)$. Set
\[
f(x) :=\frac{1}{2}\bigl(a^2+x^2\bigr), \qquad g(x) = \frac{4}{3}a^2x^2\sin^2\theta +\frac{1}{4}\bigl(a^2-x^2\bigr)^2.
\]
Then,
 \begin{equation*}
 r_{\pm}(\theta;a,x) = \sqrt{f(x) \pm \sqrt{g(x)}}.
 \end{equation*}
Note that $(iii)$, $(iv)$, are equivalent to showing that $r_{\pm}(\theta;a,x) $ is monotone in $x$, for any fixed $0< a < x$, and any $\theta$ in the corresponding parameter ranges for $r_{\pm}$.
We have that
 \begin{equation*}
 \frac{\rm d}{{\rm d}x} \sqrt{f(x) \pm \sqrt{g(x)}} =
 \frac{f'(x) \pm \frac{g'(x)}{2\sqrt{g(x)}}}{2\sqrt{f(x) \pm \sqrt{g(x)}}}.
 \end{equation*}
We must therefore show that
 \begin{align*}
 x \pm \frac{x\bigl(\frac{8}{3}a^2\sin^2\theta - \bigl(a^2-x^2\bigr)\bigr)}{2\sqrt{\frac{4}{3}a^2x^2\sin^2\theta + \frac{1}{4}(a^2-x^2)^2}} >0,
 \end{align*}
for $\sin^2\theta < 1$ in the ``$+$'' case, and $\sin^2\theta <\frac{3}{4}$ in the ``$-$'' case. Indeed, in the ``$+$'' case, since $0<x<a$,
algebraic manipulation yields that
 \begin{gather*}
\begin{split}
& 2\sqrt{\frac{4}{3}a^2x^2\sin^2\theta + \frac{1}{4}\bigl(a^2-x^2\bigr)^2} + \frac{8}{3}a^2\sin^2\theta - \bigl(a^2-x^2\bigr) \\
& \qquad{}> 2\sqrt{\frac{4}{3}a^2x^2\sin^2\theta + \frac{1}{4}\bigl(a^2-x^2\bigr)^2} + \frac{8}{3}a^2\sin^2\theta > 0,
\end{split}
 \end{gather*}
since all quantities above are positive. In the ``$-$'' case, algebraic manipulation yields the~following equivalent inequality:
 \begin{gather*}
 2\sqrt{\frac{4}{3}a^2x^2\sin^2\theta + \frac{1}{4}\bigl(a^2-x^2\bigr)^2}> \frac{8}{3}a^2\sin^2\theta - \bigl(a^2-x^2\bigr)\\
 \qquad{}\Longleftrightarrow \ 4\biggl(\frac{4}{3}a^2x^2\sin^2\theta + \frac{1}{4}\bigl(a^2-x^2\bigr)^2\biggr) > \biggl(\frac{8}{3}a^2\sin^2\theta - \bigl(a^2-x^2\bigr)\biggr)^2\\
 \qquad{}\Longleftrightarrow \ \frac{64}{9}a^4\sin^2\theta\biggl(\frac{3}{4}-\sin^2\theta\biggr) > 0.
 \end{gather*}
Since $\sin^2\theta <\frac{3}{4}$ (recall the range of $\theta$ for $r_-$), we find that the above is positive, and so{\samepage
 \begin{equation*}
 x - \frac{x\bigl(\frac{8}{3}a^2\sin^2\theta - \bigl(a^2-x^2\bigr)\bigr)}{2\sqrt{\frac{4}{3}a^2x^2\sin^2\theta + \frac{1}{4}(a^2-x^2)^2}} >0.
 \end{equation*}
It follows that the inequalities $(iii)$ and $(iv)$ hold.}

\begin{figure}[t]\centering
\includegraphics{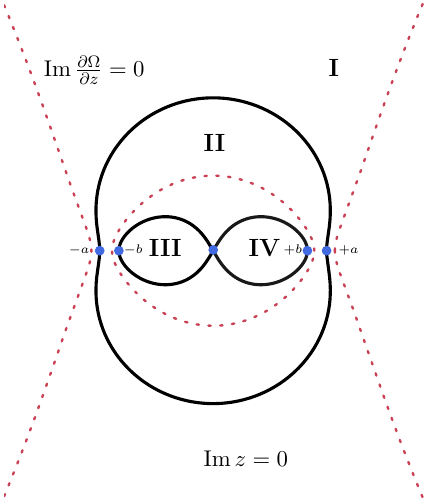}
% \begin{overpic}[scale=.4]{Pictures/LensingInequalities.pdf}
% 		 \put (65,85) {\large \bf I}
% 		 \put (40,70) {\large \bf II}
% 		 \put (29,49) {\large \bf III}
% 		 \put (49,49) {\large \bf IV}
% 		 \put (56,49.5) {\tiny $+b$}
% 		 \put (24,49.5) {\tiny $-b$}
% 		 \put (67,49.5) {\tiny $+a$}
% 		 \put (13,49.5) {\tiny $-a$}
% 		 \put (8,85) {$\operatorname{Im}\frac{\partial \Omega}{\partial z} = 0$}
% 		 \put (40,7) {$\operatorname{Im} z = 0$}
% 		\end{overpic}

 \caption{Images of the branch cuts $\operatorname{Im}z (u) = 0$ and $\operatorname{Im} \frac{\partial \Omega}{\partial z} = 0$ in the uniformizing plane, for $a=1.059$, $b=0.880$, $c=0.880$. The curves indeed do not intersect, and so the direction of steepest descent is constant along each connected component of the branch cuts.}
 \label{fig:Lensing-Inequalities}
\end{figure}

Thus, we have proven that, provided $(a,b,c) \in R = \bigl\{(a,b,c) \mid 0 < b\leq 1,\, 1 \leq a \leq b^{-1},\allowbreak {0 < c \leq b}\bigr\}$, the only intersection points of the branch cuts and the places where the direction of steepest descent of $\Omega(u)$ changes sign are at (possibly) the branch points. The branch cuts $\operatorname{Im} z = 0$ and the curves $\operatorname{Im}\frac{\partial \Omega}{\partial z} = 0$ in the uniformizing plane are shown in Figure~\ref{fig:Lensing-Inequalities} for a~particular choice of the parameters $(a,b,c)$; one can see explicitly that these curves do not intersect.
\end{proof}

This lemma is fundamental in proving the required lensing inequalities; we now proceed to the proofs of these.

%\begin{figure}
%		\begin{center}
%		 \begin{overpic}[scale=.5]{Pictures/Lensing1.pdf}
%		 \put (23,75) {$\Gamma$}
%		 \put (52,49) {$\Gamma_u$}
%		 \put (52,33) {$\Gamma_l$}
%		 \put (35,15) {$\Gamma_2$}
%		 \put (5,49) {$\Gamma_{2,u}$}
%		 \put (5,33) {$\Gamma_{2,l}$}
%		 \put (85,49) {$\Gamma_{1,u}$}
%		 \put (85,33) {$\Gamma_{1,l}$}
%		 \put (66,75) {$\Gamma_1$}
%		 \end{overpic}
%		\end{center}
% \caption{The opened lenses.}
% \label{fig:Lenses-1}
%\end{figure}

\subsection[The generic (noncritical) case: $0< b < 1$, $1 < a < b^{-1}$, $0 < c < b$]{The generic (noncritical) case: $\boldsymbol{0< b < 1}$, $\boldsymbol{1 < a < b^{-1}}$, $\boldsymbol{0 < c < b}$}
It remains to check that the $\Omega_j(z):=\Omega(u_j(z))$ have the correct sign around each of the lenses. This amounts to expanding $\Omega(u)$ around each of the
branch points, and checking that locally, the signs of the quantities we will be interested are positive. The fact that these inequalities hold
globally across each component of the support follows from the fact the intersection lemma we have just proven. We have the following proposition.

\begin{Proposition} \label{local-expansions-generic}
 Local expansions of $\Omega_j(z)$ around the branch points.
 \begin{enumerate}\itemsep=0pt
 \item[$(1)$] Around $z = \pm\alpha := z(\pm a)$
 \begin{align*}
 &\tau\Omega_1(z)= \eta_\alpha(z) + q_1 (z-\alpha)^{3/2} +\OO\bigl( (z-\alpha)^{2} \bigr),\\
 &\tau\Omega_2(z)= \eta_\alpha(z) - q_1 (z-\alpha)^{3/2} +\OO\bigl( (z-\alpha)^{2} \bigr),
 \end{align*}
 and
 \begin{align*}
 &\tau\Omega_1(z)= \begin{cases}
 \eta_{-\alpha}(z) + {\rm i}q_1 (z+\alpha)^{3/2} +\OO\bigl( (z+\alpha)^{2} \bigr), & \operatorname{Im} z >0,\\
 \eta_{-\alpha}(z) - {\rm i}q_1 (z+\alpha)^{3/2} +\OO\bigl( (z+\alpha)^{2} \bigr), & \operatorname{Im} z <0,
 \end{cases} \\
 &\tau\Omega_2(z)=
 \begin{cases}
 \eta_{-\alpha}(z) - {\rm i}q_1 (z+\alpha)^{3/2} +\OO\bigl( (z+\alpha)^{2} \bigr), & \operatorname{Im} z >0,\\
 \eta_{-\alpha}(z) + {\rm i}q_1 (z+\alpha)^{3/2} +\OO\bigl( (z+\alpha)^{2} \bigr), & \operatorname{Im} z <0,
 \end{cases}\\
 & \eta_{\pm \alpha} (z) = \Omega(\pm a) \mp \alpha\frac{a^2b^6 - 3a^2b^2 - 3b^4 - 3}{2a^2 + 6b^2}(z\mp \alpha).
 \end{align*}
Note that the constant
 \[
 q_1:=\frac{2\bigl(a^4-b^4\bigr)\bigl(a^4-1\bigr)\bigl(1-a^4b^4\bigr)}{a^{1/2}A^{3/2} (a^2-b^2)^{3/2}} >0.
 \]

 \item[$(2)$] \textit{Around $z = \pm\beta := z(\pm b)$}
 \begin{align*}
 &\tau\Omega_2(z)=
 \begin{cases}
 \eta_{\beta}(z) + {\rm i}\tilde{q}_1 (z -\beta)^{3/2} + \OO\bigl( (z-\beta)^2 \bigr), & \operatorname{Im} z > 0,\\
 \eta_{\beta}(z) - {\rm i}\tilde{q}_1(z -\beta)^{3/2} + \OO\bigl( (z-\beta)^2 \bigr),& \operatorname{Im} z < 0,\\
 \end{cases}\\
 &\tau\Omega_4(z)=
 \begin{cases}
 \eta_{\beta}(z) - {\rm i}\tilde{q}_1 (z -\beta)^{3/2} + \OO\bigl( (z-\beta)^2 \bigr), & \operatorname{Im} z > 0,\\
 \eta_{\beta}(z) + {\rm i}i\tilde{q}_1 (z -\beta)^{3/2} + \OO\bigl( (z-\beta)^2 \bigr),& \operatorname{Im} z < 0,\\
 \end{cases}
 \end{align*}
 and
 \begin{align*}
 &\tau\Omega_2(z)= \eta_{-\beta}(z) + \tilde{q}_1 (z +\beta)^{3/2} + \OO\bigl( (z+\beta)^2 \bigr),\\
 &\tau\Omega_3(z)= \eta_{-\beta}(z) - \tilde{q}_1 (z +\beta)^{3/2} + \OO\bigl( (z+\beta)^2 \bigr),\\
& \eta_{ \pm \beta}(z) = \Omega(\pm \beta) \mp \beta \frac{a^2b^6 - 3a^2b^2 - 3b^4 - 3}{2a^2 + 6b^2} (z\mp \beta).
 \end{align*}
Note that the constant
 \[
 \tilde{q}_1 := \frac{2 \bigl(a^4-b^4\bigr)\bigl(1-b^4\bigr)\bigl(1-a^4b^4\bigr)}{b^{1/2}A^{3/2} (a^2-b^2)^{3/2}} >0.
 \]
 \end{enumerate}
\end{Proposition}

 \begin{Proposition}\label{lensing-generic}
 \textit{Lensing around the cuts.}
 Let $\Omega_j(z) = \phi_j(z) + {\rm i}\psi_j(z)$, $j= 1,2,3,4$. Then, the following inequalities hold:
 \begin{enumerate}\itemsep=0pt
 \item[$(1)$] $\phi_4(z) - \phi_2(z) > 0$ in a lens around $[\beta,\infty)$,
 \item[$(2)$] $\phi_3(z) - \phi_2(z) > 0$ in a lens around $(-\infty,-\beta]$,
 \item[$(3)$] $\phi_2(z) - \phi_1(z) > 0$ in a lens around $[-\alpha,\alpha]$.
 \end{enumerate}
 \end{Proposition}

 \begin{proof}
 We have only to expand the differences $\phi_j(z)-\phi_k(z)$ around the branch points; our previous lemmas guarantee that if the correct inequality holds locally, it holds globally as well.
 For~$z$ sufficiently close to $\beta$, the difference ${\phi_4(z) - \phi_2(z) = \operatorname{Re}(\Omega_4 - \Omega_2)(z) \sim \operatorname{Re}\bigl[-2{\rm i}q (z-\beta)^{3/2}\bigr] >0}$ in the sector $0< \arg (z-\beta) < \frac{\pi}{3}$ for $|z-\beta|$ small enough.
 Now, for any $z \in (\beta,\infty)$, \smash{$u_{2,+}(z) = \overline{u_{4,-}(z)}$}, where $u_{j,\pm}(z)$ denote the continuous limits of $u_j(\zeta)$ as $\zeta \to z$ from the upper and lower half planes, respectively. Thus, since \smash{$\overline{\Omega(u)} = \Omega(\bar{u})$}, we have that
 \begin{equation*}
 \overline{\Omega(u_{2,+}(z))} = \Omega(u_{4,-}(z)),
 \end{equation*}
 and so $\operatorname{Re}(\Omega_4 - \Omega_2)(z) = 0$ for $z\in [\beta,\infty)$. Now, by definition of the sheets $2$, $4$, we have that
 \begin{equation*}
 \frac{\partial \operatorname{Re}\Omega_2}{\partial n_+} = -\frac{\partial \operatorname{Re}\Omega_4}{\partial n_-},\qquad
 \frac{\partial \operatorname{Re}\Omega_2}{\partial n_-} = -\frac{\partial \operatorname{Re}\Omega_4}{\partial n_+},
 \end{equation*}
 where \smash{$\frac{\partial}{\partial n_{\pm}}$} denote the normal derivatives in the upper/lower half planes in the $z$-coordinate, respectively (note that these normal derivatives and the one appearing in Lemma~\ref{global-inequalities} differ only by an overall positive factor $|z'(u)| >0 $).
 On the other hand, by our observation that \smash{$\overline{\Omega(u)} = \Omega(\bar{u})$}, we obtain the equalities
 \begin{equation*}
 \frac{\partial \operatorname{Re}\Omega_2}{\partial n_+} = \frac{\partial \operatorname{Re}\Omega_2}{\partial n_-},\qquad
 \frac{\partial \operatorname{Re}\Omega_4}{\partial n_+} = \frac{\partial \operatorname{Re}\Omega_4}{\partial n_-}.
 \end{equation*}
 We thus compute that
 \begin{equation*}
 \frac{\partial}{\partial n_{\pm}} [\operatorname{Re}(\Omega_4 - \Omega_2)(z)] = 2\frac{\partial}{\partial n_{\pm}}\operatorname{Re}(\Omega_4)(z).
 \end{equation*}
 Since this quantity is positive locally near $z=\beta$, Lemma~\ref{global-inequalities} allows us to conclude that \sloppy $\frac{\partial}{\partial n_{\pm}} [\operatorname{Re}(\Omega_4 - \Omega_2)(z)] >0$ for all $z \in [\beta,\infty)$.
 Therefore, we can open a lens around $[\beta,\infty)$.

 Similarly, near $z = -\beta$, again using the local expansions of Proposition~\ref{local-expansions-generic}, we have that $\phi_3(z) - \phi_2(z) = \operatorname{Re}(\Omega_3 - \Omega_2)(z) >0$ in the sector \smash{$\frac{2\pi}{3}< \arg(z+\beta) < \pi$} for $|z+\beta|$ sufficiently small. Thus, Lemma~\ref{global-inequalities} guarantees that we can open a lens around $(-\infty,\beta]$.

 Finally, near $z = +\alpha$ (respectively, $z=-\alpha$), the difference $\phi_2(z) - \phi_1(z) = \operatorname{Re}(\Omega_2 - \Omega_1)(z) >0$ for \smash{$\frac{2\pi}{3} < \arg z < \pi$}
 and $|z-\alpha|$ sufficiently small (respectively, \smash{$0< \arg (z-\beta) < \frac{\pi}{3}$} and $|z+\alpha|$ sufficiently small). Thus, we can also open a lens around the central cut $[-\alpha,\alpha]$.
 \end{proof}

\begin{Proposition}
 \textit{Inequalities off the real axis.}
 Let $\Omega_j(z) = \phi_j(z) + {\rm i}\psi_j(z)$, $j = 1,2,3,4$. Then, the following inequalities hold:
 \begin{enumerate}\itemsep=0pt
 \item[$(1)$] $\phi_2(z) - \phi_4(z) > 0$ for $z \in \Gamma_1 \cap \{\operatorname{Im} z > 0\}$,
 \item[$(2)$] $\phi_2(z) - \phi_3(z) > 0$ for $z \in \Gamma_2 \cap \{\operatorname{Im} z < 0\}$,
 \item[$(3)$] $\phi_1(z) - \phi_2(z) > 0$ for $z \in \Gamma \setminus \{\operatorname{Im} z = 0\}$.
 \end{enumerate}
\end{Proposition}
 \begin{proof}
 We prove $\phi_2(z) - \phi_4(z) > 0$ for $z \in \Gamma_1 \cap \{\operatorname{Im} z > 0\}$; the proofs of the other inequalities follow from similar
 argumentation. Using \eqref{local-expansions-generic}, we have that $\phi_2(z) - \phi_4(z) > 0$ for $|z-\beta|$ sufficiently small in the sector \smash{$\frac{2\pi}{3} < |{\arg (z-\beta)}| < \pi$}. Furthermore, at infinity, using equations~\eqref{Omega-Sheet-2-infty} and~\eqref{Omega-Sheet-4-infty}, we see that $\phi_2(z) - \phi_4(z) > 0$ for
 $|z|$ sufficiently large in the sector \smash{$\frac{3\pi}{4} < |{\arg z}| < \pi$}. Consider the domain
 \begin{equation*}
 E := \{z \mid \phi_2(z) - \phi_4(z) > 0 \},
 \end{equation*}
 by the lensing inequalities \eqref{lensing-generic}, the boundary of this domain is bounded away from the branch cuts,
 and $\phi_2(z) - \phi_4(z) = 0$ there. Since $\phi_2(z) - \phi_4(z)$ is not identically zero, the maximum principle tells us that the domain
 $E$ is necessarily unbounded, and reaches infinity in the sector $\frac{3\pi}{4} < |{\arg z}| < \pi$. Thus, we may freely redefine $\Gamma_1$ so that
 $\phi_2(z) - \phi_4(z) > 0$ along $\Gamma_1$ for all $z \in \Gamma_1 \cap \{\operatorname{Im} z > 0\}$.
 \end{proof}

\begin{Remark}[inequalities on the critical surface] In fact, all of the inequalities necessary for lensing hold on each
component of the critical surface. However, this requires a separate analysis in each case, as one must perform a local expansion
of the functions $\Omega_j(z)$ around each of the branch points, and the characteristic property of each of the components of the
critical surface, as we shall see in the next remark, is that the functions $\Omega_j(z)$ have \textit{different} expansions in these
regions in general. Since we are only interested in the genus $0$ partition function in this work, we omit the proof of these inequalities: this being said, the astute reader should have no trouble replicating the calculations performed above in each of the critical cases. We will demonstrate the proof of the relevant inequalities at the multicritical point explicitly in Part III of this work~\cite{DHL3}.
\end{Remark}

\section[The first and second transformations ${\bf Y} \mapsto {\bf X} \mapsto {\bf U}$]{The first and second transformations $\boldsymbol{{\bf Y} \mapsto {\bf X} \mapsto {\bf U}}$}\label{section4}

\subsection{Idea of the transformation}
The main idea of the transformation ${\bf Y} \mapsto {\bf X}$ is illustrated in~\cite{DK1}, who refer to an unpublished manuscript of Bertola, Harnad, and Its
as the origin of the idea. The point is that the weights appearing in the jump matrix, $f(z)$ and its derivatives, satisfy a modified form of the Pearcey differential equation:
 \begin{equation} \label{Pearcey-eq}
 \frac{t}{N^2\tau^2 q} f'''(z) + f'(z) - N\tau^2 z f(z) = 0,
 \end{equation}
whose solutions are the Pearcey-type integrals
 \begin{equation*}
 w_j(z) := \int_{\gamma_j} \exp\Biggl[ N\biggl(\tau z w \underbrace{-\frac{1}{2}w^2 - \frac{t}{4q}w^4}_{-V(q^{-1}w)} \biggr)\Biggr] {\rm d}w.
 \end{equation*}
It is also useful to notice that any solution $w(z) = w(z;\tau,t,q,N)$ to \eqref{Pearcey-eq} satisfies the partial differential equations
 \begin{align*}
 &\frac{\partial w}{\partial t}= -\frac{1}{4 N^3\tau^4} \frac{\partial^4 w}{\partial z^4} = \frac{q}{4N\tau^2 t} \frac{\partial^2 w }{\partial z^2} - \frac{qz}{4t}\frac{\partial w }{\partial z} - \frac{q}{4t} w,\\
 &\tau\frac{\partial w}{\partial \tau}= z\frac{\partial w}{\partial z}.
 \end{align*}
In the following subsection, we analyze the asymptotics of these integrals, and consider an associated Riemann--Hilbert problem for the $w_j(z)$'s, which we shall make use of in the first transformation $\YY \mapsto \XX$.

\subsection[Riemann--Hilbert problem and asymptotics for the Pearcey-type integrals $w_j(z)$]{Riemann--Hilbert problem and asymptotics \\
for the Pearcey-type integrals $\boldsymbol{w_j(z)}$}

\begin{figure}[t]\centering
\includegraphics{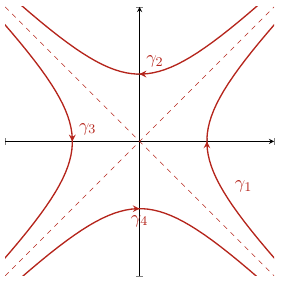}

%		\begin{tikzpicture}[scale=.8]
%			\begin{axis}[
%			axis lines=middle,
%				xticklabels=\empty,
%				yticklabels=\empty,
%				axis equal image,
% 				xmin = -2, xmax = 2,
% 				ymin = -2, ymax = 2]
%				 \addplot [BrickRed,thick,domain=-2:2,samples=50,
%					postaction={decorate},
%				 	decoration={markings, % ------
%		 				mark=at position 0.5 with {\arrow{stealth}}}							
%				 ]({cosh(x)},{sinh(x)}) node[above right,pos=.4] {$\gamma_1$};
%				 \addplot [BrickRed,thick,domain=-2:2,samples=50,
%					postaction={decorate},
%				 	decoration={markings, % ------
%		 				mark=at position 0.5 with {\arrowreversed{stealth}}}				
%				 ]({sinh(x)},{cosh(x)}) node[above right,pos=.5] {$\gamma_2$};
%				 \addplot [BrickRed,thick,domain=-2:2,samples=50,
%					postaction={decorate},
%				 	decoration={markings, % ------
%		 				mark=at position 0.5 with {\arrowreversed{stealth}}}				
%				 ]({-cosh(x)},{sinh(x)}) node[above right,pos=.5] {$\gamma_3$};
%				 \addplot [BrickRed,thick,domain=-2:2,samples=50,
%					postaction={decorate},
%				 	decoration={markings, % ------
%		 				mark=at position 0.5 with {\arrow{stealth}}}				
%				 ]({sinh(x)},{-cosh(x)}) node[below,pos=.5] {$\gamma_4$};
%				 \addplot [BrickRed,dashed,domain=-2:2,samples=50]({x},{x});
%				 \addplot [BrickRed,dashed,domain=-2:2,samples=50]({x},{-x});
%			\end{axis}
%			\end{tikzpicture}

 \caption{The contours $\gamma_j$, $j = 1,\dots,4$.} \label{Gamma-curves}
\end{figure}

We begin by utilizing classical steepest descent analysis to determine the asymptotics of each of the integrals
 \begin{equation}\label{Pearcey-int}
 w_j(z) = \int_{\gamma_j} \exp\bigl[ N(\tau z w - V(w) )\bigr] {\rm d}w,
 \end{equation}
as $z \to \infty$, where $\gamma_j$, $j = 1,\dots,4$ are one of the contours in Figure~\ref{Gamma-curves}. Before studying the asymptotics of the integrals \eqref{Pearcey-int}, we will study the solutions to the so-called saddle point equation. The saddle point equation is defined to be
 \begin{equation*}
 0 = \frac{\rm d}{{\rm d}w}\biggl[\tau z w -\frac{1}{2}w^2 -\frac{t}{4q}w^4\biggr] \ \Longleftrightarrow\ w^3 + \frac{q}{t}w -\frac{\tau q}{t}z = 0.
 \end{equation*}
The large $z$ asymptotics of the integrals \eqref{Pearcey-eq} are determined in terms of certain solutions to the saddle point equation, in a sense we shall make explicit later on.
We define $s_k(z)$, $k=1,2,3$ to be particular solutions to the saddle point equation, defined as follows. First, we define the Riemann surface of this equation
as $\mathcal{S}$, consisting of three sheets $\mathcal{S}_k$, $k=1,2,3$,
\begin{gather*}
 \mathcal{S}_{1} := \CC \setminus \bigl((-\infty,-z^*] \cup [z^*,\infty)\bigr),\\%\CC \setminus (-\infty,-z^*], \\
 \mathcal{S}_{2} := \CC \setminus [z^*,\infty),\\%\CC \setminus \bigl((-\infty,-z^*] \cup [z^*,\infty)\bigr),\\
 \mathcal{S}_{3} := \CC \setminus (-\infty,-z^*].%\CC \setminus [z^*,\infty).
\end{gather*}
Here \smash{$z^* := \pm\frac{2}{3\tau\sqrt{3}}\sqrt{\frac{q}{-t}} >0$} are the branch points of the saddle point equation.
Let $s_2(z)$ be the~unique solution of the saddle point equation which is real for $z\in \RR \setminus\{\pm z^*\}$. This solution~admits an~analytic continuation to all of $\mathcal{S}_2$. We then define $s_1(z)$, $s_3(z)$ as the analytic continuations of~$s_2(z)$ to sheets $\mathcal{S}_1$ and $\mathcal{S}_3$, respectively.
We have the following lemma, regarding the properties of the $s_k(z)$.
 \begin{Lemma}\quad
 \begin{enumerate}\itemsep=0pt
 \item[$(1)$] \textit{Boundary values of $s_k(z)$.}
 The following relations hold:
 \begin{gather*}
 s_{3,\pm}(z) = s_{1,\mp}(z),\quad z \in (-\infty,-z^{*}],\qquad s_{1,\pm}(z) = s_{2,\mp}(z),\quad z \in [z^{*},\infty).
 \end{gather*}
 \item[$(2)$] \textit{Asymptotics of $s_k(z)$.} As $z\to \infty$,
 \begin{align*}
 &s_1(z)=
 \begin{cases}
 -\omega^2 \left(\dfrac{\tau q}{-t}\right)^{1/3} z^{1/3}\bigl[1 + \OO\bigl(z^{-1/3}\bigr)\bigr], & \operatorname{Im} z >0,\\
 -\omega \left(\dfrac{\tau q}{-t}\right)^{1/3} z^{1/3}\bigl[1 + \OO\bigl(z^{-1/3}\bigr)\bigr], & \operatorname{Im} z <0,
 \end{cases}\\
 &s_2(z)= -(\dfrac{\tau q}{-t}\biggr)^{1/3} z^{1/3}\bigl[1+ \OO\bigl(z^{-1/3}\bigr)\bigr],\\
 &s_3(z)=
 \begin{cases}
 -\omega \left(\dfrac{\tau q}{-t}\right)^{1/3} z^{1/3}\bigl[1 + \OO\bigl(z^{-1/3}\bigr)\bigr], & \operatorname{Im} z >0,\\
 -\omega^2 \left(\dfrac{\tau q}{-t}\right)^{1/3} z^{1/3}\bigl[1 + \OO\bigl(z^{-1/3}\bigr)\bigr], & \operatorname{Im} z <0.
 \end{cases}
 \end{align*}
 \item[$(3)$] \textit{Symmetries of $s_k(z)$.} For each $k=1,2,3$, we have that
$s_k(-z) = -s_k(z)$, $s_k(\bar{z}) = \overline{s_k(z)}$.
 Furthermore, if $\operatorname{Re} z < 0$, then $\operatorname{Re} s_2(z)>0$.
 \end{enumerate}
 \end{Lemma}
These properties are straightforward to verify; the proof is so similar to that of~\cite[Lemmas~2.1 and~2.3]{DKM} that we omit it. For further details, we refer to this work. Consequentially, we have the following corollary.
 \begin{Corollary}
 For $s_k(z)$ defined as above, $k=1,2,3$, put
 \begin{equation*}
 \theta_k(z) := \tau z s_k(z) - V(s_k(z)),\qquad k=1,2,3.
 \end{equation*}
 $($Note that $\theta_k'(z) = \tau s_k(z)$, so that $\theta_k(z)$ are, up to a factor of $\tau$, the primitives of the functions~$s_k(z))$. Then, the functions $\theta_k(z)$ have the large-$z$ asymptotics
 \begin{align}\label{theta-z-asymp}
 &\theta_1(z) =
 \begin{cases}
 -\dfrac{3\omega^2}{4} \dfrac{\tau^{4/3}}{(-tq^{-1})^{1/3}}z^{4/3} - \dfrac{\omega}{2}\dfrac{\tau^{2/3}}{(-tq^{-1})^{2/3}}z^{2/3} \\
 \quad {}+ \dfrac{q}{6t}- \dfrac{\omega^2}{54\tau^{2/3}(-tq^{-1})^{4/3}z^{2/3}} + \OO\left(\dfrac{1}{|z|^{4/3}}\right), \qquad \operatorname{Im} z>0,\\[1.5mm]
 -\dfrac{3\omega}{4} \dfrac{\tau^{4/3}}{(-tq^{-1})^{1/3}}z^{4/3} - \dfrac{\omega^{2}}{2}\dfrac{\tau^{2/3}}{(-tq^{-1})^{2/3}}z^{2/3} \\
 \quad{}+ \dfrac{q}{6t}- \dfrac{\omega}{54\tau^{2/3}(-tq^{-1})^{4/3}z^{2/3}} + \OO\left(\dfrac{1}{|z|^{4/3}}\right), \qquad \operatorname{Im} z<0,
 \end{cases}\\
 &\theta_2(z) = -\frac{3}{4} \frac{\tau^{4/3}}{(-tq^{-1})^{1/3}}z^{4/3} - \frac{1}{2}\frac{\tau^{2/3}}{(-tq^{-1})^{2/3}}z^{2/3} + \frac{q}{6t} - \frac{1}{54\tau^{2/3}(-tq^{-1})^{4/3}z^{2/3}} + \OO\biggl(\!\dfrac{1}{|z|^{4/3}}\!\biggr),\nonumber\\
 &\theta_3(z) =
 \begin{cases}
 -\dfrac{3\omega}{4} \dfrac{\tau^{4/3}}{(-tq^{-1})^{1/3}}z^{4/3} - \dfrac{\omega^{2}}{2}\dfrac{\tau^{2/3}}{(-tq^{-1})^{2/3}}z^{2/3} \\
 \quad{}+ \dfrac{q}{6t}- \dfrac{\omega}{54\tau^{2/3}(-tq^{-1})^{4/3}z^{2/3}} + \OO\left(\dfrac{1}{|z|^{4/3}}\right), \qquad \operatorname{Im} z>0,\\[1.5mm]
 -\dfrac{3\omega^2}{4} \dfrac{\tau^{4/3}}{(-tq^{-1})^{1/3}}z^{4/3} - \dfrac{\omega}{2}\dfrac{\tau^{2/3}}{(-tq^{-1})^{2/3}}z^{2/3} \\
 \quad{} + \dfrac{q}{6t}- \dfrac{\omega^2}{54\tau^{2/3}(-tq^{-1})^{4/3}z^{2/3}} + \OO\left(\dfrac{1}{|z|^{4/3}}\right), \qquad \operatorname{Im} z<0.\nonumber
 \end{cases}
 \end{align}
 \end{Corollary}
With this information about the saddle points $s_k(z)$ in place, we are ready to state the following proposition.

 \begin{Proposition}
 Define the function $S(w)$ by
 \begin{equation*}
 S(w) := \sqrt{\frac{2\pi}{N\bigl(1+3tq^{-1}w^2\bigr)}}{\rm e}^{N(\tau z w -V(z) )}.
 \end{equation*}
 Then, the functions $w_j(z)$, $j=1,2,3,4$, defined by \eqref{Pearcey-int}, have large-$z$ asymptotics given by
 \begin{align*}
 &w_1(z)= -S(s_3(z))\bigl[1 + \OO\bigl(z^{-4/3}\bigr)\bigr],\\
 &w_2(z)=
 \begin{cases}
 S(s_1(z))\bigl[1 + \OO\bigl(z^{-4/3}\bigr)\bigr], & 0 < \arg z < \pi,\\
 S(s_3(z))\bigl[1 + \OO\bigl(z^{-4/3}\bigr)\bigr], & -\frac{\pi}{2} < \arg z < 0,\\
 S(s_2(z))\bigl[1 + \OO\bigl(z^{-4/3}\bigr)\bigr], & \pi < \arg z < -\frac{\pi}{2},
 \end{cases}\\
 &w_3(z)= -S(s_2(z))[1 + \OO\bigl(z^{-4/3}\bigr)],\\
 &w_4(z)=
 \begin{cases}
 S(s_2(z))\bigl[1 + \OO\bigl(z^{-4/3}\bigr)\bigr], & \frac{\pi}{2} < \arg z < \pi,\\
 S(s_3(z))\bigl[1 + \OO\bigl(z^{-4/3}\bigr)\bigr], & 0 < \arg z < \frac{\pi}{2},\\
 S(s_1(z))\bigl[1 + \OO\bigl(z^{-4/3}\bigr)\bigr], & -\pi < \arg z < 0.
 \end{cases}
 \end{align*}
 \end{Proposition}
 \begin{proof}
 Classical steepest descent analysis (cf., for example,~\cite{Miller}) tells us that the asymptotics of the integrals $w_j(z)$ are given by
 \begin{equation*}
 w_j(z) = \pm \sqrt{\frac{2\pi}{NV''(s_*(z))}}{\rm e}^{N(\tau z s_{*}(z) -V(s_*(z)) )}\bigl(1 + \OO\bigl(z^{-2/3}\bigr)\bigr),
 \end{equation*}
 where $*\in\{1,2,3\}$, and $s_{*}(z)$ is the dominant contributing saddle point to the integral. The proposition then follows from determining the dominant saddle in each quadrant; the sign in front of the square root is determined by the orientation of the contour $\gamma_j$ as it passes through
 the relevant saddle point.
 \end{proof}

We now define the row vectors
\begin{equation*}
 \vec{w}_j(z) := \biggl(w_j(z), \frac{w'_j(z)}{N\tau},\frac{w''_j(z)}{(N\tau)^2}\biggr),
\end{equation*}
where $'$ here denotes the derivative with respect to $z$.
We now define a $3\times 3$ matrix $\WW(z)$ as
 \begin{equation*}
 \WW(z) =
 \begin{cases}
 \begin{pmatrix}
 -\vec{w}_2(z) \\
 \hphantom{-}\vec{w}_3(z) \\
 \hphantom{-}\vec{w}_1(z)
 \end{pmatrix},& z \in\Omega_u, \\[2.5mm]
 \begin{pmatrix}
 \vec{w}_3(z) + \vec{w}_4(z) \\
 \vec{w}_3(z) \\
 \vec{w}_1(z)
 \end{pmatrix}, & z \in \Omega_c,\\[2.5mm]
 \begin{pmatrix}
 \vec{w}_4(z) \\
 \vec{w}_3(z) \\
 \vec{w}_1(z)
 \end{pmatrix}, & z \in \Omega_{\ell}.
 \end{cases}
 \end{equation*}

$\WW(z)$ is defined so that it is the unique solution to the following Riemann--Hilbert problem.

 \begin{Proposition} $\WW(z)$ is the unique solution to the following Riemann--Hilbert problem:
 \begin{enumerate}\itemsep=0pt
 \item[$(1)$] $\WW(z)$ is analytic in $\CC \setminus (\Gamma_1\cup \Gamma_2)$, with boundary values
 \begin{equation*}
 \WW_{+}(z) =
 \begin{cases}
 \begin{pmatrix}
 1 & 0 & 1 \\
 0 & 1 & 0 \\
 0 & 0 & 1
 \end{pmatrix}\WW_{-}(z) =: J_1\WW_{-}(z), & z \in \Gamma_1,\\[1.5mm]
 \begin{pmatrix}
 1 & 1 & 0 \\
 0 & 1 & 0 \\
 0 & 0 & 1
 \end{pmatrix}\WW_{-}(z)=: J_2\WW_{-}(z), & z \in \Gamma_2
 \end{cases}
 \end{equation*}
 $($note that these matrices commute, and that their product is $J_1J_2=J)$.
 \item[$(2)$] At infinity, $\WW(z)$ is normalized as
 \begin{equation*}
 \WW(z) = c_N \cdot {\rm e}^{N\Theta(z)} A(z) B(z)\hat{K}\biggl[\mathbb{I}_{3\times 3} + \OO\biggl(\frac{1}{z}\biggr)\biggr], \qquad z\to \infty,
 \end{equation*}
 where the constant $c_N:= {\rm i}\sqrt{\frac{2\pi}{3N}}{\rm e}^{N\frac{q}{6t}}$, the matrix $\Theta(z)$ is defined to be
 \begin{equation} \label{Lambda-Matrix}
 \Theta(z) =
 \begin{cases}
 {\rm diag}(\lambda_3(z), \lambda_1(z),\lambda_2(z)), & \operatorname{Im} z > 0,\\
 {\rm diag}(\lambda_2(z), \lambda_1(z),\lambda_3(z)), & \operatorname{Im} z <0,
 \end{cases} %{\rm diag}(\theta_{1}(z),\theta_{2}(z),\theta_{3}(z))
 \end{equation}
where the functions $\lambda_k(z)$ are defined by the \textit{exact} formulas
 \begin{equation*}%\label{little-theta-hat}
 \lambda_k(z) = -\frac{3\omega^{k-1}}{4} \frac{\tau^{4/3}}{(-tq^{-1})^{1/3}}z^{4/3} - \frac{\omega^{1-k}}{2}\frac{\tau^{2/3}}{(-tq^{-1})^{2/3}}z^{2/3}.
 \end{equation*}
and finally, the matrices $A(z)$, $B(z)$, and $\hat{K}$ are given by
 \begin{gather*}
 A(z) =
 \begin{cases}
 \begin{pmatrix}
 -\omega & 1 & \omega^2\\
 -1 & 1 & 1\\
 -\omega^2 & 1 & \omega
 \end{pmatrix}, & \operatorname{Im} z > 0,\\[2.5mm]
 \begin{pmatrix}
 \omega^2 & -1 & -\omega\\
 -1 & \hphantom{-}1 & \hphantom{-}1\\
 -\omega & \hphantom{-}1 & \hphantom{-}\omega^2\\
 \end{pmatrix}, & \operatorname{Im} z < 0,
 \end{cases}
 \qquad
 B(z) =
 \begin{pmatrix}
 z^{-1/3} & 0 & 0\\
 0 & 1 & 0\\
 0 & 0 & z^{1/3}
 \end{pmatrix},
 \\
 \hat{K} =
 \begin{pmatrix}
 \bigl(-tq^{-1}\bigr)^{-1/6}\tau^{-1/3} & 0 & -\frac{n+27 tq^{-1}}{54(-tq^{-1})^{13/6}\tau^{1/3}}\\[1.4mm]
 0 & \bigl(-tq^{-1}\bigr)^{-1/2} & 0\\[1.1mm]
 0 & 0 & -\bigl(-tq^{-1}\bigr)^{-5/6}\tau^{1/3}
 \end{pmatrix}.
 \end{gather*}
 \end{enumerate}
 \end{Proposition}
\begin{proof}
 The proof of this proposition is just a tedious check that the formulas we have previously derived for the asymptotics of the $w_j(z)$'s indeed guarantee that the $\WW(z)$ solves the above Riemann--Hilbert problem. Uniqueness follows from standard arguments with Morera's and Liouville's theorem. We remark that the terms decaying in $z$ as $z\to \infty$ in the functions $\theta_j(z)$ can be absorbed into the asymptotic expansion $\mathbb{I} + \OO\bigl(z^{-1}\bigr)$, by passing this part of the expansion to the right of the matrices $A(z) B(z)$. More precisely, the expansion of the Laurent series is\footnote{The regular series expansion here either by (i) calculating subleading asymptotics of the saddle point expansion to high enough order, or (ii) using the fact that the asymptotic expansion itself should satisfy the differential equation(s) \eqref{W-differentials}, and determining the coefficients
 of the subleading expansion recursively.}
 \begin{gather*}
 \mathbb{I}_{3\times 3} +
 \begin{pmatrix}
 0 & \frac{-N^3 - 72N^2t/q - 891 N t^2/q^2 - 810 t^3/q^3}{5832 N t^3/q^3\tau} & 0\\
			\frac{-N+9t/q}{54\tau t/q} & 0 & \frac{N^3 + 36 N^2 t/q - 81 N t^2/q^2 - 810 t^3/q^3}{5832 N t^3/q^3\tau}\\
			0 & \frac{-N-9t/q}{54\tau t/q} & 0
 \end{pmatrix}z^{-1}\\
 \qquad{} + \OO(z^{-2}).
\tag*{\qed}
\end{gather*}
\renewcommand{\qed}{}
\end{proof}

\begin{Remark}
 We have introduced functions $\lambda_k(z)$ in the above proposition. These functions have been chosen so that
 \begin{equation*}
 {\rm diag}(\theta_1(z),\theta_2(z),\theta_3(z)) = \Theta(z) + \OO(1), \qquad z\to \infty.
 \end{equation*}
 The reason for the introduction of these functions is to match the asymptotic theory of ordinary differential equations with rational coefficients; such an
 equation will have a large $z$ expansion consistent with the above.
\end{Remark}
We furthermore have the following proposition, which will be useful later in the computation of the $\boldsymbol{\tau}$-function.
\begin{Proposition}
 $\WW(z)$ satisfies the following differential equations:
 \begin{gather}\label{W-differentials}
 \frac{\partial \WW}{\partial z} = \WW\cdot\, \mathcal{M}^{z}(z),\qquad \frac{\partial \WW}{\partial t} = \WW\cdot\, \mathcal{M}^{t}(z), \nonumber\\
 \frac{\partial \WW}{\partial \tau} = \WW\cdot\, \mathcal{M}^{\tau}(z), \qquad \frac{\partial \WW}{\partial H} = \WW\cdot\, \mathcal{M}^{H}(z),
 \end{gather}
 where the matrices $\mathcal{M}^{z}(z)$, $\mathcal{M}^{t}(z)$, $\mathcal{M}^{\tau}(z)$, and $\mathcal{M}^{H}(z)$ are defined by
 \begin{align*}
 &\mathcal{M}^{z}(z)=
 N\tau
 \begin{pmatrix}
 0 & 0 & \tau q z/t\\
 1 & 0 & -q/t\\
 0 & 1 & 0
 \end{pmatrix},\\
 &\mathcal{M}^{t}(z)=
 \frac{N}{4t}
 \begin{pmatrix}
 -1/N & \tau q z/t & -\tau^2 q z^2/t + \frac{q}{Nt}\\
 -\tau z & -(q/t+2/N) & 2\tau q z/t\\
 1 & -\tau z & -(q/t + 3/N)
 \end{pmatrix},\\
 &\mathcal{M}^{\tau}(z)=
 Nz\begin{pmatrix}
 0 & 0 & \tau q z/t\\
 1 & 0 & -q/t\\
 0 & 1 & 0
 \end{pmatrix},\\
 &\mathcal{M}^{H}(z)=
 -\frac{N}{4}
 \begin{pmatrix}
 -1/N & \tau qz/t & -\tau^2 q z^2/t + \frac{q}{Nt}\\
 -\tau z & -(q/t+2/N) & 2\tau q z/t\\
 1 & -\tau z & -(q/t + 3/N)
 \end{pmatrix}.
 \end{align*}
\end{Proposition}
\begin{proof}
 This can be inferred immediately from the differential equation \eqref{Pearcey-eq} and the relations
 \begin{equation*}
 \frac{\partial w}{\partial t} = -\frac{{\rm e}^{-H}}{4 N^3\tau^4} \frac{\partial^4 w}{\partial z^4},\qquad \tau\frac{\partial w}{\partial \tau} = z\frac{\partial w}{\partial z},\qquad \frac{\partial w}{\partial H} = -t \frac{\partial w}{\partial t}.
 \end{equation*}
As a consistency check, one may verify that the compatibility conditions between these equations (i.e., the zero-curvature equations) hold trivially.
\end{proof}

\begin{Remark} \label{AB-matrix-remark}
 It is useful to notice that the function $A(z) B(z)$ is a solution to the following Riemann--Hilbert problem:
 $A(z) B(z)$ is analytic in $\CC\setminus\RR$, and satisfies the jump condition
 \begin{equation*}
 A_+(z) B_+(z) =
 \begin{cases}
 \begin{pmatrix}
 \hphantom{-} 0 & 0 & 1\\
 \hphantom{-} 0 & 1 & 0\\
 -1 & 0 & 0\\
 \end{pmatrix}A_-(z) B_-(z), & z > 0, \\[1.5mm]
 \begin{pmatrix}
 \hphantom{-} 0 & 1 & 0\\
 -1 & 0 & 0\\
 \hphantom{-} 0 & 0 & 1\\
 \end{pmatrix}A_-(z) B_-(z), & z < 0.
 \end{cases}
 \end{equation*}
 This is consistent with the fact that, for $z \in \RR$ sufficiently large, $\Theta(z)$ has jumps
 \begin{equation*}
 \Theta_{+}(z) =
 \begin{cases}
 \begin{pmatrix}
 \hphantom{-} 0 & 0 & 1\\
 \hphantom{-} 0 & 1 & 0\\
 -1 & 0 & 0\\
 \end{pmatrix}\Theta_{-}(z), & z > R > 0, \\[1.5mm]
 \begin{pmatrix}
 \hphantom{-} 0 & 1 & 0\\
 -1 & 0 & 0\\
 \hphantom{-} 0 & 0 & 1\\
 \end{pmatrix}\Theta_{-}(z), & z < - R < 0.
 \end{cases}
 \end{equation*}
\end{Remark}

\subsection[The transformation ${\YY} \mapsto {\XX}$]{The transformation $\boldsymbol{{\YY} \mapsto {\XX}}$}
We now define the transformation $\YY \mapsto \XX$. We set
 \begin{equation*}
 \XX(z) :=
 \begin{pmatrix}
 1 & 0 \\
 0 & c_N \hat{K}
 \end{pmatrix}
 \YY(z)
 \begin{pmatrix}
 {\rm e}^{-NV(z)} & 0 \\
 0 & \WW^{-1}(z)
 \end{pmatrix}.
 \end{equation*}
By construction, $\XX(z)$ is piecewise analytic on $\CC\setminus (\Gamma_1 \cup \Gamma_2 \cup \Gamma)$. $\XX(z)$ satisfies the following~RHP.

 \begin{Proposition}
 $\XX(z)$ is the unique solution to the following Riemann--Hilbert Problem:
 \begin{equation*}
 \XX_{+}(z) = \XX_{-}(z) \times
 \begin{cases}
 \begin{pmatrix}
 1 & 0\\
 0 & J_1^{-1}
 \end{pmatrix}, & z \in \Gamma_1,\vspace{1mm}\\
 \begin{pmatrix}
 1 & 0\\
 0 & J_2^{-1}
 \end{pmatrix}, & z \in \Gamma_2,\vspace{1mm}\\
 \begin{pmatrix}
 1 & 1 \ 0 \ 0\\
 \vec{0}_{1\times 3} & \mathbb{I}_{3\times3}
 \end{pmatrix}, & z \in \Gamma,\\
 \end{cases}
 \end{equation*}

 Subject to the normalization condition
 \begin{align}\label{X-asymptotics}
 \XX(z) =
 \biggl[\mathbb{I} + \OO\biggl(\frac{1}{z}\biggr)\biggr]\begin{pmatrix}
 1 & 0\\
 0 & B^{-1}(z) A^{-1}(z)
 \end{pmatrix}
 \begin{pmatrix}
 z^n & 0 & 0 & 0\\
 0 & z^{-n/3} & 0 & 0\\
 0 & 0 & z^{-n/3} & 0\\
 0 & 0 & 0 & z^{-n/3}
 \end{pmatrix}{\rm e}^{-N\Lambda(z)},
 \end{align}
where $\Lambda(z)$ is defined as the diagonal matrix
 \begin{equation*}
 \Lambda(z) =
 \begin{pmatrix}
 V(z) & 0\\
 0 & \Theta(z)
 \end{pmatrix},
 \end{equation*}
where $\Theta(z)$ is the diagonal $3\times 3$ matrix defined by \eqref{Lambda-Matrix}.
 \end{Proposition}

\begin{proof}
 The asymptotic condition \eqref{X-asymptotics} follows almost immediately from the definition of $\XX(z)$; along with the definition of the matrix $\WW(z)$. Indeed, this is obvious in the regions $\Omega_u$ and $\Omega_\ell$, and follows from definition of $\WW(z)$. The only detail to check
 is that the asymptotics of $\WW(z)$ in the region $\Omega_c$ are correct. We see that $\WW(z)$ in this region is obtained by adding the recessive solution $-\vec{w}_1(z)$
 to the first row; since this solution is recessive in the region $\Omega_c$, the asymptotics of $-\vec{w}_2(z) - \vec{w}_1(z)$ $(= \vec{w}_3(z)+ \vec{w}_4(z))$ are the same as the
 asymptotics of $-\vec{w}_2(z)$ there. Thus, the asymptotics of $\XX(z)$ from the
 proposition hold.

 Now, we address the jump conditions of $\XX(z)$. Since the matrix function
 \begin{equation*}
 \begin{pmatrix}
 {\rm e}^{-NV(z)} & 0 \\
 0 & \WW^{-1}(z)
 \end{pmatrix}
 \end{equation*}
 has jumps only on $\Gamma_1$, $\Gamma_2$ (arising from the jumps of $\WW(z)$), and these contours do not intersect~$\Gamma$, the first two jump conditions are clearly satisfied. It remains to check the jump of $\XX(z)$ across~$\Gamma$. The jump of $\YY(z)$ on $\Gamma$ is
 \begin{equation*}
 J_{\YY} = \mathbb{I} +
 {\rm e}^{-NV(z)}\begin{pmatrix}
 0 & f(z) & \frac{f'(z)}{N\tau} & \frac{f''(z)}{(N\tau)^2} \\
 0 & 0 & 0 & 0 \\
 0 & 0 & 0 & 0 \\
 0 & 0 & 0 & 0
 \end{pmatrix}
 =:
 \mathbb{I} + {\rm e}^{-NV(z)} \begin{pmatrix}
 0 & & \vec{f}(z) & \\
 0 & 0 & 0 & 0 \\
 0 & 0 & 0 & 0 \\
 0 & 0 & 0 & 0
 \end{pmatrix},
 \end{equation*}
 where $\vec{f}(z)$ is the row vector
 \begin{equation*}
 \vec{f}(z) := \left(
 f(z), \frac{f'(z)}{N\tau}, \frac{f''(z)}{(N\tau)^2} \right).
 \end{equation*}
 Now, since $\Gamma$ is homologically equivalent to $-\gamma_2 - \gamma_1$, we have that $\vec{L}(z) = -\vec{w}_2(z) - \vec{w}_1(z)$. Thus, the jump of $\XX(z)$ across $\Gamma$ is
 \begin{gather*}
 \begin{pmatrix}
 {\rm e}^{NV(z)} & 0 \\
 0 & \WW(z)
 \end{pmatrix}
 \left[
 \mathbb{I} + {\rm e}^{-NV(z)}
 \begin{pmatrix}
 0 & & \vec{f}(z) & \\
 0 & 0 & 0 & 0 \\
 0 & 0 & 0 & 0 \\
 0 & 0 & 0 & 0
 \end{pmatrix}\right]
 \begin{pmatrix}
 {\rm e}^{-NV(z)} & 0 \\
 0 & \WW^{-1}(z)
 \end{pmatrix}\\
 \qquad{}= \mathbb{I} +
 \begin{pmatrix}
 0 & & \vec{f}(z)\WW^{-1}(z) & \\
 0 & 0 & 0 & 0 \\
 0 & 0 & 0 & 0 \\
 0 & 0 & 0 & 0
 \end{pmatrix} = \mathbb{I} +
 \begin{pmatrix}
 0 & & (-\vec{w}_2(z) - \vec{w}_1(z))\WW^{-1}(z) & \\
 0 & 0 & 0 & 0 \\
 0 & 0 & 0 & 0 \\
 0 & 0 & 0 & 0
 \end{pmatrix} \\
 \qquad{}=
 \begin{pmatrix}
 1 & 1 & 0 & 0\\
 0 & 1 & 0 & 0 \\
 0 & 0 & 1 & 0 \\
 0 & 0 & 0 & 1
 \end{pmatrix}.
 \end{gather*}
 Here we have used the relations
 \begin{alignat*}{3}
 &\vec{w}_1(z)\WW^{-1}(z) = (0,0,1), \qquad&& \vec{w}_2(z)\WW^{-1}(z) = (-1,0,-1),&\\
 &\vec{w}_3(z)\WW^{-1}(z) = (0,1,0), \qquad&& \vec{w}_4(z)\WW^{-1}(z) = (1,-1,0),&
 \end{alignat*}
 resulting from the fact that, in a neighborhood of $\Gamma$, the matrix $\WW(z)$ admits the expression
 \begin{equation*}
 \WW(z) =
 \begin{pmatrix}
 \vec{w}_3(z) + \vec{w}_4(z) \\
 \vec{w}_3(z) \\
 \vec{w}_1(z)
 \end{pmatrix},
 \end{equation*}
 along with the identity $\WW(z) \WW^{-1}(z) = \mathbb{I}_{3\times 3}$.
\end{proof}

\begin{Remark}
 Note that, if we had equivalently chosen $\gamma_3 + \gamma_4$ as the homological representative for $\Gamma$, the same resulting jump matrix is obtained.
\end{Remark}

\subsection[The transformation $\XX \mapsto \UU$]{The transformation $\boldsymbol{\XX \mapsto \UU}$}

Define the matrix
 \begin{equation*}%\label{G-matrix}
 \GG(z) :=
 \begin{pmatrix}
 \exp[n\tau\Omega_1(z)] & 0 & 0 & 0\\
 0 & \exp[n\tau\Omega_2(z)] & 0 & 0\\
 0 & 0 & \exp[n\tau\Omega_3(z)] & 0\\
 0 & 0 & 0 & \exp[n\tau\Omega_4(z)]
 \end{pmatrix}.
 \end{equation*}
We now are ready to perform the transformation $\XX \mapsto \UU$. Set
 \begin{equation*}
 \UU(z) := [\mathbb{I} -nC_1\cdot E_{24}]{\rm e}^{-n L}\XX(z) \GG(z),
 \end{equation*}
where $\GG(z)$ is defined as above, $C_1$ is the constant appearing in the $z^{-2/3}$ term of the asymptotics of the $\Omega_j(z)$'s (cf.\ equation \eqref{C1-constant}), and $L$ is the diagonal constant (in $z$) matrix
 \begin{equation*}
 L := {\rm diag} (\ell_0, \ell_1, \ell_1, \ell_1).
 \end{equation*}
Clearly, $\UU(z)$ is analytic in $\CC \setminus (\Gamma \cup \Gamma_1 \cup \Gamma_2 )$; the goal of this subsection is to show that $\UU(z)$ is the unique solution
to its own Riemann--Hilbert problem.

 \begin{Proposition}
 The function $\UU(z)$ is the unique solution to the following Riemann--Hilbert problem:
 \begin{align*}%\label{U-jumps}
 \UU_{+}(z)={}& \UU_{-}(z) \nonumber\\
 &{\times} \begin{cases}
 \mathbb{I} - E_{24}{\rm e}^{-n\tau[\Omega_2(z) - \Omega_4(z)]},
 & z \in \Gamma_1 \cap \{\operatorname{Im} z > 0\}, \vspace{1mm}\\
 \begin{pmatrix}
 1 & 0 & 0 & \hphantom{-}0 \\
 0 & {\rm e}^{-n\tau[\Omega_{2,-}(z)-\Omega_{4,-}(z)]} & 0 & -1\\
 0 & 0 & 1 & \hphantom{-}0 \\
 0 & 0 & 0 & {\rm e}^{n\tau[\Omega_{2,-}(z)-\Omega_{4,-}(z)]}
 \end{pmatrix},
 & z \in \Gamma_1 \cap \{\operatorname{Im} z = 0\},\vspace{1mm}\\
 \mathbb{I} - E_{23}{\rm e}^{-n\tau[\Omega_2(z) - \Omega_3(z)]},
 & z \in \Gamma_2 \cap \{\operatorname{Im} z < 0\}, \vspace{2mm}\\
 \begin{pmatrix}
 1 & 0 & \hphantom{-}0 & 0 \\
 0 & {\rm e}^{-n\tau[\Omega_{2,-}(z) - \Omega_{3,-}(z)]} & -1 & 0\\
 0 & 0 & {\rm e}^{n\tau[\Omega_{2,-}(z) - \Omega_{3,-}(z)]} & 0 \\
 0 & 0 & \hphantom{-}0 & 1
 \end{pmatrix},
 & z \in \Gamma_2 \cap \{\operatorname{Im} z = 0\}, \vspace{1mm}\\
 \mathbb{I} + E_{12}{\rm e}^{-n\tau[\Omega_1(z) - \Omega_2(z)]},
 & z\in \Gamma \setminus \{\operatorname{Im} z = 0\}, \vspace{1mm}\\
 \begin{pmatrix}
 {\rm e}^{-n\tau[\Omega_{1,-}(z)-\Omega_{2,-}(z)]} & 1 & 0 & 0\\
 0 & {\rm e}^{n\tau[\Omega_{1,-}(z)-\Omega_{2,-}(z)]} & 0 & 0\\
 0 & 0 & 1 & 0 \\
 0 & 0 & 0 & 1
 \end{pmatrix},
 & z\in \Gamma \cap \{\operatorname{Im} z = 0\}.
 \end{cases}
 \end{align*}
 The asymptotics of $\UU(z)$ are given by
 \begin{equation*} %\label{U-asymptotics}
 \UU(z) =
 \biggl[\mathbb{I} + \OO\biggl(\frac{1}{z^{1/3}}\biggr)\biggr]\begin{pmatrix}
 1 & 0\\
 0 & B^{-1}(z) A^{-1}(z)
 \end{pmatrix}, \qquad z\to \infty.
 \end{equation*}
\end{Proposition}
\begin{proof}
The jump conditions satisfied by $\UU(z)$ are readily verified from the definitions of $\XX(z)$,~$\GG(z)$. Furthermore, the ``exponential asymptotics'' of $\XX(z)$
are removed by multiplication by $\GG(z)$;
comparison of formulas \eqref{Omega-Sheet-1-infty}--\eqref{Omega-Sheet-4-infty} with \eqref{X-asymptotics}, along with the explicit expressions for the asymptotics of the functions $\lambda_k(z)$ (see equation \eqref{theta-z-asymp}) shows that this is indeed the case. Indeed, we have that, as $z\to \infty$,
 \begin{align*}
 \UU(z)={}& [\mathbb{I} -nC_1 \cdot E_{24}] {\rm e}^{-n L}\bigl[\mathbb{I} + \OO\bigl(z^{-1}\bigr)\bigr]\begin{pmatrix}
 z^{n} & 0\\
 0 & z^{-n/3}\mathbb{I}_{3\times 3}
 \end{pmatrix}\\
 &\times
 \begin{pmatrix}
 1 & 0\\
 0 & B^{-1}(z) A^{-1}(z)
 \end{pmatrix}{\rm e}^{-n\hat{\Theta}(z)} \GG(z)\\
={}&[\mathbb{I} -n C_1 \cdot E_{24}]\bigl[\mathbb{I} + \OO\bigl(z^{-1}\bigr)\bigr]
 \begin{pmatrix}
 1 & 0\\
 0 & B^{-1}(z) A^{-1}(z)
 \end{pmatrix}\\
 &\times
 \begin{pmatrix}
 1 + \OO(z^{-2}) & 0 \\
 0 & \mathbb{I}_{3\times 3} + n\hat{C}z^{-2/3} + \OO(z^{-4/3})
 \end{pmatrix}.
 \end{align*}
Here $\hat{C}$ is the piecewise constant diagonal matrix
 \begin{equation*}
 \hat{C} =
 \begin{cases}
 {\rm diag}\bigl(\omega^2 C_1, C_1,\omega C_1\bigr), & \operatorname{Im} z > 0,\\
 {\rm diag}(\omega C_1, C_1,\omega^2 C_1), & \operatorname{Im} z < 0.
 \end{cases}
 \end{equation*}
If we interchange the order of the last two matrices, we obtain that, as $z\to \infty$,
 \begin{align*}
 \UU(z) &= [\mathbb{I} -n C_1\cdot E_{24}] \bigl[\mathbb{I} + \OO\bigl(z^{-1}\bigr)\bigr] \bigl[\mathbb{I} + n C_1 \cdot E_{24} + \OO\bigl(z^{-1/3}\bigr)\bigr]
 \begin{pmatrix}
 1 & 0\\
 0 & B^{-1}(z) A^{-1}(z)
 \end{pmatrix}\\
 &= \bigl[\mathbb{I} + \OO\bigl(z^{-1/3}\bigr) \bigr]
 \begin{pmatrix}
 1 & 0\\
 0 & B^{-1}(z) A^{-1}(z)
 \end{pmatrix}.
 \end{align*}

 We will analyze the jumps of this new matrix after opening lenses; this is performed in the next section.
\end{proof}

\section[The third and fourth transformations ${\bf U} \mapsto {\bf T} \mapsto {\bf S}$]{The third and fourth transformations $\boldsymbol{{\bf U} \mapsto {\bf T} \mapsto {\bf S}}$}\label{section5}
Here we perform the lensing transformations. The first lensing transformation will open lenses around the unbounded branch cuts
$(-\infty,-\beta] \cup [\beta,\infty)$; this will constitute the transformation~${\UU \mapsto \TT}$. The lens opening around the central cut
$[-\alpha,\alpha]$ is performed next; this will constitute the transformation $\TT\mapsto\boldS$. We remark here that the choice of $(a,b)$
(i.e., whether the spectral curve is critical, generic, or multicritical) is irrelevant here, as all of the lensing propositions of Section~\ref{ssec:SecondTransformation} guarantee
the same inequalities hold around the branch points. These cases will become distinguished when we later try to find a parametrix.

\subsection[The transformation ${\bf U} \mapsto {\bf T}$]{The transformation $\boldsymbol{{\bf U} \mapsto {\bf T}}$}
The opening of lenses here is based on the factorization of the jump matrix
 \begin{equation*}
 \begin{pmatrix}
 {\rm e}^{-ng_+(z)} & -1 \\
 0 & {\rm e}^{-n g_-(z)}
 \end{pmatrix}
 =
 \begin{pmatrix}
 1 & 0 \\
 -{\rm e}^{-n g_-(z)} & 1
 \end{pmatrix}
 \begin{pmatrix}
 0 & -1 \\
 1 & \hphantom{-} 0
 \end{pmatrix}
 \begin{pmatrix}
 1 & 0 \\
 -{\rm e}^{-n g_+(z)} & 1
 \end{pmatrix},
 \end{equation*}
where $g_+(z)$, $g_-(z)$ are the boundary values of one of the functions $\Omega_3(z) -\Omega_2(z)$, $\Omega_4(z) - \Omega_2(z)$ from above/below the contour.

\begin{figure}[t]\centering
\includegraphics[scale=1.1]{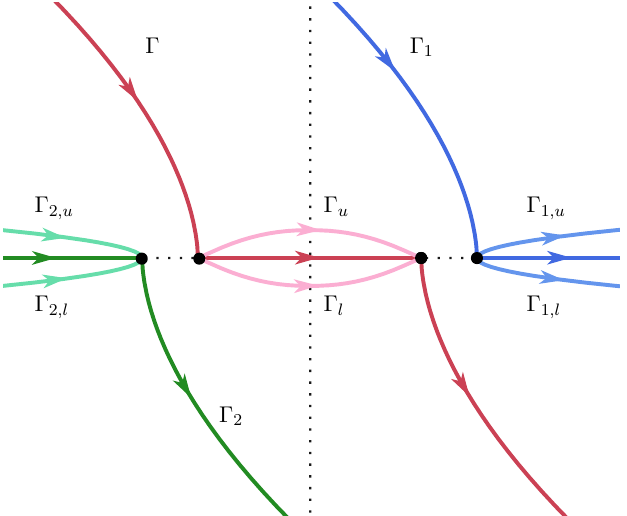}
%		 \begin{overpic}[scale=.5]{Pictures/Lensing1.pdf}
%		 \put (23,75) {$\Gamma$}
%		 \put (52,49) {$\Gamma_u$}
%		 \put (52,33) {$\Gamma_l$}
%		 \put (35,15) {$\Gamma_2$}
%		 \put (5,49) {$\Gamma_{2,u}$}
%		 \put (5,33) {$\Gamma_{2,l}$}
%		 \put (85,49) {$\Gamma_{1,u}$}
%		 \put (85,33) {$\Gamma_{1,l}$}
%		 \put (66,75) {$\Gamma_1$}
%		 \end{overpic}
 \caption{The opened lenses.} \label{fig:Lenses-1}
\end{figure}

By the lensing propositions of the previous section, there exist lens-shaped regions such as those depicted in Figure~\ref{fig:Lenses-1} around $(-\infty,-\beta]$ (respectively, $[\beta,\infty)$) such that the differences~${\operatorname{Re}[\Omega_3-\Omega_2]}$ (respectively, $\operatorname{Re}[\Omega_4-\Omega_2]$) are positive in this region. Define $\Gamma_{1,u}$, $\Gamma_{1,l}$ as the boundaries of the lensing region around $[\beta,\infty)$ in the upper and lower half planes, and similarly define $\Gamma_{2,u}$, $\Gamma_{2,l}$ as the boundaries of the lensing region around $(-\infty,-\beta]$.
The sectors enclosed by these contours are labelled as follows:
 \begin{itemize}\itemsep=0pt
 \item $\Sigma_{1,u}$ is the region enclosed by $\Gamma_{1,u}$ and $[\beta,\infty)$,
 \item $\Sigma_{1,l}$ is the region enclosed by $\Gamma_{1,l}$ and $[\beta,\infty)$,
 \item $\Sigma_{2,u}$ is the region enclosed by $\Gamma_{2,u}$ and $(-\infty,-\beta]$,
 \item $\Sigma_{2,l}$ is the region enclosed by $\Gamma_{2,l}$ and $(-\infty,-\beta]$.
 \end{itemize}
These contours are depicted in Figure~\ref{fig:Lenses-1}. Define matrices
 \begin{gather*}
 V_1(z) =
 \begin{pmatrix}
 1 & 0 & 0 & 0\\
 0 & 1 & 0 & 0\\
 0 & 0 & 1 & 0\\
 0 & -{\rm e}^{-n\tau[\Omega_4(z) - \Omega_2(z)]} & 0 & 1
 \end{pmatrix}, \\
 V_2(z) =
 \begin{pmatrix}
 1 & 0 & 0 & 0\\
 0 & 1 & 0 & 0\\
 0 & -{\rm e}^{-n\tau[\Omega_3(z) - \Omega_2(z)]} & 1 & 0\\
 0 & 0 & 0 & 1
 \end{pmatrix}.
 \end{gather*}
We define the transformation ${\bf U} \mapsto {\bf T}$ by setting
 \begin{equation*}
 \TT(z) =
 \begin{cases}
 \UU(z) V_1^{-1}(z), & z \in \Sigma_{1,u},\\
 \UU(z) V_1(z), & z \in \Sigma_{1,l},\\
 \UU(z) V_2^{-1}(z), & z \in \Sigma_{2,u},\\
 \UU(z) V_2(z), & z \in \Sigma_{2,l},\\
 \UU(z), & \text{elsewhere}.
 \end{cases}
 \end{equation*}
Clearly, $\TT(z)$ is a piecewise analytic function off of the contours $\Gamma_1$, $\Gamma_2$, $\Gamma$, $\Gamma_{1,u}$, $\Gamma_{1,l}$, $\Gamma_{2,u}$,
and~$\Gamma_{2,l}$. In fact, $\TT(z)$ is the unique solution to the following Riemann--Hilbert problem.

 \begin{Proposition}
 The function $\TT(z)$ is the unique solution to the following RHP:
 \begin{equation*}%\label{T-jumps}
 \TT_{+}(z) = \TT_{-}(z)
 \begin{cases}
 \mathbb{I} - E_{42}{\rm e}^{-n\tau[\Omega_4(z) - \Omega_2(z)]}, & z\in \Gamma_{1,u} \cup \Gamma_{1,l},\\
 \mathbb{I} - E_{32}{\rm e}^{-n\tau[\Omega_3(z) - \Omega_2(z)]}, & z\in \Gamma_{2,u} \cup \Gamma_{2,l},\\
 \begin{pmatrix}
 1 & 0 & 0 & \hphantom{-}0\\
 0 & 0 & 0 & -1\\
 0 & 0 & 1 & \hphantom{-}0\\
 0 & 1 & 0 & \hphantom{-}0
 \end{pmatrix}, & z\in [\beta,\infty),\vspace{1mm}\\
 \begin{pmatrix}
 1 & 0 & \hphantom{-}0 & 0\\
 0 & 0 & -1 & 0\\
 0 & 1 & \hphantom{-}0 & 0\\
 0 & 0 & \hphantom{-}0 & 1
 \end{pmatrix}, & z\in (-\infty,-\beta],\\
 \mathbb{I} - E_{24}{\rm e}^{-n\tau[\Omega_2(z) - \Omega_4(z)]},
 & z \in \Gamma_1 \cap \{\operatorname{Im} z > 0\},\\
 \mathbb{I} - E_{23}{\rm e}^{-n\tau[\Omega_2(z) - \Omega_3(z)]},
 & z \in \Gamma_2 \cap \{\operatorname{Im} z < 0\},\\
 \mathbb{I} + E_{12}{\rm e}^{-n\tau[\Omega_1(z) - \Omega_2(z)]},
 & z\in \Gamma \setminus [-\alpha,\alpha],\\
 \vspace{2mm}
 \begin{pmatrix}
 {\rm e}^{-n\tau[\Omega_{1,-}(z)-\Omega_{2,-}(z)]} & 1 & 0 & 0 \\
 0 & {\rm e}^{n\tau[\Omega_{1,-}(z)-\Omega_{2,-}(z)]} & 0 & 0\\
 0 & 0 & 1 & 0 \\
 0 & 0 & 0 & 1
 \end{pmatrix},
 & z\in [-\alpha,\alpha].
 \end{cases}
 \end{equation*}
 Furthermore, $\TT(z)$ has asymptotics
\begin{equation*}%\label{T-asymptotics}
 \TT(z) =
 \bigl[\mathbb{I} + \OO \bigl(z^{-1/3}\bigr)\bigr]
 \begin{pmatrix}
 1 & 0\\
 0 & B^{-1}(z) A^{-1}(z)
 \end{pmatrix}, \qquad z\to \infty.
\end{equation*}
 \end{Proposition}
\begin{proof}
 The proof of this proposition follows immediately from the definition of $\TT(z)$.
\end{proof}

\subsection[The transformation ${\bf T} \mapsto {\bf S}$]{The transformation $\boldsymbol{{\bf T} \mapsto {\bf S}}$}
We now open the lens around the segment $[-\alpha,\alpha]$. This is based on the following factorization of the jump matrix:
\begin{align*}
 &\begin{pmatrix}
 {\rm e}^{-n\tau[\Omega_{2,+}(z)-\Omega_{1,+}(z)]} & 1 & 0 & 0 \\
 0 & {\rm e}^{-n\tau[\Omega_{2,-}(z)-\Omega_{1,-}(z)]} & 0 & 0\\
 0 & 0 & 1 & 0 \\
 0 & 0 & 0 & 1
 \end{pmatrix}\\
 &\qquad{}=\begin{pmatrix}
 1 & 0 & 0 & 0\\
 {\rm e}^{-n\tau[\Omega_{2,-}(z)-\Omega_{1,-}(z)]} & 1 & 0 & 0\\
 0 & 0 & 1 & 0 \\
 0 & 0 & 0 & 1
 \end{pmatrix}\!
 \begin{pmatrix}
 \hphantom{-} 0 & 1 & 0 & 0\\
 -1 & 0 & 0 & 0\\
 \hphantom{-} 0 & 0 & 1 & 0 \\
 \hphantom{-} 0 & 0 & 0 & 1
 \end{pmatrix}\\
 & \qquad\quad\times
 \begin{pmatrix}
 1 & 0 & 0 & 0\\
 {\rm e}^{-n\tau[\Omega_{2,+}(z)-\Omega_{1,+}(z)]} & 1 & 0 & 0\\
 0 & 0 & 1 & 0 \\
 0 & 0 & 0 & 1
 \end{pmatrix}.
\end{align*}
By the lensing propositions of Section~\ref{section3}, there exists a lens-shaped region around $[-\alpha,\alpha]$ such that \[
\operatorname{Re}[\Omega_2 -\Omega_1](z) >0.
\]
 Define contours $\Gamma_u$, $\Gamma_l$ as the boundaries of this lens-shaped region in the upper and lower half planes, respectively. Further, put $\Sigma_u$, $\Sigma_l$ to be the regions enclosed by $[-\alpha,\alpha]$ and $\Gamma_u$, $\Gamma_l$, respectively.
Define the invertible matrix $V_0(z)$ in the lensed region $\Sigma_u \cup \Sigma_l$~by\looseness=-1
 \begin{equation*}
 V_0(z) := \begin{pmatrix}
 1 & 0 & 0 & 0\\
 {\rm e}^{-n\tau[\Omega_{2}(z)-\Omega_{1}(z)]} & 1 & 0 & 0\\
 0 & 0 & 1 & 0 \\
 0 & 0 & 0 & 1
 \end{pmatrix}.
 \end{equation*}

We define the piecewise analytic function $\boldS(z)$ by
 \begin{equation*}
 \boldS(z) :=
 \begin{cases}
 \TT(z)V_0^{-1}(z), & z\in\Sigma_{l},\\
 \TT(z)V_0(z), & z\in\Sigma_{u},\\
 \TT(z), & \text{otherwise}.
 \end{cases}
 \end{equation*}
In this case, $\boldS(z)$ is the unique solution to the following RHP.

 \begin{Proposition} %\label{lensing-proposition}
 The function $\boldS(z)$ is the unique solution to the following Riemann--Hilbert problem:
 $\boldS(z)$ is piecewise analytic off the contours
 \[
 \Gamma_1 , \ \Gamma_2, \ \Gamma, \ \Gamma_{1,u}, \ \Gamma_{1,l}, \ \Gamma_{2,u}, \ \Gamma_{2,l}, \ \Gamma_l, \ \Gamma_u,
 \]
 satisfying the jump condition
 \begin{equation*} %\label{S-jumps}
 \boldS_{+}(z) = \boldS_{-}(z)
 \begin{cases}
 \mathbb{I} - E_{42}{\rm e}^{-n\tau[\Omega_4(z) - \Omega_2(z)]}, & z\in \Gamma_{1,u} \cup \Gamma_{1,l},\\
 \mathbb{I} - E_{32}{\rm e}^{-n\tau[\Omega_3(z) - \Omega_2(z)]}, & z\in \Gamma_{2,u} \cup \Gamma_{2,l},\\
 \begin{pmatrix}
 1 & 0 & 0 & \hphantom{-}0\\
 0 & 0 & 0 & -1\\
 0 & 0 & 1 & \hphantom{-}0\\
 0 & 1 & 0 & \hphantom{-}0
 \end{pmatrix}, & z\in [\beta,\infty),\vspace{1mm}\\
 \begin{pmatrix}
 1 & 0 & \hphantom{-}0 & 0\\
 0 & 0 & -1 & 0\\
 0 & 1 & \hphantom{-}0 & 0\\
 0 & 0 & \hphantom{-}0 & 1
 \end{pmatrix}, & z\in (-\infty,-\beta],\vspace{1mm}\\
 \begin{pmatrix}
 \hphantom{-}0 & 1 & 0 & 0\\
 -1 & 0 & 0 & 0\\
 \hphantom{-}0 & 0 & 1 & 0 \\
 \hphantom{-}0 & 0 & 0 & 1
 \end{pmatrix}, & z\in [-\alpha,\alpha],\\
 \mathbb{I} - E_{24}{\rm e}^{-n\tau[\Omega_2(z) - \Omega_4(z)]},
 & z \in \Gamma_1 \cap \{\operatorname{Im} z > 0\},\\
 \mathbb{I} - E_{23}{\rm e}^{-n\tau[\Omega_2(z) - \Omega_3(z)]},
 & z \in \Gamma_2 \cap \{\operatorname{Im} z < 0\},\\
 \mathbb{I} + E_{12}{\rm e}^{-n\tau[\Omega_1(z) - \Omega_2(z)]},
 & z\in \Gamma \setminus [-\alpha,\alpha],\\
 \mathbb{I} + E_{21}{\rm e}^{-n\tau[\Omega_{2}(z)-\Omega_{1}(z)]}, & z\in \Gamma_{u}\cup\Gamma_{l}.\\
 \end{cases}
 \end{equation*}
Furthermore, $\boldS(z)$ has asymptotics
\begin{equation*} %\label{S-asymptotics}
 \boldS(z) =
 \biggl[\mathbb{I} + \OO\bigl(z^{-1}\bigr)\biggr]
 \begin{pmatrix}
 1 & 0\\
 0 & B^{-1}(z) A^{-1}(z)
 \end{pmatrix},
 \qquad z \to \infty .
 \end{equation*}
 \end{Proposition}
\begin{proof}
 Again, the proof of this proposition follows immediately from the definition of $\boldS(z)$. The fact that the stronger condition
 \begin{equation*}
 \boldS(z) =
 \biggl[\mathbb{I} + \OO\bigl(z^{-1}\bigr)\biggr]
 \begin{pmatrix}
 1 & 0\\
 0 & B^{-1}(z) A^{-1}(z)
 \end{pmatrix},
 \qquad z \to \infty,
 \end{equation*}
 holds is due to the fact that the jumps of $A(z)B(z)$ match the jumps of $\boldS(z)$ at infinity, as per Remark~\ref{AB-matrix-remark}. Thus,
 $\OO\bigl(z^{-1/3}\bigr)$ can be replaced with $\OO\bigl(z^{-1}\bigr)$ in the asymptotics of $\boldS(z)$.
\end{proof}

\section[Construction of parametrices and the transformation ${\bf S} \mapsto {\bf R}$]{Construction of parametrices \\ and the transformation $\boldsymbol{{\bf S} \mapsto {\bf R}}$}\label{section6}
All of the jumps of $\boldS$ are either constant, or exponentially small. Our next task is to try and eliminate these constant jumps. We will accomplish this task by searching for an approximate solution, called the \textit{global parametrix} to the Riemann--Hilbert problem at hand; this
approximate solution will match the constant jumps of $\boldS$ exactly, but the difference of jumps near the branch points will be ``bad'', requiring us to find local approximations to the RHP (local parametrices) there. Aside from the proofs of the lensing inequalities, this is really the first place we will see a difference between the multicritical, critical, and generic (non-critical) cases.

\subsection{Global parametrix}
If we ignore the exponentially small jumps of $\boldS(z)$, we obtain the following model RHP for a~${4\times 4}$ matrix-valued function $M(z)$:
\begin{equation}\label{Model-RHP-a}
 \begin{cases}
 \text{$M$ is analytic in $\CC\setminus\RR$},\\
 M_+(z) = M_-(z)
 \begin{pmatrix}
 1 & 0 & 0 & 0\\
 0 & 0 & -1 & 0\\
 0 & 1 & 0 & 0\\
 0 & 0 & 0 & 1\\
 \end{pmatrix}, & z \in (-\infty,-\beta],\vspace{1mm}\\
 M_+(z) = M_-(z)
 \begin{pmatrix}
 0 & 1 & 0 & 0\\
 -1 & 0 & 0 & 0\\
 0 & 0 & 1 & 0\\
 0 & 0 & 0 & 1\\
 \end{pmatrix}, & z \in [-\alpha,\alpha],\vspace{1mm}\\
 M_+(z) = M_-(z)
 \begin{pmatrix}
 1 & 0 & 0 & 0\\
 0 & 0 & 0 & -1\\
 0 & 0 & 1 & 0\\
 0 & 1 & 0 & 0\\
 \end{pmatrix}, & z \in [\beta,\infty),\vspace{1mm}\\
 M(z) =
 \bigl[\mathbb{I} + \OO\bigl(z^{-1}\bigr)\bigr]
 \begin{pmatrix}
 1 & 0\\
 0 & B^{-1}(z) A^{-1}(z)
 \end{pmatrix},
 & z \to \infty.
 \end{cases}
\end{equation}
In general, solutions to \eqref{Model-RHP-a} will not be unique. Uniqueness can be guaranteed by imposing additionally that
\begin{gather}
 M(z) = \OO\bigl( \bigl(z\mp \alpha\bigr)^{-1/4} \bigr), \qquad z\to \pm \alpha,\nonumber\\
 M(z) = \OO\bigl( \bigl(z\mp \beta\bigr)^{-1/4} \bigr), \qquad z\to \pm \beta.\label{global-parametrix-constraint}
\end{gather}
Then, from the usual Liouville argument, it is clear that if a solution to \eqref{Model-RHP-a} (with the constraint~\eqref{global-parametrix-constraint} imposed) exists, it is unique. We show that a solution exists by direct construction.

\begin{figure}[t]\centering
\includegraphics{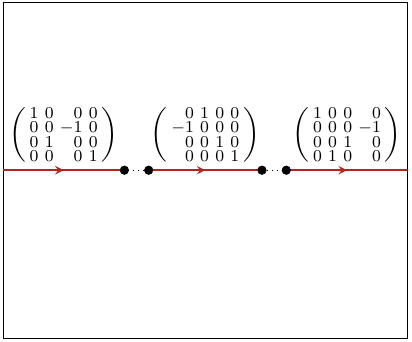}

 \caption{The jumps of the global parametrix $M(z)$ on $\RR$. Note that, when $a,b\to 1$, branch points merge, and the cuts cover the whole real line.}
 %\label{fig:Global-Parametrix}
\end{figure}

 \begin{Proposition}
 The solution to the global parametrix is given by
 \begin{equation*}
 M_{jk}(z) = \begin{cases}
 \psi_j(u_k(z)), & \operatorname{Im} z >0,\\
 \psi_j(u_\ell(z))S_{\ell k}, & \operatorname{Im} z <0,\\
 \end{cases}
 \end{equation*}
 $($here, summation over $\ell$ is implied$)$ where $u_k(z)$ is the restriction of the uniformizing coordinate to the $k^{\rm th}$ sheet, and the functions $\psi_j(u)$ are given by
 \begin{alignat*}{3}
 &\psi_1\bigl(u\bigr) = \frac{u^2}{\sqrt{\bigl(u^2-b^2\bigr)\bigl(u^2-a^2\bigr)}}, \qquad && \psi_2\bigl(u\bigr) = \frac{3^{2/3}A^{1/3}}{a^{1/3}b^{1/3} }\frac{2a^2b^2-3\bigl(a^2+b^2\bigr)u^2}{18u\sqrt{\bigl(u^2-b^2\bigr)\bigl(u^2-a^2\bigr)}}& \\
 &\psi_3\bigl(u\bigr) = \frac{ab}{3\sqrt{\bigl(u^2-b^2\bigr)\bigl(u^2-a^2\bigr)}}, \qquad && \psi_4\bigl(u\bigr) = \frac{a^{1/3}b^{1/3} }{3^{2/3}A^{1/3} }\frac{u}{\sqrt{\bigl(u^2-b^2\bigr)\bigl(u^2-a^2\bigr)}},&
 \end{alignat*}
 and $S = {\rm diag}(1, -1, 1, 1) = [\sigma_3 \oplus \mathbb{I}_2]$.
 \end{Proposition}

\begin{proof}
Let us assume we are in either the generic (noncritical) or critical case, we have that $\alpha = \alpha(a,b) < \beta(a,b) = \beta$.
Following~\cite{DK1}, we will solve find it convenient to solve this problem in the uniformizing coordinate $z = z(u)$:
 \begin{equation*}
 z(u) = A(a,b)\biggl(u + \frac{a^2+b^2}{u} - \frac{a^2b^2}{3u^3}\biggr).
 \end{equation*}
Consider the general row vector
 \begin{equation*}
 \vec{\psi}(z) =
 \begin{cases}
 [f^u(u_1(z)), g^u(u_2(z)), h^u(u_3(z)), k^u(u_4(z)) ], & \operatorname{Im} z>0,\\
 [f^l(u_1(z)), g^l(u_2(z)), h^l(u_3(z)), k^l(u_4(z)) ], & \operatorname{Im} z<0.
 \end{cases}
 \end{equation*}
where $f^u$, $g^u$, $h^u$, $k^u$ \big(respectively, $f^l$, $g^l$, $h^l$, $k^l$\big) are analytic functions away from the real axis, to be determined. Suppose \smash{$\vec{\psi}(z)$} satisfies the jumps of the
global parametrix. Now, consider the analytic continuation of $\vec{\psi}$
through $(-\infty,-\beta]$: upon continuation, $u_2(z)$ and $u_3(z)$ are interchanged. On the other hand, the analytic continuation is determined by the
jump condition of the global parametrix. This leads to the constraint
 \begin{gather*}
 [f^u(u_1(z)), g^u(u_3(z)),h^u(u_2(z)),k^u(u_4(z))]
 = \bigl[f^l(u_1(z)), h^l(u_3(z)),-g^l(u_2(z)),k^l(u_4(z))\bigr],
 \end{gather*}
for $z\in (-\infty,-\beta]$. In particular, this implies the equalities $f^u = f^l$, $k^u=k^l$, $g^u = h^l$, and $h^u = -g^l$. Similar analysis on the other cuts yields the further compatibility conditions
 $g^u = k^l$, $k^u = -g^l$, $f^u = -g^l$, $g^u = f^l$.
Thus, the global parametrix depends on only on the unknown function: $f^u =: \psi$. Therefore,
 \begin{equation*}
 \vec{\psi}(z) =
 \begin{cases}
 [\psi(u_1(z)), \psi(u_2(z)), \psi(u_3(z)), \psi(u_4(z)) ], & \operatorname{Im} z>0,\\
 [\psi(u_1(z)), -\psi(u_2(z)), \psi(u_3(z)), \psi(u_4(z)) ], & \operatorname{Im} z<0.
 \end{cases}
 \end{equation*}

We have thus shown that, in the upper half plane,
 \begin{equation*}
 \vec{\psi}(z) = [\psi(u_1(z)), \psi(u_2(z)), \psi(u_3(z)), \psi(u_4(z)) ],
 \end{equation*}
where $\vec{\psi}(z)$ defines a (possibly multivalued) analytic function in the uniformizing plane, which is defined by its components $\psi(u_j(z))$ in each of the images of the sheets $j = 1,2,3,4$. Suppose $\vec{\psi}(z)$ satisfies the jump conditions of \eqref{Model-RHP-a}.
Let us compute the monodromy of $\psi$ around each of the branch points. In the uniformizing coordinate, the branch points are mapped on
to $\pm a$ and $\pm b$ \big(Note: provided we are away from the multi-critical point, we have that $0 < b < 1 \leq a \leq b^{-1}$.\big) By direct computation using the Riemann--Hilbert problem, we find that $\psi(u)$ has square-root singularities at each of the branch points $\pm a$, $\pm b$,
and possibly a pole singularity at $u=0$. The form of $\psi(u)$ is then fixed to be
 \begin{equation*}
 \psi(u) = \frac{p_3(u)}{u\sqrt{(u^2-a^2)(u^2-b^2)}},
 \end{equation*}
where $p_3(u)$ is some polynomial of degree $\leq 3$, and the positive branch cut of the square root is chosen, with branch cuts on the intervals $[-a,-b]\cup [b,a]$.
The form of $\psi(u)$ is fixed by the following constraints. For any choice of $p_3(u)$, $\psi_k(u)$ is analytic in $\CC \setminus \gamma$, where $\gamma$ is the image of the cuts in the uniformizing plane, and satisfies the following conditions:
\begin{alignat*}{3}
 &\psi_+(u) = -\psi_-(u),\qquad&& u \in \gamma,&\\
 &\psi(u) = \OO(1),\qquad&& u \to \infty,&\\
 &\psi(u) = \OO\bigl(u^{-1}\bigr),\qquad&& u \to 0,&\\
 &\psi(u) = \OO\bigl( (u\mp a)^{-1/2}\bigr),\qquad&& u \to \pm a,&\\
 &\psi(u) = \OO\bigl( (u\mp b)^{-1/2}\bigr),\qquad&& u \to \pm b.&
\end{alignat*}

These conditions guarantee that the entries of $M(z)$ are $\OO(1)$ as $z\to \infty$ on the first sheet, and $\OO\bigl(z^{1/3}\bigr)$ on sheets $2$--$4$. It also guarantees that $M(z)$ has the correct jumps.
For a given row, we can use the normalization condition of the Riemann--Hilbert problem to determine the coefficients of $p_3(u)$.

Setting $p_3(u) = c_0 + c_1 u + c_2 u^2 + c_3u^3$, then we have the following large $z$ expansions of the
functions $\psi(u_j(z))$:
 \begin{align*}
 &\psi (u_1(z)) = c_3 + \frac{c_2 A}{z} + \frac{\bigl((a^2+b^2)c_3 + 2c_1\bigr)A^2}{2z^2} + \frac{(3(a^2 + b^2)c_2 + 2c_0)A^3}{2z^3} + \OO\bigl(z^{-4}\bigr),\\
 &\psi (u_2(z)) =
 \begin{cases}
 -\dfrac{ 3^{\frac{1}{3}} c_0}{(ab)^{\frac{5}{3}}} \dfrac{\omega^2z^{\frac{1}{3}}}{A^{\frac{1}{3}}} + \dfrac{c_1}{ab} - \dfrac{2a^2b^2c_2 + 3c_0(a^2+b^2)}{3^{\frac{1}{3}}(ab)^{\frac{7}{3}}} \dfrac{\omega A^{\frac{1}{3}}}{ z^{\frac{1}{3}}}&\\
 \quad{} + \dfrac{3^{\frac{1}{3}}(2a^2b^2c_3 - (a^2 + b^2)c_1)}{6a^{\frac{5}{3}}b^{\frac{5}{3}}} \dfrac{\omega^2A^{\frac{2}{3}}}{z^{\frac{2}{3}}} + \OO\bigl(z^{-1}\bigr), &\operatorname{Im}z >0,\\
 -\dfrac{ 3^{\frac{1}{3}} c_0}{(ab)^{\frac{5}{3}}} \dfrac{\omega z^{\frac{1}{3}}}{A^{\frac{1}{3}}} + \dfrac{c_1}{ab} - \dfrac{2a^2b^2c_2 + 3c_0(a^2+b^2)}{3^{\frac{1}{3}}(ab)^{\frac{7}{3}}} \dfrac{\omega^2A^{\frac{1}{3}}}{ z^{\frac{1}{3}}} &\\
 \quad{}+ \dfrac{3^{\frac{1}{3}}(2a^2b^2c_3 - (a^2 + b^2)c_1)}{6a^{\frac{5}{3}}b^{\frac{5}{3}}} \dfrac{\omega A^{\frac{2}{3}}}{z^{\frac{2}{3}}} + \OO\bigl(z^{-1}\bigr), &\operatorname{Im}z <0,
 \end{cases}\\
 &\psi(u_3(z)) = -\frac{ 3^{\frac{1}{3}} c_0}{(ab)^{\frac{5}{3}}} \frac{z^{\frac{1}{3}}}{A^{\frac{1}{3}}} + \frac{c_1}{ab} - \frac{2a^2b^2c_2 + 3c_0(a^2+b^2)}{3^{\frac{1}{3}}(ab)^{\frac{7}{3}}} \frac{A^{\frac{1}{3}}}{z^{\frac{1}{3}}}\\
 &\hphantom{\psi(u_3(z)) =}{}
 + \frac{3^{\frac{1}{3}}(2a^2b^2c_3 - (a^2 + b^2)c_1)}{6a^{\frac{5}{3}}b^{\frac{5}{3}}} \frac{A^{\frac{2}{3}}}{z^{\frac{2}{3}}} + \OO\bigl(z^{-1}\bigr),\\
 &\psi(u_4(z)) =
 \begin{cases}
 -\dfrac{ 3^{\frac{1}{3}} c_0}{(ab)^{\frac{5}{3}}} \dfrac{\omega z^{\frac{1}{3}}}{A^{\frac{1}{3}}} + \dfrac{c_1}{ab} - \dfrac{2a^2b^2c_2 + 3c_0(a^2+b^2)}{3^{\frac{1}{3}}(ab)^{\frac{7}{3}}} \dfrac{\omega^2A^{\frac{1}{3}}}{ z^{\frac{1}{3}}} &\\
 \quad{}+ \dfrac{3^{\frac{1}{3}}(2a^2b^2c_3 - (a^2 + b^2)c_1)}{6a^{\frac{5}{3}}b^{\frac{5}{3}}} \dfrac{\omega A^{\frac{2}{3}}}{z^{\frac{2}{3}}} + \OO\bigl(z^{-1}\bigr), &\operatorname{Im}z >0,\\
 -\dfrac{ 3^{\frac{1}{3}} c_0}{(ab)^{\frac{5}{3}}} \dfrac{\omega^2z^{\frac{1}{3}}}{A^{\frac{1}{3}}} + \dfrac{c_1}{ab} - \dfrac{2a^2b^2c_2 + 3c_0(a^2+b^2)}{3^{\frac{1}{3}}(ab)^{\frac{7}{3}}} \dfrac{\omega A^{\frac{1}{3}}}{ z^{\frac{1}{3}}} &\\
 \quad{}+ \dfrac{3^{\frac{1}{3}}(2a^2b^2c_3 - (a^2 + b^2)c_1)}{6a^{\frac{5}{3}}b^{\frac{5}{3}}} \dfrac{\omega^2A^{\frac{2}{3}}}{z^{\frac{2}{3}}} + \OO\bigl(z^{-1}\bigr), &\operatorname{Im}z <0,
 \end{cases}
 \end{align*}

Since \smash{$\bigl(\begin{smallmatrix}
 1 & 0\\
 0 & A(z)B(z)
 \end{smallmatrix}\bigr)M(z)$} has no jumps near infinity, it admits the expansion
 \begin{equation*}
 \begin{pmatrix}
 1 & 0\\
 0 & A(z)B(z)
 \end{pmatrix} M(z) = \mathbb{I} + \OO\bigl(z^{-1}\bigr), z\to \infty.
 \end{equation*}
This fact allows us to determine the constants $\{c_i\}$ for each row; the rest of the proof follows from direct computation. A similar analysis in the lower half plane may be performed, and the same result for $\psi(z)$ is obtained. The expression in the lower half plane can also be obtained by analytic continuation of the solution in the upper half plane, in accordance with the jump conditions of the global parametrix.
\end{proof}

\begin{Remark}
 So far, we have ignored the multicritical case, when $a=b=c=1$. In this case, the branch points in both the spectral and uniformizing planes merge into a pair of branch points, say, $\pm \alpha$. If we follow the same procedure as before, we find that any row vector ${\psi(u) = [\psi(u_1(z)), \dots, \psi(u_4(z))]}$ has no monodromy, and thus defines a meromorphic function on the spectral curve. In fact, if we complete the calculations, we find that the solution to the global parametrix in the multicritical case is just the degeneration of the general global parametrix as~${a,b,c\to 1}$. Direct inspection of the general global parametrix shows us that~${(a,b,c) = (1,1,1)}$ is the only point in $R$ where the rows of the global parametrix are single-valued in the uniformizing plane. The same is true when we approach a point on the curves $\gamma_b$; this is beyond the scope of the present work.
\end{Remark}
The next proposition will become useful when we study the local parametrix problem.
\begin{Proposition}\label{global-parametrix-prop}
 $M(z)$ satisfies the symmetry condition
 \begin{equation} \label{global-parametrix-symmetry}
 M(-z) = \left[\sigma_3 \oplus \sigma_3 \right] M(z) [\sigma_3 \oplus \sigma_1].
 \end{equation}
\end{Proposition}
\begin{proof}
 First, note that the previous proposition yields that
 \begin{equation*}
 \psi_j(-u) = -\psi_j(u), \quad j \mod 2 \equiv 0, \qquad \psi_j(-u) = \psi_j(u), \quad j \mod 2 \equiv 1.
 \end{equation*}
 Furthermore, the uniformizing coordinates $u_j(z)$ satisfy the following relations (cf.\ Appendix~\ref{Appendix-A} for details):
 \begin{alignat*}{3}
 &u_1(-z) = -u_1(z), \qquad&& u_2(-z) = -u_2(z),&\\
 &u_3(-z) = -u_4(z), \qquad&& u_4(-z) = -u_3(z).&
 \end{alignat*}
 Thus, we can compute entrywise the transformation $M(-z)$. For example, if $z$ belongs to the upper half plane, then $-z$ is in the lower half plane, and we obtain that
 \begin{align*}
 &M_{11}(-z)= \psi_1(u_1(-z)) = \psi_1(-u_1(z)) = \psi_1(u_1(z)),\\
 &M_{22}(-z)= -\psi_2(u_2(-z)) = -\psi_2(-u_2(z)) = \psi_2(u_2(z)),\\
 &M_{33}(-z)= \psi_3(u_3(-z)) = \psi_3(-u_4(z)) = \psi_4(u_4(z)).
 \end{align*}
 Note that $\operatorname{Im} z > 0$ became relevant in the calculation for the 2-2 entry.
 Calculation of the remaining entries yields equation \eqref{global-parametrix-symmetry}.
\end{proof}
\begin{Remark}
 Note here that the right multiplication $[\sigma_3 \oplus \sigma_1] = [\sigma_3\oplus \mathbb{I}_2][\mathbb{I}_2\oplus \sigma_1]$ has two purposes: the
block $[\mathbb{I}_2\oplus \sigma_1] = \hat{\sigma}_{34}$ interchanges of columns $3$ and $4$, and the block $[\sigma_3\oplus \mathbb{I}_2] = S$ accounts
for the change in sign of the global parametrix upon moving from the upper to the lower half plane (and vice versa).
\end{Remark}

\begin{Remark}
 The final step in the steepest descent analysis is to calculate the local parametrices near the branch points, where the global parametrix is a bad approximation to $\boldS(z)$. In the next section, many of the jumps will involve differences of the $\Omega_j(z)$'s. For this reason, we introduce the notation
 \begin{equation*}
 \delta \Omega_{ij}(z) := \Omega_i(z) - \Omega_j(z),
 \end{equation*}
 for $i,j\in \{1,\dots,4\}$.
\end{Remark}

We have now found an approximate solution to the Riemann--Hilbert problem for $\boldS(z)$. Indeed, if we consider the matrix
 \begin{equation*}
 \boldR_{\rm out}(z) := \boldS(z)M^{-1}(z),
 \end{equation*}
we see that $\boldR_{\rm out}(z) \to \mathbb{I}$ as $z\to \infty$. Furthermore, the jumps of $\boldR_{\rm out}(z)$ are all exponentially small (in $n$), with one exception: near the branch points $\pm \alpha$, $\pm \beta$, the jumps are not close to the identity, as $n\to \infty$. Therefore, we must try to find a better approximation of $\boldS(z)$ near the branch points.

\begin{Remark} %\label{remark:localparametrices}
Here is the first place in the Riemann--Hilbert analysis where we will see explicitly that the generic, critical, and multicritical cases differ. In this work, we are interested only in the generic situation. In Part III, we will address the multicritical situation; the other critical cases we hope to study in a later work. We record here how the local parametrices for each of these cases should work out. One should compare this to the remarks in Section~\ref{phase-transitions-section}.
 \begin{enumerate}\itemsep=0pt
 \item[$(1)$] \textit{The generic case: Airy parametrices.}
 For $(\tau,t,H)$ off the critical surface, the situation is generic, and the behavior of the $\delta\Omega_{ij}(z)$'s near the branch points is
 \begin{equation*}
 \delta\Omega_{ij}(z) \sim (z\pm \alpha)^{3/2} \qquad \text{\big(resp.\ $\sim (z\pm \beta)^{3/2}$\big)},
 \end{equation*}
 cf.\ Proposition~\ref{local-expansions-generic}. By now, it is well-established in the literature that this behavior leads to Airy-type parametrices at each of the branch points. Since this computation is familiar, we omit the explicit calculation of the all of the parametrices here. We present the calculation of the local parametrix at $z=-\beta$ in the next subsection.

 \item[$(2)$] \textit{$(\tau,t,H)\in \gamma_{{\rm low},0}$: Painlev\'{e}~I.}
 For $(\tau,t,H)$ on the low-temperature critical curve (i.e., the curve defined by the equation \smash{$t=-\frac{1}{12}+\frac{2}{9}\tau^2$}, $0<\tau<\frac{1}{4}, H=0$),
 the behavior of the $\delta \Omega_{ij}(z)$'s near each of the four branch points $\pm \alpha$, $\pm \beta$, is
 \begin{equation*}
 \delta \Omega_{ij}(z) \sim (z\pm \alpha)^{5/2} \qquad \text{\big(resp.\ $\sim (z\pm \beta)^{5/2}$\big)}.
 \end{equation*}
 It is also widely acknowledged in the literature (cf., for example,~\cite{DK0}) that such local behavior leads to a Painlev\'{e}~I-type Riemann--Hilbert
 problem, which has been the subject of intensive study~\cite{Kapaev,Kapaev-Kitaev}. We expect that the same analysis applies to the situation at hand,
 and leads to a description of the partition function in terms of the same solution to Painlev\'{e}~I that appears in the description of the critical
 $1$-matrix model~\cite{DK0, FIK1,FIK2}. Since we are mainly interested in the behavior of the partition function in the generic case and at the multicritical point, we omit the explicit calculation of the parametrices here.

 \item[$(3)$] \textit{$(\tau,t,H)\in \gamma_{{\rm high},0} \cup S_{{\rm high}}$: Painlev\'{e}~I and Airy.}
 For $(\tau,t,H)$ on the high-temperature critical surface, the behavior of the $\delta \Omega_{ij}(z)$'s the branch points is
 \begin{equation*}
 \delta \Omega_{ij}(z) \sim (z\pm \alpha)^{5/2}, \qquad \delta \Omega_{ij}(z) \sim (z\pm \beta)^{3/2},
 \end{equation*}
 as $z\to \pm \alpha$, $z\to \pm \beta$, respectively. This indicates that the appropriate local
 parametrices to use near $z=\pm \beta$ are Airy-type parametrices, by our previous commentary for the generic situation. Similarly, we expect that near
 the branch points $z=\pm \alpha$, one should use Painlev\'{e}~I parametrices. Thus, the local parametrix structure on the high-temperature critical curve
 is a mix of the low-temperature critical and generic situations. On~the~support of the main cut, the density is still critical, and Painlev\'{e}~I
 parametrices are needed; however, the other cuts cease to be ``critical'', as they were in the low-temperature regime.
 We again omit the explicit calculation of the parametrices here, as our main interests lie elsewhere.

 \item[$(4)$] \textit{$(\tau,t,H)\in S_{{\rm low}}$: Airy and Painlev\'{e}~I.}
 Similarly to the high-temperature surface, on the low-temperature surface both Airy and Painlev\'{e}~I-type parametrices appear. In this case, the behavior of the $\delta \Omega_{ij}(z)$'s the branch points is
 \begin{equation*}
 \delta \Omega_{ij}(z) \sim (z\pm \alpha)^{3/2}, \qquad \delta \Omega_{ij}(z) \sim (z\pm \beta)^{5/2}.
 \end{equation*}
 Thus, one must construct Airy-type parametrices at $z=\pm \alpha$, and Painlev\'{e}~I parametrices near
 $z=\pm\beta$.

 \item[$(5)$] \textit{$(\tau,t,H)\in \gamma_b$: New Painlev\'{e}~I parametrix.}
 On the critical curve $\gamma_b$ (the interface of the surfaces $S_{{\rm low}}$ and $S_{{\rm high}}$), the branch points at $z=+\alpha,+\beta$
 (respectively, $z=-\alpha,-\beta$) merge, and one obtains a pair of cubic branch points $z=\pm \alpha$, at which the $\delta\Omega_{ij}(z)$'s
 behave as
 \begin{equation*}
 \delta \Omega_{ij}(z) \sim (z\pm \alpha)^{5/3}.
 \end{equation*}
 To the best of our knowledge, an appropriate model Riemann--Hilbert problem for such a parametrix has not yet appeared in the literature. Some
 preliminary calculations (see also Part II~\cite{DHL2}) show that the ``right'' parametrix is a $3\times 3$ RHP for a function $\Psi(\xi;x)$,
 which has asymptotics
 \begin{equation*}
 \Psi(\xi;x) = g(\xi)\biggl[\mathbb{I} + \frac{\Psi_{1}(x)}{\xi^{1/3}} + \OO\biggl(\frac{1}{\xi^{2/3}}\biggr)\biggr]{\rm e}^{\Theta(\xi;x)},
 \end{equation*}
 where $g(\xi) = \OO\bigl(\xi^{1/3}\bigr)$, $\Theta(\xi;x)$ is a diagonal matrix whose entries are analytic continuations of the function \smash{$\theta_{1}(\xi;x) = \frac{3}{5}\xi^{5/3} + x\xi^{1/3}$}, and the jumps of $\Psi$ appearing only from the Stokes phenomenon. Some calculation reveals
 that this model Riemann--Hilbert problem is related to the Painlev\'{e}~I equation, in that $\frac{\rm d}{{\rm d}x}\left[\Psi_{1}(x)\right]_{11} = u(x)$, and $u(x)$ solves a scaled version of the Painlev\'{e}~I equation. The associated Lax pair \textit{has} appeared before in the literature~\cite{JKT}; one might hope to relate this parametrix to the ``standard'' $2\times 2$ parametrix for Painlev\'{e}~I via the
 techniques of Liechty and Wang~\cite{LW}, who studied a similar relation between two instances of Painlev\'{e}~II parametrices. Both
 seem to be instances of the so-called $(p,q)\leftrightarrow (q,p)$ duality of string equations~\cite{FKN,GGPZ} (for~PII, $q=2$, $p=4$, and for PI, $q=2$, $p=3$).
 It would be interesting to investigate this correspondence in more detail. For now, we leave this problem open.

 \item[$(6)$] \textit{$(\tau,t,H) = \bigl(\frac{1}{4},-\frac{5}{72},0\bigr)$: A new critical phenomenon.}
 Finally, if we are at the multicritical point, we again find that the branch points at $z=+\alpha,+\beta$
 (respectively, $z=-\alpha,-\beta$) merge, and one obtains a pair of cubic branch points $z=\pm \alpha$, at which the $\delta\Omega_{ij}(z)$'s
 behave as
 \begin{equation*}
 \delta \Omega_{ij}(z) \sim (z\pm \alpha)^{7/3}.
 \end{equation*}
 This requires the construction of a new $3\times 3$ parametrix, which we discuss fully in Part~II~\cite{DHL2}. This parametrix is connected to the $(3,4)$ string equation~\cite{Douglas1}, and is meant to describe the Ising phase transition coupled to gravity, cf.~\cite{KB,DS,GM1,GM2}. We fully investigate how the multicritical partition function of the 2-matrix model studied in this work is connected
 to this equation in Part III of this work~\cite{DHL3}.
 \end{enumerate}
\end{Remark}

\subsection{Calculation of the local parametrices}
We now construct the local parametrices. Define four discs
 \begin{equation*}
 D_{\pm\alpha} := \{z\in\CC \mid |z\mp\alpha| < \epsilon\},\qquad D_{\pm \beta} := \{z\in\CC \mid |z\mp\beta| < \epsilon\},
 \end{equation*}
where $\epsilon>0$ is a radius, to be determined. Our new parametrix $M^{\infty}(z)$ will be defined as
 \begin{equation*}
 M^{\infty}(z) =
 \begin{cases}
 M(z), & z\in \CC \setminus (D_{\pm\alpha} \cup D_{\pm\beta}),\\
 P^{(\pm\alpha)}(z), & z\in D_{\pm\alpha},\\
 P^{(\pm\beta)}(z), & z\in D_{\pm\beta}.
 \end{cases}
 \end{equation*}
The functions $P^{(\pm \alpha)}(z)$ will be chosen to so that the following conditions are met:
 \begin{enumerate}\itemsep=0pt
 \item[$(1)$] $P^{(\pm \alpha)}(z)$ matches the jumps of $\boldS(z)$ \textit{exactly} in the discs $D_{\pm \alpha}$,
 \item[$(2)$] $P^{(\pm \alpha)}(z) = \bigl[\mathbb{I} + \OO\bigl(n^{-\delta}\bigr)\bigr] M(z)$, for $z\in D_{\pm \alpha}$, $n\to \infty$ for some $\delta >0$.
 \end{enumerate}
The functions $P^{(\pm \beta)}(z)$ should be chosen so that identical statements to the above hold, with~$\alpha$ replaced with $\beta$.
In the case of generic values of the parameters, one should expect that near the branch points $z=\pm\alpha,\pm \beta$, an Airy-type parametrix can be used to describe the local behavior of the Riemann--Hilbert problem at hand. This is a common calculation in the literature (cf., for example,~\cite{DKMVZ}).
Before calculating the parametrices, it is useful to notice that we only have to calculate two of the parametrices; the others can be calculated
by symmetry. This statement is summarized in the following lemma.

\begin{Lemma}\label{Parametrix-symmetry-prop}
 Suppose $P^{(+\beta)}(z)$ satisfies conditions $(1)$ and $(2)$ above. Then, $P^{(-\beta)}(z)$ can be constructed as
 \begin{equation*}
 P^{(-\beta)}(z) = [\sigma_3\oplus \sigma_3]P^{(+\beta)}(-z)[\sigma_3 \oplus \sigma_1].
 \end{equation*}
\end{Lemma}
\begin{proof}
 Let us verify condition $(2)$ first. Indeed, if $z$ belongs to a sufficiently small neighborhood of $z = -\beta$, by Proposition
~\ref{global-parametrix-prop}, we have that
 \begin{align*}
 [\sigma_3\oplus \sigma_3]P^{(+\beta)}(-z)[\sigma_3 \oplus \sigma_1] &= [\sigma_3\oplus \sigma_3]\bigl[\mathbb{I} + \OO\bigl(n^{-\delta}\bigr)\bigr] M(-z)[\sigma_3 \oplus \sigma_1]\\
 &=\bigl[\mathbb{I} + \OO\bigl(n^{-\delta}\bigr)\bigr]M(z),
 \end{align*}
 where $M(z)$ above denotes the local expansion of $M(z)$ about $z=-\beta$. Now, let us check condition $(1)$ By direct calculation,
 \begin{align*}
 P^{(-\beta)}_{+}(z) &= [\sigma_3\oplus \sigma_3]P^{(+\beta)}_{+}(-z)[\sigma_3 \oplus \sigma_1]\\
 &= [\sigma_3\oplus \sigma_3]P^{(+\beta)}_{-}(-z) J_{P}(-z)[\sigma_3 \oplus \sigma_1]\\
 &= P^{(-\beta)}_-(z) [\sigma_3 \oplus \sigma_1] J_{P}(-z)[\sigma_3 \oplus \sigma_1],
 \end{align*}
 where $J_P(z)$ denotes the jumps of $P^{(+\beta)}(z)$. We must verify that $[\sigma_3 \oplus \sigma_1]J_{P}(-z)[\sigma_3 \oplus \sigma_1]$ agrees with the
 jumps of $\boldS(z)$ in a neighborhood of $z=-\beta$; this is done by direct calculation. For example, for $z \in (-\infty,-\beta]\cap D_{-\alpha}$,
 $-z \in [\beta,\infty) \cap D_{+\alpha}$. Furthermore, we have that
 \begin{equation*}
 [\sigma_3 \oplus \sigma_1]J_{P}(-z)[\sigma_3 \oplus \sigma_1] = [\sigma_3 \oplus \sigma_1]
 \begin{pmatrix}
 1 & 0 & 0 & \hphantom{-} 0\\
 0 & 0 & 0 & -1\\
 0 & 0 & 1 & \hphantom{-} 0\\
 0 & 1 & 0 & \hphantom{-} 0
 \end{pmatrix}[\sigma_3 \oplus \sigma_1] =
 \begin{pmatrix}
 1 & \hphantom{-} 0 & 0 & 0\\
 0 & \hphantom{-} 0 & 1 & 0\\
 0 & -1 & 0 & 0\\
 0 & \hphantom{-} 0 & 0 & 1
 \end{pmatrix},
 \end{equation*}
 which indeed agrees with the jump of $\boldS(z)$ there, upon reorientation of the jump contour. Similarly, if $z \in \Gamma_2 \cap \{\operatorname{Im} z < 0\} \cap D_-$, then $-z \in \Gamma_1 \cap \{\operatorname{Im} z > 0\} \cap D_+$, and
 \begin{align*}
 [\sigma_3 \oplus \sigma_1]J_{P}(-z)[\sigma_3 \oplus \sigma_1] &= [\sigma_3 \oplus \sigma_1]\bigl[\mathbb{I} - E_{24}{\rm e}^{-n\tau \delta\Omega_{24}(-z)}\bigr][\sigma_3 \oplus \sigma_1]\\
 &= \mathbb{I} + E_{23}{\rm e}^{-n\tau \delta\Omega_{24}(-z)} = \mathbb{I} + E_{23}{\rm e}^{-n\tau \delta\Omega_{23}(z)},
 \end{align*}
by the symmetry of the functions $\Omega_j(z)$ (and again reorientation of the jump contour). The rest of the calculations are similar, so we omit them.
\end{proof}

We state without proof the same lemma for the parametrices at $z=\pm \alpha$.

\begin{Lemma}%\label{Parametrix-symmetry-prop-2}
 Suppose $P^{(+\alpha)}(z)$ satisfies conditions $(1)$ and $(2)$ above. Then, $P^{(-\alpha)}(z)$ can be constructed as
$ P^{(-\alpha)}(z) = [\sigma_3\oplus \sigma_3]P^{(+\alpha)}(-z)[\sigma_3 \oplus \sigma_1]$.
\end{Lemma}

Thus, it is sufficient to calculate the parametrix at $z=\alpha$, $z=\beta$ only, and use the above symmetry relations
as the definition of $P^{(-\alpha)}(z)$, $P^{(-\beta)}(z)$.

\begin{figure}[t]
\centering
\includegraphics{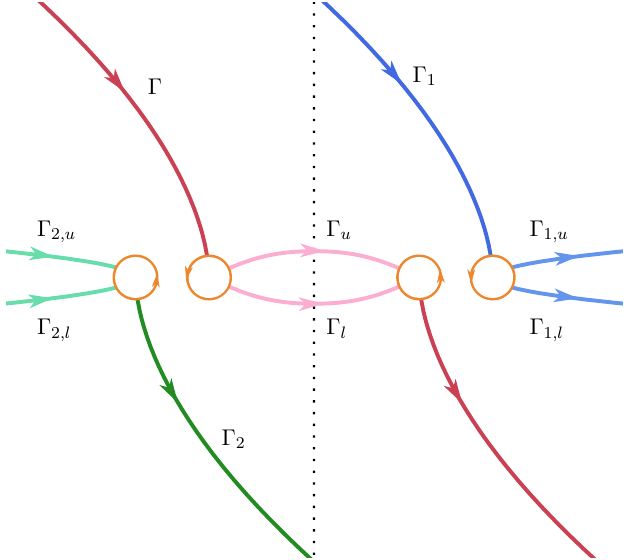}
%		 \begin{overpic}[scale=.5]{Pictures/Lensing2.pdf}
%		 \put (23,75) {$\Gamma$}
%		 \put (52,52) {$\Gamma_u$}
%		 \put (52,36) {$\Gamma_l$}
%		 \put (35,18) {$\Gamma_2$}
%		 \put (5,52) {$\Gamma_{2,u}$}
%		 \put (5,36) {$\Gamma_{2,l}$}
%		 \put (85,52) {$\Gamma_{1,u}$}
%		 \put (85,36) {$\Gamma_{1,l}$}
%		 \put (66,77) {$\Gamma_1$}
%		 \end{overpic}

 \caption{The opened lenses after the final transformation.}
 %\label{fig:Lenses-2}
\end{figure}

For brevity, we only compute explicitly the parametrix at $z=-\beta$. The rest of the local parametrices are calculated in an identical manner.

Let $\mathcal{A}(\zeta)$ denote the unique solution to the following $2\times 2$ Riemann--Hilbert problem:
 \begin{gather}
 \mathcal{A}_+(\zeta) = \mathcal{A}_-(\zeta)\times
 \begin{cases}
 \begin{pmatrix}1 & 1\\ 0 & 1\end{pmatrix}, & \arg \zeta = 0,\vspace{1mm}\\
 \begin{pmatrix}1 & 0\\ 1 & 1\end{pmatrix}, & \arg \zeta = \pm 2\pi/3,\vspace{1mm}\\
 \begin{pmatrix}\hphantom{-}0 & 1\\ -1 & 0\end{pmatrix}, & \arg \zeta = \pi.
 \end{cases}\nonumber\\
 \mathcal{A}(\zeta) = \bigl[\mathbb{I} + \OO\bigl(\zeta^{-1}\bigr)\bigr]\zeta^{-\frac{1}{4}\sigma_3} U {\rm e}^{-\frac{2}{3}\zeta^{3/2}\sigma_3}, \qquad \zeta\to \infty,\label{Airy-RHP}
 \end{gather}
where here \smash{$U = \frac{1}{\sqrt{2}}\begin{psmallmatrix}1 &{\rm i}\\ {\rm i} & 1\end{psmallmatrix}$}. The solution to the RHP \eqref{Airy-RHP} in the sector
$0 < \arg \zeta < 2\pi/3$ is given~by
 \begin{equation*}
 \mathcal{A}(\zeta) = \sqrt{2\pi}
 \begin{pmatrix}
 \hphantom{-}\Ai(\zeta) &-\omega^2 \Ai(\omega^2\zeta)\\
 -{\rm i}\Ai'(\zeta)& {\rm i}\omega \Ai'(\omega^2\zeta)
 \end{pmatrix}.
 \end{equation*}
Here $\Ai(\zeta)$ denotes the usual Airy function; the solution to the RHP \eqref{Airy-RHP} in the other sectors can be obtained by analytic continuation, and
application of the jump condition.

Define the conformal mapping
 \begin{equation*}
 \zeta(z) := \biggl[\frac{3}{4}n\tau \delta\Omega_{23}(z)\biggr]^{2/3}.
 \end{equation*}
Note that $\zeta(z)$ is univalent in a neighborhood of $z=-\beta$, since $\delta\Omega_{23}(z) = \text{const}\cdot (z+\beta)^{3/2}[1+ \OO(z+\beta)]$.
Finally, set
 \begin{equation*}
 \hat{P}(z) := \mathcal{A}(\zeta(z)) {\rm e}^{\frac{2}{3}\zeta(z)^{3/2}\sigma_3}\sigma_3.
 \end{equation*}
We redefine the jump contours of $\mathcal{A}(\zeta)$ so that the jump contours of~$\hat{P}(z)$ align with those of~$\boldS(z)$, in a neighborhood of $z=-\beta$.
The following lemma then holds.

\begin{Lemma}
 The jumps of $1\oplus \hat{P}(z) \oplus 1$ and $\boldS(z)$ agree, in a neighborhood of $z=-\beta$.
\end{Lemma}
 \begin{proof}
 This is an immediate consequence of the fact that multiplication of the matrix $\mathcal{A}(\zeta(z))$ by \smash{$\mathfrak{p}:= {\rm e}^{\frac{3}{2}\zeta(z)^{3/2}\sigma_3}\sigma_3$} conjugates its jumps by this factor; in other words, $J_{\hat{P}}(z) = \mathfrak{p}^{-1}J_{\mathcal{A}}(z)\mathfrak{p}$.
 \end{proof}

Now, consider $2\times 2$ matrix-valued function
 \begin{equation*}
 \tilde{\mathfrak{m}}(z) := (z+\beta)^{-\frac{1}{4}\sigma_3} U \sigma_3.
 \end{equation*}
It follows immediately that the jump of $1\oplus \tilde{\mathfrak{m}} \oplus 1$ and the jump of the global parametrix~$M(z)$ match in a neighborhood of $z\hspace{-0.25pt}=\hspace{-0.25pt}-\beta$.
Therefore, one can locally represent the global parametrix~as%
\begin{equation*}
 M(z) = \mathfrak{H}(z) [1\oplus \tilde{\mathfrak{m}}(z) \oplus 1],
\end{equation*}
for some locally holomorphic matrix function $\mathfrak{H}(z)$. Furthermore, since $\zeta(z) = n^{2/3} (z+\beta) g(z)$, for some non-vanishing analytic function $g(z)$ in a neighborhood of $z=-\beta$, by a slight re-definition of $\mathfrak{H}(z)$, we can rewrite $M(z)$ as
 \begin{equation*}
 M(z) = \tilde{\mathfrak{H}}(z) \bigl[1\oplus n^{\frac{1}{6}\sigma_3}[\zeta(z)]^{-\frac{1}{4}\sigma_3} U \sigma_3 \oplus 1 \bigr],
 \end{equation*}
where $\tilde{\mathfrak{H}}(z)$ is some $n$-independent invertible function, analytic in a neighborhood of $z=-\beta$.

We now define the parametrix in a sufficiently small disc about $z=-\beta$ as
 \begin{equation*}
 P^{(-\beta)}(z) := M(z) \bigl[1\oplus \sigma_3 U^{-1}[\zeta(z)]^{\frac{1}{4}\sigma_3} \oplus 1 \bigr] \hat{P}(z).
 \end{equation*}
We remark that $P^{(-\beta)}(z)$ and $\hat{P}(z)$ have identical jumps inside the disc at $z=-\beta$, since
\begin{equation*}
M(z) \bigl[1\oplus \sigma_3 U^{-1}[\zeta(z)]^{\frac{1}{4}\sigma_3} \oplus 1 \bigr]
\end{equation*}
has no jumps there, by our previous considerations. The parametrix at $z=+\beta$ can be constructed using our previous symmetry considerations, cf.\ Lemma~\ref{Parametrix-symmetry-prop}. One can similarly construct parametrices at $P^{(\pm\alpha)}(z)$, by replacing in the definition of $\hat{P}(z)$: (1)~the conformal
map~$\zeta(z)$ with \smash{$\zeta(z) := \bigl[\frac{3}{4}n\tau \delta\Omega_{24}(z) \bigr]{}^{2/3}$}, (2)~conjugating by $\hat{\sigma}_{34}$, and (3)~readjusting the jump contours, if needed. Such a construction is identical to that of $P_{-\beta}(z)$, and so we omit the explicit calculation.

Finally, set
 \begin{equation*}
 {\bf R}(z) =
 \begin{cases}
 \boldS(z) M^{-1}(z), & \text{$z$ outside the discs at $z=\pm\alpha$, $z=\pm\beta$},\\
 \boldS(z) \bigl(P^{(\pm\alpha)} \bigr)^{-1}(z), & \text{$z$ inside the discs at $z=\pm\alpha$},\\
 \boldS(z) \bigl(P^{(\pm\beta)} \bigr)^{-1}(z), & \text{$z$ inside the discs at $z=\pm\beta$}.
 \end{cases}
 \end{equation*}
We have the following proposition.

\begin{Proposition}
 On the boundary of the discs at $z=\pm\alpha,\pm\beta$,
 \begin{equation*}
 {\bf R}_+(z) = {\bf R}_-(z) \bigl[\mathbb{I} + \OO \bigl(n^{-1}\bigr) \bigr], \qquad n\to\infty.
 \end{equation*}
\end{Proposition}
\begin{proof}
 We furnish a proof only on the boundary of the disc $z=-\beta$. Let us assume the boundary of the disc at $z=-\beta$ is oriented counterclockwise. Then,
 \begin{align*}
 {\bf R}_+(z) &{}= \boldS(z) \bigl(P^{(-\beta)}\bigr)^{-1}(z) = \boldS(z)M^{-1}(z)M(z) \bigl(P^{(-\beta)}\bigr)^{-1}(z) \\
 &{}= {\bf R}_-(z)M(z) \bigl(P^{(-\beta)}\bigr)^{-1}(z).
 \end{align*}
Now, we have that
 \begin{gather*}
 M(z) \bigl(P^{(-\beta)}\bigr)^{-1}(z)\\
 \qquad{}= M(z) [1\oplus \hat{P}^{-1}(z)\oplus 1] \bigl[1\oplus [\zeta(z)]^{-\frac{1}{4}\sigma_3}U \oplus 1\bigr] M^{-1}(z)\\
 \qquad{}= M(z)
 \underbrace{\bigl[1\oplus \sigma_3 {\rm e}^{-\frac{2}{3}\zeta(z)^{3/2}}\mathcal{A}^{-1}(\zeta(z))\oplus 1\bigr]}_{1\oplus \hat{P}^{-1}(z)\oplus 1}
 \bigl[1\oplus [\zeta(z)]^{-\frac{1}{4}\sigma_3}U\sigma_3 \oplus 1\bigr]\\
 \qquad\quad{}\times\underbrace{\bigl[1\oplus \sigma_3 U^{-1} [\zeta(z)]^{\frac{1}{4}\sigma_3}n^{-\frac{1}{6}\sigma_3}\oplus 1\bigr] \tilde{\mathfrak{H}}^{-1}(z)}_{M^{-1}(z)}\\
 \qquad{}=\underbrace{\tilde{\mathfrak{H}}(z) \bigl[1\oplus n^{\frac{1}{6}\sigma_3}[\zeta(z)]^{-\frac{1}{4}\sigma_3} U \sigma_3\oplus 1\bigr]}_{M(z)}
 \bigl[1\oplus \sigma_3 {\rm e}^{-\frac{2}{3}\zeta(z)^{3/2}}\mathcal{A}^{-1}(\zeta(z))\oplus 1\bigr]\\
 \qquad\quad{}\times \bigl[1\oplus n^{-\frac{1}{6}\sigma_3}\oplus 1\bigr] \tilde{\mathfrak{H}}^{-1}(z)\\
 \qquad{}=\tilde{\mathfrak{H}}(z) \bigl[1\oplus n^{\frac{1}{6}\sigma_3}[\zeta(z)]^{-\frac{1}{4}\sigma_3} U \sigma_3\oplus 1\bigr]
 \bigl[1\oplus \sigma_3 U^{-1}[\zeta(z)]^{\frac{1}{4}\sigma_3}\bigl[\mathbb{I}_{2\times 2} + \OO\bigl(\zeta^{-1}\bigr)\bigr]\oplus 1\bigr]\\
 \qquad\quad{}\times\bigl[1\oplus n^{-\frac{1}{6}\sigma_3}\oplus 1\bigr] \tilde{\mathfrak{H}}^{-1}(z)\\
 \qquad{}=\tilde{\mathfrak{H}}(z) \bigl[1\oplus n^{\frac{1}{6}\sigma_3}\oplus 1\bigr] \bigl[1\oplus\bigl[\mathbb{I}_{2\times 2} + \OO\bigl(\zeta^{-1}\bigr)\bigr]\oplus 1\bigr]\bigl[1\oplus n^{-\frac{1}{6}\sigma_3}\oplus 1\bigr] \tilde{\mathfrak{H}}^{-1}(z).
 \end{gather*}
 Since $\OO\bigl(\zeta^{-1}\bigr) = \OO\bigl(n^{-2/3}\bigr)$ in the $z$-plane, by crude estimation we see that the above jump is at worst
 \begin{equation*}
 \bigl[1\oplus n^{\frac{1}{6}\sigma_3}\oplus 1\bigr] \bigl[1\oplus\bigl[\mathbb{I}_{2\times 2} + \OO\bigl(\zeta^{-1}\bigr)\bigr]\oplus 1\bigr]\bigl[1\oplus n^{-\frac{1}{6}\sigma_3}\oplus 1\bigr] = \mathbb{I} + \OO\bigl(n^{-1/3}\bigr), \qquad n\to \infty
 \end{equation*}
 (conjugation by an $n$-independent holomorphic matrix does not change the error). In fact, as~is typical for the Airy parametrix, a more detailed expansion of the above product reveals that cancellations occur, and one finds that indeed
 \begin{equation*}
 \boldR_-(z)^{-1}\boldR_+(z) = M(z) \bigl(P^{(-\beta)}\bigr)^{-1}(z) = \mathbb{I} + \OO\bigl(n^{-1}\bigr),\qquad n\to \infty.
 \end{equation*}
 Thus, we have proven what is desired.
\end{proof}

Aside from the jumps on the local discs, all other jumps of $\boldR(z)$ are exponentially close to the identity matrix as $n\to \infty$, as we have already seen; furthermore, this bound improves as~${z\to \infty}$. If we let $\Gamma_{\boldR}$ denote the union of all of the jump contours of $\boldR(z)$, and $J_{\boldR}(z)\colon \Gamma_{\boldR} \to \CC$ denote the jumps of $\boldR(z)$, then we have that
 \begin{equation*}
 J_{\boldR}(z) = \mathbb{I} + \OO\bigl({\rm e}^{-cn(|z|+1)}\bigr), \qquad n\to\infty,
 \end{equation*}
uniformly for $z\in \Gamma_{\boldR} \setminus \left(D_{\pm\alpha}\cup D_{\pm\beta}\right)$, for some positive constant $c>0$. Using similar arguments as in~\cite{DK1,DKM}, we can conclude the following.

\begin{Proposition}
 There exists a constant $C>0$ such that, for every $n$, and uniformly for ${z\in \CC\setminus \Gamma_{\boldR}}$,
 \begin{equation*}
 ||\boldR(z)-\mathbb{I}|| \leq \frac{C}{n(1+|z|)}.
 \end{equation*}
\end{Proposition}
This allows us to conclude that $\boldR(z)$ admits a large $n$ asymptotic expansion in powers of~$n^{-1}$, cf.~\cite[Theorem 7.10]{DKMVZ}. Since all of the Riemann--Hilbert transformations we made were invertible, by tracing back these transformations we can obtain a large $n$ asymptotic expansion of $\YY(z)$. This concludes the steepest descent analysis.

\section{Proof of main theorem and concluding remarks}\label{section7}
\subsection{Calculation of the free energy}
We are now in a position where we are able to calculate the asymptotics of the partition function. As noted in the introduction, the partition function for the 2-matrix model can be written in terms of an isomonodromic $\boldsymbol{\tau}$ function, as per~\cite{Bertola-Marchal}. Explicitly,
the $\boldsymbol{\tau}$-function is expressible in terms of the solution of the Riemann--Hilbert problem for $\YY(z)$. Since we have succeeded in
finding the asymptotics of $\YY(z)$, we in turn can produce an asymptotic expression for the partition function. The expression for the $\boldsymbol{\tau}$-differential is
 \begin{gather}
 {\rm d}\log \boldsymbol{\tau}_n(\tau,t) = \left\langle \YY^{-1}\YY'
 \begin{pmatrix}
 0 & 0\\
 0 & \frac{\partial \WW} {\partial \tau} \WW^{-1}
 \end{pmatrix}\right\rangle {\rm d}\tau
 + \left\langle \YY^{-1}\YY' \begin{pmatrix}
 0 & 0\\
 0 & \frac{\partial \WW }{\partial T} \WW^{-1}
 \end{pmatrix}\right\rangle {\rm d}T \nonumber\\
 \hphantom{{\rm d}\log \boldsymbol{\tau}_n(\tau,t) = }{}
 + \left\langle \YY^{-1}\YY' \begin{pmatrix}
 0 & 0\\
 0 & \frac{\partial \WW }{\partial \bar{T}} \WW^{-1}
 \end{pmatrix}\right\rangle {\rm d}\bar{T},\label{tau-diff-1}
 \end{gather}
where $'$ denotes the derivative with respect to the spectral variable $z$, $\WW$ is the matrix which appears in the first transformation, $T := qt$, $\bar{T} = t/q$, and
for a given matrix-valued $1$-form~$A(z)$, we have introduced the notation
 \begin{equation*}
 \langle A(z) \rangle := \Res_{z=\infty} \tr \left[ A(z) \right].
 \end{equation*}
The partition function's differential is then
 \begin{equation*}
 {\rm d}\frac{\log Z_{n}(\tau,t,q;N)}{\tau^{\frac{-n(n-1)}{2}}C_{n,N}} = {\rm d}\log\biggl[\biggl(\frac{\tau}{t^2}\biggr)^{\frac{n}{2}(\frac{n}{3}-1)} \boldsymbol{\tau}_n \biggr],
 \end{equation*}
 and the free energy can be written as
 \begin{equation*}
 {\rm d}\log \frac{Z_{n}(\tau,t,q;N)}{Z_{n}(\tau,0,0;N)} = {\rm d}\log \left[\frac{\bigl(1-\tau^2\bigr)^{n^2/2}}{\tau^{n^2/3}t^{n(\frac{n}{3}-1)}}\boldsymbol{\tau}_n\right].
 \end{equation*}
The derivations of the above formulae for the $\boldsymbol{\tau}$-function and partition function are sketched in Appendix~\ref{Appendix-B}. Combined with the asymptotic formulae of the previous sections, these formulae allow us to finally prove the main theorem. Let us recall the main theorem (this is identical to Theorem~\ref{main-theorem}).\looseness=-1

\begin{Theorem*}
 Let $(\tau,t,H)$ belong to the region $\mathcal D$. Then, as $n\to \infty$,
 \begin{align*}
 F(\tau,t,H):={}&\lim_{n\to \infty} \frac{1}{n^2}\log \frac{ Z_{n}(\tau,t,H;n) }{ Z_{n}(\tau,0,0;n)} \\
={}& \frac{3}{4} + \frac{1}{2}\log \frac{\bigl(1-\tau^2\bigr)\sigma(\tau,t,H)}{-3t} - \int_{0}^{\sigma(\tau,t,H)}\biggl(\lambda(u)-\frac{1}{2}\lambda(u)^2\biggr) \frac{{\rm d}u}{u},
\end{align*}
 where $\lambda(u)$ is the rational function
 \begin{equation*}
 \lambda(u) = -\frac{1}{t}\biggl[\frac{1}{9}\tau^2u\bigl(u^2-3\bigr) + \frac{1}{3}\frac{u}{(u+1)^2} - \frac{2}{3}\biggl(\frac{u}{u^2-1}\biggr)^2[\cosh H -1]\biggr],
 \end{equation*}
 and $\sigma(\tau,t,H)$ from Definition~$\ref{def:sigmaDtcrx}$.
 \end{Theorem*}

\begin{proof}
 Let $\hat{\sigma} := {\rm diag}(1,-1/3,-1/3,-1/3)$, and set
 \begin{align*}
 &\YY^{(1)} := \lim_{z\to \infty} z\bigl[\YY(z)z^{-n\hat{\sigma}} - \mathbb{I}\bigr],\qquad
 \YY^{(2)} := \lim_{z\to \infty} z^2\biggl[\YY(z)z^{-n\hat{\sigma}} - \mathbb{I} - \frac{\YY^{(1)}}{z}\biggr].
 \end{align*}
 Using the expression \eqref{tau-diff-1}, and the fact that the matrices $\mathcal{M}^{\bullet}$, $\bullet \in \{\tau,t,q\}$, are all polynomials of degree at most $2$ in $z$, we can derive an explicit expression for the tau function in terms of the entries of $\YY(z)$:
 \begin{gather*}
 \varrho_{\tau} = -\frac{\tau q n}{t}\biggl[\YY^{(1)}_{12}\YY^{(1)}_{4 1} + \YY^{(1)}_{2 2}\YY^{(1)}_{4 2} + \YY^{(1)}_{3 2}\YY^{(1)}_{4 3} + \YY^{(1)}_{42}\YY^{(1)}_{4 4} - 2\YY^{(2)}_{4 2}+ \frac{1}{\tau}\YY^{(1)}_{4 3}\\
 \hphantom{\varrho_{\tau} = -\frac{\tau q n}{t}\biggl[}{}
 - \frac{t}{\tau q}\YY^{(1)}_{34} - \frac{t}{\tau q}\YY^{(1)}_{2 3} \biggr],\\
 \varrho_t = -\frac{\tau}{4t}\varrho_{\tau} + \frac{n\tau q}{4t^2}\biggl[\YY^{(1)}_{43} + \YY^{(1)}_{32} - \frac{2n}{3\tau} - \frac{2t}{\tau q}\biggr], \qquad
 \varrho_q = -\frac{t}{q}\varrho_t.
 \end{gather*}
 The above three formulae tell us that we must compute the asymptotics of $\YY(z)$ to two subleading orders in $z$ at infinity. In an appropriately chosen sector (chosen so that we are outside of the lenses) at infinity, the asymptotics of $\YY(z)$ we have calculated are given by
 \begin{gather} \label{Y-asymptotics-final}
 \YY(z) =
 \!\begin{pmatrix}
 1 & 0\\
 0 & c_N^{-1} \hat{K}^{-1}
 \end{pmatrix}\!
 {\rm e}^{nL} [\mathbb{I} + n C_1 \cdot E_{24}] \bigl[\mathbb{I} + \OO\bigl(n^{-2}\bigr) \bigr] M(z) \GG^{-1}(z)
 \!\begin{pmatrix}
 {\rm e}^{n V(z)} & 0\\
 0 & \WW(z)
 \end{pmatrix}\!.\!\!
 \end{gather}
 This implies that we must compute:
 \begin{itemize}\itemsep=0pt\samepage
 \item the regular expansion of $M(z) B^{-1}(z)A^{-1}(z)$ at infinity to order $z^{-2}$,
 \item the expansion of the effective potentials $\Omega_j(z)$ to second order, $j=1,2,3,4$; in particular, we write
 \begin{align*}
 \tau \Omega_3(z) ={}&{-}\frac{3}{4}\frac{\tau^{4/3}}{(-t/q)^{1/3}}z^{4/3} - \frac{1}{2}\frac{\tau^{2/3}}{(-t/q)^{2/3}}z^{2/3} + \frac{1}{3}\log z + \ell_1\\
 &{}+ \frac{C_1}{z^{2/3}} + \frac{C_2}{z^{4/3}} + \frac{C_3}{z^2} + \OO\bigl(z^{-8/3}\bigr), \qquad z\to \infty,
 \end{align*}
 with similar expressions for $\Omega_2(z)$, $\Omega_4(z)$,
 \item the regular part of the expansion of $\WW(z)$, up to order $z^{-2}$.
 \end{itemize}

 The explicit expressions for $C_1$, $C_2$ are given as
 \begin{align*}
 C_1(a,b,c)={}& \frac{c^{3/2}\tau^{7/3}}{18b^{3/2}(-t)^{4/3}q^{1/6}}\bigl(3a^8b^4c^2 + 3a^6b^6c^2 + 3a^8b^2 + 3a^6b^4 + 3a^6b^2c^2\nonumber\\
 &\hphantom{\frac{c^{3/2}\tau^{7/3}}{18b^{3/2}(-t)^{4/3}q^{1/6}}\bigl(}{}
 + 3a^4b^6 + 3a^4b^4c^2 + 3a^2b^6c^2 - a^6 - 9a^4b^2 - 9a^2b^4 - b^6\bigr),\\
 C_2(a,b,c)={}& {-}\frac{c^2 \tau^{8/3}}{54 b^2 (-t)^{5/3} q^{1/3}} \bigl( 6a^{10}b^4c^2 + 9a^8b^6c^2 + 6a^6b^8c^2 + 2a^{10}b^2 - 6a^8b^4 + 2a^8b^2c^2 \nonumber\\
 &\hphantom{{-}\frac{c^2 \tau^{8/3}}{54 b^2 (-t)^{5/3} q^{1/3}} \bigl(}{}
 -6a^6b^6 - 6a^6b^4c^2 + 2a^4b^8 - 6a^4b^6c^2 + 2a^2b^8c^2 - a^8\\
 &\hphantom{{-}\frac{c^2 \tau^{8/3}}{54 b^2 (-t)^{5/3} q^{1/3}} \bigl(}{}
 - 10a^6b^2 - 12a^4b^4 - 10a^2b^6 - b^8\bigr).
 \end{align*}
 With these calculations in hand, one may insert the asymptotic expansions derived above into~\eqref{Y-asymptotics-final}, and furthermore insert these expressions into the expressions of
 $\rho_{\tau}$, $\rho_t$, and $\rho_q$, to obtain
 \begin{align*}
 &\frac{1}{n^2}\varrho_{\tau}= \frac{2}{3}\frac{C_1}{\tau^{1/3}} \Bigl(-\frac{q}{t}\Bigr)^{2/3} + 4 C_2\Bigl(-\frac{\tau q}{t}\Bigr)^{1/3} + \OO\bigl(n^{-2}\bigr),\\
 &\frac{1}{n^2}\varrho_t= -\frac{\tau}{4t}\varrho_\tau +\frac{2}{\tau^{1/3}}C_1 -\frac{q^2}{27t^3} - \frac{2q}{3t^2} + \OO\bigl(n^{-2}\bigr),\\
 &\frac{1}{n^2}\varrho_q= -\frac{t}{q}\varrho_t + \OO\bigl(n^{-2}\bigr).
 \end{align*}
 Now, $\tau$, $t$, $q$, $C_1$, and $C_2$ all have explicit expressions in terms of $a$, $b$, $c$, and so we can compute the tau differential in these coordinates. We thus obtain an expression for ${\rm d}F$. This expression is exact,
 and can thus be integrated. Since we know the value of this expression at $(\tau,0,0)$ (namely, $F(\tau,0,0)\equiv 0$), we obtain an exact expression for the function $F(\tau,t,H)$. The final result is yielded by comparing the result of the previously outlined calculation to the expression~\eqref{Kazakov-Free-Energy}.
\end{proof}

\subsection{Concluding remarks}
This work is the first in our series of papers on the Ising transition on the $2$-matrix model: the study of the free energy of this
model at the multicritical point will be the subject of~\cite{DHL3, DHL2}. These works are perhaps one of the main reasons the $2$-matrix
model merits study, and will hopefully shed light on minimal matter coupled to gravity. Aside from this work, there are still many other
interesting questions we did not pursue in this work which absolutely merit further study. We discuss some of these problems below.

\textit{Genus expansion of the free energy.} One of the more pressing questions we did not
address is the conjectured genus expansion of the free energy~\cite{PGZ, Eynard}. It is conjectured that the following~${N\to \infty}$ asymptotic expansion holds, in the genus $0$ region of phase space we considered in this work:
 \begin{equation*}
 \frac{1}{N^2} \log \frac{Z_N(\tau,t,q;N)}{Z_N(\tau,0,0;N)} \sim F(\tau,t,q) + \sum_{g=1}^{\infty} \frac{F_g(\tau,t,q)}{N^{2g}},
 \end{equation*}
where $F_g(\tau,t,q)$ is the analog of \eqref{Ising-random-partition-generating} (see also equation \eqref{Ising-random-partition-n}), but with the sum taken over all connected, $4$-regular, genus $g$ maps. We fully expect that this is the case, and this result should follow from a more careful look at the Riemann--Hilbert analysis performed in this work. Concretely, there have been several works which
give formulae for the genus $1$ contribution to the free energy $F_1(\tau,t,H)$~\cite{AAD,Eynard}. It is of therefore of interest to confirm if our results agree with the findings of these works.

\textit{Critical partition function and Painlev\'{e}~I.} One corollary of the famous Yang--Lee theorem~\mbox{\cite{YangLee2, YangLee1}} is that
the partition function for the Ising model should be an analytic function of all of its parameters, away from the segment $H=0$, $[0,T_{\rm c})$,
where $T_{\rm c}$ is the critical temperature. Indeed, this formally seems to be the case for the model studied here, even when coupled
to gravity. However, the spectral curve seemingly undergoes a phase transition on the critical surface for $H\neq 0$, characterized by
the critical curves $\gamma_b$ depicted in Figure~\ref{fig:PhaseDiagram3D}. This transition is characterized by a merging of the
measures $\mu_1$ and $\mu_2$ discussed in Remark~\ref{phase-transitions-section}. This does not necessarily imply that the partition
function itself undergoes a transition, but certainly closer attention to this case is needed. Some preliminary analysis reveals the
appearance of a~special $3\times 3$ parametrix, which involves Painlev\'{e}~I. The Lax pair associated to this parametrix has
appeared in the literature~\cite{JKT}, but the parametrix itself requires a more careful analysis, especially if one wants to identify
the particular solution appearing when one approaches $\gamma_b$. More generally, this parametrix, along with the $(3,4)$ parametrix
discussed in Part II of this work~\cite{DHL2}, belong to a hierarchy of string equations of type $(3,p)$, where $p = 3k+1, 3k+2$. A~more comprehensive study of these equations, as well as the more general $(q,p)$ string equations, is needed, if one is interested in completing the picture of minimal models coupled to gravity.

\textit{Yang--Lee zeros of the partition function.} Another tantalizing feature of the model we studied comes from the works of~\cite{AAM1,AAM2, Staudacher}. It is well-known that the Ising model~\cite{YangLee2, YangLee1} that the partition function for the
Ising model, as a function of the complexified parameter ${\rm e}^H$, has all of its zeros on the unit circle. This is often thought of as
the mechanism through which the ferromagnetic phase transition occurs, and is comes with its own associated minimal CFT
(the Yang--Lee edge, the so-called ``$(2,5)$'' minimal model, cf.~\cite{Cardy}). However, it is far from obvious that sums of
such partition functions should also exhibit this feature. Nevertheless, the numerical results of~\cite{AAM1,AAM2,AAD, Staudacher}
seem to indicate that this is indeed the case. A proper study of this phenomenon would require the study of the partition
function defined by \eqref{eq:twomatrixmodel} for \textit{complex} values of the parameters, as one would need to study this model
with imaginary external field ($H\to {\rm i}H$). A~drawback of our current approach is that the construction of the spectral curve
relied on the ansatz that the associated ``$S$-curves'' all lie on the real axis; it is not clear that this remains true when
the parameters of the model are complexified (indeed, this is \textit{not} the case for the $1$-matrix model, cf., for example, \cite{Bertola-Tovbis,BGM}). It would be interesting if one could characterize these $S$-curves in terms of some
canonical quadratic differential, as has been performed for some closely related matrix models (see~\cite{KS, Rakhmanov} for
this construction pertaining to the $1$-matrix model with/without hard walls, and~\cite{MFS1,MFS2,MFS3} for the construction
related to the external source model).

\appendix

\section{Expansion of the uniformizing coordinate \\ near the branch points} \label{Appendix-A}
\begin{figure}[t]
 \centering
\includegraphics{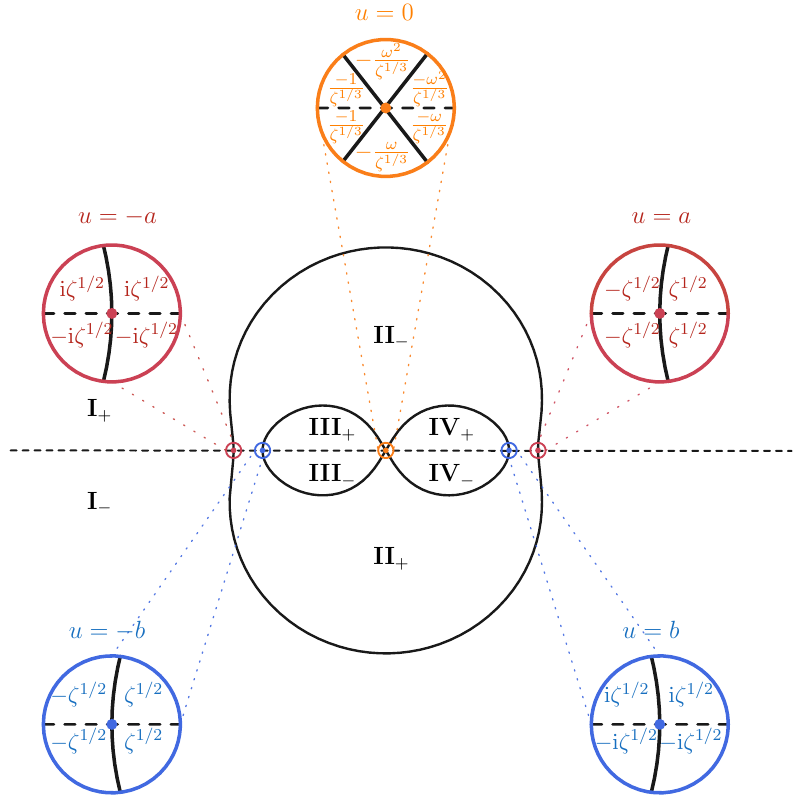}

 \caption{The uniformizing plane for generic values of the parameters $(a,b,c)$. The leading asymptotic behavior of $u_j(z)$ on each sheet is given in the appropriate local coordinate. For example, in a neighborhood of $u=a$, the local coordinate $\zeta = {\rm const}\cdot (z-\beta)$, and $u_2(z) \sim -\zeta^{1/2} \bigl[1 + \OO\bigl(\zeta^{1/2}\bigr)\bigr]$, whereas $u_3(z) \sim \zeta^{1/2} \bigl[1 + \OO\bigl(\zeta^{1/2}\bigr)\bigr]$. }
 \label{fig:Uniformizing-Plane}
\end{figure}

Here we list the relevant expansions of the uniformizing coordinate on each sheet of the spectral curve. We have included Figure~\ref{fig:Uniformizing-Plane}, which depicts the leading order asymptotics of the uniformizing coordinate near each of the branch points, for the convenience of the reader. The~expansions at infinity hold for all $(a,b,c) \in R = \{0<b\leq 1,\, 0< c\leq b,\, 1 \leq a \leq b^{-1}\}$, and so we list them first. Let $A= A(a,b,c)$. Then,
\begin{align*}
 &u_1(z)= \frac{z}{A} - \frac{\bigl(a^2+b^2\bigr)A}{z} - \frac{\bigl(\frac{5}{3}a^2b^2 + a^4 + b^4\bigr)A^3}{z^3} - \frac{\bigl(\frac{14}{3}\bigl(a^4b^2 + a^2b^4\bigr) + 2a^6+ 2b^6\bigr)A^5}{z^5}\\
 & \hphantom{u_1(z)=}{}
 + \OO\bigl(z^{-7}\bigr),\\
 &u_2(z)=
 \begin{cases}
 -\left(\dfrac{a^2b^2}{3}\right)^{1/3} \dfrac{\omega A^{1/3}}{z^{1/3}} + \dfrac{(a^2+b^2)A}{3z} &\\
 \quad{}- \left(\dfrac{3}{a^2b^2}\right)^{1/3} \dfrac{\omega^2(a^4+a^2b^2+b^4)A^{5/3}}{9z^{5/3}} + \OO\bigl(z^{-7/3}\bigr), & \operatorname{Im}z >0,\\[1mm]
 -\left(\dfrac{a^2b^2}{3}\right)^{1/3} \dfrac{\omega^2 A^{1/3}}{z^{1/3}} + \dfrac{(a^2+b^2)A}{3z} &\\
 \quad{} - \left(\dfrac{3}{a^2b^2}\right)^{1/3} \dfrac{\omega(a^4+a^2b^2+b^4)A^{5/3}}{9z^{5/3}} + \OO\bigl(z^{-7/3}\bigr), & \operatorname{Im}z <0,
 \end{cases}\\
 &u_3(z)= -\biggl(\frac{a^2b^2}{3}\biggr)^{1/3} \frac{A^{1/3}}{z^{1/3}} + \frac{(a^2+b^2)A}{3z} - \biggl(\frac{3}{a^2b^2}\biggr)^{1/3} \frac{(a^4+a^2b^2+b^4)A^{5/3}}{9z^{5/3}} + \OO\bigl(z^{-7/3}\bigr),\\
 &u_4(z)=
 \begin{cases}
 -\left(\dfrac{a^2b^2}{3}\right)^{1/3} \dfrac{\omega^2 A^{1/3}}{z^{1/3}} + \dfrac{(a^2+b^2)A}{3z} &\\
 \quad{}- \left(\dfrac{3}{a^2b^2}\right)^{1/3} \dfrac{\omega(a^4+a^2b^2+b^4)A^{5/3}}{9z^{5/3}} + \OO\bigl(z^{-7/3}\bigr), & \operatorname{Im}z >0,\\[1mm]
 -\left(\dfrac{a^2b^2}{3}\right)^{1/3} \dfrac{\omega A^{1/3}}{z^{1/3}} + \dfrac{(a^2+b^2)A}{3z} &\\
 \quad{}- \left(\dfrac{3}{a^2b^2}\right)^{1/3} \dfrac{\omega^2(a^4+a^2b^2+b^4)A^{5/3}}{9z^{5/3}} + \OO\bigl(z^{-7/3}\bigr), & \operatorname{Im}z <0,
 \end{cases}
\end{align*}

On the other hand, the local expansions of the uniformizing coordinate differ in the noncritical/critical cases and multicritical cases. We indicate the behavior of the uniformization coordinate around these branch points here.

\subsection[Expansion of the uniformizing coordinate in the generic and critical cases]{Expansion of the uniformizing coordinate\\ in the generic and critical cases}
All of the following expansions are valid for $0 < b < 1$, $1\leq a\leq b$ (and also $0<c\leq b$, but this parameter does not play a role). We also let $A:=A(a,b,c) >0$ be as in \eqref{A-param}.
 \begin{itemize}\itemsep=0pt
 \item \textit{Expansion at $z= + \alpha$.}
 As $z\to \alpha$, letting $\zeta = z-\alpha$,
 \begin{align*}
 &u_1(z)= a +\frac{a^{3/2}}{A^{\frac{1}{2}}(a^2-b^2)^{\frac{1}{2}}}\zeta^{\frac{1}{2}} + \frac{a^2\bigl(3a^2-7b^2\bigr)}{6A(a^2-b^2)^2} \zeta + \frac{a^{5/2}C_2}{72A^{\frac{3}{2}}(a^2-b^2)^{\frac{7}{2}}}\zeta^{\frac{3}{2}} + \OO\bigl(\zeta^2\bigr),\\
 &u_2(z)= a-\frac{a^{3/2}}{A^{\frac{1}{2}}(a^2-b^2)^{\frac{1}{2}}}\zeta^{\frac{1}{2}} + \frac{a^2\bigl(3a^2-7b^2\bigr)}{6A(a^2-b^2)^2} \zeta - \frac{a^{5/2}C_2}{72A^{\frac{3}{2}}(a^2-b^2)^{\frac{7}{2}}}\zeta^{\frac{3}{2}} + \OO\bigl(\zeta^2\bigr),
 \end{align*}
 and $u_3(z)$, $u_4(z)$ have regular expansions. Here
$C_2:=C_2(a,b,c)>0$.

 \item \textit{Expansion at $z= - \alpha$.}
 As $z\to -\alpha$, letting $\zeta = z+\alpha$,
 \begin{align*}
 &u_1(z)=
 \begin{cases}
 -a +\dfrac{{\rm i}a^{3/2}}{A^{\frac{1}{2}}(a^2-b^2)^{\frac{1}{2}}}\zeta^{\frac{1}{2}} + \dfrac{a^2(3a^2-7b^2)}{6A(a^2-b^2)^2} \zeta &\\
 \quad{} + \dfrac{{\rm i}a^{5/2}C_2}{72A^{\frac{3}{2}}(a^2-b^2)^{\frac{7}{2}}}\zeta^{\frac{3}{2}} + \OO\bigl(\zeta^2\bigr), & \operatorname{Im} \zeta >0,\\[2mm]
 -a -\dfrac{{\rm i}a^{3/2}}{A^{\frac{1}{2}}(a^2-b^2)^{\frac{1}{2}}}\zeta^{\frac{1}{2}} + \dfrac{a^2(3a^2-7b^2)}{6A(a^2-b^2)^2} \zeta &\\
 \quad{} - \dfrac{{\rm i}a^{5/2}C_2}{72A^{\frac{3}{2}}(a^2-b^2)^{\frac{7}{2}}}\zeta^{\frac{3}{2}} + \OO\bigl(\zeta^2\bigr), &\operatorname{Im} \zeta <0,
 \end{cases}\\
 &u_2(z)=
 \begin{cases}
 -a -\dfrac{{\rm i}a^{3/2}}{A^{\frac{1}{2}}(a^2-b^2)^{\frac{1}{2}}}\zeta^{\frac{1}{2}} + \dfrac{a^2(3a^2-7b^2)}{6A(a^2-b^2)^2} \zeta &\\
 \quad{} - \dfrac{{\rm i}a^{5/2}C_2}{72A^{\frac{3}{2}}(a^2-b^2)^{\frac{7}{2}}}\zeta^{\frac{3}{2}} + \OO\bigl(\zeta^2\bigr), & \operatorname{Im} \zeta >0,\\[2mm]
 -a +\dfrac{{\rm i}a^{3/2}}{A^{\frac{1}{2}}(a^2-b^2)^{\frac{1}{2}}}\zeta^{\frac{1}{2}} + \dfrac{a^2(3a^2-7b^2)}{6A(a^2-b^2)^2} \zeta & \\
 \quad{}+ \dfrac{{\rm i}a^{5/2}C_2}{72A^{\frac{3}{2}}(a^2-b^2)^{\frac{7}{2}}}\zeta^{\frac{3}{2}} + \OO\bigl(\zeta^2\bigr), &\operatorname{Im}\zeta <0,
 \end{cases}
 \end{align*}
 and $u_3(z)$, $u_4(z)$ have regular expansions. Here
$C_2:=C_2(a,b,c)>0$.
 \end{itemize}
 The expansions at $z=\pm \beta$ can be obtained in a similar manner, by exchanging the roles of~$a$,~$b$ in the expansions. We obtain that
 \begin{itemize}\itemsep=0pt
 \item \textit{Expansion at $z= +\beta$.}
 As $z\to \beta$, letting $\zeta = z-\beta$,
 \begin{align*}
 &u_2(z)=
 \begin{cases}
 b+ \dfrac{{\rm i}b^{3/2}}{A^{\frac{1}{2}}(a^2-b^2)^{\frac{1}{2}}}\zeta^{1/2} -\dfrac{(7a^2 - 3b^2)b^2}{6A(a^2 - b^2)^2} \zeta &\\
 \quad{} + \dfrac{{\rm i}b^{5/2}\tilde{C}_2}{72A^{\frac{5}{2}}(a^2-b^2)^{\frac{7}{2}}}\zeta^{\frac{3}{2}} + \OO\bigl(\zeta^2\bigr), & \operatorname{Im}\zeta > 0, \\[2mm]
 b -\dfrac{{\rm i}b^{3/2}}{A^{\frac{1}{2}}(a^2-b^2)^{\frac{1}{2}}}\zeta^{1/2} -\dfrac{(7a^2 - 3b^2)b^2}{6A(a^2 - b^2)^2} \zeta & \\
 \quad {} - \dfrac{{\rm i}b^{5/2}\tilde{C}_2}{72A^{\frac{5}{2}}(a^2-b^2)^{\frac{7}{2}}}\zeta^{\frac{3}{2}} + \OO\bigl(\zeta^2\bigr), & \operatorname{Im}\zeta < 0,
 \end{cases}\\
 &u_4(z)=
 \begin{cases}
 b -\dfrac{{\rm i}b^{3/2}}{A^{\frac{1}{2}}(a^2-b^2)^{\frac{1}{2}}}\zeta^{1/2} -\dfrac{(7a^2 - 3b^2)b^2}{6A(a^2 - b^2)^2} \zeta &\\
 \quad {}- \dfrac{{\rm i}b^{5/2}\tilde{C}_2}{72A^{\frac{5}{2}}(a^2-b^2)^{\frac{7}{2}}}\zeta^{\frac{3}{2}} + \OO\bigl(\zeta^2\bigr), & \operatorname{Im}\zeta > 0, \\[2mm]
 b + \dfrac{{\rm i}b^{3/2}}{A^{\frac{1}{2}}(a^2-b^2)^{\frac{1}{2}}}\zeta^{1/2} -\dfrac{(7a^2 - 3b^2)b^2}{6A(a^2 - b^2)^2} \zeta & \\
 \quad{} + \dfrac{{\rm i}b^{5/2}\tilde{C}_2}{72A^{\frac{5}{2}}(a^2-b^2)^{\frac{7}{2}}}\zeta^{\frac{3}{2}} + \OO\bigl(\zeta^2\bigr), & \operatorname{Im}\zeta < 0,
 \end{cases}
 \end{align*}
 and $u_1(z)$, $u_3(z)$ have regular expansions. Here
$\tilde{C}_2 := \tilde{C}_2(a,b,c)>0$.

 \item \textit{Expansion at $z= - \beta$.}
 As $z\to -\beta$, letting $\zeta = z+\beta$,
 \begin{align*}
 &u_2(z)= -b-\frac{b^{3/2}}{A^{\frac{1}{2}}(a^2-b^2)^{\frac{1}{2}}}\zeta^{1/2} -\frac{\bigl(7a^2 - 3b^2\bigr)b^2}{6A(a^2 - b^2)^2} \zeta - \frac{b^{5/2}\tilde{C}_2}{72A^{\frac{5}{2}}(a^2-b^2)^{\frac{7}{2}}}\zeta^{\frac{3}{2}} + \OO\bigl(\zeta^2\bigr),\\
 &u_3(z)= -b +\frac{b^{3/2}}{A^{\frac{1}{2}}(a^2-b^2)^{\frac{1}{2}}}\zeta^{1/2} -\frac{\bigl(7a^2 - 3b^2\bigr)b^2}{6A(a^2 - b^2)^2} \zeta + \frac{b^{5/2}\tilde{C}_2}{72A^{\frac{5}{2}}(a^2-b^2)^{\frac{7}{2}}}\zeta^{\frac{3}{2}} + \OO\bigl(\zeta^2\bigr),
 \end{align*}
 and $u_1(z)$, $u_4(z)$ have regular expansions. Here
 \[
 \tilde{C}_2 := \tilde{C}_2(a,b)= 101a^4 - 30a^2b^2 + 9b^4 = C_2(b,a).
 \]
 \end{itemize}

\subsection[Expansions of the uniformizing coordinate at the multicritical point and $\gamma_b$]{Expansions of the uniformizing coordinate\\ at the multicritical point and $\boldsymbol{\gamma_b}$}
\begin{itemize}\itemsep=0pt
 \item \textit{Expansion at $z= +\alpha$}. Let $\zeta:=z-\alpha$, $A = A(1,1,c)>0$; as $\zeta\to 0$, we have the expansions:
 \begin{align*}
 &u_1(z) = 1+ \biggl(\frac{3}{4A}\biggr)^{1/3}\!\zeta^{1/3} + \frac{3}{4}\biggl(\frac{3}{4A}\biggr)^{2/3}\!\zeta^{2/3} + \frac{21}{64A}\!\zeta + \frac{37}{192}\biggl(\frac{3}{4A}\biggr)^{4/3}\zeta^{4/3} + \OO\bigl(\zeta^{5/3}\bigr),\\
 &u_2(z) =
 \begin{cases}
 1+ \left(\dfrac{3}{4A}\right)^{1/3}\omega^2\zeta^{1/3} + \dfrac{3}{4}\left(\dfrac{3}{4A}\right)^{2/3}\omega\zeta^{2/3} + \dfrac{21}{64A}\zeta&\\
 \quad{}+ \dfrac{37}{192}\left(\dfrac{3}{4A}\right)^{4/3}\omega^2\zeta^{4/3} + \OO\bigl(\zeta^{5/3}\bigr),
 & \operatorname{Im}\zeta >0,\\[1mm]
 1+ \left(\dfrac{3}{4A}\right)^{1/3}\omega\zeta^{1/3} + \dfrac{3}{4}\left(\dfrac{3}{4A}\right)^{2/3}\omega^2\zeta^{2/3} + \dfrac{21}{64A}\zeta&\\
 \quad{}+ \dfrac{37}{192}\left(\dfrac{3}{4A}\right)^{4/3}\omega\zeta^{4/3} + \OO\bigl(\zeta^{5/3}\bigr), & \operatorname{Im}\zeta <0,
 \end{cases}\\
 &u_3(z) =-\frac{1}{3} + \frac{1}{64A}\zeta - \frac{27}{16384A^2}\zeta^2 + \frac{891}{4194304A^3}\zeta^3 + \OO\bigl(\zeta^4\bigr),\\
 &u_4(z) =
 \begin{cases}
 1+ \left(\dfrac{3}{4A}\right)^{1/3}\omega\zeta^{1/3} + \dfrac{3}{4}\left(\dfrac{3}{4A}\right)^{2/3}\omega^2\zeta^{2/3}+ \dfrac{21}{64A}\zeta&\\
 \quad{}+ \dfrac{37}{192}\left(\dfrac{3}{4A}\right)^{4/3}\omega\zeta^{4/3} + \OO\bigl(\zeta^{5/3}\bigr), & \operatorname{Im}\zeta >0,\\[0.8mm]
 1+ \left(\dfrac{3}{4A}\right)^{1/3}\omega^2\zeta^{1/3} + \dfrac{3}{4}\left(\dfrac{3}{4A}\right)^{2/3}\omega\zeta^{2/3}+ \dfrac{21}{64A}\zeta&\\
 \quad{}+ \dfrac{37}{192}\left(\dfrac{3}{4A}\right)^{4/3}\omega^2\zeta^{4/3} + \OO\bigl(\zeta^{5/3}\bigr), & \operatorname{Im}\zeta <0.
 \end{cases}
 \end{align*}
 \item \textit{Expansion at $z= -\alpha$}. Let $\zeta:=z+\alpha$, $A= A(1,1,c)>0$; as $\zeta\to 0$, we have the expansions:
 \begin{align*}
 &u_1(z) =
 \begin{cases}
 -1+ \left(\dfrac{3}{4A}\right)^{1/3}\omega\zeta^{1/3} - \dfrac{3}{4}\left(\dfrac{3}{4A}\right)^{2/3}\omega^2\zeta^{2/3}+ \dfrac{21}{64A}\zeta &\\
 \quad{} - \dfrac{37}{192}\left(\dfrac{3}{4A}\right)^{4/3}\omega\zeta^{4/3} + \OO\bigl(\zeta^{5/3}\bigr), & \operatorname{Im} \zeta >0,\\[1mm]
 -1+ \left(\dfrac{3}{4A}\right)^{1/3}\omega^2\zeta^{1/3} - \dfrac{3}{4}\left(\dfrac{3}{4A}\right)^{2/3}\omega\zeta^{2/3}+ \dfrac{21}{64A}\zeta &\\
 \quad{} - \dfrac{37}{192}\left(\dfrac{3}{4A}\right)^{4/3}\omega^2\zeta^{4/3} + \OO\bigl(\zeta^{5/3}\bigr), & \operatorname{Im} \zeta <0,
 \end{cases}\\
 &u_2(z) =
 \begin{cases}
 -1+ \left(\dfrac{3}{4A}\right)^{1/3}\omega^2\zeta^{1/3} - \dfrac{3}{4}\left(\dfrac{3}{4A}\right)^{2/3}\omega\zeta^{2/3}+ \dfrac{21}{64A}\zeta &\\
 \quad{} - \dfrac{37}{192}\left(\dfrac{3}{4A}\right)^{4/3}\omega^2\zeta^{4/3} + \OO\bigl(\zeta^{5/3}\bigr),
 & \operatorname{Im} \zeta >0,\\[1mm]
 -1+ \left(\dfrac{3}{4A}\right)^{1/3}\omega\zeta^{1/3} - \dfrac{3}{4}\left(\dfrac{3}{4A}\right)^{2/3}\omega^2\zeta^{2/3}+ \dfrac{21}{64A}\zeta &\\
 \quad{}- \dfrac{37}{192}\left(\dfrac{3}{4A}\right)^{4/3}\omega\zeta^{4/3} + \OO\bigl(\zeta^{5/3}\bigr), & \operatorname{Im} \zeta <0,
 \end{cases}\\
 &u_3(z) = -1+ \biggl(\frac{3}{4A}\biggr)^{1/3}\! \zeta^{1/3}\! - \frac{3}{4}\biggl(\frac{3}{4A}\biggr)^{2/3}\! \zeta^{2/3}\! + \frac{21}{64A} \zeta - \frac{37}{192}\biggl(\frac{3}{4A}\biggr)^{4/3}\zeta^{4/3}\! + \OO\bigl(\zeta^{5/3}\bigr),\\
 &u_4(z) = \frac{1}{3} + \frac{1}{64A}\zeta + \frac{27}{16384A^2}\zeta^2 + \frac{891}{4194304A^3}\zeta^3 + \OO\bigl(\zeta^4\bigr).
 \end{align*}
\end{itemize}

We also note the following symmetry properties of these coordinates:
 \begin{Proposition}
 The expansions $u_j(z)$ and $u_j(-z)$, $j=1,2,3,4$ are related by
 \begin{alignat*}{3}
 &u_1(-z) = -u_1(z), \qquad&& u_2(-z) = -u_2(z),&\\
 &u_3(-z) = -u_4(z), \qquad&& u_4(-z) = -u_3(z).&
 \end{alignat*}
 \end{Proposition}
\begin{proof}
 This proposition can be observed directly from the expansions of $u_j(z)$ given above.
\end{proof}

\section{The isomonodromic tau function} \label{Appendix-B}
 The partition function of the $2$-matrix model was identified with an isomonodromic $\boldsymbol{\tau}$-function by Bertola and Marchal,
 cf.~\cite{Bertola-Marchal}. Their derivation is relatively straightforward, and applies almost directly to our situation. However, some of
 the details of the calculation are different enough that the proof merits discussion. We present the proof of the fact
 that the partition function for the $2$-matrix model is an isomonodromic $\boldsymbol{\tau}$-function here. The proof mirrors almost directly that of
 Bertola and Marchal's; one should consult their work and references therein for further details and commentary.
 For sake of readability, let us introduce the notation, for a given matrix-valued $1$-form $A(z)$,
 \begin{equation*}
 \langle A(z) \rangle := \Res_{z=\infty} \tr [ A(z)].
 \end{equation*}
 Recall that, in general, the $n^{\rm th}$ monic biorthogonal polynomial $p_n(z)$ with respect to the weight\footnote{Here we take $N >0$ as a free parameter, in general different from the index of the biorthogonal polynomial. We will later set $N=n$.}
 \begin{equation*}
 {\rm e}^{N W(z,w)} := {\rm e}^{N[\tau zw - V(z;T) - V(w;\bar{T})]}
 \end{equation*}
 \big(Here \smash{$V(z;T):= \frac{1}{2}z^2 + \frac{T}{4}z^4$}, and we take $T$, $\bar{T}$ as independent parameters in general\big) on contour(s) $(\Gamma,\Gamma)$
 is given in terms of the solution to the following Riemann--Hilbert problem:
 \begin{enumerate}\itemsep=0pt
 \item[$(1)$] $\YY_{n}(z)$ is a piecewise analytic function in $\CC\setminus \Gamma$.
 \item[$(2)$] $\YY_n(z)$ has boundary values
 \begin{gather*}
 \YY_{n,+}(z) = \YY_{n,-}(z) {\rm e}^{-NV(z)}\!\biggl[ \mathbb{I} + w(z)E_{12} + \frac{1}{N\tau} w'(z)E_{13} + \frac{1}{N^2\tau^2} w''(z)E_{14} \biggr],\quad z\in \Gamma,
 \end{gather*}
 where \smash{$w(z) :=\int_{\Gamma}{\rm e}^{N [\tau zw - V(w,\bar{T})]} {\rm d}w$} (note that in the general situation, the integration contour $\Gamma$ is allowed to differ from the jump contour $\Gamma$; in this work, this distinction is irrelevant).
 \item[$(3)$] As $z\to \infty$,
 \begin{gather*}
 \YY_n(z) = \biggl[\mathbb{I} + \frac{\YY_{n}^{(1)}}{z} + \frac{\YY_{n}^{(2)}}{z^2} + \OO\biggl(\frac{1}{z^3}\biggr)\biggr]
 \begin{pmatrix}
 z^n & 0 & 0\\
 0 & z^{-m_n - 1}\mathbb{I}_{r_n} & 0\\
 0 & 0 & z^{-m_n}\mathbb{I}_{3-r_n}\\
 \end{pmatrix},
 \end{gather*}
 where $m_n\in \mathbb{N}$, $r_n \in \{0,1,2\}$ are such that $n = 3 m_n + r_n$.
 \end{enumerate}
Note that, when $n = 3k$ is a multiple of $3$, the above coincides with the Riemann Hilbert problem analyzed in the present work.
If we put $h_n\bigl(\tau,T,\bar{T};N\bigr)$ to be the norming constant for the monic $n^{\rm th}$-order polynomials,
 \begin{equation*}
 \int_{\Gamma}\int_{\Gamma} p_n(z)q_m(w) {\rm e}^{N W(z,w)}{\rm d}z{\rm d}w = h_n\bigl(\tau,T,\bar{T};N\bigr)\delta_{nm},
 \end{equation*}
then the \textit{partition function} of this model is defined to be
 \begin{equation*}
 Z_n\bigl(\tau,T,\bar{T};N\bigr) := \prod_{k=0}^{n-1}h_n\bigl(\tau,T,\bar{T};N\bigr).
 \end{equation*}
We now summarize some of the basic results pertaining to this RHP which we will need in our analysis of the $\tau$-differential.

The solution to this Riemann--Hilbert problem is given explicitly,
 \begin{equation*}
 \YY_n(z) =
 \begin{pmatrix}
 p_n(z) & C_\Gamma[p_n w](z) & \frac{1}{N\tau} C_\Gamma[p_n w'](z) & \frac{1}{N^2\tau^2} C_\Gamma[p_n w''](z)\\[0.9mm]
 Q_{n-1}(z) & C_\Gamma[Q_{n-1} w](z) & \frac{1}{N\tau} C_\Gamma[Q_{n-1} w'](z) & \frac{1}{N^2\tau^2} C_\Gamma[Q_{n-1} w''](z)\\[0.9mm]
 Q_{n-2}(z) & C_\Gamma[Q_{n-2} w](z) & \frac{1}{N\tau} C_\Gamma[Q_{n-2} w'](z) & \frac{1}{N^2\tau^2} C_\Gamma[Q_{n-2} w''](z)\\[0.9mm]
 Q_{n-3}(z) & C_\Gamma[Q_{n-3} w](z) & \frac{1}{N\tau} C_\Gamma[Q_{n-3} w'](z) & \frac{1}{N^2\tau^2} C_\Gamma[Q_{n-3} w''](z)
 \end{pmatrix},
 \end{equation*}
where $Q_{n-1}$, $Q_{n-2}$, and $Q_{n-3}$ are some appropriately chosen polynomials of degrees $n-1$, $n-2$, and $n-3$, respectively, $p_n(z)$ is the
$n^{\rm th}$ monic biorthogonal polynomial, and
 \begin{equation*}
 C_{\Gamma}[f](z) := \frac{1}{2\pi {\rm i}} \int_{\Gamma} \frac{f(x)}{x-z} {\rm d}z
 \end{equation*}
denotes the Cauchy transform with respect to the contour $\Gamma$. We can also relate the Riemann--Hilbert problem for $\YY_n(z)$ to the Riemann--Hilbert
problem for $\YY_{n+1}(z)$ by means of a \textit{raising operator}
 \begin{equation*}
 \YY_{n+1}(z) = R_n(z) \YY_n(z),
 \end{equation*}
where \smash{$R_n(z) := R^{(1)}_n z + R^{(0)}_n$} is a degree $1$ matrix-valued polynomial in $z$. The existence of $R_n(z)$ follows immediately from the fact that
$\YY_{n+1}(z)\YY_n^{-1}(z)$ has no jumps, and thus extends to an entire function. The asymptotics of $\YY_{n+1}(z)\YY_n^{-1}(z)$ uniquely fix the form of
$R_n(z)$. Setting $\alpha_0 := r_N + 1$, we have that \smash{$R^{(1)}_n = E_{11}$}, whereas the matrix \smash{$\bigl(R^{(0)}_n\bigr)_{jk}$} has entries as given in the following table:
 \begin{equation*}%\renewcommand{\arraystretch}{1.5}
 \begin{tabular}{c| c c c }
 & $k=\alpha_0$ & $k = 1$ & $k\neq 1, \alpha_0$\\
 \hline
 $j=\alpha_0$ & $\frac{-(\YY_{n}^{(2)})_{\alpha_0,1} +\sum_{\ell\neq \alpha_0} (\YY_{n}^{(1)})_{\alpha_0,\ell} \bigl(\YY_{n}^{(1)}\bigr)_{\ell,1}}{(\YY_{n,1})_{\alpha_0,1}}$ & $-\bigl(\YY_{n}^{(1)}\bigr)_{\alpha_0,1}$ & $-\bigl(\YY_{n}^{(1)}\bigr)_{\alpha_0,k}$ \vspace{1mm}\\
 $j=1$ & $\frac{1}{(\YY_{n}^{(1)})_{\alpha_0,1}}$ & $0$ & $0$ \vspace{1mm}\\
 $j\neq 1, \alpha_0$ & $\frac{(\YY_{n}^{(1)})_{j,1}}{(\YY_{n}^{(1)})_{\alpha_0,1}}$ & $0$ & $\delta_{jk}$ \\
 \end{tabular}
 \end{equation*}
This implies that the matrix $R_n(z)$ is determined entirely in terms of $\YY_n(z)$.

 The isomonodromic $\boldsymbol{\tau}$-differential corresponding to $\YY := \YY_n$ is defined to be
 \begin{equation*}
 {\rm d}\log \boldsymbol{\tau}_n := \bigl\langle \YY_n^{-1}\YY_n' {\rm d}\hat{\WW} \hat{\WW}^{-1}\bigr\rangle,
 \end{equation*}
 where $\hat{\WW}(z)$ is essentially the augmented $\WW$-matrix from the first transformation
 \begin{equation*}
 \hat{\WW}(z) :=
 \begin{pmatrix}
 {\rm e}^{-NV(z)} & 0 \\
 0 & \WW^{-1}(z)
 \end{pmatrix},
 \end{equation*}
 with the only difference here being that \smash{$V(z) = \frac{1}{2}z^2 + \frac{\bar{T}}{4}z^4$}. For now, we treat the parameter~$N$ in the matrix \smash{$\hat{\WW}$} as a \textit{fixed} parameter independent of the index of the polynomial $n$; we shall later set $N = n$. Since multiplication by \smash{$\hat{\WW}$} yields a constant jump RHP, we have the following~formulae for the differential of \smash{${\rm d}\hat{\WW} \hat{\WW}^{-1}$}:
 renders $\YY$
 \begin{equation*}
 {\rm d}\hat{\WW} \hat{\WW}^{-1} =
 \begin{pmatrix}
 0 & 0\\
 0 & -\WW^{-1}\frac{\partial \WW }{\partial \tau}
 \end{pmatrix} {\rm d}\tau +
 \begin{pmatrix}
 0 & 0\\
 0 & -\WW^{-1}\frac{\partial\WW }{\partial T}
 \end{pmatrix} {\rm d}T
 + \begin{pmatrix}
 -Nz^4/4 & 0\\
 0 & 0
 \end{pmatrix} {\rm d}\bar{T}.
 \end{equation*}

In~\cite{Bertola-Marchal}, the parameters of the isomonodromic $\boldsymbol{\tau}$-differential come from the coefficients of the potential (as opposed to
our case, where one of the parameters is the coefficient of the interaction term $XY$; namely, $\tau$), and the
definition of \smash{${\rm d}\hat{\WW} \hat{\WW}^{-1}$} is slightly different. However, it is only important to the proof that \smash{$\hat{\WW}$} renders the jumps of $\YY$ constant, which we have already seen (indeed, this was the point of the first transformation $\YY\mapsto\XX$).

The biorthogonal polynomials double as a particular sequence of multiple orthogonal polynomials. Summarizing the arguments of~\cite{Bertola-Marchal}, one can use the sequence of raising operators arising from the multiple orthogonality to produce the next \textit{biorthogonal polynomial} in the sequence. Let us
denote this raising operator generically by $R_n(z)$. $R_n(z)$ is defined so that
 \begin{equation}\label{raising-op}
 \YY_{n+1}(z) = R_{n}(z) \YY_n(z).
 \end{equation}
Generically, $R_n(z)$ is a degree $1$ polynomial in $z$; its inverse is also a degree $1$ polynomial in $z$. We have the following proposition.
\begin{Proposition}
 \begin{equation*}
 {\rm d}\log\frac{\boldsymbol{\tau}_{n+1}}{\boldsymbol{\tau}_{n}} = -\bigl\langle R_n^{-1} R_n' {\rm d}\YY_n \YY^{-1}_n\bigr\rangle.
 \end{equation*}
\end{Proposition}
 \begin{proof}
 By equation \eqref{raising-op}, we have that
 \begin{equation*}
 \YY_{n+1}^{-1} \YY_{n+1}' = \YY_n^{-1} R_n^{-1} R_n'\YY_n + \YY_n^{-1}\YY_n'.
 \end{equation*}
 Thus, the quotient of $\boldsymbol{\tau}$ differentials is
 \begin{align*}
 \begin{split}
 {\rm d}\log\frac{\boldsymbol{\tau}_{n+1}}{\boldsymbol{\tau}_{n}} &= \bigl\langle \bigl[\YY_n^{-1} R_n^{-1} R_n'\YY_n + \YY_n^{-1}\YY_n'\bigr] {\rm d}\hat{\WW}\hat{\WW}^{-1}\bigr\rangle - \bigl\langle \YY_n^{-1}\YY_n' {\rm d}\hat{\WW} \hat{\WW}^{-1} \bigr\rangle\\
 &=\bigl\langle\YY_n^{-1} R_n^{-1} R_n'\YY_n {\rm d}\hat{\WW}\hat{\WW}^{-1}\bigr\rangle.
 \end{split}
 \end{align*}
 Now, recall that $\YY_n \WW$ has constant jumps, and so the differential ${\rm d}[\YY_n \WW]\WW^{-1} \YY_n^{-1}$ has coefficients which
 are polynomial in $z$, by the standard Liouville argument. This statement can be rewritten as
 \begin{equation*}
 \YY_n {\rm d}\hat{\WW}\hat{\WW}^{-1} \YY_n^{-1} = {\rm d}\YY_n \YY_n^{-1} + \text{polynomial}.
 \end{equation*}
 Inserting the above into our expression for the $\boldsymbol{\tau}$-quotient, we obtain that
 \begin{equation*}
 {\rm d}\log\frac{\boldsymbol{\tau}_{n+1}}{\boldsymbol{\tau}_{n}} = \bigl\langle R_n^{-1}R_n'[\text{polynomial}] - R_n^{-1}R_n' {\rm d}\YY_n\YY_n^{-1}\bigr\rangle;
 \end{equation*}
 since $R_n^{-1}R_n'$ is a polynomial in $z$, the first term is a polynomial in $z$, and thus has no residues at infinity. This completes the proof.
 \end{proof}

We can use the explicit form of the raising operators obtained before to get an expression for the $\tau$-differential in terms of the coefficients of
$\YY_n(z)$. The exact expression is summarized by the following proposition.
\begin{Proposition}
 The ratio of consecutive $\boldsymbol{\tau}$ differentials, up to multiplication by a function independent of the isomonodromic times, is given by
 \begin{equation*}
 \frac{\boldsymbol{\tau}_{n+1}}{\boldsymbol{\tau}_n} = \bigl(\YY_{n}^{(1)}\bigr)_{1,\alpha_0}.
 \end{equation*}
\end{Proposition}
\begin{proof}
 This follows immediately from inspection of the previous proposition, and the explicit form of the matrices $R_n(z)$. For details, see~\cite{Bertola-Marchal}.
\end{proof}

Furthermore, we can relate the coefficients $\bigl(\YY_{n}^{(1)}\bigr)_{1,\alpha_0}$ to the biorthogonality coefficients $h_n$.
\begin{Proposition}
 The matrix coefficient $\bigl(\YY_{n}^{(1)}\bigr)_{1,\alpha_0}$ is given in terms of the $n^{\rm th}$ normalizing constant of the biorthogonal polynomials:
 \begin{equation*}
 \bigl(\YY_{n}^{(1)}\bigr)_{1,\alpha_0} = \biggl(\frac{T}{\tau}\biggr)^S h_n,
 \end{equation*}
 where $S\in \mathbb{N}$, $\alpha_0 \in \{0,1,2\}$ are such that $n = 3S + \alpha_0 -1$.
\end{Proposition}

\begin{Proposition}
 The isomonodromic $\boldsymbol{\tau}$-function $\boldsymbol{\tau}_n$ is related to the partition function $Z_n(\tau,t,\allowbreak H;N)$ for $n$ a multiple of $3$ by
 the formula
 \begin{equation*}
 Z_n\bigl(\tau,T,\bar{T};N\bigr) = \Bigl(\frac{\tau}{T}\Bigr)^{\frac{n}{2}(\frac{n}{3}-1)} \boldsymbol{\tau}_n.
 \end{equation*}
\end{Proposition}
\begin{Remark}
 We remark that there is a minor error in the statement of the above proposition in~\cite{Bertola-Marchal}. The power of $(T/\tau)$ should be inverse to what it reads in their Theorem~3.4 (see page~17). The proof is otherwise correct, and there are no essential changes to the results otherwise.
\end{Remark}

Thus, we have an explicit expression for the partition function in terms of quantities we can calculate, by making the change of variables $T = q t$, $\bar{T} = q^{-1} t$. However, the calculation of the differential in the variable $\bar{T}$ is computationally difficult, as it would require us to calculate the expansion of $\YY$ to 4 subleading terms. However, we can use the fact that the partition function is symmetric in $T$, $\bar{T}$; in other words,
 \begin{equation*}
 {\rm d} \log Z_n\bigl(\tau,T,\bar{T};N\bigr) = {\rm d} \log Z_n\bigl(\tau,\bar{T},T;N\bigr).
 \end{equation*}
If we set
 \begin{align*}
 &\varpi_{\tau}\bigl(\tau,T,\bar{T},n;N\bigr) := \bigl\langle \YY_n^{-1}\YY_n' \hat{\WW}_\tau \hat{\WW}^{-1}\bigr\rangle,\qquad
 \varpi_{T}\bigl(\tau,T,\bar{T},n;N\bigr) := \bigl\langle \YY_n^{-1}\YY_n' \hat{\WW}_T \hat{\WW}^{-1}\bigr\rangle,
 \end{align*}
then we have the following proposition, which follows immediately from the symmetry
\[
\log Z_n(T,\bar{T}) = \log Z_n(\bar{T},T),
\] and our previous calculations.

\begin{Proposition}
 The partition function is given by the expression
 \begin{gather*}
 {\rm d} \log Z_n = {\rm d}\log\Bigl(\frac{\tau}{T\bar{T}}\Bigr)^{\frac{n}{2}(\frac{n}{3}-1)} + \varpi_{\tau}\bigl(\tau,T,\bar{T},n;N\bigr) {\rm d}\tau +
 \varpi_{T}\bigl(\tau,T,\bar{T},n;N\bigr) {\rm d}T\\
 \hphantom{{\rm d} \log Z_n =}{}
 + \varpi_{T}\bigl(\tau,\bar{T},T,n;N\bigr) {\rm d}\bar{T}.
 \end{gather*}
\end{Proposition}

\section[Explicit formulae for the spectral curve and the critical surface]{Explicit formulae for the spectral curve \\ and the critical surface}\label{Critical-Surface-Appendix}
Here we present explicit formulae for the spectral curve $\mathfrak{S}(X,Y) = 0$, and the critical surface $\Xi(\tau,t,q) = 0$. Both of these formulae are rather longer and not particularly enlightening. We~have therefore relinquished their presentation in the main text and placed them here, for the sake of completeness.

\subsection{Implicit formula for the spectral curve}
One can eliminate the parameter $u$ from the pair of functions $(X(u),Y(u))$ to obtain an implicit formula for the Riemann surface these functions parametrize
 \begin{equation*}
 \mathfrak{S}(X(u),Y(u)) = 0.
 \end{equation*}
$\mathfrak{S}(X,Y)$ is a degree $6$ polynomial in $X$ and $Y$, with rational coefficients in the variables $\tau$, $t$, $q$, and $\sigma$ (recall that $\sigma$ is defined as a special solution to the algebraic equation \eqref{critical-surface-ideal}). Explicitly, this polynomial is
 \begin{align*}
 \mathfrak{S}(X,Y)={}& \tau q X^4 + \tau q^{-1}Y^4 -tX^3Y^3 - qX^3Y - q^{-1}Y^3X + t\tau^{-1}X^2Y^2\\
 &{}{+}\, \mathfrak{s}_2(\sigma;\tau,t,q) X^2 + \mathfrak{s}_2\bigl(\sigma;\tau,t,q^{-1}\bigr) Y^2 + \mathfrak{s}_1(\sigma;\tau,t,q) XY + \mathfrak{s}_0(\sigma;\tau,t,q).\nonumber
 \end{align*}
where $\mathfrak{s}_i(\sigma;\tau,t,q)$ are given by
 \begin{gather*}
 \mathfrak{s}_2(\sigma;\tau,t,q) = \frac{1-6\sigma-3\sigma^2}{27\tau t (\sigma+1)^3} - \frac{1}{27\tau t}\bigl(\sigma^{3} \tau^{2}+3 \sigma^{2} \tau^{2}-9 \sigma \tau^{2}-27 \tau^{2}+1\bigr) \nonumber\\
 \hphantom{\mathfrak{s}_2(\sigma;\tau,t,q) =}{}
 -\frac{(q-1)\sigma}{27\tau t(\sigma^2 - 1)^3}\bigl[\bigl(9 \tau^{2}-9\bigr)+9(q+1) \sigma +\bigl(4q^{-1}-\bigl(28 \tau^{2}+6\bigr) \bigr) \sigma^{2} \\
 \hphantom{\mathfrak{s}_2(\sigma;\tau,t,q) =}\qquad{}
 -6(q+1 ) \sigma^{3}+\bigl(30 \tau^{2}+3\bigr)\sigma^{4}+(q+1) \sigma^{5}-12\sigma^{6} \tau^{2}+\sigma^{8} \tau^{2}\bigr]\nonumber,\\
 \mathfrak{s}_1(\sigma;\tau,t,q) = -\frac{8 \bigl(5 \sigma +3\bigr)}{81 (\sigma +1)^{2} t} -\frac{\sigma^{6} \tau^{2}-15 \sigma^{4} \tau^{2}+27 \sigma^{2} \tau^{2}-3 \sigma^{2}+243 \tau^{2}+24 \sigma +171}{243 t}\nonumber\\
 \hphantom{\mathfrak{s}_1(\sigma;\tau,t,q) =}{}
 -\frac{4\sigma^3\bigl(\sigma^2+3\bigr)}{81t(\sigma^2 - 1)^2}\bigl[q+q^{-1}-2\bigr],\\
 \mathfrak{s}_0(\sigma;\tau,t,q) = -\frac{\sigma}{19683t^2\tau q^2(\sigma^2 - 1)^4}\biggl[15 q \sigma^{6} \tau^{2}+9 q \sigma^{5} t +\bigl(12-111 \tau^{2}\bigr) q\sigma^{4} \nonumber \\
 \hphantom{\mathfrak{s}_0(\sigma;\tau,t,q) =}\qquad{}
 +15\biggl(q^{2}-\frac{6}{5}t q +1\biggr) \sigma^{3}+\bigl(177 \tau^{2}-15\bigr) q \sigma^{2}-54\biggl(q^{2}-\frac{1}{6}t q +1\biggr) \sigma\nonumber\\
 \hphantom{\mathfrak{s}_0(\sigma;\tau,t,q) =}\qquad{}
 + 81\bigl(1-\tau^{2}\bigr) q\biggr]^2.
 \end{gather*}

\subsection{Implicit formula for the critical surface}
Recall that the critical surface is characterized by the vanishing of the discriminant of the polynomial $\mathfrak{I}(\sigma;\tau,t,q)$, defined in
equation \eqref{critical-surface-ideal}. In other words, let $\Xi(\tau,t,q)$ denote this discriminant. Then, we have that
 \begin{equation*}
 (\tau,t,q) \in S_{L} \cup S_{H} \ \Longrightarrow \ \Xi(\tau,t,q) = 0.
 \end{equation*}
Explicitly, we can write this curve as $\bigl(C = \frac{1}{2}\bigl(q+q^{-1}\bigr) = \cosh(H)\bigr)$
 \begin{align*}
 \mathcal{J}(\tau,t,C) :={}& 32400000 t \tau^{4}C^5 + 1350000 \tau^{4} \bigl(\tau^{4}+288 t^{2}-2 \tau^{2}+1\bigr)C^4\\
 &{}{+}\, 36000 t \tau^{2} \bigl(3160 \tau^{6}+28026 t^{2} \tau^{2}-2745 \tau^{4}+4374 t^{2}-1215 \tau^{2}\bigr)C^3\\
 &{}{+}\, \bigl(3840000 \tau^{12}-423014400 t^{2} \tau^{8}-10560000 \tau^{10}-1700611200 t^{4} \tau^{4}\\
 &{}\, \quad-235612800 t^{2} \tau^{6}+ 7440000 \tau^{8}-306110016 t^{6}+2173003200 t^{4} \tau^{2}\\
 &{}\, \quad-666646200 t^{2} \tau^{4}+1440000 \tau^{6}+ 25223400 t^{2} \tau^{2}-2160000 \tau^{4}\bigr)C^2\\
 &{}{+}\,\bigl(-10174464 t \tau^{12}+403107840 t^{3} \tau^{8}+60549120 t \tau^{10}-3809369088 t^{5} \tau^{4}\\
 &{}\, \quad+2205895680 t^{3} \tau^{6}
 -197475840 t \tau^{8}-3673320192 t^{7}+8865853056 t^{5} \tau^{2}\\
 &{}\, \quad-2979218880 t^{3} \tau^{4}+147277440 t \tau^{6}-25509168 t^{5}
 +249930360 t^{3} \tau^{2}\\
 &{}\, \quad-1451520 t \tau^{4}+1275264 t \tau^{2}\bigr)C\\
 &{}{-}\,65536 \tau^{16}+5308416 t^{2} \tau^{12}+1277952 \tau^{14}-161243136 t^{4} \tau^{8}-29859840 t^{2} \tau^{10}\\
 &{}{-}\,7827456 \tau^{12}
 +2176782336 t^{6} \tau^{4}-362797056 t^{4} \tau^{6}-21772800 t^{2} \tau^{8}\\
 &{}{+}\, 15861760 \tau^{10}-11019960576 t^{8}+8979227136 t^{6} \tau^{2}
 -5048925696 t^{4} \tau^{4}\\
 &{}{+}\,806993280 t^{2} \tau^{6}-12437760 \tau^{8}-153055008 t^{6}+583036704 t^{4} \tau^{2}\\
 &{}{+}\,42729120 t^{2} \tau^{4}+2624256 \tau^{6}-531441 t^{4}+6601824 t^{2} \tau^{2}\\
 &{}{+}\,546048 \tau^{4}+20736 \tau^{2} = 0.
 \end{align*}
One can readily check that $\vec{r}_{{\rm low}}(b,c)$, $\vec{r}_{{\rm high}}(b,c)$ (defined below), indeed both parametrize the above algebraic equation.

\section{Parametrization of the phase space} \label{app:param}
Here we discuss the parametrization of $D_q$ by \eqref{tau-param}--\eqref{q-param} in more detail and prove Proposition~\ref{prop:phase}.

\subsection{The critical surfaces}
Let us start by studying the critical surfaces defined in Definition~\ref{defn:criticalsurfaces}.
Note that, by definition, the low and high temperature curves are given by the pair of parametric equations $\vec{r}_{\bullet}(b,c) = \langle \tau_{\bullet}(b,c), t_{\bullet}(b,c), q_{\bullet}(b,c)\rangle$, $\bullet \in \{{\rm low}, {\rm high}\}$, defined by
 \begin{align*}
 &\vec{r}_{{\rm low}}(b,c):= \Biggl\langle \frac{b^3}{\sqrt{N_1(b,c)N_2(b,c)}},-\frac{bc\bigl(b^6c^2 + \frac{4}{3}b^2c^2 + 1\bigr)}{3N_1(b,c)N_2(b,c)},\frac{bcN_2(b,c)}{N_1(b,c)}\Biggr\rangle,\\
 &\vec{r}_{{\rm high}}(b,c):= \Biggl\langle \frac{1}{\sqrt{M_1(b,c)M_2(b,c)}},-\frac{bc\bigl(4b^2c^2 + 3b^2 + 3c^2\bigr)}{9M_1(b,c)M_2(b,c)},\frac{cM_2(b,c)}{bM_1(b,c)}\Biggr\rangle,
 \end{align*}
respectively,
where
 \begin{alignat*}{3}
 &N_1(b,c)=2b^4c^2 + b^2 + c^2, \qquad&& N_2(b,c) = b^4 + b^2c^2 + 2,&\\
 &M_1(b,c)=b^2c^2 + 2c^2 + 1, \qquad&& M_2(b,c) =b^2c^2 + 2b^2 + 1.&
 \end{alignat*}
In both cases, the parameter range for $(b,c)$ is $0 < b < 1$, $0< c < b$. The critical curve $\gamma_b$, defined as the boundary connecting the surfaces $S_{{\rm low}}$ and $S_{{\rm high}}$, is given parametrically as
 \begin{equation*}
 \vec{r}(c) = \Biggl\langle\frac{1}{\sqrt{(3c^2+1)(c^2+3)}},-\frac{c\bigl(7c^2 + 3\bigr)}{9(c^2+3)(3c^2+1)},\frac{c\bigl(c^2 + 3\bigr)}{3c^2+1}\Biggr\rangle, \qquad 0 < c < 1,
 \end{equation*}
 which results from the limit $b\to 1$ in either of the critical surfaces $\vec{r}_{{\rm low}}$, $\vec{r}_{{\rm high}}$.

Note that by changing variables from $(\tau,t,q)$ to $\bigl(\tau^2,t,q\bigr)$ the parametrizations become rational and thus the surface $S_{{\rm low}}$ and $S_{{\rm high}}$ are algebraic.
\begin{Lemma} \label{lem:surface=graph}
 The union $S_{{\rm low}}\cup S_{{\rm high}}$ of the low- and high-temperature critical surface is the graph of a function of $0<\tau<1$ and $0<q<1$.
\end{Lemma}
\begin{proof}
It suffices to show that the projection of $S_{{\rm high}} \cup S_{{\rm low}}\cup \gamma_b$ to the $t=0$ plane is a bijection onto the square defined by $0 < \tau <1$ and $0 < q < 1$.

 The projection of the surface $S_{{\rm high}}$ to the $t=0$ plane is obtained by setting $t=0$ in its parameterization. The parameterization becomes singular near $b=c=0$, but this can be resolved by setting $\tilde c=c/b$, leading to the map
 \begin{equation} \label{eq:projection}
(\tau,q)= \Biggl(\frac{1}{\sqrt{(1+ 2\tilde c^2 b^2 + \tilde c^2b^4)(1+2b^2 +\tilde c^2b^4)}},
 \frac{\tilde c \bigl(1+2b^2 +\tilde c^2b^4\bigr)}{1+ 2\tilde c^2 b^2 + \tilde c^2b^4}
\Biggr).
\end{equation}
 Note that this defines a smooth map on $\mathbb R^2$, but we are mainly interested in its restriction to the closure of the parameter space given by unit square $0\leq b\leq 1$ and $0\leq \tilde c<1$. The map $(b,\tilde c) \mapsto (\tau,q)$ is differentiable and the Jacobian never vanishes in the open square. This implies that it is locally a diffeomorphism. The image of the boundary is given by the following observations:
\begin{itemize}\itemsep=0pt
\item For $b=0$, we find $\tau=1$ and $0\leq q \leq 1$.

\item For $b=1$, we obtain the projection of curve $\gamma_b$ to the $t=0$ plane.

\item For $\tilde c=0$, we find $q=0$ and $1/\sqrt{3} \leq \tau \leq 1$.

\item For $\tilde c=1$, we find $q=1$ and $1/4 \leq \tau \leq 1$.
\end{itemize}

 It is also easy to verify that the map \eqref{eq:projection} provides a bijection from the boundary of the unit square to the boundary of this region.

 Concluding, \eqref{eq:projection} is a continuous map from the closed unit square, that is locally diffeomorphic and maps the boundary bijective to the boundary of the image. Since $\phi$ is continuous and the unit square is compact, it is also a proper map (pre-image of compacts sets are compact). By furthering invoking the Hadamard--Caccioppoli theorem~\cite{KrantzParks}, one find that the map defines a~homeomorphism, from the closed unit disk onto its image. Moreover, the image of the boundary divides $\mathbb R^2$ into two components and the image equals the bounded component. In other words, the projection of $S_{{\rm high}}$ is the region enclosed by $\tau=1$, $q=0$, $q=1$ and the projection of $\gamma_b$ to the $t=0$ plane.

 In a similar fashion, one can show that the projection of $S_{{\rm low}}$ is the region enclosed by $\tau=0$, $q=0$, $q=1$ and the projection of $\gamma_b$ to the $t=0$ plane. The projections together thus fill out the unit square bijectively and this is proves the statement.
 \end{proof}

\subsection{Proof of Proposition~\ref{prop:phase}}
\begin{proof}
We will characterize the image of the parametrization inspired by the following simple principle: if $D$ is an open set with compact closure and $\phi\colon \overline D\to \mathbb R^n$ is a continuous function such that $\phi(D)$ open, then $\partial \phi(D) \subset\phi(\partial D)$. Since the Jacobian \eqref{eq:jacobian} does not vanish in the interior of $R$, the map $\Pi$ maps $R^\circ$ to an open subset of $\mathbb R^3$. However, $R$ is not compact and $\Pi$ is not continuous for $b\to 0$. Nevertheless, as we will show below by a limiting procedure, the boundary $\partial \Pi(R)$ can still be found from the behavior of $\Pi$ near the boundary~$\partial R$.

For $\eps>0$, define
$ R^{\eps}=R \cap \{ b > \eps\}.$
Then the closure of $R^{\eps}$ is compact and $\Pi$ is continuous up to the boundary. The boundary of $\Pi(R^{\eps})$ consists of the images of the boundary part of~$R^{\eps}$ given by respectively $a=b^{-1}$, $a=1$, $c=b$, $c=0$ and $b=\eps$. The first two conditions, lead to the critical surfaces $S_{{\rm low}}$ and $S_{{\rm high}}$ (note that for $\eps \downarrow 0$ this will only trace our part of these surface but when $\eps \downarrow 0$ we retrieve the entire surfaces again). The condition $c=b$ gives us (part~of) the plane $q=1$. The condition $c=0$ turns out to be obsolete as it leads to both~${t=0}$ and $q=0$ and this is the part where the critical surfaces meet the $t=0$ plane. So it remains to investigate the behavior of the image of $b=\eps$ when $\eps \downarrow 0$. Note that the image of $b=\eps$ under $\Pi$ gives a smooth surface. We claim that for $\eps \downarrow 0$ this surface will converge to parts of the planes $\tau=0$
 and $t=0$.

To study the $\eps \downarrow 0$ limit we start, as in the proof of Lemma~\ref{lem:surface=graph}, by setting
$c=\tilde c b$,
and consider the parametrization by $a$, $b$ and $\tilde c$. Now $0\leq \tilde c \leq 1$ and the limit $b= \eps\downarrow 0$ simply gives
\begin{equation*} %\label{eq:abct=0}
 \tau(a,b,c)\to 1/a^2, \qquad t(a,b, c)\to 0,\qquad q(a,b,c) \to \tilde c.
\end{equation*}
 One can readily verify that this parameterizes the $t=0$ wall of the phase space for $0 < q < 1$:
 \begin{equation*} %\label{eq:boundaryt=0}
 \{(\tau,t,q)\mid t = 0,\, 0 < \tau < 1,\, 0 < q < 1\}.
 \end{equation*}
The $\tau=0$ plane can obtained in a similar way. To this end, set
$
 a=\tilde a/b$, $\tilde c=c/b$,
 with $0<\tilde a\leq 1$ and $0\leq \tilde c \leq 1$.
 Then for $b=\eps\downarrow 0$ we find
 \begin{equation*}
 \tau( a,b, c)\to 0,\qquad t( a,b, c) \to -\frac{\tilde a^2 \tilde c}{3(1+\tilde a^2)(1+\tilde a^2 \tilde c^2)},\qquad
 q( a,b, c) \to \frac{\tilde c\bigl(
 1+\tilde a^2\bigr)}{1+\tilde a^2\tilde c^2}.
 \end{equation*}
 One can readily verify that the above parametrizes the following set:
 \begin{equation} \label{eq:tau0wall}
 \biggl\{(\tau,t,q)\mid \tau = 0,\, -\frac{1}{12}q<t<0,\, 0 < q < 1\biggr\}.
 \end{equation}
 Equality $t=-q/12$ in this limiting procedure is obtained by setting $\tilde a=1$ and thus $a=b^{-1}$. Therefore, \eqref{eq:tau0wall} is the part of the plane $\tau=0$ enclosed by $t=0$, $q=0$, $q=1$ and the part of the boundary of $S_{{\rm low}}$ that is in the $\tau=0$ plane.

 Concluding, after taking the limit $\eps \downarrow 0$, we find that $\Pi$ maps $R^{\circ}$ to the open region in $\mathbb R^3$ for which the boundary consists of the critical surfaces $S_{{\rm low}} \cup S_{{\rm high}}\cup \gamma_b$ and the four planes defined by respectively $q=0$, $q=1$, $t=0$ and $\tau=0$.

Next we show that $\Pi(R)=D$. Note that, by Lemma~\ref{lem:surface=graph}, we can write
\[
\Pi(R)=\{(t,\tau,q) \mid 0<\tau<1,\, q \in \mathbb R,\, t_0(\tau,q)< t <0\},
\]
for some function $t_0$. It remains to show that $t_0(\tau,q)=t_{\rm cr}(\tau,q)$. We first note that it by a~simple but tedious computation one can show that $\sigma=a^2bc$ indeed solves \eqref{critical-surface-ideal}, for~$\tau$,~$t$,~$q$ as defined in \eqref{tau-param}--\eqref{q-param}. It is also clear that $\sigma$ depends continuously on $a$, $b$, $c$ in the interior of $R$ and for~${b\to 0}$ we find $\sigma\to 0$. Moreover, since the Jacobian of $\Pi$ is not vanishing in the interior of $R$, we see that $\sigma$ is even differentiable as a function of $t$. We thus proved that $\sigma(t)=a^2bc$ for $t_0 \leq t <0$. Finally, at the boundary the Jacobian vanishes and $\sigma(t)=a^2bc$ is no longer differentiable as a function of $t$ and thus no longer analytic in $t$. By definition, we must have~${t_{\rm cr}=t_0}$. This finishes the proof.
\end{proof}

\section[Calculations from physics: the genus $0$ partition function]{Calculations from physics: the genus $\boldsymbol{0}$ partition function}
In this appendix, we study the genus $0$ free energy, which is defined in terms of the partition function
\begin{align} \label{C2:2-matrix-partition function}
 Z_n(\tau,t,H;N) := \int_{\mathcal{H}_n}\!\int_{\mathcal{H}_n}\!\! \exp \biggl\{\tr \biggl[\tau XY - \frac{1}{2}X^2 -\frac{1}{2}Y^2 - \frac{t{\rm e}^H}{4N} X^4 - \frac{t{\rm e}^{-H}}{4N}Y^4\biggr]\!\biggr\} {\rm d}X{\rm d}Y.\!\!
 \end{align}
The (genus $0$) free energy is then defined as the ratio
 \begin{equation*}
 F(\tau,t,H) = \lim_{N\to \infty}\frac{1}{N^2}\log \frac{Z_N(\tau,t,H;N)}{Z_N(\tau,0,0;N)}.
 \end{equation*}
More precisely, we reproduce the following results from the physics literature:
 \begin{enumerate}\itemsep=0pt
 \item[$(1)$] $Z_n(\tau,t,H;N)$ is a generating function for partition functions of the Ising model on genus $g$ $4$-regular graphs, in a sense we shall make precise. This is the result of Kazakov (for $H=0$) and Kazakov and Boulatov ($H\neq 0$)~\cite{Kazakov2,Kazakov1}. This is presented in Appendix~\ref{Genus0-A}.
 \item[$(2)$] The large $N$ limit of the genus $0$ partition function admits an exact expression in terms of the solution to a certain implicit equation. This
 is also the result of~\cite{Kazakov2,Kazakov1}. This is presented in Appendix~\ref{Genus0-B}.
 \item[$(3)$] We check that the results of these calculations indeed agree. We calculate the first few terms of this generating function by hand, and show that this result agrees with the exact formula from~\cite{Kazakov2,Kazakov1}. This is presented in Appendix~\ref{Genus0-C}.
 \item[$(4)$] The asymptotics of the Taylor coefficients of $\sigma(\tau,t,H)$ are derived, and explicitly so for the special case $H=0$, which is again a result of~\cite{Kazakov1}.
 \end{enumerate}
We present these results in more detail here, as the calculations are somewhat involved and the methods from the physics literature are perhaps unfamiliar to the target audience of this work. This appendix is meant to hopefully clarify these results; we do not claim any originality here. We also stress that these methods are formal: they are best thought of as guiding principles.

\subsection{Wick expansion of the partition function}\label{Genus0-A}
In this appendix, we replicate in detail the Wick expansion of the free energy, and show that it
is a formal generating function for the Ising model on random $4$-regular genus $g$ graphs.
We begin by considering the Gaussian model:
 \begin{equation}\label{Gaussian-measure}
 {\rm d}\mathbb{P}(X,Y) = \frac{1}{Z_n(\tau,0,0;N)}\exp \biggl\{N\tr\biggl[\tau XY - \frac{1}{2}X^2 -\frac{1}{2}Y^2\biggr]\biggr\} {\rm d}X{\rm d}Y.
 \end{equation}
Let $\langle\cdot\rangle_n$ denote the expected value with respect to this measure. Then, one can directly compute the following.
\begin{Lemma}The matrix elements $X_{ij}$, $Y_{ij}$
have the following means and covariances:
 \begin{alignat}{3}
 &\langle X_{ij} \rangle_n = 0, && \langle Y_{ij} \rangle_n = 0,&\label{2-point-function-2-matrix}\\
 &\langle X_{ij}X_{k\ell}\rangle_n = \langle Y_{ij}Y_{k\ell}\rangle_n = \frac{1}{N(1-\tau^2)} \delta_{i\ell}\delta_{jk},\qquad &&
 \langle X_{ij}Y_{k\ell}\rangle_n =\langle Y_{ij}X_{k\ell}\rangle_n = \frac{\tau}{N(1-\tau^2)} \delta_{i\ell}\delta_{jk}.&\nonumber
 \end{alignat}
\end{Lemma}
\begin{proof}
 We only sketch the idea of the proof here. The main point is that one can explicitly compute the moment generating function of this model.
 Let $J, K \in \mathcal{H}_n$ be a fixed Hermitian matrices, and consider the quantity
 \begin{equation*}
 f(J, K) := \langle \exp (\tr J X + K Y) \rangle_n,
 \end{equation*}
 where the expected value is taken with respect to the measure \eqref{Gaussian-measure}. By the usual trick of completing the square, one can compute that
 \begin{equation*}
 f(J,K) = \exp\biggl\{\frac{1}{N(1-\tau^2)}\tr\biggl[\frac{1}{2}J^2 + \frac{1}{2}K^2 -\tau JK\biggr]\biggr\}.
 \end{equation*}
 Taking derivatives with respect to the appropriate matrix elements of $J$, $K$ yields \eqref{2-point-function-2-matrix}.
\end{proof}

Now, since the entries of both $X$, $Y$ are all centered Gaussian variables, we can apply Wick's theorem to calculate the expected value of any higher order moment in terms of the covariances~\eqref{2-point-function-2-matrix}. Let us now explain how to calculate expected values of the form
 \begin{equation}\label{wick-expected-value}
 \Biggl\langle \prod_{p=1}^k \tr X^{i_p} \prod_{q=1}^\ell \tr Y^{j_q}\Biggr\rangle_n,
 \end{equation}
where $(i_1,\dots,i_p)$ and $(j_1,\dots,j_p)$ are sequences of positive integers. The main takeaway will be that one can interpret this expected value digramatically; we shall now explain this procedure. Let \smash{$K = \sum_{p=1}^k i_p$}, \smash{$L = \sum_{q=1}^\ell j_q$}, and $E = K+L$. We assume that $E$ is even; otherwise, this expected value is automatically zero. We can apply Wick's theorem to \eqref{wick-expected-value}, and convert this expected value into a sum over all possible pairings of the indices:
 \begin{equation*}
 \Biggl\langle \prod_{p=1}^k \tr X^{i_p} \prod_{q=1}^\ell \tr Y^{j_q}\Biggr\rangle_n = \sum_{I} \sum_{\pi \in \Pi_{E}}\prod_{\{(r_a,r_b),(r_c,r_d)\} \in \pi} \langle Z_{r_a r_b} Z_{r_c r_d}\rangle_n,
 \end{equation*}
where $\Pi_{E}$ denotes the set of pairings of $E$ elements, the sum $I$ runs over all of the indices of the traces $\tr X^{i_p}$, $p = 1,\dots,k$, and $\tr Y^{j_q}$, $q = 1,\dots,j$, and
 \begin{gather*}
 Z_{r_a r_b} \!:=\!
 \begin{cases}
 X_{r_a r_b}, & (r_a,r_b) \text{ belongs to the index set of one of the traces $\tr X^{i_p}$, $p\! =\! 1,\dots,k$},\\
 Y_{r_a r_b}, & (r_a,r_b) \text{ belongs to the index set of one of the traces $\tr Y^{i_q}$, $q\! =\! 1,\dots,j$}.
 \end{cases}
 \end{gather*}
Interchanging the order of summation, we see that each pairing $\pi$ contributing a factor of
 \begin{equation*}
 w[R_{\pi}] := \sum_{I}\prod_{\{(r_a,r_b),(r_c,r_d)\} \in \pi} \langle Z_{r_a r_b} Z_{r_c r_d}\rangle_n
 \end{equation*}
to the expected value. To each such pairing $\pi$, we associate a 2-colored ribbon graph $R_\pi$ as follows:
 \begin{enumerate}\itemsep=0pt
 \item[$(1)$] Draw $k$ $X$-colored vertices, each with degree $i_p$, $p=1,\dots,k$, with half-edges labelled by the indices of the trace of the corresponding vertex.
 \item[$(2)$] Draw $\ell$ $Y$-colored vertices, each with degree $j_q$, $q=1,\dots,\ell$, with half-edges labelled by the indices of the trace of the corresponding vertex.
 \item[$(3)$] Pair the half-edges according to the pairing prescribed by $\pi$.
 \end{enumerate}
The resulting diagram is called a $2$-colored ribbon graph. We will sometimes denote a $2$-colored ribbon graph (associated to a pairing $\pi$)
on $V:= k+\ell$ vertices, with $k$ vertices of the first color, and $\ell := V-k$ of the second color, by $R_{\pi}(k,\ell)$. The \textit{size} of such a ribbon graph is the number of vertices, denoted $|R_{\pi}(k,\ell)| = V$. We call such a coloring a \textit{coloring of type $(k,\ell)$}.
The contribution of such a graph to the Wick sum we will denote by $w[R_{\pi}(k,\ell)]$, or by $w[R_{\pi}]$, when the coloring type is
clear.

Let us compute more explicitly the contribution to the sum of a particular pairing (ribbon graph) $\pi$ ($R_{\pi}$).
The ribbon graph associated to the pairing $\pi$ will by construction have $V$ vertices, and $E$ edges, by the handshaking lemma.
Each face in the resulting ribbon graph corresponds to a collection of indices which will be identified by Wick's theorem, since \smash{$\langle Z_{ab}Z_{cd}\rangle_n = \frac{\star}{n(1-\tau^2)}\delta_{ad}\delta_{bc}$}, where $Z$ is either $X$, $Y$, and $\star$ is either $1$ or $\tau$. Thus,
the number of independent summation indices $I'$ left over after accounting for the identified indices is the same as the number of faces, which
we shall denote by $F$. The contribution will be multiplied by a factor of $\tau^D$, where $D$ counts
the number of edges connecting $X$-colored vertices to $Y$-colored vertices. Thus, we see that $w[R_{\pi}]$ is (upon identifying
$N := n$):
 \begin{equation*}
 w[R_{\pi}] = \frac{1}{n^E} \biggl(\frac{1}{1-\tau^2}\biggr)^E n^F \tau^D = \biggl(\frac{1}{1-\tau^2}\biggr)^E n^{\chi(R_{\pi}) -V} \tau^D.
 \end{equation*}
Let $U$ denote the number of edges in $R_{\pi}$ connecting $X$-colored vertices to $X$-colored vertices and $Y$-colored vertices to
$Y$-colored vertices.
Then, we have the equality $D+U = E$, the total number of edges in $R_{\pi}$. If we define $S := D - U$, then $D = \frac{1}{2}(E+S)$, and
we can express the weight $w[R_{\pi}]$ finally as
 \begin{equation}\label{C2:colored-weight-expression}
 w[R_{\pi}] = \biggl(\frac{\tau^{1/2}}{1-\tau^2}\biggr)^E n^{\chi(R_{\pi}) -V}\tau^{\frac{1}{2}S}.
 \end{equation}
Note that, since $E$ is always even, as the graph comes from a pairing, so there is no problem defining the square root.
What we have proven is the following proposition:

\begin{Proposition}
 Let $X$, $Y$ be Gaussian random matrices, distributed according to the measure~\eqref{Gaussian-measure}, with parameter $N := n$. Let
 $(i_1,\dots,i_k)$, $(j_1,\dots,j_\ell)$ be a $k$- $($respectively, $\ell)$-tuple of positive integers, and set $V := k+\ell$. Furthermore, put
 \smash{$E :=\frac{1}{2}\bigl( \sum_p i_p + \sum_{q} j_q\bigr)$}. Then,
 \begin{equation*}
 n^{V}\Biggl\langle \prod_{p=1}^k \tr X^{i_p}\prod_{q = 1}^\ell \tr Y^{j_q}\Biggr\rangle_n
 = \biggl(\frac{\tau^{1/2}}{1-\tau^2}\biggr)^E\sum_{g\geq 0} \frac{e_g(\tau)}{n^{2g-2}},
 \end{equation*}
 where $e_g(\tau)$ is a polynomial in $\tau$, defined by
 \begin{equation*}
 e_g(\tau) = \sum_{R_{\pi} } \tau^{\frac{1}{2}S(R_{\pi})},
 \end{equation*}
 where the sum runs over all genus $g$ unordered $(k,\ell)$-colored ribbon graphs, and $S(R_{\pi})$ is the number of edges in $R_{\pi}$ between
 like-colored vertices minus the number of edges between unalike vertices.
\end{Proposition}

Now, let us apply this proposition to the matrix integral \eqref{C2:2-matrix-partition function}.
Expanding $\exp \bigl[-\frac{t}{4N} \tr X^4 \bigr]$, $\exp \bigl[-\frac{t}{4N} \tr Y^4 \bigr]$ as series in $t$, we obtain that
 \begin{gather*}
 \exp \biggl[-\frac{t}{4N} \tr X^4 \biggr]\exp \biggl[-\frac{t}{4N} \tr Y^4 \biggr]\\
 \qquad{}= \sum_{V = 0}^{\infty}\frac{(-t/4N)^V}{V!}\sum_{k+j = V} \frac{V!}{k!j!} \bigl(\tr X^4\bigr)^k \bigl(\tr Y^4\bigr)^j {\rm e}^{(k-j)H}.
 \end{gather*}
If we insert this expression into equation \eqref{C2:2-matrix-partition function}, and divide by the Gaussian partition function
$Z_n(\tau,0,0;N)$, we obtain that
 \begin{equation*}
 \frac{Z_n(\tau,t,H;N)}{Z_n(\tau,0,0;N)} = \sum_{M = 0}^{\infty}\frac{(-t/4N)^V}{V!}\sum_{k+j = V} \frac{V!}{k!j!} \bigl\langle \bigl(\tr X^4\bigr)^k \bigl(\tr Y^4\bigr)^j \bigr\rangle_n {\rm e}^{(k-j)H},
 \end{equation*}
where the expected value here is taken with respect to the Gaussian measure \eqref{Gaussian-measure}.
It is important to notice that, provided $k+j = V$, \smash{$\frac{V!}{k!j!}$} counts the number of 2-colorings of $V$ objects, with $k$ of the first color
and $j$ of the second. Expanding the expected value \smash{$\bigl\langle \bigl(\tr X^4\bigr)^k \bigl(\tr Y^4\bigr)^j \bigr\rangle_n$} using Wick's theorem and the diagrammatic rules
we established, we obtain that
 \begin{equation*}
 \frac{Z_n(\tau,t,H;N)}{Z_n(\tau,t,H;N)} = \sum_{V = 0}^{\infty}\frac{(-t/4N)^V}{V!}\sum_{k+j = V} \frac{V!}{k!j!} \sum_{|R_{\pi}(k,j)| = V} w [R_{\pi}(k,j)]{\rm e}^{(k-j)H},
 \end{equation*}
where the innermost sum runs over all $2$-colored ribbon graphs on $V$ vertices with $k$ edges of the first color and $j$ of the second, $k+j = V$.
The colorings in this case are \textit{unlabelled}, in the sense that any graph of a fixed type with coloring $(k,j)$ are considered to be the same.
However, we can count \textit{labelled} colorings by noticing that
 \begin{enumerate}\itemsep=0pt
 \item[$(1)$] If we color any labelled ribbon graph $R_{\pi}$ in two different ways with precisely $k$ vertices of the first color and $j$ of the second, the resulting weights these diagrams contribute to the sum are identical.
 \item[$(2)$] If $k+j = V$, there are precisely \smash{$\frac{V!}{k!j!}$} possible colorings of type $(k,j)$ on a given graph on~$V$ vertices.
 \end{enumerate}
Thus, we see that inner sum over $k$, $j$ can be interpreted as a sum over the distinct possible colorings of the vertices, and we have that
 \begin{equation*}
 \frac{Z_n(\tau,t,H;N)}{Z_n(\tau,0,0;N)} = \sum_{V = 0}^{\infty}\frac{(-t/4N)^V}{V!}\sum_{|R_{\pi}| = V} w_h[R_{\pi}],
 \end{equation*}
where we have defined a modified weight function
 \begin{equation*}
 w_h[R_{\pi}] := w[R_{\pi}]{\rm e}^{(k-j)H},
 \end{equation*}
and the internal sum now runs over all \textit{labelled, $2$-colored} ribbon graphs on $V$-vertices, with \textit{any} coloring scheme.
For our purposes, it is useful switch the order of summation over colorings and graphs, i.e., to write the above as
 \begin{equation*}
 \frac{Z_n(\tau,t,H;N)}{Z_n(\tau,0,0;N)} = \sum_{V = 0}^{\infty}\frac{(-t/4N)^V}{V!}\sum_{\substack{\text{ribbon}\\\text{graphs}\\ R_{\pi} }}\sum_{\substack{\text{colorings of} \\ R_{\pi} }} w_h[R_{\pi}] {\rm e}^{(k-j)H},
 \end{equation*}
where the sum
 \begin{equation*}
 \sum_{\substack{\text{ribbon}\\\text{graphs}\\ R_{\pi} }}
 \end{equation*}
is a sum over all \textit{uncolored, labelled} ribbon graphs on $V$ vertices, and
 \begin{equation*}
 \sum_{\substack{\text{colorings of} \\ R_{\pi} }}
 \end{equation*}
runs over all possible $2$-colorings of $R_{\pi}$. Now, using the formula for the weights $w[R_{\pi}]$ of colored ribbon graphs
we derived earlier (cf.\ equation \eqref{C2:colored-weight-expression}), and using the fact that $E = 2V$ by the handshaking lemma, we have that
(again putting $N := n$):
 \begin{equation*}
 \frac{Z_n(\tau,t,H;n)}{Z_n(\tau,0,0;n)} = \sum_{V = 0}^{\infty}\frac{1}{V!} \biggl(\frac{-t\tau}{4(1-\tau^2)^2}\biggr)^V\sum_{\substack{\text{ribbon}\\\text{graphs}\\ R_{\pi} }} n^{\chi(R_{\pi})}\sum_{\substack{\text{colorings of} \\ R_{\pi} }} \tau^{\frac{1}{2} S(R_{\pi})} {\rm e}^{(k-j)H}.
 \end{equation*}
For a fixed graph $R_{\pi}$, this sum is nothing but the partition function for the Ising model on this graph, with the parameter identification
$\tau = {\rm e}^{-2\beta}$, $H = \beta h$. Thus, we have shown that
 \begin{equation*}
 \frac{Z_n(\tau,t,H;n)}{Z_n(\tau,0,0;n)} = \sum_{V = 0}^{\infty}\frac{1}{V!} \biggl(\frac{-t\tau}{4(1-\tau^2)^2}\biggr)^V\sum_{\substack{\text{ribbon}\\\text{graphs}\\ R_{\pi} }} n^{\chi(R_{\pi})} Z_{R_{\pi}}(\beta,h).
 \end{equation*}
A general principle in the theory of generating functions is the following. Suppose we have an exponential generating function which counts the number of labelled objects. Then, its logarithm counts the number of \textit{connected} objects of the same kind. For more details pertaining to this fact, one may consult, for example, the book~\cite[Section~5]{Stanley}, instead a sum over \textit{connected} ribbon graphs:
 \begin{equation*}
 \log \frac{Z_n(\tau,t,H;n)}{Z_n(\tau,0,0;n)} = \sum_{V = 1}^{\infty}\frac{1}{V!}\biggl(\frac{-t\tau}{(1-\tau^2)^2}\biggr)^V\sum_{\substack{\text{ribbon}\\\text{graphs}\\ R_{\pi} }} {}^{'} n^{\chi(R_{\pi})} Z_{R_{\pi}}(\beta,h),
 \end{equation*}
where $'$ denotes the sum only over connected diagrams, and $\beta$, $h$ are related to $\tau$, $H$ via the formulae
 \begin{equation*}
 \tau = {\rm e}^{-2\beta}, \qquad H = \beta h.
 \end{equation*}
In particular, if we divide through by $n^2$, and take a limit as $n\to \infty$, we obtain that
 \begin{gather}\label{generating-function-genus-0}
 F(\tau,t,H) := \lim_{n\to\infty} \frac{1}{n^2}\log \frac{Z_n(\tau,t,H;n)}{Z_n(\tau,0,0;n)} = \sum_{V = 1}^{\infty}\frac{1}{V!} \biggl(\frac{-t\tau}{4(1-\tau^2)^2}\biggr)^V\sum_{\substack{\text{ribbon}\\\text{graphs}\\ R_{\pi} }} {}^{'} Z_{R_{\pi}}(\beta,h),
 \end{gather}
where the internal sum runs over all connected, $4$-valent planar ribbon graphs on $V$ vertices.

\subsection[Formal calculation of the genus 0 partition function]{Formal calculation of the genus $\boldsymbol{0}$ partition function} \label{Genus0-B}
Here we calculate the genus $0$ partition function in the limit as $N:=n$ tends to infinity.
Define the function \smash{$V(z;t) :=\frac{1}{2} z^2 + \frac{t}{4}z^4$}, and put
 \begin{equation*}
 W(x,y) := \tau xy - V\bigl(x,t{\rm e}^{H}\bigr) - V\bigl(y,t{\rm e}^{-H}\bigr).
 \end{equation*}
We consider the family of biorthogonal polynomials defined by the formula
 \begin{equation*}
 h_k(\tau,t,H;N)\delta_{kj} = h_k\delta_{kj} = \iint P_k(x) Q_k(y) {\rm e}^{NW(x,y)} {\rm d}x{\rm d}y,
 \end{equation*}
where the integration is carried out over an appropriately chosen contour, so that the above integral makes sense. The genus zero generating function for
the Ising model on random spherical quadrangulations is then
 \begin{equation*}
 F(\tau,t,H) := \lim_{N \to \infty} \frac{1}{N^2} \log \prod_{k=0}^N \frac{h_k(\tau,t,H;N)}{h_k(\tau,0,0;N)}.
 \end{equation*}
In order to calculate $F(\tau,t,H)$, we must first study some properties of the corresponding biorthogonal polynomials. We have the following lemma.
\begin{Lemma}
 The polynomials $\{P_k(x)\}$, $\{Q_k(y)\}$ satisfy the recursion relations
 \begin{align}
 &xP_k(x)= P_{k+1}(x) + R_k P_{k-1}(x) + S_k P_{k-3}(x), \label{A}\\
 &yQ_k(y)= Q_{k+1}(y) + \tilde{R}_k Q_{k-1}(y) + \tilde{S}_k Q_{k-3}(y).\label{B}
 \end{align}
\end{Lemma}
\begin{proof}
 We sketch the proof of \eqref{A}. Obviously, \smash{$xP_k(x) = P_{k+1}(x) + \sum_{j=0}^k c_{jk} P_j(x)$}. Multiplying this relation by $Q_{\ell}(y)$ and integrating against the measure ${\rm e}^{NW(x,y)}{\rm d}x{\rm d}y$, one obtains
 \begin{equation*}
 c_{k\ell}h_{\ell} =\iint xP_k(x) Q_{\ell}(y) {\rm e}^{NW(x,y)}{\rm d}x{\rm d}y.
 \end{equation*}
 Using the identity \smash{$x {\rm e}^{NW(x,y)} = \frac{1}{N \tau}\bigl(\frac{\partial}{\partial y} {\rm e}^{NW(x,y)}\bigr) + \frac{1}{\tau}\bigl(y + t{\rm e}^{-H}y^3\bigr)$}, one sees
 that
 \begin{equation*}
 c_{k\ell}h_{\ell} = \iint P_k(x) Q_{\ell}(y)\biggl[\frac{1}{N \tau}\biggl(\frac{\partial}{\partial y} {\rm e}^{NW(x,y)}\biggr) + \frac{1}{\tau}\bigl(y + t{\rm e}^{-H}y^3\bigr)\biggr]{\rm d}x{\rm d}y.
 \end{equation*}
 Since $\ell < k+1$, the first term vanishes identically, since integration by parts yields the integrand as \smash{$-P_k(x) Q_{\ell}'(y){\rm e}^{NW(x,y)}$},
 and $Q_{\ell}'(y)$ is a polynomial of degree less than $k$. Similarly, last two terms can only be nonzero if $\ell = k,k-1,k-3$; the $k$ term is also
 zero, by the symmetry of the integration measure upon the change of variables $(x,y) \to (-x,-y)$. The result \eqref{B} is obtained in an identical manner.
\end{proof}

We now show that the recursion coefficients satisfy additional relations.
\begin{Lemma}
 Define $f_k := h_k/h_{k-1}$. The coefficients $R_k$, $S_k$, $\tilde{R}_k$, $\tilde{S}_k$ satisfy the following relations:
 \begin{align}
 &\tau S_k= t {\rm e}^{-H} f_k f_{k-1} f_{k-2},\label{S1}\\
 &\tau R_k= f_k\bigl[1 + t{\rm e}^{-H}\bigl(\tilde{R}_{k+1} + \tilde{R}_k + \tilde{R}_{k-1}\bigr)\bigr], \label{string1-A}\\
 &\frac{k}{N}= -\tau f_k + \tilde{R}_k + t{\rm e}^{-H}\bigl[\tilde{S}_{k+2} + \tilde{S}_{k+1} + \tilde{S}_{k} + \tilde{R}_k\bigl(\tilde{R}_{k+1} + \tilde{R}_{k} + \tilde{R}_{k-1}\bigr)\bigr],\\
 &\tau \tilde{S}_k= t {\rm e}^{H} f_k f_{k-1} f_{k-2},\\
 &\tau \tilde{R}_k= f_k\bigl[1 + t{\rm e}^{H}\bigl({R}_{k+1} + {R}_k + {R}_{k-1}\bigr)\bigr], \label{string1-B}\\
 &\frac{k}{N}= -\tau f_k + {R}_k + t{\rm e}^{H}\bigl[{S}_{k+2} + {S}_{k+1} + {S}_{k} + {R}_k\bigl({R}_{k+1} + {R}_{k} + {R}_{k-1}\bigr)\bigr]\label{f2}.
 \end{align}
\end{Lemma}
 \begin{proof}
 We only prove the first relation; the last three are proven identically. Multiplying equation \eqref{A} by $\tau Q_{k-3}(y)$, and integrating with respect to the measure ${\rm e}^{NW(x,y)}{\rm d}x{\rm d}y$, we find~that
 \begin{equation*}
 \tau S_k h_{k-3} = \iint \tau xP_k(x) Q_{k-3}(y) {\rm e}^{NW(x,y)}{\rm d}x{\rm d}y.
 \end{equation*}
 Using the identity \smash{$\tau x {\rm e}^{NW(x,y)} = \frac{1}{N}\bigl(\frac{\partial}{\partial y} {\rm e}^{NW(x,y)}\bigr) + \bigl(y + t{\rm e}^{-H}y^3\bigr)$}, we see that
 \begin{equation*}
 \tau S_k h_{k-3} = \iint \tau P_k(x) Q_{k-3}(y) \biggl[\frac{1}{N}\biggl(\frac{\partial}{\partial y} {\rm e}^{NW(x,y)}\biggr) + \bigl(y + t{\rm e}^{-H}y^3\bigr)\biggr]{\rm d}x{\rm d}y.
 \end{equation*}
 The first term drops out upon integrating by parts; similarly, since $\deg [ y Q_{k-3}] = k-2$, the second term also evaluates to $0$. On the
 last term, we use iteratively the recursion formula~\eqref{B}:
 \begin{align*}
 \tau S_k h_{k-3} &{}= t {\rm e}^{-H} \iint P_k(x) \left[Q_k(y) + \text{lower degree polynomials }\right]{\rm e}^{NW(x,y)}{\rm d}x{\rm d}y \\
 &{}= t {\rm e}^{-H} h_k.
\tag*{\qed}
\end{align*}
\renewcommand{\qed}{}
\end{proof}

In planar limit, we assume the scaling $\frac{k}{N} \to \lambda$, $f_k \to f(\lambda)$, and so on. The above formulas~\eqref{S1}--\eqref{f2} imply that, in the limit
as $N\to \infty$,
 \begin{align}
 &\tau S(\lambda)= t {\rm e}^{-H} f(\lambda)^3,\nonumber\\
 &\tau R(\lambda)= f(\lambda)\bigl[1 + 3t{\rm e}^{-H}\tilde{R}(\lambda)\bigr],\nonumber\\
 &\lambda= -\tau f(\lambda) + \tilde{R}(\lambda) + 3t{\rm e}^{-H}\bigl[\tilde{S}(\lambda) + \tilde{R}(\lambda)^2\bigr], \label{C}\\
 &\tau \tilde{S}(\lambda)= t {\rm e}^{H} f(\lambda)^3,\nonumber\\
 &\tau \tilde{R}(\lambda)= f(\lambda)\bigl[1 + 3t{\rm e}^{H}{R}(\lambda)\bigr],\nonumber\\
 &\lambda= -\tau f(\lambda) + {R}(\lambda) + 3t{\rm e}^{H}\bigl[S(\lambda) + R(\lambda)^2\bigr],\label{D}
 \end{align}
and these relations we can solve for $S(\lambda)$, $R(\lambda)$, $\tilde{R}(\lambda)$, and $\tilde{S}(\lambda)$ in terms of $f(\lambda)$,
and the~param\-e\-ters~$\tau$, $t$, $H$:
 \begin{alignat*}{3}
 &S(\lambda)= \frac{t{\rm e}^{-H}}{\tau} f(\lambda)^3,\qquad&& \tilde{S}(\lambda) = \frac{t{\rm e}^{H}}{\tau} f(\lambda)^3,&\\
 &R(\lambda)= \frac{\tau + 3t{\rm e}^{-H} f(\lambda)}{\tau^2 - 9t^2f(\lambda)^2}f(\lambda),\qquad&&
 \tilde{R}(\lambda) = \frac{\tau + 3t{\rm e}^{H} f(\lambda)}{\tau^2 - 9t^2f(\lambda)^2}f(\lambda).&
 \end{alignat*}
These formulae can then be subsequently substituted into either equations \eqref{C} or \eqref{D} (they yield the same equation in the end) to obtain an
implicit formula for the function $f(\lambda) = f(\lambda;\tau,t,H)$:
 \begin{align}
 &\lambda= -\tau f(\lambda) + \frac{\tau + 3t{\rm e}^{-H} f(\lambda)}{\tau^2 - 9t^2f(\lambda)^2}f(\lambda) + \frac{3t{\rm e}^{H}\bigl(\tau + 3t{\rm e}^{-H} f(\lambda)\bigr)^2}{(\tau^2 - 9t^2f(\lambda)^2)^2}f(\lambda)^2 + \frac{3t^2f(\lambda)^3}{\tau}\label{f-formula}\\
 &\qquad{}\Longleftrightarrow \lambda = -\tau f(\lambda) + \frac{3t^2}{\tau}f(\lambda)^3 + \frac{\tau f(\lambda)}{(\tau - 3tf(\lambda))^2}+\frac{6\tau^2tf(\lambda)^2}{(\tau^2-9t^2f(\lambda)^2)^2}[\cosh H -1].\nonumber
 \end{align}
In particular, when $t=H=0$, we obtain the solution
 \begin{equation*}
 f(\lambda;\tau,0,0) = \frac{\tau \lambda}{1-\tau^2}.
 \end{equation*}

Returning again to the calculation of the function $F(\tau,t,H)$, note that, by the definition of the variables $f_k$, \smash{$\prod_{k=0}^N h_k = h_0^N\prod_{k=1}^N f_k^{N-k}$}, and so
 \begin{equation*}
 \frac{1}{N^2} \log \prod_{k=0}^N \frac{h_k(\tau,t,H;N)}{h_k(\tau,0,0;N)} = \frac{1}{N} \log \frac{h_0(\tau,t,H;N)}{h_0(\tau,0,0;N)} + \sum_{k=1}^N \biggl(1-\frac{k}{N}\biggr) \log \frac{f_k(\tau,t,H;N)}{f_k(\tau,0,0;N)} \frac{1}{N}.
 \end{equation*}
Taking the limit as $N\to \infty$, the first term vanishes (the argument of the logarithm is bounded), and the second term approximates a
Riemann integral; we thus obtain the formula
 \begin{align}
 F(\tau,t,H) &= \lim_{N \to \infty} \frac{1}{N^2} \log \prod_{k=0}^N \frac{h_k(\tau,t,H;N)}{h_k(\tau,0,0;N)} = \int_{0}^{1}(1-\lambda) \log\frac{f(\lambda;\tau,t,H)}{f(\lambda;\tau,0,0)} {\rm d}\lambda \nonumber \\
 &= \frac{3}{4} - \frac{1}{2}\log \frac{\tau}{1-\tau^2} + \int_{0}^{1}(1-\lambda) \log f(\lambda;\tau,t,H) {\rm d}\lambda \label{Free2}.
 \end{align}
This, along with the implicit formula \eqref{f-formula}, are enough (in principle) to determine the function $F(\tau,t,H)$. Integrating the above formula by parts,
 \begin{align*}
 F(\tau,t,H)={}& \frac{3}{4} - \frac{1}{2}\log \frac{\tau}{1-\tau^2} + \biggl[\biggl(\lambda-\frac{1}{2}\lambda^2\biggr)\log f(\lambda;\tau,t,H)\biggr]_{0}^{1}\\
 &{}{-}\,\int_{0}^{1} \biggl(\lambda-\frac{1}{2}\lambda^2\biggr) \frac{f'(\lambda;\tau,t,H)}{f(\lambda;\tau,t,H)} {\rm d}\lambda\\
={}& \frac{3}{4} - \frac{1}{2}\log \frac{\tau}{1-\tau^2} + \frac{1}{2}\log f(1;\tau,t,H) - \int_{0}^{1}\biggl(\lambda-\frac{1}{2}\lambda^2\biggr)\frac{f'(\lambda;\tau,t,H)}{f(\lambda;\tau,t,H)} {\rm d}\lambda
 \end{align*}
(Here we have used the fact that $f(0;\tau,t,H) \equiv 0$). Making the change of variables $u = f(\lambda;\tau,t,H)$,
\smash{$\frac{{\rm d}u}{u} = \frac{f'}{f}{\rm d}\lambda$},
 \begin{align*}
 &F(\tau,t,H)= \frac{3}{4} - \frac{1}{2}\log \frac{\tau}{1-\tau^2} + \frac{1}{2}\log f(1;\tau,t,H) - \int_{0}^{f(1;\tau,t,H)}\biggl(\tilde{\lambda}(u)-\frac{1}{2}\tilde{\lambda}(u)^2\biggr) \frac{{\rm d}u}{u},
 \end{align*}
where $\tilde{\lambda}(u)$ is defined by use of the implicit equation \eqref{f-formula}:
 \begin{equation*}
 \tilde{\lambda}(u) = -\tau u + \frac{3t^2}{\tau}u^3 + \frac{\tau u}{(\tau - 3tu)^2}+\frac{6\tau^2tu^2}{(\tau^2-9t^2u^2)^2}[\cosh H -1].
 \end{equation*}
The integration can be carried out directly, since $\lambda(u)$ is a rational function. Finally, defining
 \begin{equation*}
 \sigma(\tau,t,H) := -\frac{3t}{\tau}f(1;\tau,t,H),
 \end{equation*}
and making an appropriate change of variables, we obtain the following Proposition:
 \begin{Proposition}
 The genus $0$ partition function admits the expression
 \begin{equation*}
 F(\tau,t,H) = \frac{3}{4} + \frac{1}{2}\log \frac{\bigl(1-\tau^2\bigr)\sigma(\tau,t,H)}{-3t} - \int_{0}^{\sigma(\tau,t,H)}\biggl(\lambda(u)-\frac{1}{2}\lambda(u)^2\biggr) \frac{{\rm d}u}{u},
 \end{equation*}
 where $\lambda(u)$ is the rational function
 \begin{equation*}
 \lambda(u) = -\frac{1}{t}\biggl[\frac{1}{9}\tau^2u\bigl(u^2-3\bigr) + \frac{1}{3}\frac{u}{(u+1)^2} - \frac{2}{3}\biggl(\frac{u}{u^2-1}\biggr)^2[\cosh H -1]\biggr]
 \end{equation*}
 and $\sigma = \sigma(\tau,t,H)$ is defined implicitly by the equation
 \begin{equation*}
 t = -\frac{1}{9}\tau^2\sigma \bigl(\sigma^2 - 3\bigr) -\frac{1}{3}\frac{\sigma}{(1+\sigma)^2} + \frac{2}{3}\biggl(\frac{\sigma}{1-\sigma^2}\biggr)^2[\cosh H -1].
 \end{equation*}
 \end{Proposition}
This is the result originally obtained by Kazakov ($H=0$, cf.~\cite{Kazakov1}) and Boulatov ($H\neq 0$, cf.~\cite{Kazakov2}). If one expands
the function $F(\tau,t,H)$ as a series in $t$, one obtains that
 \begin{align} \label{Kazakov-F}
 &F(\tau,t,H)= \textcolor{teal}{4\tau^{-1}\cosh{H}}\biggl(\frac{-t\tau}{4(1-\tau^2)^2}\biggr) + \textcolor{teal}{\bigl(8\tau^2 + 64 +72\tau^{-2}\cosh(2H)\bigr)}\frac{1}{2}\biggl(\frac{-t\tau}{4(1-\tau^2)^2}\biggr)^2 \nonumber \\
 &\qquad{}+\textcolor{teal}{\biggl(\frac{3456}{\tau^3}\cosh(H)\bigl(2\cosh(2H)+\tau^4+2\tau^2-1\bigr)\biggr)}\frac{1}{6}\biggl(\frac{-t\tau}{4(1-\tau^2)^2}\biggr)^3 + \OO\bigl(t^4\bigr).
 \end{align}

\subsection{Comparison with explicit calculation} \label{Genus0-C}
On the other hand, we can compute the first few contributions to the genus $0$ free energy by hand.
The following graphs (shown in Figure~\ref{tab:Genus0-contributions}) contribute at genus zero:
\begin{figure}[t]
 \centering
 \includegraphics{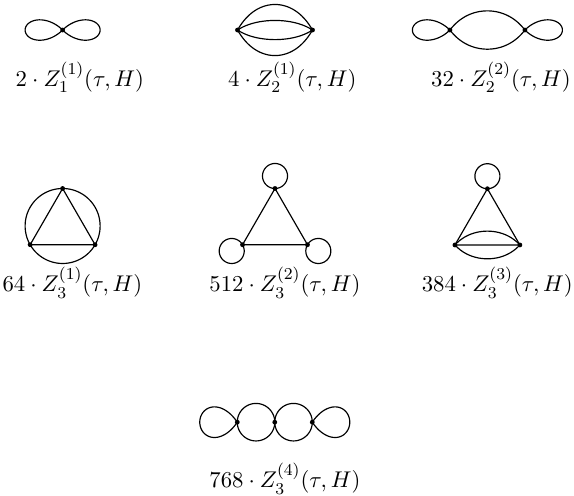}
% \begin{tabular}{c c c}
% \begin{overpic}[scale=.15]{Pictures/1-vertex.pdf}
% \put (25,20) {$2 \cdot Z_1^{(1)}(\tau,H)$}
% \end{overpic} & \begin{overpic}[scale=.15]{Pictures/2-vertex-type1.pdf}
% \put (25,20) {$4 \cdot Z_2^{(1)}(\tau,H)$}
% \end{overpic} & \begin{overpic}[scale=.15]{Pictures/2-vertex-type2.pdf}
% \put (20,20) {$32 \cdot Z_2^{(2)}(\tau,H)$}
% \end{overpic} \\
% \begin{overpic}[scale=.15]{Pictures/3-vertex-type1.pdf}
% \put (18,15) {$64 \cdot Z_3^{(1)}(\tau,H)$}
% \end{overpic} & \begin{overpic}[scale=.15]{Pictures/3-vertex-type2.pdf}
% \put (15,15) {$512 \cdot Z_3^{(2)}(\tau,H)$}
% \end{overpic} & \begin{overpic}[scale=.15]{Pictures/3-vertex-type3.pdf}
% \put (15,15) {$384 \cdot Z_3^{(3)}(\tau,H)$}
% \end{overpic}\\
% & \begin{overpic}[scale=.15]{Pictures/3-vertex-type4.pdf}
% \put (15,15) {$768 \cdot Z_3^{(4)}(\tau,H)$}
% \end{overpic} &
% \end{tabular}
 \caption{The isomorphism classes of graphs contributing at genus $0$ to the free energy, up to $V=3$ vertices. Each isomorphism class of graph contributes a factor of $N_i \cdot Z_V^{(i)}(\tau,H)$.}
 \label{tab:Genus0-contributions}
\end{figure}
\begin{enumerate}\setlength{\leftskip}{0.35cm}\itemsep=0pt
 \item[$V=1$.] For one vertex, there is one type of graph contributing, shown in the figure. The partition function for the Ising model on this graph is, in the variables $\tau$, $H$,
 \begin{equation*}
 Z_1^{(1)}(\tau,H) = 2\tau^{-1}\cosh H.
 \end{equation*}
 \item[$V=2$.] For two vertices, there are two nonisomorphic types of graphs contributing, shown in the figure. Their corresponding partition functions are
 \begin{equation*}
 Z_2^{(1)}(\tau,H) = 2\tau^{-2} \cosh(2H) + 2\tau^2, \qquad Z_2^{(2)}(\tau,H) = 2\tau^{-2}\cosh(2H) + 2.
 \end{equation*}
 \item[$V=3$.] There are four nonisomorphic types of graphs contributing, shown in the figure. Their partition functions are
 \begin{align*}
 &Z_3^{(1)}(\tau,H)= 2\tau^{-3}\cosh(3H)+6\tau\cosh(H),\nonumber\\
 &Z_3^{(2)}(\tau,H)= 2\tau^{-3}\cosh(3H) + 2\tau^{-1}\cosh(H) + 4\tau \cosh(H),\\
 &Z_3^{(3)}(\tau,H)= 2\tau^{-3}\cosh(3H)+6\tau^{-1}\cosh(H), \nonumber\\
 &Z_3^{(4)}(\tau,H)= 2\tau^{-3}\cosh(3H) + 4\tau^{-1}\cosh(H) + 2\tau \cosh(H).\nonumber
 \end{align*}
\end{enumerate}

Observe that we can write the genus zero free energy \eqref{generating-function-genus-0} equivalently as
 \begin{equation*}
 F(\tau,t,H) = \sum_{V=0}^{\infty} \textcolor{teal}{\biggl(\sum_{|G| = V} N_G Z_G(\tau,H) \biggr)}\frac{1}{V!}\biggl(\frac{-t\tau}{4(1-\tau^2)^2}\biggr)^V,
 \end{equation*}
where the internal sum is taken over \textit{isomorphism classes} of graphs $G$ on $V$ vertices, and $N_G$ counts the number of labelled graphs isomorphic to $G$ when one ``forgets''
the labelling. In this way, one can explicitly compute that
 \begin{gather*}
 F(\tau,t,H) = \textcolor{teal}{\bigl[2\cdot Z_1^{(1)}(\tau,H)\bigr]}\biggl(\frac{-t\tau}{4(1-\tau^2)^2}\biggr) \\
 \hphantom{F(\tau,t,H) =}{}
 + \textcolor{teal}{\bigl[4 \cdot Z_2^{(1)}(\tau,H) + 32\cdot Z_2^{(2)}(\tau,H)\bigr]}\frac{1}{2}\biggl(\frac{-t\tau}{4(1-\tau^2)^2}\biggr)^2\\
 \hphantom{F(\tau,t,H) =}{}
 + \textcolor{teal}{\bigl[64\cdot Z_3^{(1)}(\tau,H) + 512\cdot Z_3^{(2)}(\tau,H) + 384\cdot Z_3^{(3)}(\tau,H) + 768\cdot Z_3^{(4)}(\tau,H)\bigr]}\\
 \hphantom{F(\tau,t,H) =}{}\quad
 \times\frac{1}{6}\biggl(\frac{-t\tau}{4(1-\tau^2)^2}\biggr)^3+ \OO\bigl(t^4\bigr)\\
 \hphantom{F(\tau,t,H)}{}=
 \textcolor{teal}{4\tau^{-1}\cosh{H}}\biggl(\frac{-t\tau}{4(1-\tau^2)^2}\biggr) + \textcolor{teal}{\bigl(8\tau^2 + 64+72\tau^{-2}\cosh(2H)\bigr)}\frac{1}{2}\biggl(\frac{-t\tau}{4(1-\tau^2)^2}\biggr)^2\\
 \hphantom{F(\tau,t,H) =}{}
 \qquad{}+ \textcolor{teal}{\biggl(\frac{3456}{\tau^3}\cosh(H)\bigl(2\cosh(2H)+\tau^4+2\tau^2-1\bigr)\!\biggr)}\frac{1}{6}\biggl(\frac{-t\tau}{4(1-\tau^2)^2}\biggr)^3\!\! + \OO\bigl(t^4\bigr),
 \end{gather*}
which is in agreement with the formula \eqref{Kazakov-F} originally obtained by Kazakov.

\begin{Remark} \textit{High/low temperature limits of the free energy.}
 As a final remark, let us show that in the limit $\tau,H\to 0$, the free energy of the quartic-two matrix model reduces to twice the free energy of the
 quartic $1$-matrix model. Intuitively, one should expect this result because $\tau$ acts as a coupling parameter between the matrices $X$ and $Y$; when
 $\tau,H \to 0$, we obtain two independent copies of the quartic $1$-matrix model. Recall that the free energy for the quartic $1$-matrix model is
 \begin{equation*}
 \mathcal{F}(t) :=\lim_{N\to\infty}\log\frac{Z^{(1)}_{N}(t;N)}{Z^{(1)}_{N}(0;N)},
 \end{equation*}
 where $Z^{(1)}_{n}(t;N)$ is given by the formula
 \begin{equation*}
 Z^{(1)}_{n}(t;N) = \int_{\mathcal{H}_n} \exp\biggl[-N\tr\biggl(\frac{1}{2}X^2 + \frac{t}{4}X^4\biggr) \biggr]{\rm d}X.
 \end{equation*}
 Explicitly, one may evaluate $\mathcal{F}(t)$ to be (cf., for example,~\cite{BIZ}, or the more recent~\cite{BGM})
 \begin{align*}
 &\mathcal{F}(t)= \int_{0}^{1}(1-\lambda)\log \frac{-1 + \sqrt{1+12t\lambda}}{6t\lambda}{\rm d}\lambda.
 \end{align*}
 On the other hand, let us return to the formula for the partition function of the $2$-matrix model (see the first expression in equation \eqref{Free2}):
 \begin{align*}
 F(\tau,t,H) &= \int_{0}^{1}(1-\lambda) \log\frac{f(\lambda;\tau,t,H)}{f(\lambda;\tau,0,0)} {\rm d}\lambda\\
 &=\int_{0}^{1}(1-\lambda) \log\frac{\bigl(1-\tau^2\bigr)f(\lambda;\tau,t,H)}{\tau \lambda} {\rm d}\lambda,
 \end{align*}
 where $f(\lambda;\tau,t,H)$ is defined by the implicit equation
 \begin{equation}\label{implicit-eq-final}
 \lambda = -\tau f(\lambda) + \frac{\tau + 3t{\rm e}^{-H} f(\lambda)}{\tau^2 - 9t^2f(\lambda)^2}f(\lambda) + \frac{3t{\rm e}^{H}\bigl(\tau + 3t{\rm e}^{-H} f(\lambda)\bigr)^2}{(\tau^2 - 9t^2f(\lambda)^2)^2}f(\lambda)^2 + \frac{3t^2f(\lambda)^3}{\tau}.
 \end{equation}
 Define the new variable \smash{$\tilde{f}(\lambda;\tau,t,H) := \frac{(1-\tau^2)}{\tau}f(\lambda;\tau,t,H)$}. The reason for this change of variables is that \smash{$\tilde{f}(\lambda;\tau,t,H)\to \lambda$ when $t\to 0$}, i.e., the limit function is independent of $\tau$. Under this change of variables, equation \eqref{implicit-eq-final} becomes
 \begin{equation*}
 0 = \lambda - \frac{\bigl(1+3t{\rm e}^{-H}\tilde{f}\bigr)\tilde{f}}{1-9t^2\tilde{f}^2} - \frac{3t{\rm e}^H\bigl(1+3t{\rm e}^{-H}\tilde{f}\bigr)^2}{\bigl(1-9t^2\tilde{f}^2\bigr)^2} + \OO\bigl(\tau^2\bigr).
 \end{equation*}
 In the limit as $\tau,H\to 0$, one sees that
 \begin{equation*}
 0 = \lambda - \frac{\bigl(1+3t\tilde{f}\bigr)\tilde{f}}{1-9t^2\tilde{f}^2} - \frac{3t\bigl(1+3t\tilde{f}\bigr)^2}{\bigl(1-9t^2\tilde{f}^2\bigr)^2}.
 \end{equation*}
 This equation can be solved for $\tilde{f}$:
 \begin{equation*}
 \tilde{f}(t) = \frac{1 + 6t\lambda -\sqrt{1 + 12t\lambda}}{18t^2\lambda},
 \end{equation*}
 where we have taken the branch of the solution which tends to $\lambda$ as $t\to 0$.
 Inserting this expression into the expression for the free energy of the $2$-matrix model,
 \begin{align*}
 \lim_{\tau,H\to 0} F(\tau,t,H) &= \lim_{\tau,H\to 0} \int_{0}^{1}(1-\lambda) \log\frac{\tilde{f}(\tau,t,H)}{\lambda} {\rm d}\lambda\\
 &= \int_{0}^{1}(1-\lambda) \log\frac{1 + 6t\lambda -\sqrt{1 + 12t\lambda}}{18t^2\lambda^2} {\rm d}\lambda\\
 &= \int_{0}^{1}(1-\lambda) \log\biggl(\frac{-1 + \sqrt{1+12t\lambda}}{6t\lambda}\biggr)^2 {\rm d}\lambda\\
 &=2\int_{0}^{1}(1-\lambda) \log\frac{-1 + \sqrt{1+12t\lambda}}{6t\lambda} {\rm d}\lambda\\
 &=2\mathcal{F}(t).
 \end{align*}
 More generally, we have the following limits:
 \begin{align*}
 &\lim_{H\to 0}\lim_{\tau\to 0} F\bigl(\tau,\bigl(1-\tau^2\bigr)t,H\bigr)= \lim_{H\to 0}\bigl[\mathcal{F}\bigl({\rm e}^{H}t\bigr) + \mathcal{F}\bigl({\rm e}^{-H}t\bigr)\bigr] = 2\mathcal{F}(t),\\
 &\lim_{H\to 0}\lim_{\tau\to 1} F\bigl(\tau,\bigl(1-\tau^2\bigr)t,H\bigr)= \lim_{H\to 0}\bigl[\mathcal{F}\bigl(2{\rm e}^{H}t\bigr) + \mathcal{F}\bigl(2{\rm e}^{-H}t\bigr)\bigr] = 2\mathcal{F}(2t).
 \end{align*}
 The first limit follows from our previous considerations, without setting $H=0$. The replacement of $t$ by $\bigl(1-\tau^2\bigr)t$ here is natural from the point
 of view of the graphical expansion of the $2$-matrix model. The second limit follows from identical calculations. These limits correspond to the low and high temperature limits of the Ising model. As originally pointed out by Kazakov~\cite{Kazakov1}, the factor of $2$ appearing in the $\tau\to 1$ limit comes from the fact that, for a graph $G$ on $V$ vertices, the partition function for the Ising model on that graph \big(recalling that $\tau = {\rm e}^{-2\beta}$\big)
 \begin{equation*}
 \lim_{\beta\to 0}Z_G(\beta,h) \to 2^V.
 \end{equation*}
 Thus, in the limit as $\tau\to 1$, each planar, $4$-regular graph on $V$ vertices should contributes the same factor of $2^V$ to the free energy.
\end{Remark}

\subsection[Large $V$ asymptotics of $\sigma(\tau,t,H)$]{Large $\boldsymbol{V}$ asymptotics of $\boldsymbol{\sigma(\tau,t,H)}$}
Here we calculate the large $V$ asymptotics of the implicitly defined function $\sigma(\tau,t,H)$. The reason for these calculations is the following: in the works~\cite{Kazakov2,Kazakov1}, it is argued that the asymptotics of the Taylor coefficients of $F(\tau,t,H)$ are dominated by the nearest critical point to the origin of the equation $\mathfrak{I}(\sigma;\tau,t,H) = 0$. The authors make the assertion that the Taylor coefficients of~$F$ behave like $F_V \sim {\rm const}\cdot V^\alpha[g(\tau,H)]^V$; it is eventually the expression $g(\tau,H)$ that they are interested in. Indeed, calculation of $g(\tau,H)$ is equivalent to finding the radius of convergence of~$F$, considered
as an analytic function of $t$. Since $F$ is defined in terms of $\sigma(\tau,t,H)$, it is therefore plausible that this radius of convergence should coincide with the radius of convergence of~$\sigma(\tau,t,H)$ as a function of $t$. However, we were unfortunately unable to show
that this was the case, as it is not obvious that terms such as \smash{$\frac{1}{(\sigma+1)^3}$}, \smash{$\log \frac{(1-\tau^2)\sigma }{-3t}$} do
not alter the radius of convergence of $F$ (the remaining terms in the expression for $F$ are polynomial in $\sigma$, and thus have the same radius of convergence as that of $\sigma$). We therefore instead have opted to simply investigate the Taylor coefficients of $\sigma(\tau,t,H)$ as a function of $t$, and take the assumptions Kazakov makes about the implication of this function as fact.

Let us now set about calculating the Taylor coefficients of $\sigma(\tau,t,H)$ as a function of $t$. We do so explicitly in the case when
$H=0$, and also provide an expansion which holds as $H\to 0$. Recall that $\sigma = \sigma(\tau,t,H)$ is defined by the equation
 \begin{equation*}
 \mathfrak{I}(\sigma;\tau,t,H) = -\frac{1}{9}\tau^2\sigma \bigl(\sigma^2 - 3\bigr) -\frac{1}{3}\frac{\sigma}{(1+\sigma)^2} + \frac{2}{3}\biggl(\frac{\sigma}{1-\sigma^2}\biggr)^2[\cosh H -1] - t = 0.
 \end{equation*}
When $H=0$, the last term vanishes identically. Note that, for any fixed $\tau$, $H$, the above implicit equation is of special form: it
is actually an equation defining an \textit{inverse} function to $t(\sigma)$.
We recall the following theorem, due to Lagrange (cf., for instance,~\cite[p.~149, Exercise~25]{WW}):
 \begin{Theorem}[Lagrange inversion formula]
 Suppose $t = G(\sigma)$ is analytic in a neighborhood of $\sigma= 0$, and suppose $G(0) = 0$, $G'(0) \neq 0$. Then, an inverse function $\sigma = \sigma(t)$ exists in a neighborhood of $t = 0$, and is given by the power series
 \begin{equation*}
 \sigma(t) = \sum_{V=1}^{\infty} \sigma_V \frac{t^V}{V!},
 \end{equation*}
 where
 \begin{equation}\label{sigma-coeff}
 \sigma_V = \lim_{\sigma\to 0}\frac{{\rm d}^{V-1}}{{\rm d}\sigma^{V-1}} \biggl(\frac{\sigma^V}{G(\sigma)^V}\biggr).
 \end{equation}
 \end{Theorem}
The proof of this theorem follows almost immediately from the Cauchy integral formula. In~our situation, for any fixed $0<\tau<1$,
and $H\in \RR$, we have that
 \begin{equation*}
 \mathfrak{I}(0;\tau,t,H) = 0,\qquad \mathfrak{I}_\sigma(0;\tau,t,H) = \frac{1}{3}\bigl(\tau^2-1\bigr) \neq 0,
 \end{equation*}
and so we can develop a Taylor series expansion of $\sigma(\tau,t,H)$ about $t=0$, by taking $G(\sigma)$ to be
 \begin{equation*}
 G(\sigma) := \mathfrak{I}(\sigma;\tau,0,H) = -\frac{1}{9}\tau^2\sigma \bigl(\sigma^2 - 3\bigr) -\frac{1}{3}\frac{\sigma}{(1+\sigma)^2} + \frac{2}{3}\biggl(\frac{\sigma}{1-\sigma^2}\biggr)^2[\cosh H -1].
 \end{equation*}
By the Cauchy integral formula, along with formula \eqref{sigma-coeff}, we can write
 \begin{equation}\label{sigma-int}
 \sigma_V(\tau,H) = \frac{1}{2\pi {\rm i}}\oint_{C_0} \frac{{\rm d}\zeta}{\mathfrak{I}(\zeta;\tau,0,H)^V} = \frac{1}{2\pi {\rm i}}\oint_{C_0}{\rm e}^{-V\log\mathfrak{I}(\zeta;\tau,0,H)}{\rm d}\zeta,
 \end{equation}
where $C_0$ is a sufficiently small positively oriented circle enclosing the origin. We are interested in the large-$V$ asymptotics of
$\sigma_V(\tau,H)$. When $H=0$, and $\tau \neq \frac{1}{4}$, these asymptotics can be calculated explicitly using classical
steepest descent analysis (cf., for example,~\cite{Miller}). When~${H=0}$, the saddle points of the integrand are
 \begin{equation*}
 \zeta = 1,\qquad -1 \pm \tau^{-1/2},\qquad -1 \pm {\rm i}\tau^{-1/2}.
 \end{equation*}
The dominant saddle point $\zeta^*$ is the one nearest to the origin. Thus, we see that
 \begin{equation*}
 \zeta^*(\tau) =
 \begin{cases}
 1, & 0 < \tau < \frac{1}{4},\\
 \tau^{-1/2} - 1, & \frac{1}{4} < \tau < 1.
 \end{cases}
 \end{equation*}
(Note that this change in the dominant saddle point is ultimately the source of the phase transition). It follows that
 \begin{equation*}
 \sigma_V(\tau,0) = \frac{\pm}{2\pi {\rm i}}\sqrt{\frac{2\pi}{V[-\log\mathfrak{I}(\zeta^*;\tau,0,0)]_{\zeta\zeta}}}\biggl[\frac{1}{\mathfrak{I}(\zeta^*;\tau,0,0)}\biggr]^V\bigl[1 + \OO\bigl(V^{-1/2}\bigr)\bigr],
 \end{equation*}
where the sign $\pm$ is chosen appropriately in accordance with the contour of steepest descent. This addresses the asymptotics of
$\sigma_V$ away from the critical point \smash{$\tau = \frac{1}{4}$}. In fact, we can actually say more. We summarize the above result, as well as an extension of it to a neighborhood of the critical point, in the following proposition:
\begin{Proposition} Define $t_{{\rm low}}(\tau) := -\frac{1}{12}+\frac{2}{9}\tau^2$,
$t_{{\rm high}}(\tau) = -\frac{2}{9} \sqrt{\tau}(\sqrt{\tau} - 1)^2(\sqrt{\tau} +2)$.
As $V\to \infty$, for any fixed $0 < \tau < 1$, $\tau \neq \frac{1}{4}$,
 \begin{equation}\label{sigma-asymptotics-nofield}
 \sigma_V(\tau,0) =
 \begin{cases}
 \dfrac{1}{\sqrt{3\pi V}}\sqrt{\dfrac{8\tau^2-3}{16\tau^2-1}} t_{{\rm low}}(\tau)^{-V}\bigl[1 + \OO\bigl(V^{-1/2}\bigr)\bigr], & 0 < \tau < \frac{1}{4},\\[1.5mm]
 \dfrac{(\sqrt{\tau}-1)}{2\sqrt{3\pi V}}\sqrt{\dfrac{2+\sqrt{\tau}}{\tau (2\sqrt{\tau}-1)}}t_{{\rm high}}(\tau)^{-V}\bigl[1 + \OO\bigl(V^{-1/2}\bigr)\bigr], & \frac{1}{4} < \tau < 1.
 \end{cases}
 \end{equation}
If $\tau$ belongs to a sufficiently small neighborhood of $\tau = \frac{1}{4}$, then we have the \textit{uniform in $\tau$} asymptotic
 \begin{gather*}
 \sigma_V(\tau,0) = [t_{{\rm low}}(\tau)t_{{\rm high}}(\tau)]^{-V/2}\biggl[\frac{a_0(\tau)}{V^{1/3}}\Ai(V^{2/3}s(\tau)) + \frac{b_0(\tau)}{V^{2/3}}\Ai'(V^{2/3}s(\tau))\biggr]\\
 \hphantom{\sigma_V(\tau,0) =}{}
 \times \bigl[1 + \OO\bigl(V^{-1/3}\bigr)\bigr],
 \end{gather*}
Where $\Ai(z)$ denotes the Airy function, $s(\tau) \geq 0$ is the continuous function
 \begin{equation}\label{s-definition}
 s(\tau) = \frac{3}{4}\biggl|\log\biggl[\frac{t_{{\rm low}}(\tau)}{t_{{\rm high}}(\tau)}\biggr]\biggr|^{2/3},
 \end{equation}
satisfying $s\bigl(\frac{1}{4}\bigr) = 0$, and $a_0(\tau)$, $b_0(\tau)$ are given by
 \begin{align*}
 &a_0(\tau)=\frac{s(\tau)^{1/4}}{2}\Biggl[\frac{2}{\sqrt{3}}\sqrt{\frac{3-8\tau^2}{1-16\tau^2}} + \sqrt{\frac{(1-\sqrt{\tau})^2(2+\sqrt{\tau})}{3\tau(1-2\sqrt{\tau})}}\Biggr],\\
 &b_0(\tau)=\frac{s(\tau)^{-1/4}}{2}\Biggl[\frac{2}{\sqrt{3}}\sqrt{\frac{3-8\tau^2}{1-16\tau^2}} - \sqrt{\frac{(1-\sqrt{\tau})^2(2+\sqrt{\tau})}{3\tau(1-2\sqrt{\tau})}}\Biggr].
 \end{align*}
\end{Proposition}

\begin{proof}
 As discussed above, when $\tau\neq \frac{1}{4}$, all of the saddle points are simple, and classical steepest descent analysis
 yields the expansion \eqref{sigma-asymptotics-nofield}. However, when $\tau\to \frac{1}{4}$, the saddle points $\zeta = 1$,
 $\zeta = -1 + \tau^{-1/2}$ coalesce, and the expansions in the low and high temperature regimes are no longer valid. Indeed,
 upon direct inspection of the expansions \eqref{sigma-asymptotics-nofield}, one sees that the leading terms in
 both expressions diverge as $\tau\to \frac{1}{4}$. In principle, one can find an asymptotic expansion which holds precisely at
 $\tau = \frac{1}{4}$. However, we can find a more robust expansion, which holds uniformly for $\tau$ sufficiently close to
 $\frac{1}{4}$; this kind of asymptotic expansion was first demonstrated in~\cite{CFU}. The proof relies on the construction
 of a family of conformal maps $\zeta \mapsto u(\zeta;\tau)$, which satisfy the algebraic equation
 \begin{equation}\label{conformal-map-eq}
 -\log \mathfrak{I}(\zeta;\tau,0,0) = \frac{1}{3} u^3 -s(\tau) u + C(\tau).
 \end{equation}
 Here $s$, $C$ are determined by the equations
 \begin{gather*}
 -\log \mathfrak{I}(1;\tau,0,0) = -\frac{2}{3}s(\tau)^{3/2} + C(\tau),\\
 -\log \mathfrak{I}\bigl(-1+\tau^{-1/2};\tau,0,0\bigr) = + \frac{2}{3}s(\tau)^{3/2} + C(\tau).
 \end{gather*}
 \big(note that these conditions guarantee that $\frac{{\rm d}u}{{\rm d}\zeta}\neq 0,\infty$\big). One finds that
 \begin{align*}
 &C(\tau) = -\frac{1}{2}\log[t_{{\rm low}}(\tau)t_{{\rm high}}(\tau)],\\
 &s(\tau) = \biggl((-1)^{\mathbf{1}_{\{\tau>\frac{1}{4}\}}}\frac{3}{4}\log\biggl[\frac{t_{{\rm low}}(\tau)}{t_{{\rm high}}(\tau)}\biggr]\biggr)^{2/3}.
 \end{align*}
\big(Note that \smash{$\frac{t_{{\rm low}}}{t_{{\rm high}}} > 1$} for $0 < \tau < \frac{1}{4}$, and is smaller than $1$ for $\frac{1}{4} < \tau < 1$, and
so the above expression for $S$ is equivalent to \eqref{s-definition}\big). Provided that $\tau$ is sufficiently close to $\frac{1}{4}$, $C(\tau)$, $s(\tau)$ are positive and real-valued.
 Following~\cite{CFU}, we can then rewrite the integral \eqref{sigma-int} as
 \begin{equation*} % RS: можливо воно й правильно, але на всяк випадок виділила
 \sigma_V(\tau,0) = \frac{{\rm e}^{V C(\tau)}}{2\pi {\rm i}}\oint_{\tilde{C}} {\rm e}^{V\left[\frac{1}{3} u^3 -s(\tau) u\right]} \frac{{\rm d}\zeta}{{\rm d}u} {\rm d}u,
 \end{equation*}
 where $\tilde{C}$ is the image of the circle $C$ under the conformal map $u(\zeta)$. Now, we can express $\frac{{\rm d}\zeta}{{\rm d}u}$ as
 \begin{equation*}
 \frac{{\rm d}\zeta}{{\rm d}u} = \sum a_k(\tau)\bigl(u^2-s(\tau)\bigr)^k + \sum b_k(\tau)u\bigl(u^2-s(\tau)\bigr)^k;
 \end{equation*}
 the only terms we calculate explicitly are $a_0(\tau)$, $b_0(\tau)$, as they will be the dominant terms in the expansion. Differentiating \eqref{conformal-map-eq}, and expanding using L'H\^{o}pital's rule, one can determine~that
 \begin{align*}
 &\biggl(\frac{{\rm d} \zeta}{{\rm d} u}\bigg|_{u = s^{1/2}}\biggr)^2= \frac{4}{3}\frac{3-8\tau^2}{1-16\tau^2}s(\tau)^{1/2},\\
 &\biggl(\frac{{\rm d} \zeta}{{\rm d} u}\bigg|_{u = -s^{1/2}}\biggr)^2= \frac{(1-\sqrt{\tau})^2(2+\sqrt{\tau})}{3\tau(1-2\sqrt{\tau})}s(\tau)^{1/2},
 \end{align*}
 and so these derivatives are determined up to a sign, which can be determined for instance by requiring the asymptotic expansion we are developing match with classical steepest descent results precisely at $\tau = \frac{1}{4}$. This determines $a_0(\tau)$, $b_0(\tau)$
 to be
 \begin{align*}
 &a_0(\tau)=\frac{s(\tau)^{1/4}}{2}\Biggl[\frac{2}{\sqrt{3}}\sqrt{\frac{3-8\tau^2}{1-16\tau^2}} + \sqrt{\frac{(1-\sqrt{\tau})^2(2+\sqrt{\tau})}{3\tau(1-2\sqrt{\tau})}}\Biggr],\\
 &b_0(\tau)=\frac{s(\tau)^{-1/4}}{2}\Biggl[\frac{2}{\sqrt{3}}\sqrt{\frac{3-8\tau^2}{1-16\tau^2}} - \sqrt{\frac{(1-\sqrt{\tau})^2(2+\sqrt{\tau})}{3\tau(1-2\sqrt{\tau})}}\Biggr].
 \end{align*}
 These functions are indeed continuous, as one can readily compute the left/right limits as $\tau\to\frac{1}{4}$.
 One finally obtains that
 \begin{align*}
 \sigma_V(\tau,0) &\!=\! {\rm e}^{VC(\tau)}\!\biggl[\frac{a_0(\tau)}{V^{1/3}}\Ai\bigl(V^{2/3}s(\tau)\bigr) + \frac{b_0(\tau)}{V^{2/3}}\Ai'\bigl(V^{2/3}s(\tau)\bigr)\!\biggr]\!\bigl[1 + \OO\bigl(V^{-1/3}\bigr)\bigr]\\
 &\!=\![t_{{\rm low}}(\tau)t_{{\rm high}}(\tau)]^{-V/2}\!\biggl[\frac{a_0(\tau)}{V^{1/3}}\Ai\bigl(V^{2/3}s(\tau)\bigr) + \frac{b_0(\tau)}{V^{2/3}}\Ai'\bigl(V^{2/3}s(\tau)\bigr)\!\biggr]\!\bigl[1 + \OO\bigl(V^{-1/3}\bigr)\bigr],
 \end{align*}
 as desired. As a consistency check, for $\tau\neq \frac{1}{4}$, if one further replaces $\Ai(V^{2/3}s(\tau))$ with its asymptotic expansion for large $V$ (note that $s(\tau) \geq 0$, by construction), one can match the formulae \eqref{sigma-asymptotics-nofield}.
\end{proof}

\begin{Remark}
 We can also extend the some of the above calculations by continuity to the case of nonzero $H$, provided $H$ is sufficiently small. However, in a neighborhood of the critical point $\tau=\frac{1}{4}$, $H=0$, the above calculations do not apply, as we shall now explain.

 Provided that $\tau\neq \frac{1}{4}$, one can continue the dominant saddle point $\zeta^*$ to nonzero $H$ in an~unambiguous manner, and the dominant saddle remains simple for sufficiently small $H$. For~${\frac{1}{4}< \tau < 1}$,
 recall that the dominant saddle point was $\zeta^* = -1 + \tau^{-1/2}$. In this case, $\zeta^*$ is an analytic function of $H$ in a neighborhood of $H=0$, and admits the expansion
 \begin{equation*}
 \zeta^*(\tau,H) = -1 + \tau^{-1/2} - \frac{(1-\sqrt{\tau})(1+2\sqrt{\tau}+2\tau)}{2(1-2\sqrt{\tau})^4}H^2 + \OO\bigl(H^4\bigr).
 \end{equation*}
 By continuity, this is the nearest saddle point to the origin when $H$ is sufficiently small, for any fixed \smash{$\frac{1}{4}< \tau < 1$}.
 The situation is slightly more complicated when \smash{$0 < \tau < \frac{1}{4}$}. In this case, one sees that the saddle point at $\zeta^* = 1$ is actually the result of $4$ simple saddle points for nonzero~$H$. These saddle points do \textit{not} depend analytically on $H$ in a neighborhood of $H=0$, and instead have a series expansion in powers of $|H|^{1/2}$. One finds that
 \begin{equation*}
 \zeta = 1 \pm \frac{\sqrt{2}}{(1-16\tau^2)^{1/4}}|H|^{1/2} + \OO(|H|),\qquad 1 \pm \frac{{\rm i}\sqrt{2}}{(1-16\tau^2)^{1/4}}|H|^{1/2} + \OO(|H|).
 \end{equation*}
 These saddle points are continuous in $H$; the one nearest to the origin for $|H|>0$ is therefore
 \begin{equation*}
 \zeta^* = 1-\frac{\sqrt{2}}{(1-16\tau^2)^{1/4}}|H|^{1/2} + \OO(|H|).
 \end{equation*}
 When $H\neq 0$, it is apparent that the saddle point at $\zeta = 1$ split into a quadruple of saddle points when $H\neq 0$,
 and for $0 < \tau < \frac{1}{4}$. This is also the case when $\frac{1}{4} < \tau < 1$, although the saddle point is irrelevant for the
 calculations at hand. However, when $\tau = \frac{1}{4}$, the saddle points at $\zeta = 1$, $\zeta = -1 + \tau^{-1/2}$ merge; when the
 external field $H$ is ``turned on'' at this point, one finds that the saddle point at $\zeta= 1$ splits into \textit{five} distinct
 saddle points, each which has the expansion
 \begin{equation*}
 \zeta_j = 1 - 2^{1/5} \mathfrak{s}^j |H|^{2/5} + \OO\bigl(|H|^{4/5}\bigr),\qquad j=0,\dots,4,
 \end{equation*}
 where \smash{$\mathfrak{s} :={\rm e}^{\frac{2\pi {\rm i}}{5}}$} is a $5^{\rm th}$ root of unity. If one is interested in asymptotics precisely at $\tau = \frac{1}{4}$ (as~one might be, if attempting to calculate the critical exponent $\delta$ from the Landau theory, cf.~\cite{Kazakov2}), then one
 recognizes $\zeta_0$ as the dominant saddle point, and can proceed to use classical steepest descent to calculate the asymptotics of
 \smash{$\sigma_V\bigl(\frac{1}{4},H\bigr)$} there. A more interesting calculation, which is for the moment out of reach, would be to obtain \textit{uniform}
 asymptotics in a neighborhood of the critical point $\tau =\frac{1}{4}$, $H=0$, as we have done in the case when we ignored the parameter~$H$.
 This would require an analysis similar to that of~\cite{CFU}, but instead for a merging of $5$ branch points simultaneously. Similarly to
 how the uniform asymptotics arising from the merging of \textit{two} branch points was written in terms of solutions to a \textit{second order} equation (the Airy equation), one should expect that the asymptotics arising from $5$ merging branch points should be characterized in terms of a degree $5$ equation, whose solutions are integrals of the form
 \begin{equation*}
 P(x,a,b,c;\gamma) := \frac{1}{2\pi {\rm i}}\int_{\gamma}{\rm e}^{\frac{1}{6}\zeta^{6} + a\zeta^4 + b\zeta^3 + c\zeta^2 - x\zeta} {\rm d}\zeta,
 \end{equation*}
 where the parameters $a$, $b$, $c$, $x$ are functions of $\tau$, $H$, chosen so that all tend to $0$ as $\tau\to\frac{1}{4}$, $H\to 0$. One must construct an associated conformal map, which is by no means a simple task. We hope to pursue this analysis in a later work.
\end{Remark}

\subsection*{Acknowledgements}
MD was supported by the Swedish Research Council (VR), grant no. 2021-06015, and the European Research
Council (ERC), Grant Agreement No.~101002013. NH was supported by the European Research
Council (ERC), Grant Agreement No.~101002013. Part of the work was completed while NH was at the University of South Florida. We would like to thank an anonymous referee, who pointed out a number of results in the literature which rigorously derive the genus zero free energy using combinatorial methods. We would also like to thank all of the referees, whose valuable comments helped improve the overall presentation of this manuscript.

\LastPageEnding

\end{document}